\documentclass[twocolumn]{aastex701}
\usepackage{amsmath}
\usepackage{bigints}
\usepackage{booktabs}
\usepackage{hyperref}
\usepackage{enumitem}

\begin{document}

\title{Optimizing Kilonova Searches: A Case Study of the Type IIb SN\,2025ulz in the Localization Volume of the Low-Significance Gravitational Wave Event S250818k}

\newcommand{\LCO}{\affiliation{Las Cumbres Observatory, 6740 Cortona Drive, Suite 102, Goleta, CA 93117-5575, USA}}
\newcommand{\UCSB}{\affiliation{Department of Physics, University of California, Santa Barbara, CA 93106-9530, USA}}
\newcommand{\KITP}{\affiliation{Kavli Institute for Theoretical Physics, University of California, Santa Barbara, CA 93106-4030, USA}}
\newcommand{\UCD}{\affiliation{Department of Physics, University of California, 1 Shields Avenue, Davis, CA 95616-5270, USA}}
\newcommand{\WIS}{\affiliation{Department of Particle Physics and Astrophysics, Weizmann Institute of Science, 76100 Rehovot, Israel}}
\newcommand{\OKC}{\affiliation{Oskar Klein Centre, Department of Astronomy, Stockholm University, Albanova University Centre, SE-106 91 Stockholm, Sweden}}
\newcommand{\OAPD}{\affiliation{INAF-Osservatorio Astronomico di Padova, Vicolo dell'Osservatorio 5, I-35122 Padova, Italy}}
\newcommand{\Caltech}{\affiliation{Cahill Center for Astronomy and Astrophysics, California Institute of Technology, Mail Code 249-17, Pasadena, CA 91125, USA}}
\newcommand{\GSFC}{\affiliation{Astrophysics Science Division, NASA Goddard Space Flight Center, Mail Code 661, Greenbelt, MD 20771, USA}}
\newcommand{\UMD}{\affiliation{Joint Space-Science Institute, University of Maryland, College Park, MD 20742, USA}}
\newcommand{\UCB}{\affiliation{Department of Astronomy, University of California, Berkeley, CA 94720-3411, USA}}
\newcommand{\TTU}{\affiliation{Department of Physics, Texas Tech University, Box 41051, Lubbock, TX 79409-1051, USA}}
\newcommand{\STScI}{\affiliation{Space Telescope Science Institute, 3700 San Martin Drive, Baltimore, MD 21218, USA}}
\newcommand{\UT}{\affiliation{Department of Astronomy, The University of Texas at Austin, 2515 Speedway, Stop C1400, Austin, TX 78712, USA}}
\newcommand{\IoA}{\affiliation{Institute of Astronomy, University of Cambridge, Madingley Road, Cambridge CB3 0HA, UK}}
\newcommand{\QUB}{\affiliation{Astrophysics Research Centre, School of Mathematics and Physics, Queen's University Belfast, Belfast BT7 1NN, UK}}
\newcommand{\IPAC}{\affiliation{Spitzer Science Center, California Institute of Technology, Pasadena, CA 91125, USA}}
\newcommand{\JPL}{\affiliation{Jet Propulsion Laboratory, California Institute of Technology, 4800 Oak Grove Dr, Pasadena, CA 91109, USA}}
\newcommand{\Southampton}{\affiliation{Department of Physics and Astronomy, University of Southampton, Southampton SO17 1BJ, UK}}
\newcommand{\LANL}{\affiliation{Space and Remote Sensing, MS B244, Los Alamos National Laboratory, Los Alamos, NM 87545, USA}}
\newcommand{\Tsinghua}{\affiliation{Physics Department and Tsinghua Center for Astrophysics, Tsinghua University, Beijing, 100084, People's Republic of China}}
\newcommand{\NAOC}{\affiliation{National Astronomical Observatory of China, Chinese Academy of Sciences, Beijing, 100012, People's Republic of China}}
\newcommand{\Itagaki}{\affiliation{Itagaki Astronomical Observatory, Yamagata 990-2492, Japan}}
\newcommand{\Einstein}{\altaffiliation{Einstein Fellow}}
\newcommand{\Hubble}{\altaffiliation{Hubble Fellow}}
\newcommand{\CfA}{\affiliation{Center for Astrophysics \textbar{} Harvard \& Smithsonian, 60 Garden Street, Cambridge, MA 02138-1516, USA}}
\newcommand{\UA}{\affiliation{Department of Astronomy and Steward Observatory, University of Arizona, 933 North Cherry Avenue, Tucson, AZ 85721-0065, USA}}
\newcommand{\MPA}{\affiliation{Max-Planck-Institut f\"ur Astrophysik, Karl-Schwarzschild-Stra\ss e 1, D-85748 Garching, Germany}}
\newcommand{\DSFP}{\altaffiliation{LSST-DA Data Science Fellow}}
\newcommand{\catalyst}{\altaffiliation{LSST-DA Catalyst Fellow}}
\newcommand{\HCO}{\affiliation{Harvard College Observatory, 60 Garden Street, Cambridge, MA 02138-1516, USA}}
\newcommand{\Carnegie}{\affiliation{Observatories of the Carnegie Institute for Science, 813 Santa Barbara Street, Pasadena, CA 91101-1232, USA}}
\newcommand{\TAU}{\affiliation{School of Physics and Astronomy, Tel Aviv University, Tel Aviv 69978, Israel}}
\newcommand{\Edinburgh}{\affiliation{Institute for Astronomy, University of Edinburgh, Royal Observatory, Blackford Hill EH9 3HJ, UK}}
\newcommand{\Birmingham}{\affiliation{Birmingham Institute for Gravitational Wave Astronomy and School of Physics and Astronomy, University of Birmingham, Birmingham B15 2TT, UK}}
\newcommand{\CIERA}{\affiliation{Center for Interdisciplinary Exploration and Research in Astrophysics and Department of Physics and Astronomy, \\Northwestern University, 1800 Sherman Ave., 8th Floor, Evanston, IL 60201, USA}}
\newcommand{\Bath}{\affiliation{Department of Physics, University of Bath, Claverton Down, Bath BA2 7AY, UK}}
\newcommand{\CTIO}{\affiliation{Cerro Tololo Inter-American Observatory, National Optical Astronomy Observatory, Casilla 603, La Serena, Chile}}
\newcommand{\Potsdam}{\affiliation{Institut f\"ur Physik und Astronomie, Universit\"at Potsdam, Haus 28, Karl-Liebknecht-Str. 24/25, D-14476 Potsdam-Golm, Germany}}
\newcommand{\INPE}{\affiliation{Instituto Nacional de Pesquisas Espaciais, Avenida dos Astronautas 1758, 12227-010, S\~ao Jos\'e dos Campos -- SP, Brazil}}
\newcommand{\UNC}{\affiliation{Department of Physics and Astronomy, University of North Carolina, 120 East Cameron Avenue, Chapel Hill, NC 27599, USA}}
\newcommand{\Ohio}{\affiliation{Astrophysical Institute, Department of Physics and Astronomy, 251B Clippinger Lab, Ohio University, Athens, OH 45701-2942, USA}}
\newcommand{\AAS}{\affiliation{American Astronomical Society, 1667 K~Street NW, Suite 800, Washington, DC 20006-1681, USA}}
\newcommand{\MMT}{\affiliation{MMT and Steward Observatories, University of Arizona, 933 North Cherry Avenue, Tucson, AZ 85721-0065, USA}}
\newcommand{\Geneva}{\affiliation{ISDC, Department of Astronomy, University of Geneva, Chemin d'\'Ecogia, 16 CH-1290 Versoix, Switzerland}}
\newcommand{\Steward}{\affiliation{Steward Observatory, University of Arizona, 933 North Cherry Avenue, Tucson, AZ 85721, USA}}
\newcommand{\Leiden}{\affiliation{Leiden Observatory, Leiden University, PO Box 9513, 2300 RA Leiden, The Netherlands}}
\newcommand{\PSU}{\affiliation{Department of Astronomy \& Astrophysics, The Pennsylvania State University, University Park, PA 16802, USA}}
\newcommand{\PSUa}{\affiliation{Department of Astronomy \& Astrophysics, The Pennsylvania State University, University Park, PA 16802, USA}}
\newcommand{\PSUb}{\affiliation{Institute for Computational \& Data Sciences, The Pennsylvania State University, University Park, PA 16802, USA}}
\newcommand{\PSUc}{\affiliation{Institute for Gravitation and the Cosmos, The Pennsylvania State University, University Park, PA 16802, USA}}
\newcommand{\IAIFI}{\affiliation{The NSF AI Institute for Artificial Intelligence and Fundamental Interactions, USA}}
\newcommand{\JHU}{\affiliation{Department of Physics and Astronomy, Johns Hopkins University, 3400 North Charles Street, Baltimore, MD 21218, USA}}
\newcommand{\Utah}{\affiliation{Department of Physics \& Astronomy, University of Utah, Salt Lake City, UT 84112, USA}}
\newcommand{\UIUC}{\affiliation{Department of Astronomy, University of Illinois, 1002 W. Green St., Urbana, IL 61801, USA}}
\newcommand{\Maryland}{\affiliation{Department of Astronomy, University of Maryland, College Park, MD 20742-2421, USA}}
\newcommand{\keck}{\affiliation{W.~M.~Keck Observatory, 65-1120 M\=amalahoa Highway, Kamuela, HI 96y43-8431, USA}}
\newcommand{\cbpf}{\affiliation{Laboratório de Inteligência Artificial, Centro Brasileiro de Pesquisas Físicas, 138 Rua Dr. Xavier Sigaud 150, CEP 22290-180, 139 Rio de Janeiro, RJ, Brazil}}
\newcommand{\UFRJ}{\affiliation{Instituto de Física, Universidade Federal do Rio de Janeiro (UFRJ), Caixa Postal 68528, 21941-972 Rio de Janeiro, Brazil}}
\newcommand{\Monash}{\affiliation{School of Physics and Astronomy, Monash University, Clayton, Victoria 3800, Australia}}
\newcommand{\UCSD}{\affiliation{Department of Astronomy \& Astrophysics, University of California, San Diego, 9500 Gilman Drive, MC 0424, La Jolla, CA 92093-0424, USA}}
\newcommand{\Northwestern}{\affiliation{Department of Physics and Astronomy, Northwestern University, Evanston, IL 60208, USA}}
\author[orcid=0000-0003-4537-3575, gname=Noah, sname=Franz]{Noah Franz}
\UA
\email[show]{nfranz@arizona.edu}

\author[0000-0001-8073-8731, gname=Bhagya, sname=Subrayan]{Bhagya Subrayan}
\email{bsubrayan@arizona.edu}
\UA

\author[0000-0002-5740-7747, gname=Charles, sname=Kilpatrick]{Charles~D.~Kilpatrick}
\email{ckilpatrick@northwestern.edu}
\CIERA

\author[0000-0002-0832-2974, gname=Griffin, sname=Hosseinzadeh]{Griffin Hosseinzadeh}
\email{ghosseinzadeh@ucsd.edu}
\UCSD

\author[0000-0003-4102-380X, gname=David, sname=Sand]{David J. Sand}
\email{dsand@arizona.edu}
\UA

\author[0000-0002-8297-2473, gname=Kate, sname=Alexander]{Kate D. Alexander}
\email{kdalexander@arizona.edu}
\UA

\author[0000-0002-7374-935X, gname=Wen-fai, sname=Fong]{Wen-fai Fong}
\email{wfong@northwestern.edu}
\CIERA
\Northwestern

\author[0000-0003-0528-202X, gname=Collin, sname=Christy]{Collin T. Christy}
\email{collinchristy@arizona.edu}
\UA

\author[orcid=0000-0002-0744-0047, gname=Jeniveve, sname=Pearson]{Jeniveve Pearson}
\email{jenivevepearson@arizona.edu}
\UA

\author[0000-0003-1792-2338, gname=Tanmoy, sname=Laskar]{Tanmoy Laskar}
\email{tanmoy.laskar@utah.edu}
\Utah

\author[0000-0002-9454-1742, gname=Brian, sname=Hsu]{Brian Hsu}
\email{bhsu@arizona.edu}
\UA

\author[0000-0002-9267-6213, gname=Jillian, sname=Rastinejad]{Jillian Rastinejad}
\email{jcrastin@umd.edu}
\altaffiliation{NHFP Einstein Fellow}
\Maryland

\author[0000-0001-9589-3793, gname=Michael, sname=Lundquist]{Michael J. Lundquist}
\email{mlundquist@keck.hawaii.edu}
\keck

\author[0000-0002-9392-9681, gname=Edo, sname=Berger]{Edo Berger}
\email{eberger@cfa.harvard.edu}
\CfA

\author[0000-0002-4924-444X, gname=Azalee, sname=Bostroem]{K. Azalee Bostroem}
\email{bostroem@arizona.edu}
\UA
\catalyst

\author[0000-0003-4383-2969, gname=Clecio, sname=Bom]{Clecio R. Bom}
\email{clecio@debom.com.br}
\cbpf

\author[0000-0001-8833-474X, gname=Phelipe, sname=Darc]{Phelipe Darc}
\email{phelipedarc@gmail.com}
\cbpf

\author[0000-0003-0685-3621, gname=Mark, sname=Gurwell]{Mark Gurwell}
\email{mgurwell@cfa.harvard.edu}
\CfA

\author[0000-0002-2184-4646, gname=Shelbi, sname=Schimpf]{Shelbi Hostler Schimpf}
\email{shostler@cfa.harvard.edu}
\CfA

\author[0000-0002-3490-146X, gname=Garrett, sname=Keating]{Garrett K. Keating}
\email{garrett.keating@cfa.harvard.edu}
\CfA

\author[orcid=0009-0006-0647-636X, gname=Phillip, sname=Noel]{Phillip Noel}
\email{phillipnoel@arizona.edu}
\UA

\author[0000-0003-4175-4960, gname=Conor, sname=Ransome]{Conor Ransome}
\email{cransome@arizona.edu}
\UA

\author[0000-0002-1407-7944, gname=Ramprasad, sname=Rao]{Ramprasad Rao}
\email{rrao@cfa.harvard.edu}
\CfA

\author[0000-0003-3402-6164, gname=Luidhy, sname=Santana-Silva]{Luidhy Santana-Silva}
\email{luidhysantana@gmail.com}
\cbpf

\author[0000-0002-1420-3584, gname=Souza, sname=Santos]{A. Souza Santos}
\email{asantos.astro@gmail.com}
\cbpf

\author[0000-0002-4022-1874, gname=Manisha, sname=Shrestha]{Manisha Shrestha}
\email{manisha.shrestha@monash.edu}
\Monash

\author[0000-0002-4989-6253, gname=Ramya, sname=Anche]{Ramya Anche}
\email{ramyaanche@arizona.edu}
\UA

\author[0000-0003-0123-0062, gname=Jennifer, sname=Andrews]{Jennifer E. Andrews}
\email{jennifer.andrews@noirlab.edu}
\affiliation{Gemini Observatory/NSF's NOIRLab, 670 N. A'ohoku Place, Hilo, HI 96720, USA}

\author[0000-0002-2724-8298, gname=Sanchayeeta, sname=Borthakur]{Sanchayeeta Borthakur}
\email{sanch@asu.edu}
\affiliation{School of Earth and Space Exploration, Arizona State University, P.O. Box 871404, Tempe, AZ 85287-1404, USA}

\author[0000-0002-9110-6673, gname=Nathaniel, sname=Butler]{Nathaniel R. Butler}
\email{natbutler@asu.edu}
\affiliation{School of Earth and Space Exploration, Arizona State University, P.O. Box 871404, Tempe, AZ 85287-1404, USA}

\author[0000-0001-5126-6237, gname=Deanne, sname=Coppejans]{Deanne L. Coppejans}
\email{deanne.coppejans@warwick.ac.uk}
\affiliation{Department of Physics, University of Warwick, Coventry CV4 7AL, UK}

\author{Philip N Daly}
\email{pndaly@arizona.edu}
\UA

\author[0000-0003-2594-8052, gname=Kathryne, sname=Daniel]{Kathryne J. Daniel}
\email{kjdaniel@arizona.edu}
\UA

\author[0000-0001-7626-9629, gname=Paul, sname=Duffell]{Paul C. Duffell}
\email{pduffell@purdue.edu}
\affiliation{Department of Physics and Astronomy, Purdue University, 525 Northwestern Avenue, West Lafayette, IN 47907, USA}

\author[0000-0003-0307-9984, gname=Tarraneh, sname=Eftekhari]{Tarraneh Eftekhari}
\email{teftekhari@northwestern.edu}
\CIERA

\author[0000-0002-8925-057X, gname=Carl, sname=Fields]{Carl E. Fields}
\email{carlnotsagan@arizona.edu}
\UA

\author[0000-0003-4906-8447, gname=Alexander, sname=Gagliano]{Alexander T. Gagliano}
\email{gaglian2@mit.edu}
\affiliation{The NSF AI Institute for Artificial Intelligence and Fundamental Interactions}
\CfA
\affiliation{Department of Physics and Kavli Institute for Astrophysics and Space Research, Massachusetts Institute of Technology, 77 Massachusetts Avenue, Cambridge, MA 02139, USA}

\author[0000-0001-7946-1034, gname=Walter, sname=Golay]{Walter W. Golay}
\email{wgolay@cfa.harvard.edu}
\CfA

\author[0000-0002-2215-1841, gname=Aldana, sname=Grichener]{Aldana Grichener}
\email{agrichener@arizona.edu}
\UA

\author[0000-0002-3131-7372, gname=Erika, sname=Hamden]{Erika T. Hamden}
\email{hamden@arizona.edu}
\UA

\author[0000-0002-1125-9187, gname=Daichi, sname=Hiramatsu]{Daichi Hiramatsu}
\email{dhiramatsu@ufl.edu}
\affiliation{Department of Astronomy, University of Florida, 211 Bryant Space Science Center, Gainesville, FL 32611-2055 USA}

\author[0000-0003-0871-4641, gname=Harsh, sname=Kumar]{Harsh Kumar}
\email{harsh.kumar@cfa.harvard.edu}
\CfA
\affiliation{The NSF AI Institute for Artificial Intelligence and Fundamental Interactions, USA.}

\author[0000-0003-0547-6158, gname=Vikram, sname=Manikantan]{Vikram Manikantan}
\email{vik@arizona.edu}
\UA

\author[0000-0003-4768-7586, gname=Raffaella, sname=Margutti]{Raffaella Margutti}
\email{rmargutti@berkeley.edu}
\affiliation{Department of Astronomy, University of California, Berkeley, CA 94720-3411, USA}
\affiliation{Berkeley Center for Multi-messenger Research on Astrophysical Transients and Outreach (Multi-RAPTOR), University of California,12
Berkeley, CA 94720-3411, USA}
\affiliation{Department of Physics, University of California, 366 Physics North MC 7300, Berkeley, CA 94720, USA}

\author[0000-0002-8099-9023, gname=Vasileios, sname=Paschalidis]{Vasileios Paschalidis}
\email{vpaschal@arizona.edu}
\UA
\affiliation{Department of Physics, University of Arizona, Tucson, AZ 85721, US}

\author[0000-0001-8340-3486, gname=Kerry, sname=Paterson]{Kerry Paterson}
\email{paterson@mpia.de}
\affiliation{Max-Planck-Institut f\"ur Astronomie, K\"onigstuhl 17, 69117 Heidelberg, Germany}

\author[0000-0002-5060-3673, gname=Daniel, sname=Reichart]{Daniel E. Reichart}
\email{reichart@physics.unc.edu}
\affiliation{Department of Physics and Astronomy, University of North Carolina at Chapel Hill, Campus Box 3255, Chapel Hill, NC 27599-3255}

\author[0000-0002-6718-9472, gname=Mathieu, sname=Renzo]{Mathieu Renzo}
\email{mrenzo@ariona.edu}
\UA

\author[0009-0009-7182-6022, gname=Kali, sname=Salmas]{Kali Salmas}
\email{ksalmas@mmto.org}
\affiliation{MMT and Steward Observatory, University of Arizona, 933 North Cherry Avenue, Tucson, AZ 85721-0065, USA}

\author[0000-0001-9915-8147, gname=Genevieve, sname=Schroeder]{Genevieve Schroeder}
\email{gms279@cornell.edu}
\affiliation{Department of Astronomy, Cornell University, Ithaca, NY 14853, USA}

\author[0000-0001-5510-2424, gname=Nathan, sname=Smith]{Nathan Smith}
\email{nathans@as.arizona.edu}
\UA

\author[0000-0002-0956-7949, gname=Kristine, sname=Spekkens]{Kristine Spekkens}
\email{kristine.spekkens@queensu.ca}
\affiliation{Department of Physics, Engineering Physics and Astronomy, Queen's University, Kingston, ON, K7L 2E1, Canada}

\author[0000-0002-1468-9668, gname=Jay, sname=Strader]{Jay Strader}
\email{straderj@msu.edu}
\affiliation{Center for Data Intensive and Time Domain Astronomy, Department of Physics and Astronomy, Michigan State University, East Lansing, MI 48824, USA}

\author[0000-0003-4580-3790, gname=David, sname=Trilling]{David E. Trilling}
\email{david.trilling@nau.edu}
\affiliation{Department of Astronomy and Planetary Science, Northern Arizona University, PO Box 6010, Flagstaff, AZ 86011 USA}

\author[0000-0001-7815-7604, gname=Nicholas, sname=Vieira]{Nicholas Vieira}
\email{nicholas.vieira@northwestern.edu}
\CIERA

\author[0000-0001-6065-7483, gname=Benjamin, sname=Weiner]{Benjamin Weiner}
\email{bjw@mmto.org}
\affiliation{MMT and Steward Observatory, University of Arizona, 933 North Cherry Avenue, Tucson, AZ  85721-0065, USA}

\author[0000-0003-3734-3587, gname=Peter, sname=Williams]{Peter K. G. Williams}
\email{pwilliams@cfa.harvard.edu}
\CfA

\newcommand{\todo}{\textcolor{red}{TODO: }\textcolor{red}}
\newcommand{\note}{\textcolor{blue}{NOTE: }\textcolor{blue}}
\setcounter{footnote}{0}

\begin{abstract}
Kilonovae, the ultraviolet/optical/infrared counterparts to binary neutron star mergers, are an exceptionally rare class of transients. Optical follow-up campaigns are plagued by \added{contaminating transients, which may mimic kilonovae, but do not receive sufficient observations to measure the full photometric evolution}. In this work, we present an analysis of the multi-wavelength dataset of supernova (SN) 2025ulz, a proposed kilonova candidate following the low-significance detection of gravitational waves originating from the potential binary neutron star merger S250818k.  Despite an early rapid decline in brightness, our multi-wavelength observations of SN\,2025ulz reveal that it is a type IIb supernova. As part of this analysis, we demonstrate the capabilities of a novel quantitative scoring algorithm to determine the likelihood that a transient candidate is a kilonova, based primarily on its 3D location and light curve evolution. We also apply our scoring algorithm to other transient candidates in the localization volume of S250818k and find that, at all times after the discovery of SN\,2025ulz, there are $\geq 4$ candidates with a score \added{comparable to} SN\,2025ulz\added{, indicating that the kilonova search may have benefited from the additional follow-up of other candidates}. During future kilonova searches, this type of scoring algorithm will be useful to rule out contaminating transients in real time, optimizing the use of valuable telescope resources.
\end{abstract}


\defcitealias{ligo_scientific_collaboration_and_virgo_collaboration_gw170817_2017}{LIGO/Virgo Collaboration (2017)}
\defcitealias{abbott_multi-messenger_2017}{LIGO/Virgo Collaboration et al. 2017a}
\defcitealias{abbott_gravitational_2017}{LIGO/Virgo Collaboration et al. 2017b}
\defcitealias{the_ligo_scientific_collaboration_and_the_virgo_collaboration_gw170817_2018}{LIGO/Virgo Collaboration 2018}
\defcitealias{gcn41437}{LVK Collaboration 2025a}
\defcitealias{gcn41440}{LVK Collaboration 2025b}

\section{Introduction}
The detection of gravitational waves (GWs) associated with the binary neutron star (BNS) merger GW\,170817 by the \citetalias{ligo_scientific_collaboration_and_virgo_collaboration_gw170817_2017}, and the subsequent discovery of GW170817's electromagnetic counterparts \citepalias{abbott_multi-messenger_2017} GRB\,170817A \citep{savchenko_integral_2017, goldstein_ordinary_2017} and AT\,2017gfo \citep{arcavi_optical_2017, coulter_swope_2017, lipunov_master_2017, smartt_kilonova_2017, soares-santos_electromagnetic_2017, tanvir_emergence_2017, valenti_discovery_2017}, opened a powerful new window on our Universe. The unambiguous association of the short gamma-ray burst (SGRB) GRB\,170817A and kilonova (KN) AT\,2017gfo \citepalias{abbott_multi-messenger_2017} provided unique insights into the relativistic engine driving SGRBs (\citetalias{abbott_gravitational_2017}; \citealt{margutti_electromagnetic_2017, alexander_electromagnetic_2017, fong_electromagnetic_2017, savchenko_integral_2017, murguia-berthier_fate_2021}), the properties of neutron stars (\textit{e.g.}, their equation of state; \citetalias{the_ligo_scientific_collaboration_and_the_virgo_collaboration_gw170817_2018}; \citealt{radice_gw170817_2018}), and the formation of $r$-process elements in the ejecta of BNS mergers \citep{arcavi_optical_2017, cowperthwaite_electromagnetic_2017, tanvir_emergence_2017, drout_light_2017, kasliwal_illuminating_2017, kilpatrick_electromagnetic_2017, smartt_kilonova_2017}.  

Theoretical models \citep[\textit{e.g.},][]{kasen_kilonova_2015, kasen_origin_2017, metzger_kilonovae_2017, metzger_fragmentation_2024, chen_gravitational_2025}, combined with observations of AT\,2017gfo, demonstrate that photometric observations can typically discriminate between KNe and other transient types. After a BNS merger occurs, $r$-process nucleosynthesis in the neutron-rich ejecta forms heavy $r$-process elements, including lanthanides and actinides \citep[][]{Metzger10}. The radioactive decay of the heavy elements heats the ejecta, resulting in a faint and rapidly fading (few days to a week) transient whose spectral energy distribution (SED) primarily peaks in the near-infrared \citep{li_transient_1998, barnes_effect_2013, tanaka_radiative_2013, metzger_kilonovae_2017}. If the merger remnant launches outflows, it may (primarily depending on the neutron mass in the ejecta) inject energy and result in a re-brightening of the light curve in the infrared, while only resulting in a ``plateau'' in the optical \citep{kasen_kilonova_2015}. 

Beyond GW\,170817, searches for an electromagnetic counterpart associated with a GW signal have been unsuccessful \citep{Andreoni+19,Coughlin+19,Dobie+19,Goldstein+19,Gomez+19,Hosseinzadeh+19,Lundquist+19,Andreoni+20,Antier+20a,Antier+20b,Ackley+20,Garcia+20,Gompertz+20,Kasliwal+20,Morgan+20,Pozanenko+20,Thakur+20,Vieira+20,Watson+20,Anand+21,Alexander+21,Becerra+21,Bhakta+21,Chang+21,de_Wet+21,Dichiara+21,Dobie+21,kilpatrick_gravity_2021,Oates+21,Ohgami+21,Paterson+21,tucker_soargoodman_2022,mcbrien_ps15cey_2021,deJaeger+22,rastinejad_systematic_2022,ahumada_searching_2024,Pillas25, hu_kilonova_2025, frostig_winter_2025, ahumada_ligovirgokagra_2025, paek_gecko_2025, darc_long-term_2025, gillanders_pan-starrs_2025}. During these searches, there are often many transients that are spatially and temporally co-located with a given GW event, a subset of which are new, extragalactic events. Some of these contaminants are impostors, which have light curves that are initially faint and rapidly declining, appearing similar to a KN for several days, but eventually evolving away from expectations \citep{mcbrien_ps15cey_2021, barna_iib_2025}. Quickly ruling out interloping transients in real time will help the community focus their follow-up efforts on more promising candidates.

\begin{figure*}[!ht]
    \centering
    \includegraphics[width=\linewidth]{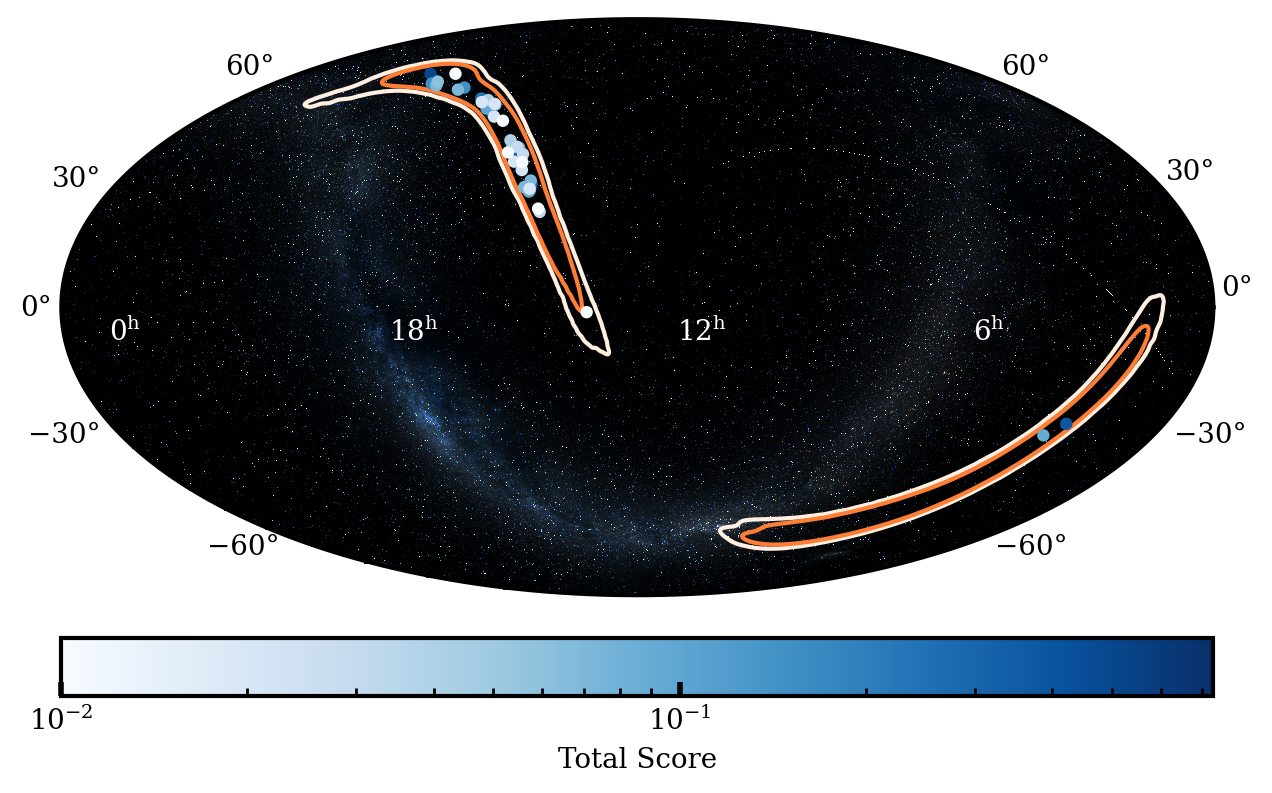}
    \caption{The IGWN (International Gravitational Wave Network) skymap credibility contours (orange: 50\%, white: 90\%) of S250818k, and the locations of all 121 candidates we found within the 95\% credibility region from the Transient Name Server (TNS). The color of the point represents the overall score from our vetting, which is described in \autoref{sec:search}.}
    \label{fig:skymap}
\end{figure*}

Due to their initial rapid decline, type IIb supernovae (SNe IIb) are common KN impostors \citep{barna_iib_2025}. SNe IIb are characterized by weak hydrogen absorption lines at early times ($\sim$weeks to months after explosion) and likely originate from a progenitor star with a partially stripped hydrogen envelope \citep{podsiadlowski_progenitor_1993, woosley_sn_1994, richmond_ubvri_1994, matheson_detailed_2000, elmhamdi_hydrogen_2006, chevalier_type_2010}. Importantly, SNe IIb can feature a double-peaked light curve in the ultraviolet, optical, and infrared \citep[\textit{e.g.,}][]{2002AJ....123..753M}. The first peak ($\sim$few days after explosion) is dominated by shock breakout and subsequent cooling \citep{richmond_ubvri_1994, arcavi_constraints_2017, das_probing_2023}. The initial shock cooling lasts $\sim$ 5--7 days, after which the radioactive decay of $^{56}$Ni heats the ejecta and dominates the observed emission, powering the second peak \citep[\textit{e.g.}, ][]{benson_light_1994}. In addition to their initial rapid decline, this re-brightening may appear similar to the ``plateau'' phase of a KN light curve. But, conventional KN models are unable to reproduce a significant re-brightening in the optical for more than a few days \citep{kasen_kilonova_2015, kasen_origin_2017}.

In this paper, we focus on electromagnetic candidates potentially associated with the sub-threshold GW event S250818k, a GW event with a 29\% probability of being a BNS merger and a 71\% probability of being of terrestrial origin \citepalias{gcn41437,gcn41440}. In particular, we present our observations of SN\,2025ulz\footnote{We consistently refer to the transient as ``SN\,2025ulz'', rather than ``AT\,2025ulz'', since this is the official IAU name on the Transient Name Server.}, including new observational evidence that SN\,2025ulz is a type IIb SN and not a KN counterpart to S250818k. We find that both the photometric evolution and spectrum are consistent with other SNe\,IIb and inconsistent with both AT\,2017gfo and a suite of KNe models. 

We also present a novel algorithm for quantifying the likelihood that a specific electromagnetic counterpart is associated with a BNS or Neutron Star--Black Hole (NSBH) GW event. We apply this algorithm to SN\,2025ulz and the 120 other transients discovered (1) within one week of the putative merger and (2) within the 95\% localization region of S250818k. We find $\sim 1-2$ candidates have a photometric evolution roughly consistent with a KN. But, these $\sim 2$ candidates only have $2-5$ publicly available photometry points, making it unclear if any of them are true KNe. Through this analysis, we demonstrate the current capabilities of this quantitative scoring algorithm. In the future, we will continue development of this algorithm and make it available as part of a publicly available web application.

The paper is organized as follows. Section \ref{sec:discovery} describes the discovery and initial public observations of S250818k and SN\,2025ulz. Section \ref{sec:obs} provides the details of our follow-up observations, spanning ultraviolet to radio wavelengths. In Section \ref{sec:analysis_discussion}, we analyze our observations and discuss the implications for the interpretation of SN\,2025ulz as an SN\,IIb. In Section \ref{sec:search} we describe our search for additional KN candidates associated with S250818k. Finally, we summarize and conclude in Section \ref{sec:conclusions}. Throughout this work we assume a flat $\Lambda {\rm CDM}$ cosmology with ${\rm H_0 = 69.6 ~km~s^{-1}~Mpc^{-1}}$ \citep{freedman_calibration_2020}.

\begin{table*}[]
    \centering
    \caption{SN\,2025ulz Properties}
    \begin{tabular}{llr}
        \hline
        Property & Value & Source \\
        \hline
         RA & 15h51m54.201s & \citet{2025TNSTR3264....1S} \\
         Dec & +30d54m08.67s & \citet{2025TNSTR3264....1S} \\
         Redshift & 0.0848 &  \citet{2025GCN.41436....1K} \\
         E(B-V)$_{\rm MW}$ & 0.0243 mag & \citet{schlegel_maps_1998,schlafly_measuring_2011} \\
         Host SFR & $\sim 1~{\rm M_\odot~yr}^{-1}$ & \citet{jones_blast_2024}; This work (\autoref{sec:host}) \\
         Envelope Mass (RSG) & $0.027^{+0.009}_{-0.007} ~{\rm M_\odot}$ & This work (\autoref{sec:early-lc-model}) \\
         Progenitor Radius (RSG) & $0.6^{+1.6}_{-0.4} \times 10^{13} ~{\rm cm}$& This work (\autoref{sec:early-lc-model}) \\
         Envelope Mass (BSG) & $0.14^{+0.08}_{-0.04} ~{\rm M_\odot}$& This work (\autoref{sec:early-lc-model}) \\
         Progenitor Radius (BSG) & $0.5^{+2.0}_{-0.3}\times10^{13}~{\rm cm}$& This work (\autoref{sec:early-lc-model}) \\
         \hline
    \end{tabular}
    \label{tab:ulz_props}
\end{table*}

\section{Discovery of S250818\lowercase{k} and SN\,2025\lowercase{ulz}} \label{sec:discovery}

The GW event S250818k was initially discovered by the LIGO/Virgo/KAGRA (LVK) Collaboration as a sub-threshold event detected at 2025-08-18 01:20:19 UTC (GPS time: 1439515224.03) and reported at 2025-08-18 01:20:35 UTC.  Preliminary classification of the event from the {\tt pycbc} pipeline yielded a 29\% probability of being a BNS merger and a 71\% probability of being a Terrestrial event \citep{gcn41437}. This initial classification also yielded a 50th (90th) percentile confidence sky localization that spanned 205 deg$^{2}$ (786 deg$^{2}$) with a posterior luminosity distance of 259$\pm$62 Mpc. S250818k has a false alarm rate, also estimated by the \texttt{pycbc} online analysis pipeline \citep{pycbc}, of 6.81$\times$10$^{-8}$~Hz (1 per 170~days), close to the reported LVK threshold for significant events after accounting for the trials factor\footnote{See description in \url{https://emfollow.docs.ligo.org/userguide/analysis/index.html\#alert-threshold-trial-factor}.}\added{\citep[e.g., see][for a full description of these parameters]{chaudhary_low-latency_2024}}.  We note that the probability that the system had a neutron star ({\tt HasNS}) and the probability that the system ejected a non-zero amount of neutron star matter ({\tt HasRemnant}) were both estimated to be 1.0, under the assumption that the event was non-terrestrial.

The revised S250818k event analysis (\textit{i.e.}, the ``initial'' analysis), provided on 2025-08-20 04:52:51 UTC ($\sim48$ hours after the first alert), and the S250818k ``update'' analysis, provided on 2025-08-20 09:57:08 UTC, changed these parameters slightly \citepalias[see, \textit{e.g.},][]{gcn41437,gcn41440}.  Given the short gap between these updates, we provide only the latter here, which we use for the majority of our analysis. The final event classification changed only marginally and remained consistent with a 29\% probability of being a BNS merger and 71\% probability of being Terrestrial. Similarly, the false alarm rate remained at 6.81$\times$10$^{-8}$~Hz.  The final {\tt Bilby} offline analysis map expanded the localization slightly to a 50th (90th) percentile sky localization of 276~deg$^{2}$ (949~deg$^{2}$) with a corresponding all-sky distance constraint of 237$\pm$62~Mpc (\autoref{fig:skymap}). Finally, we note that the {\tt Bilby} analysis revised the {\tt HasNS} and {\tt HasRemnant} estimates down to 0.8 each, while the chirp mass estimate placed $>$99\% of the probability within the (0.1, 0.87) $M_{\odot}$ bin.

On 2025 August 18 17:01:30 UTC, $\sim16$ hours after the initial GW alert, \citet{2025GCN.41414....1S} reported the discovery of a potential electromagnetic counterpart to S250818k: SN\,2025ulz. Following this at $\sim2.6$ days after the discovery of S250818k, \citet{2025GCN.41436....1K} reported that the most likely host galaxy of SN\,2025ulz has a spectroscopic redshift $z=0.0848$ (D$\approx$360 Mpc)\added{. This distance is} somewhat consistent with the distance to S250818k\added{ along the line of sight to SN\,2025ulz, with a $\approx68\%$ joint probability of the transient and GW distance distributions (see subsections \ref{sec:3d_score} \& \ref{sec:25ulz-score-time})}. This motivated a global follow-up campaign with facilities covering the X-ray through radio wavelengths \citep[\textit{e.g.}, ][ in the optical]{gillanders_pan-starrs_2025}. Subsequent messages posted on the General Coordinates Network (GCN) over the next $\sim3$ days reported photometry consistent with a fast-fading red transient with a similar decay rate to AT\,2017gfo \citep{2025GCN.41433....1H}, but $\sim1.5$ orders of magnitude more luminous. The source was never detected in the X-ray \citep{2025GCN.41433....1H, 2025GCN.41410....1N, 2025GCN.41528....1B, 2025GCN.41460....1L}. A source was detected in the radio by the MeerKAT telescope \citep{2025GCN.41500....1B, 2025GCN.41594....1B} but this emission was diffuse and later shown to likely be associated with the host galaxy \citep{2025GCN.41666....1R, 2025GCN.42032....1B}. The source was not detected with any other radio telescope \citep{2025GCN.41464....1R, 2025GCN.41542....1R, 2025GCN.41577....1B}. The basic properties of SN\,2025ulz are summarized in \autoref{tab:ulz_props}. 

On 2025-08-23 15:18:41 UTC, $\sim 5$ days after the discovery of S250818k, \citet{2025GCN.41507....1F} reported a re-brightening in the $i$-band light curve of SN\,2025ulz, and subsequent GCNs confirmed this finding \citep{2025GCN.41518....1A, 2025GCN.41540....1G, 2025GCN.41544....1B}, making the light curve more consistent with a SN IIb rather than AT\,2017gfo or other plausible KN emission. Soon after, a new spectrum confirmed the presence of a broad ($v \sim 15{,}600~{\rm km~s}^{-1}$) P-Cygni H$\alpha$ line, consistent with other SN\,IIb at this phase \citep{2025GCN.41532....1B}. Despite this, SN\,2025ulz continued to be observed and the exact origin remains hotly debated \citep[\textit{e.g.},][]{2025GCN.41538....1K, gillanders_pan-starrs_2025}. \citet{gillanders_pan-starrs_2025} searched the S250818k localization region with Pan-STARRS and found no promising candidate KNe, concluding that SN\,2025ulz is qualitatively consistent with a SN classification and incompatible with KN models. 

\section{Multiwavelength Follow-up Observations of SN\,2025\lowercase{ulz}} \label{sec:obs}
Given the strong temporal and spatial association of SN\,2025ulz with S250818k, we observed it from the ultraviolet to radio wavelengths. In the following subsections we describe these previously unreported observations. All optical photometry and spectra will be made public on WISeREP \citep{2012PASP..124..668Y}, and the radio observations and optical photometry will be made public, jointly, on the Open mulTiwavelength Transient Event Repository \citep[OTTER]{franz_open_2025}. All photometry is also provided in \autoref{app:data}.

\subsection{Public Photometry}
We gather the publicly available data reported on the GCN and the TNS for SN\,2025ulz. This includes the ultraviolet, optical, infrared, and radio observations reported in GCN circulars (with the exception of the MeerKAT data, which we reanalyze ourselves; Section \ref{sec:radio-obs}). We also include the photometry from \citet{gillanders_pan-starrs_2025}.

Using the methodology described in \citet{2024ApJ...964...35H}, we also query the Asteroid Terrestrial-impact Last Alert System (ATLAS; \citealt{2018PASP..130f4505T}) forced photometry server \citep{2020PASP..132h5002S,2021TNSAN...7....1S} at the coordinates of SN\,2025ulz from $\sim 180$ days prior to $\sim 20$ days after it was discovered. ATLAS forced photometry subtractions are based on difference images, and are subject to false detections attributed to bad subtractions, potentially due to artifacts in template images. Therefore, for SN\,2025ulz, we consider an ATLAS forced photometry observation to be a detection if it (1) is significant at the $5\sigma$ level and (2) has at least 2 other $5\sigma$ detections within a 10-day window.

\subsection{Optical Photometry}

\begin{figure}
    \centering
    \includegraphics[width=\linewidth]{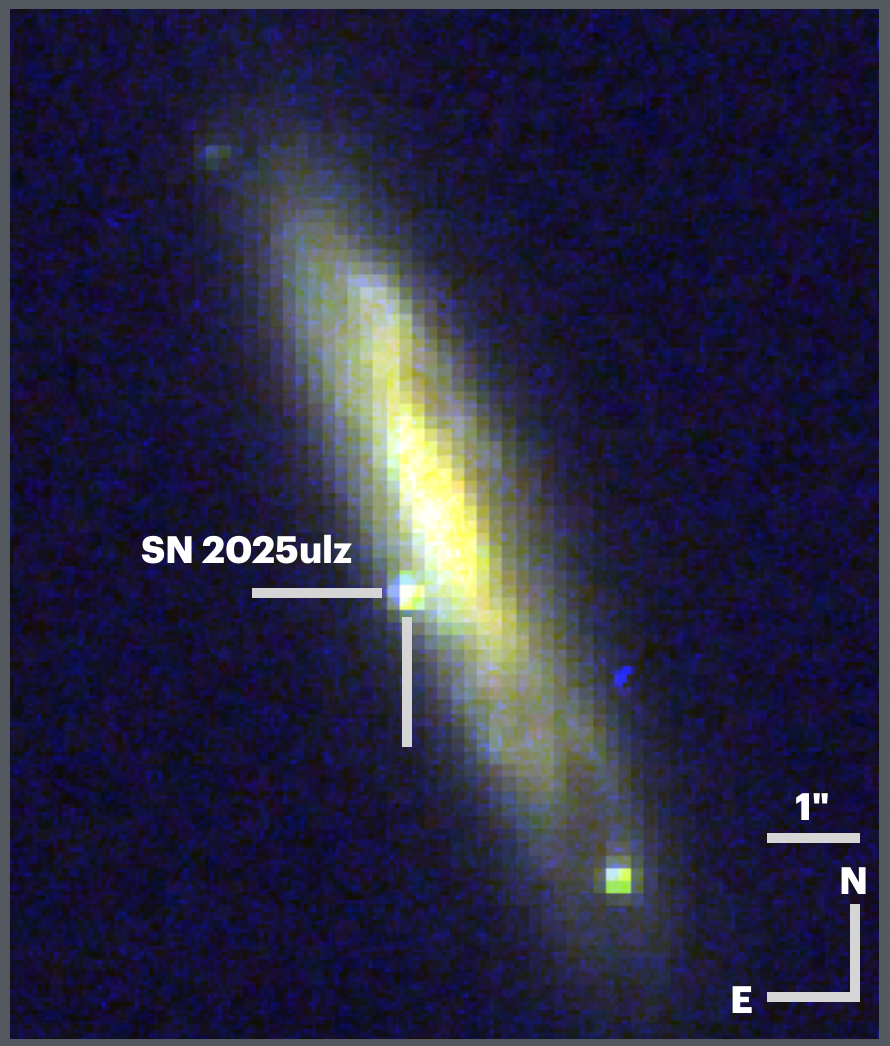}
    \caption{A false color image of the HST WFC3 observations of SN\,2025ulz.  We include the F606W image from 26 Aug (blue; $\delta t \sim 8$ days), the F110W image from 27 Aug (green; $\delta t \sim 9$ days), and the F160W image from 28 Aug 2025 (red; $\delta t \sim 10$ days). We highlight the position of SN\,2025ulz showing its close in projection to the core of its host galaxy.}
    \label{fig:hst-img}
\end{figure}

\begin{figure*}
    \centering
    \includegraphics[width=\linewidth]{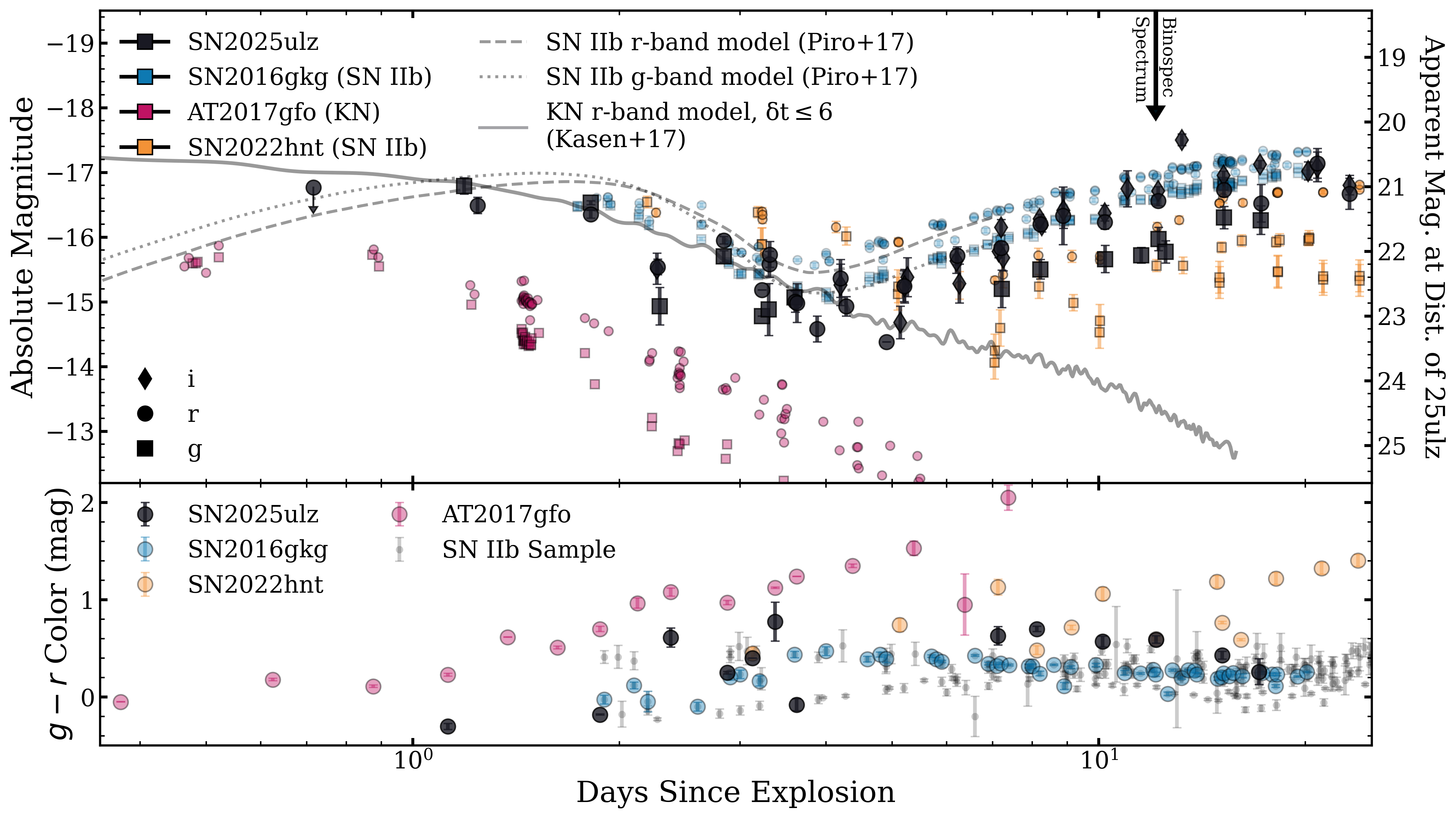}
    \caption{\textit{Top}: A comparison of the $gri$ light curves for SN\,2025ulz (black), SN\,2016gkg \citep[blue; ][]{tartaglia_progenitor_2017}, SN\,2022hnt \citep[orange; ][]{farah_when_2025}, and AT\,2017gfo \citep[red; ][]{villar_combined_2017}. For SN\,2025ulz, We choose an explosion date of MJD=60904 based on our shock cooling modeling (see \autoref{sec:early-lc-model}). The best fit (\textit{i.e.}, lowest sum of the residual squared) $r$-band KN model \citep[from][]{kasen_origin_2017} to SN\,2025ulz for the early time ($\delta t \leq 6$ days) light curve decline is shown as a solid line. We also show the $r$- (black dashed line) and $g$-band (black dotted line) models for SN\,2016gkg from \citet{piro_numerically_2017}. At phase $\delta t \lesssim 6$ days all four observed light curves appear to redden as they fade rapidly at around the same rate. At $\delta t \gtrsim 6$ days, the light curves of SN\,2025ulz and the other SNe IIb begin to re-brighten while the KN light curve continues to fade rapidly. \textit{Bottom}: The $g-r$ color evolution of SN\,2025ulz as compared to AT\,2017gfo, SN\,2016gkg, SN\,2022hnt, and 7 other SNe\,IIb (SN\,1993J, \citealt{richmond_ubvri_1994}; SN\,2008ax, \citealt{Pastorello2008}; SN\,2011dh, \citealt{Arcavi2011dh}; SN\,2011ei, \citealt{Dan2013}; SN\,2011fu, \citealt{Morales2015_2011fu}; SN\,2013df, \citealt{morales2014_2013df}; SN\,2024uwq, \citealt{Subrayan_2024uwq}). The color evolution of SN\,2025ulz is very similar to the rest of the SNe\,IIb population.}
    \label{fig:optical-lc}
\end{figure*}

The site of SN\,2025ulz was observed over multiple epochs with the {\it Hubble Space Telescope} ({\it HST}) using the Wide Field Camera 3 (WFC3) as part of GO-17450 and GO-17805 (PI: Troja).  The counterpart and its host galaxy were imaged in the F336W, F606W, F110W, and F160W bands from 22--28 August 2025.  We downloaded all {\it HST} images of SN\,2025ulz from the Mikulski Archive for Space Telescopes (MAST) and processed them with {\tt hst123} \citep[see][]{hst123,Kilpatrick_2019yvr}, including frame-to-frame alignment with {\tt Tweakreg} and image co-addition with {\tt astrodrizzle} \citep{drizzlepac}, and photometry in the calibrated image frames with {\tt dolphot} \citep{dolphot}.  We obtain detections of SN\,2025ulz in every epoch and band, and we provide the corresponding photometry in \autoref{tab:phot} (\autoref{app:data}). A false color image of a subset of the {\it HST} imaging from 26--28 August 2025 in the F606W, F110W, and F160W bands is in Figure \ref{fig:hst-img}.

We observed SN\,2025ulz in $ri$-bands on 22 August 2025 and $iz$-bands on 30 August 2025 using the imaging mode of the Binospec instrument \citep{Fabricant2019} on the 6.5-m MMT telescope located on Mt. Hopkins in Arizona, USA.  We processed all imaging using the {\tt POTPyRI} pipeline\footnote{\url{https://github.com/CIERA-Transients/POTPyRI}} \citep[see, \textit{e.g.},][]{Dong24,CHIME_FRB_2025}. We calibrate photometric zeropoints using stars in the Panoramic Survey Telescope and Rapid Response System \citep[Pan-STARRS;][]{PanSTARRS} data release (DR) 1 photometric survey. We derive an effective point spread function (ePSF) model for each processed image by fitting bright, isolated stars with the \texttt{EPSFBuilder} tool from the \texttt{photutils} package in \texttt{Astropy}. We then perform PSF-fitting at the location of SN~2025ulz, as well as a set of 20 or more stars spread throughout the image. A first-order two-dimensional polynomial is included in the PSF fitting to account for any spatially varying background and to avoid over-fitting of the stars. To estimate the uncertainty of each flux measurement, we set the statistical uncertainty per pixel using the RMS error of the fit residuals scaled by a factor of the square root of the reduced $\chi^2$. We then multiply this value by the number of `noise pixels' of the ePSF\footnote{A derivation of this quantity by F. Masci can be found at \url{http://web.ipac.caltech.edu/staff/fmasci/home/mystats/noisepix_specs.pdf}}. We use the set of Pan-STARRS calibration stars to derive aperture corrections ($\lesssim$0.1 mag in all filters) to scale PSF-fitting magnitudes to the images' photometric zeropoints. For the total uncertainty in our reported magnitudes, we report the statistical flux uncertainty summed in quadrature with the RMS error of the stars used in the zeropoint and ePSF aperture correction. Despite the large FOV of Binospec, the limited number of isolated stars in the Pan-STARRS catalog results in the zeropoint RMS dominating the reported error. For photometry taken on 22 August 2025, we report $5\sigma$ limits. The limiting magnitude is calculated by placing 50 random background apertures within 70 pixels from the target position and taking the standard deviation of the aperture values. We then repeat this ten times and choose $\sigma$ to be the average of the ten runs. Finally, we convert the $5\sigma$ flux in counts to magnitude using the photometric zeropoint derived from the calibration stars. 

We observed SN\,2025ulz on 28 August 2025 with the Goodman High-throughput Spectrograph \citep{Clemens04} in imaging mode on the Southern Astrophysical Research Telescope (SOAR) on Cerro Pachon, Chile.  We used $riz$ bands with a single 100\,s frame and 3$\times$200\,s images in each band.  All SOAR/Goodman imaging data were processed using {\tt photpipe} \citep{Rest05} using methods described in \citet{Rastinejad25}.  We calibrated all data using bias and dome flat-field frames obtained on the same night and instrumental configuration and aligned to the {\it Gaia} DR3 astrometric frame \citep{gaiadr3} using point-like astrometric standards.  We then performed PSF photometry in each frame using a custom version of {\tt dophot} \citep{dophot} and calibrated the photometry using Pan-STARRS DR2 $riz$ photometric standards in each image \citep{Flewelling20}.  We then resampled each image to a common, undistorted astrometric frame and stacked images in each band with {\tt Swarp} \citep{swarp}.  SN\,2025ulz is located close to the nucleus of its host galaxy, and so we subtracted Pan-STARRS 3$\pi$ images in each band using {\tt hotpants} \citep{hotpants}.  We detect a faint residual in $i$-band in the subtracted images and we place 3$\sigma$ upper limits on the presence of a counterpart in $r$~and $z$ bands using forced aperture photometry within 2.5$\times$ the full-width at half maximum (FWHM) of the difference image PSF and at the location of SN\,2025ulz.

We observed SN\,2025ulz over multiple epochs starting on 21 August 2025 with the T80N-Cam on the Observatorio Astrof\'isico de Javalambre (OAJ) 83\,cm telescope located at the Astrophysical Observatory in Teruel, Spain.  We observed in $r$ and $i$-bands with 2$\times$300\,s in $r$-band and 4$\times$300\,s in $i$-band.  All imaging was processed following the same procedures as with the SOAR imaging, including Pan-STARRS subtractions, using the S-PLUS Transient Extension Program (STEP) pipeline, as described in \citet{Santos24}.  We report detections and 3$\sigma$ upper limits on our T80S observations using a methodology similar to our SOAR imaging.

The $gri$ light curve of SN\,2025ulz is shown in \autoref{fig:optical-lc}, with both our photometry and publicly available photometry. Our other observations are used for modeling and provided in \autoref{tab:phot}.

\subsection{Optical Spectra}

\begin{figure*}
    \centering
    \includegraphics[width=\linewidth]{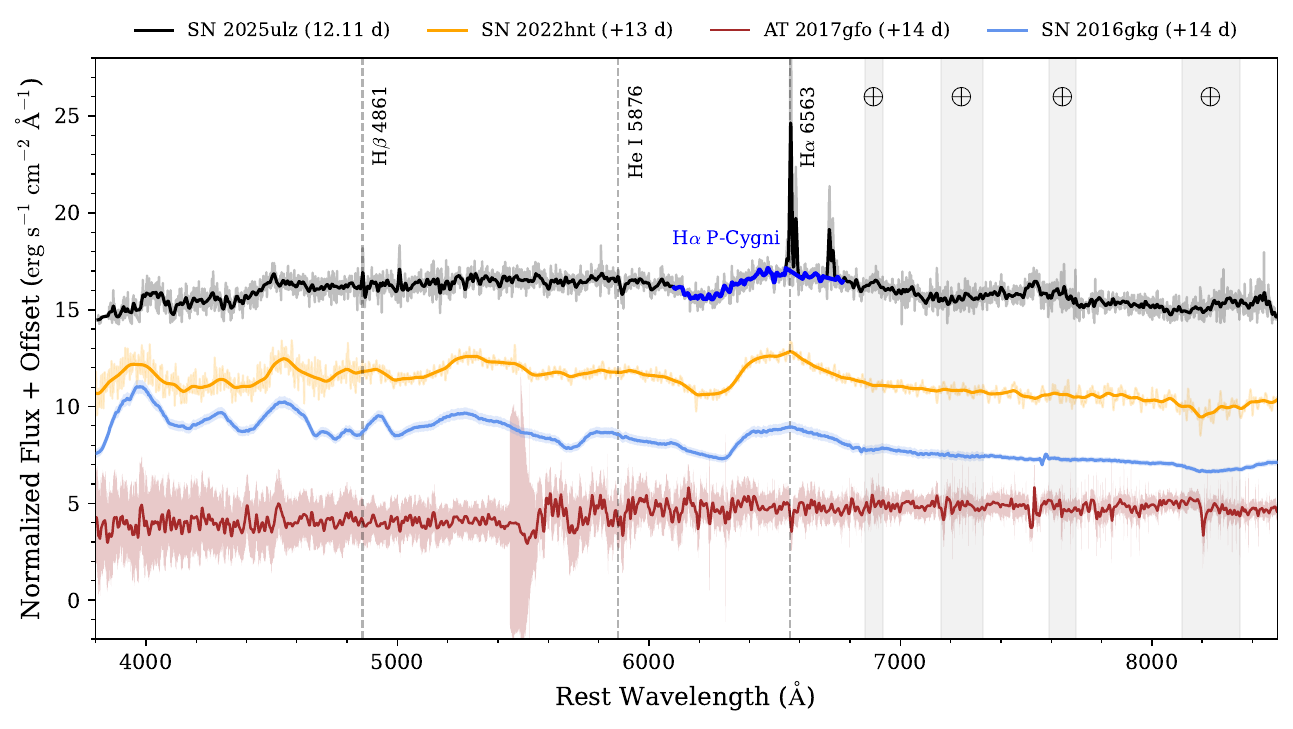}
    \caption{Comparison of the MMT optical spectrum of SN\,2025ulz (black) to the type IIb SN\,2022hnt \citep[orange;][]{farah_when_2025}, SN\,2016gkg \citep[sky blue;][]{tartaglia_progenitor_2017} and the kilonova AT\,2017gfo \citep[red; ][]{pian_spectroscopic_2017}, all at similar phases. The low opacity colored regions behind the spectra show the 1-sigma uncertainty and the grey vertical strips are regions of telluric features. The spectral features from SN\,2025ulz appear most similar to SN\,2022hnt and do not show any similarity with AT\,2017gfo. Most notably, the broad P-Cygni H$\alpha$ feature highlighted in dark blue, present in both SN\,2025ulz and SN\,2022hnt, is not expected, nor seen, in the KN spectrum.}
    \label{fig:binospec}
\end{figure*}

We observed SN\,2025ulz and its host galaxy SDSS\,J155154.16+305409.3 starting on 21 August 2025 with the Low-Resolution Imaging Spectrograph \citep[LRIS;][]{lris} on the Keck-I telescope located on Maunakea, Hawaii, USA.  We observed with the 400/3400 grism on the blue side and 400/8500 grating on the red side in conjunction with the d560 dichroic and the 1.0\arcsec\ long slit, covering a continuous wavelength range of approximately 3000--10500~\AA.  In this mode, we observed for 5$\times$930\,s and 5$\times$900\,s on the blue and red arms, respectively, although we cut the final blue exposure to 628\,s and red exposure to 646.6\,s due to poor atmospheric transparency.  We processed all spectra using {\tt PypeIt} \citep{pypeit:joss_pub,pypeit:zenodo}, calibrating the science spectra with dome flat exposures and obtaining a wavelength solution with arc lamp exposures obtained on the same night and instrumental configuration.  We extracted a one-dimensional spectrum located around the position of the host galaxy and performed flux calibration with a spectrum of BD+33d2642 obtained on the same night, with telluric corrections derived from atmospheric grids of Maunakea. We then scale the continuum flux to the Pan-STARRS photometry, which scales the line flux appropriately.

Additionally, we observed SN~2025ulz on 30 August 2025 ($\delta t \sim 12$ days post-discovery) using Binospec on MMT. This spectrum was taken with an exposure time of 5$\times$900 s using a 1.0\arcsec\ slit and the 270 lines/mm grating centered at 6800~\AA\ for a total wavelength range of roughly 3800-9200~\AA. This spectrum was triggered using the {\tt PyMMT} package \citep{Shrestha2024} as part of the Searches After Gravitational waves Using ARizona Observatories program \citep[SAGUARO;][]{2019ApJ...881L..26L, 2021ApJ...912..128P, 2022ApJ...927...50R, 2024ApJ...964...35H}. Processing of the 2D Binospec spectrum, including flat-fielding, sky subtraction, and wavelength and flux calibration, were done using the Binospec IDL pipeline \citep{Kansky2019}. The 1D spectrum was then extracted using standard IRAF techniques \citep{Tody1986, Tody1993}. A careful telluric correction was performed using the IRAF \texttt{telluric}\footnote{\url{https://iraf.readthedocs.io/en/doc-autoupdate/tasks/noao/imred/specred/telluric.html}} task, by creating an atmospheric model using the standard star (G191-B2B). We removed the intrinsic stellar features from the standard and the resulting telluric model was scaled, wavelength-shifted and divided to get the telluric corrected science spectrum \citep{telluric_1}. The reduced Binospec spectrum is shown in \autoref{fig:binospec}, alongside similar phase spectra of the type IIb SN~2022hnt \citep{farah_when_2025}, SN~2016gkg \citep{tartaglia_progenitor_2017}, and the kilonova AT~2017gfo \citep{pian_spectroscopic_2017}.

\subsection{Radio Observations}\label{sec:radio-obs}

On 26 August 2025, we observed SN\,2025ulz with the upgraded Giant Metrewave Radio Telescope (uGMRT) under program 48\_120 (PI: Laskar). Observations were taken using the band 5 receiver ($\nu \approx 1.26$ GHz) with a total observation length of 106 minutes. We used 3C286 and J1609+266 as the flux and phase calibrators, respectively. The data were reduced and imaged using standard practices in the Common Astronomy Software Applications version 6.5.4 \citep[CASAv6.5.4;][]{mcmullin_casa_2007, casa_team_casa_2022}, including two rounds of self-calibration. This observation had a beam size of $\sim 2.5''$ and a root mean square (RMS) error of $\sim 27~\mu{\rm Jy}$, and no evidence of emission at the location of SN\,2025ulz.

We observed SN\,2025ulz with the Very Large Array (VLA) over 2 epochs on 22 and 30 August 2025 (VLA configuration B$\rightarrow$C) under program 22A-417 (PI: Alexander). The first epoch spanned 1 hour and was at C-band ($\nu \approx 6$ GHz). In the first epoch 3C48 and J1602+3326 were used as the flux and phase calibrators, respectively. The second epoch spanned 2 hours and was at S-, C-, and X-bands ($\nu \approx 3,\,6,\,10$ GHz, respectively). During the second epoch, 3C286 and J1602+3326 were used as the flux and phase calibrators respectively. All data were reduced and imaged using standard practices in CASAv6.5.4. All epochs had a beam size of $\sim 0.5''-1.5''$ and the image RMS was $\sim (6 - 10)~\mu{\rm Jy}$, with no evidence of emission.

SN\,2025ulz was observed for three epochs using shared time on the South African MeerKAT radio telescope from programs MKT-24113/MKT-24127/MKT-24270 (PI: Bruni/Alexander/Mooley; triggered by PI Bruni under MKT-24113) on 21 and 28 August 2025 and 13 September 2025. The first two epochs used the MeerKAT S-band receiver ($\nu \approx 3$ GHz) and the third epoch used the UHF-, L-, and S-band receivers ($\nu \approx 0.81,\,1.28,\,3$ GHz, respectively). The MeerKAT observations were reduced and imaged by the SARAO (South African Radio Astronomy Observatory) pipeline. This resulted in a detection of diffuse resolved (\textit{i.e.}, larger than the synthesized beam size) emission consistent with the location of the transient host galaxy, with a major beam size of $\sim 5''-20''$ (although the beam is highly elongated, \textit{e.g.,} \autoref{tab:radio}) and an image RMS $\sim(4-45)~\mu{\rm Jy/beam}$. Note that we know that the emission must be mostly diffuse because it is not detected in the deeper, higher resolution radio images from the VLA, indicating that the emission is resolved out. Given the previous disagreement about the transient nature of the flux density measured from these observations \citep{2025GCN.41500....1B, 2025GCN.41594....1B, 2025GCN.41666....1R, 2025GCN.42032....1B}, we re-extract a flux from the images using both a 2D Gaussian fixed to the size of the beam and a 2D Gaussian with a variable size. We then convert all flux density values to units of $\mu$Jy/beam and find $F_{\nu} \approx 59 - 185~\mu{\rm Jy/beam}$ that is dependent only on the frequency of the observation, not the time. Therefore, regardless of the fitting method used, the detections are consistent within $1\sigma$ with non-varying emission.

We observed SN\,2025ulz with the Northern Extended Millimetre Array (NOEMA) 3mm receiver under program S25CS (PI: Laskar) on 22 August 2025. The observation used 3C345 and J1600+335 as the flux and phase calibrator, respectively. The data were reduced in GILDAS\footnote{\href{https://www.iram.fr/IRAMFR/GILDAS/}{https://www.iram.fr/IRAMFR/GILDAS/}}, following standard calibration procedures, and imaged using CASAv6.5.4. This resulted in a beam size of $\sim 3''-4''$, an image RMS of $\sim 21~\mu{\rm Jy}$, and no evidence of emission.

Finally, we observed SN\,2025ulz at 225.5 GHz with the Submillimeter Array (SMA) on 21 August 2025 under the POETS (Pursuit of Extragalactic Transients with the SMA) program (PI: Berger). The data were reduced and imaged using the COMPASS (Calibrator Observations for Measuring the Performance of Array Sensitivity and Stability) pipeline resulting in an RMS of $250 \mu {\rm Jy}$, a beam size of $\sim 3''$, and no evidence of emission.  

We do not detect any radio emission from SN\,2025ulz, and we are confident that the diffuse emission detected by MeerKAT is from the host galaxy rather than the transient (also see \autoref{sec:host}). For flux and limit extraction from all radio data (except those from the SMA, for which we use COMPASS), we fit a 2D Gaussian fixed to the synthesized beam size using the CASA task {\tt imfit} at the location of SN\,2025ulz. All limits are three times the RMS noise in a nearby empty portion of the sky. Our radio observations are summarized in \autoref{tab:radio}.

\section{Analysis \& Discussion of SN\,2025\lowercase{ulz}} \label{sec:analysis_discussion}
\subsection{The Light Curve and Spectrum are Consistent with SNe IIb}

\autoref{fig:optical-lc} shows the $gri$ optical light curve of SN\,2025ulz compared to the $gr$ light curves of KN AT\,2017gfo \citep{villar_combined_2017} and two SNe IIb: SN\,2022hnt \citep{farah_when_2025} and SN\,2016gkg \citep{tartaglia_progenitor_2017}. For reference, we show the numerical model fits to SN\,2016gkg from \citet{piro_numerically_2017}. For SN\,2025ulz we include the $i$-band to fill in the gaps and guide the eye, as it evolves similarly to the $r$-band light curve. As shown in \autoref{fig:optical-lc}, SN\,2025ulz follows a similar photometric evolution to both SN\,2016gkg and SN\,2022hnt, with all three showing a reddening $\sim 2$ days after explosion and a re-brightening $\sim 5-7$ days after explosion. In addition, \autoref{fig:optical-lc} shows that, at all times, the $g-r$ color evolution of SN\,2025ulz is comparable to that of other SNe\,IIb. Given that this re-brightening is inconsistent with KN models, the overall light curve of SN\,2025ulz is more similar to that of an SN\,IIb than a KN.

\autoref{fig:binospec} shows the optical spectrum of SN\,2025ulz from our MMT Binospec observation at phase $\sim12$ days post-discovery. We also show optical spectra of SN\,2022hnt \citep{farah_when_2025}, SN\,2016gkg \citep{tartaglia_progenitor_2017}  and AT\,2017gfo \citep{pian_spectroscopic_2017} for comparison. SN\,2025ulz shows a broad P-Cygni H$\alpha$ line, which we attribute to the SN, with a narrow H$\alpha$ component, which we attribute to the host galaxy. SN\,2022hnt shows a P-Cygni H$\alpha$ line with an absorption component with a similar velocity ($v \sim 13,000$ km~s$^{-1}$) to SN\,2025ulz ($v\sim15,600$ km~s$^{-1}$) at a similar phase. In contrast, AT\,2017gfo shows no strong spectral features at this phase, including a lack of any Hydrogen. Furthermore, KNe are relativistic transients with extremely fast outflows \citep[dynamical ejecta velocities $v_{\rm ej} \sim (0.1-0.3)c$; ][]{kasen_origin_2017}, generally expected to be faster than the velocity of the hydrogen line in the SN\,2025ulz spectrum. Therefore, based on the broad P-Cygni H$\alpha$ line, the spectrum of SN\,2025ulz is consistent with an SN\,IIb classification and inconsistent with a KN.

\subsection{Modeling the Light Curve as a KN}\label{sec:early-lc-model}
For completeness, we attempt to fit the SN\,2025ulz light curve as a conventional KN (\textit{i.e.}, AT\,2017gfo-like) using the models from \citet{kasen_origin_2017}. The \citet{kasen_origin_2017} model grid produces spectral evolution models for KNe spanning dynamical ejecta velocities $v_{\rm ej}\approx(0.03-0.4)c$, ejecta mass $M=(0.001-0.1)M_\odot$, and lanthanide fraction $X_{\rm lan} = 10^{-9}-10^{-1}$. We take their spectral model data and integrate the simulated spectrum convolved with the SDSS $r$-band transmission function to extract a simulated light curve for each model. For each model in their grid, we downsample the simulated light curve to our temporal coverage and compute the sum of the square residual ($\Sigma \epsilon^2$) with respect to our $r$-band light curve. \added{We note that, since we are not interpolating over the \citet{kasen_origin_2017} grid, this approach only produces approximations of the KN parameters. Exact KN parameters could be derived from a more detailed modeling of the SN\,2025ulz light curve using existing Bayesian fitting codes (e.g., \texttt{MOSFiT}, \citealt{guillochon_mosfit_2018}; \texttt{gwemopt}, \citealt{Coughlin:2018lta, Coughlin:2019qkn}; \texttt{nmma}, \citealt{Pang:2022rzc}; \texttt{redback} \citealt{sarin_redback_2024}), something that is out of the scope of this work.}

We then find the best-fitting model for both the early light curve decline ($\leq 6$ days since explosion) and the entire light curve. The early light curve fit has a low $\Sigma \epsilon^2=1.66$ and the full light curve fit has a poor $\Sigma \epsilon^2=118.55$. As a result, in \autoref{fig:optical-lc} we show the best-fit model (lowest $\Sigma \epsilon^2$) for the early light curve as a solid line. This demonstrates that the early light curve can be fit well with a conventional KN model, but not the full light curve. The early time best-fit model has an ejecta mass of $M_{\rm ej} \approx 0.05~{\rm M_\odot}$, outflow velocity of $v_{\rm ej} \approx 0.2c$, and a lanthanide mass fraction $X_{\rm lan} \approx 10^{-2}$. In contrast, the best-fit model for the entire light curve, which is an extremely poor fit, has $M_{\rm ej} \approx 0.01~{\rm M_\odot}$, $v_{\rm ej} \approx 0.03c$, and $X_{\rm lan}  10^{-9}$.

Based on this analysis, no model in the \citet{kasen_origin_2017} grid is able to reproduce the luminosity of the second optical peak while simultaneously capturing the early time light curve features. This is expected given that the optical luminosity of the first peak requires a high $X_{\rm lan}$ while the optical luminosity of the second optical peak requires a small $X_{\rm lan}$ \citep{kasen_kilonova_2015}. Furthermore, even though \citet{kasen_kilonova_2015} predict a second peak in the {\it infrared}, we are not aware of any conventional KN model, irrespective of the viewing angle, that has a second {\it optical} light curve peak more luminous than its initial peak (as is seen in SN\,2025ulz).

We briefly consider, and subsequently disfavor, more extreme and irregular KN models. One model for GW signals from a subsolar-mass (SSM) BNS merger is the fragmentation of a collapsar disk into multiple, low-mass NSs that later merge \citep{metzger_fragmentation_2024, chen_gravitational_2025}. Under this model, we would expect an SN followed by multiple GW signals, with the last signal consistent with a NSBH/BNS merger (almost certainly with a chirp mass $\gtrsim 1 M_\odot$) that occurs $\sim$hours-days after the initial GW signal \citep{metzger_fragmentation_2024}. In the case of S250818k, there was only one (low-significance) GW signal reported from its localization region \citep[\textit{e.g.}, as was noted in ][]{gillanders_pan-starrs_2025}. However, it is possible, given the model parameters, that LVK only detected the strongest of the many predicted signals. While we can not entirely rule out this model, we strongly disfavor it given the overwhelming observational similarities between SN\,2025ulz and other prototypical SN\,IIb, making the SN explanation considerably more plausible. 

Related, the SN\,IIb SN\,2025uso was discovered $\sim 2$ days after S2501818k and is in the localization volume of the event \citep{2025TNSTR3300....1C}. Based on publicly available light curve information, the explosion date of this event is roughly consistent with the S2501818k merger date ($\delta t \sim -1$ days, similar to our computed explosion date for SN\,2025ulz). Under the \citet{metzger_fragmentation_2024} model for KN from SSM BNS mergers, SN\,2025uso should be considered an equally likely candidate KN associated with S250818k. However, we rule this out for the same reasons as SN\,2025ulz; the simpler, prototypical SN explanation is much more plausible. 
   
Another possible KN model is an extreme eccentric encounter between a NS and a BH. Under this model, the NS initially approaches on an orbit grazing the tidal radius of the BH and then on the next orbit is fully disrupted. This, in theory, can explain two electromagnetic flares (at the pericenter of both orbits) with only one GW signal (at the time of the full disruption of the NS; \citealt{east_eccentric_2015}). This model is rather unlikely, as (1) it would require a very extreme initial orbit to explain the timescales of the two flares in the optical light curve; (2) the International Gravitational Wave Network (IGWN) alert had a $\sim 0\%$ probability that this was a NSBH merger; (3) the {\it maximum} chirp mass is 0.87 M$_{\odot}$, indicating a hypothetical black hole mass $\lesssim 1 {\rm M_\odot}$ and making this scenario extremely unlikely; and (4) it predicts the formation of an accretion disk, rather than the expanding outflow that is required to produce the P-Cygni H$\alpha$ profile in our optical spectrum. Therefore, we also rule out this model.

\subsection{Early Shock Cooling Light Curve Modeling}
\begin{figure*}
    \centering
    \includegraphics[width=\linewidth]{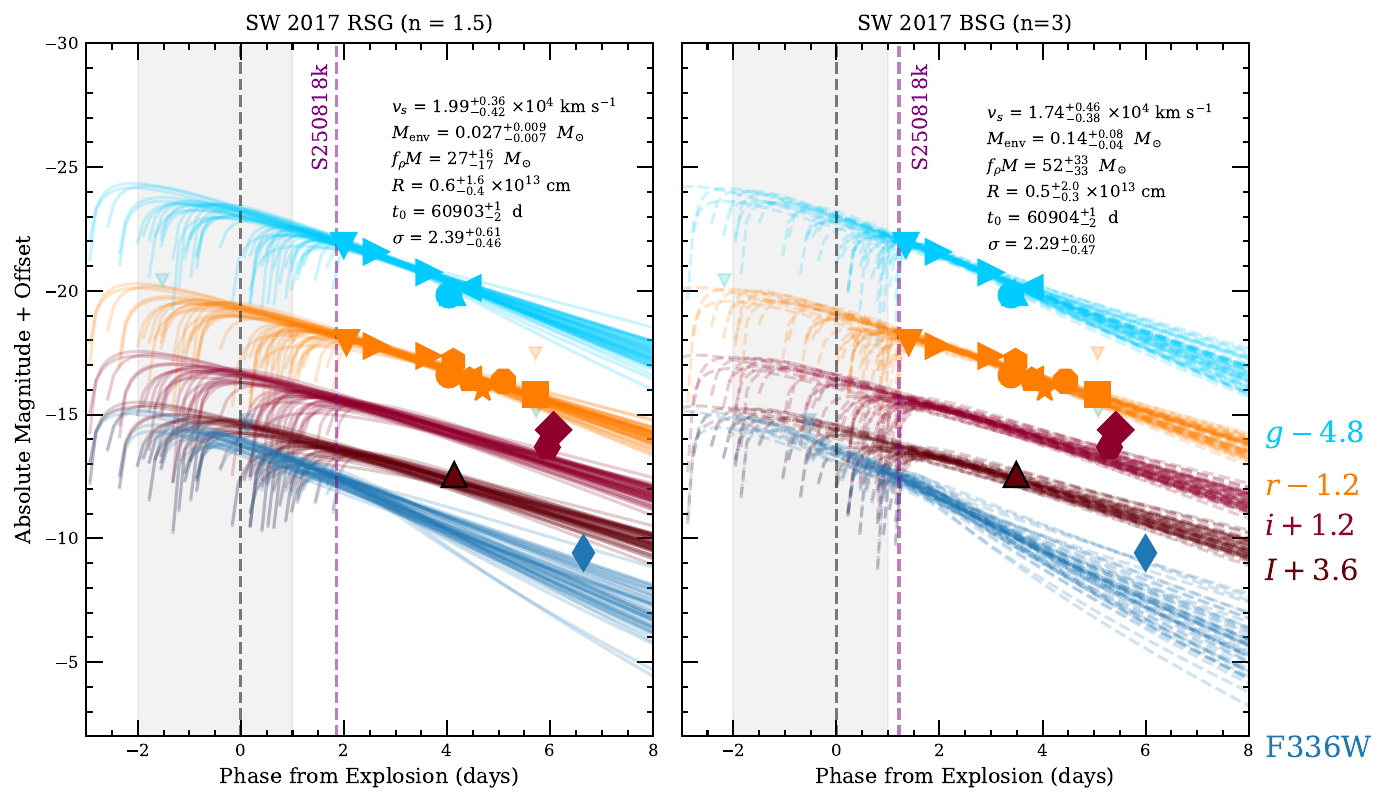}
    \caption{Fits of the $griI$ and HST F336W (which is comparable to Swift/UVOT $U$-band) light curves of SN\,2025ulz with a shock cooling modeling \citep{SW2017}, assuming two polytropic indices: $n=3/2$ (left) and $n=3$ (right). The best-fit explosion date, with uncertainties, and the detection date of S250818k are shown as grey and purple dashed lines, respectively.}
    \label{fig:sw_rsg_bsg}
\end{figure*}

\begin{figure}
    \centering
    \includegraphics[width=\linewidth]{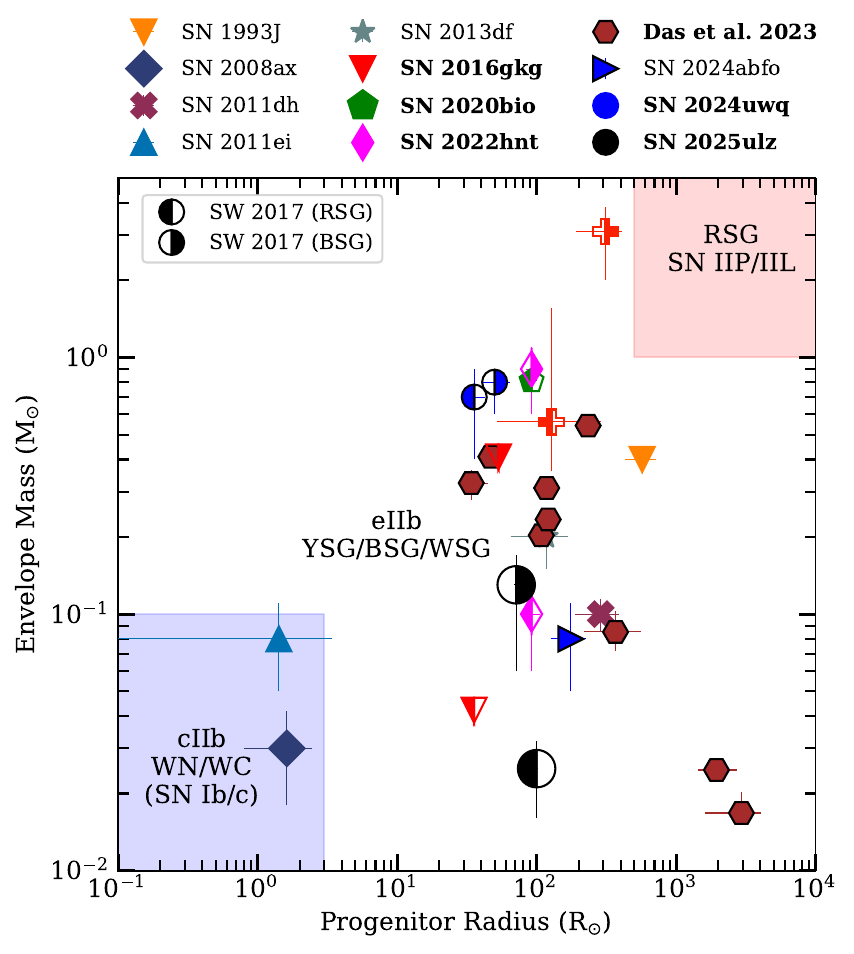}
    \caption{The shock cooling model envelope mass vs. progenitor radius for SN\,2025ulz as compared to a population of SNe IIb. SN\,2025ulz lies in a similar region of parameter space to the other SNe IIb and is in nearly the same region as SN\,2016gkg using the same model. SNe highlighted in bold have values derived from similar shock-cooling modeling, while the others adopt values from alternate methods. Points with the right half filled are from the \citet{SW2017} BSG model and points with the left half filled are from the \citet{SW2017} RSG model. Published values for other SNe in this figure are obtained from: SN~1993J: \citealt{Woosley1994a}; SN~2008ax: \citealt{Pastorello2008,CS2010}; SN~2011dh \citealt{B2012}; SN~2011ei: \citealt{Dan2013}; SN~2011fu: \citealt{Morales2015_2011fu}; SN~2013df: \citealt{morales2014_2013df}; SN~2016gkg: \citealt{Arcavi2017}; SN~2017jgh: \citealt{Armstrong_2017jgh}; SN~2020bio: \citealt{Pellegrino_2023_SN2020bio}; SN~2022hnt: \citealt{F2025_2022hnt}; SN~2024uwq: \citealt{Subrayan_2024uwq},
    SN~2024abfo: \citealt{Regutti2025}; and
    \citealt{Das2023}.}
    \label{fig:menv_r_phase_space}
\end{figure}

Shock cooling in type IIb SNe occurs when the explosion shock wave deposits energy into the progenitor’s extended, partially stripped hydrogen envelope, which then radiates its residual thermal energy as it rapidly expands and cools in the subsequent hours to days following shock breakout \citep{R1994, Woosley1994a, SW2017, Arcavi2017}. We fit the early ($\delta t \lesssim 7$ days since discovery) light curve dataset of SN~2025ulz to semi-analytic shock cooling models described by \citet{SW2017}, using a Markov Chain Monte Carlo (MCMC) routine implemented in the Light Curve Fitting package \citep{hosseinzadeh_2024_11405219}. We follow the same methods for the early shock cooling light curve modeling for type IIb as described in \citet{Subrayan_2024uwq}. Two polytropic indices (\(n=3/2\) and \(n=3\)) were considered to represent convective (Red Supergiant; RSG) and radiative (Blue Supergiant; BSG) progenitor envelopes. The MCMC simultaneously constrains the progenitor radius (\(R\)), shock velocity (\(v_s\)), and envelope mass (\(M_{\rm env}\)), while an intrinsic scatter term (\(\sigma\)) scales the observational errors by \(\sqrt{1+\sigma^2}\) to account for additional variance. A scaled ejecta mass parameter was also included but remains largely unconstrained due to its minimal impact on the early light curve. The best fit models and the derived parameters for RSG and BSG progenitors are shown in Figure \ref{fig:sw_rsg_bsg}.

The early-time light curve of SN~2025ulz is well described by the semi-analytic shock cooling models of \citet{SW2017}, with the inferred envelope mass \(M_{\rm env}\) and progenitor radius \(R\) falling within the parameter space spanned by known type~IIb events in the literature, as shown in \autoref{fig:menv_r_phase_space}. The best-fit values of SN~2025ulz lie closest to SN~2022hnt and SN~2016gkg. This is also consistent with the spectral similarity between the two SNe at $\sim +$12 days (see \autoref{fig:binospec}). The inferred \(M_{\rm env}\) and \(R\) for SN~2025ulz also match the theoretical ranges predicted by \citet{Yoon2017} and \citet{S2020} for RSG progenitors with residual, thin hydrogen envelopes, typical of SN~IIb that display P-Cygni H profiles in their spectra. We also note that the modeling yields an explosion epoch that consistently precedes the reported subthreshold GW event. However, this model can overestimate the rise time of the SN and may produce explosion epochs before the true explosion date \citep{pearson_circumstellar_2023}.

\subsection{Host Galaxy Properties Reveal a Star-Forming Galaxy}\label{sec:host}
The star formation rate (SFR) of a host galaxy can provide important contextual information for the classification of transients. For example, due to the short lifetimes of their high mass progenitors, core collapse SNe (CCSNe) are expected to occur at a higher rate in galaxies with active star formation. We use the results from \texttt{Blast}\footnote{\url{https://blast.ncsa.illinois.edu/}}, a web service and automated galaxy spectral energy distribution modeling software, to infer the physical properties of SN\,2025ulz's host galaxy \citep{jones_blast_2024}. For SN\,2025ulz, \texttt{Blast} finds a stellar mass $M_* = 1.23^{+0.21}_{-0.15} \times 10^{10} ~{\rm M_\odot}$, ${\rm SFR} = 0.298^{+0.244}_{-0.143} ~{\rm M_\odot~yr}^{-1}$, and a specific SFR ${\rm sSFR} = 2.37^{+2.02}_{-1.19} \times 10^{-11} ~{\rm yr^{-1}}$. This SFR and host galaxy mass are consistent with the broader population of type IIb SNe \citep{qin_linking_2024, nugent_characterizing_2025}. We now explore other probes of star formation rate including H$\alpha$, which probes star formation in the last $10$ Myr, and the diffuse radio emission observed in MeerKAT, which probes the last $150$ Myr \citep{kennicutt_star_2012}.

\begin{figure}
    \centering
    \includegraphics[width=\linewidth]{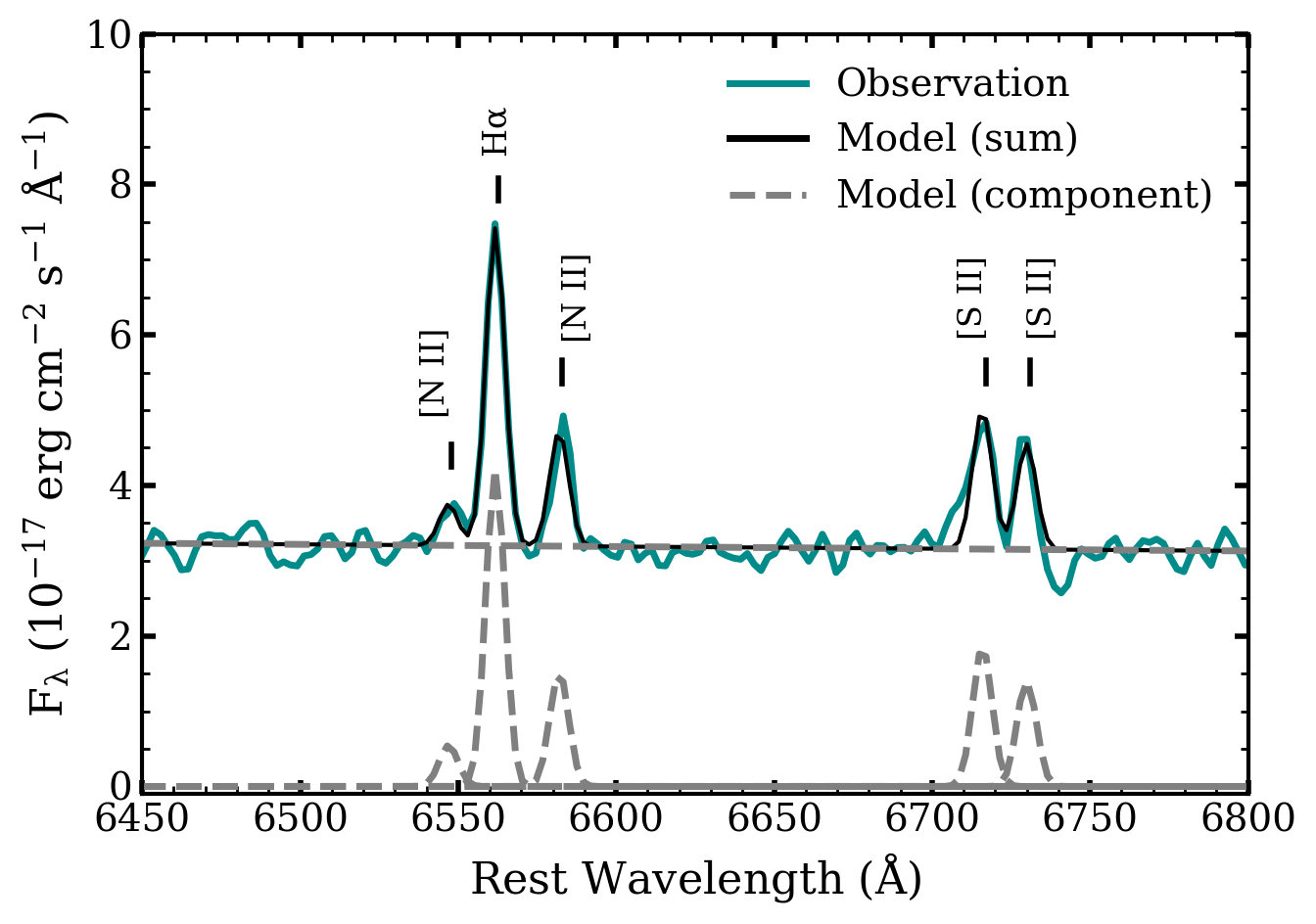}
    \caption{Model of our Keck spectrum that is dominated by emission from the SN\,2025ulz host. We use this model to derive a star formation rate for the host galaxy for comparison with our radio derived star formation rate.}
    \label{fig:keck-spec-model}
\end{figure}

Since no emission at the location of SN\,2025ulz was visible in our Keck spectrum, we instead extracted a host galaxy spectrum. We first estimate the host extinction $E(B-V)$ using the Balmer decrement. We model both the H$\alpha$ and H$\beta$ lines with a Gaussian plus a linear continuum\footnote{Additionally, we add a [\ion{N}{2}] and [\ion{S}{2}] Gaussian component to our H$\alpha$ model.}. The model of H$\alpha$ is shown in \autoref{fig:keck-spec-model}. We integrate both the H$\alpha$ and H$\beta$ best fit models to obtain the line flux. From the H$\alpha$/H$\beta$ ratio we find a host extinction of $E(B-V)\approx0.49$ mag using the relations from \citet{dominguez_dust_2013}. This extinction is for the entire host galaxy and is likely overestimated along the line of sight to SN\,2025ulz, as its location is outside the plane of the host galaxy and we do not detect \ion{Na}{1}D in the MMT Binospec spectrum of the transient.  

As a check on the \texttt{Blast} values, we compute the SFR and sSFR from H$\alpha$. By assuming a Milky Way $R_V=3.1$, we correct the H$\alpha$ line flux for both host and Milky Way dust extinction using the \texttt{extinction} Python package. We apply the relation from \citet{kennicutt_star_1998} to obtain a host galaxy star formation rate ${\rm SFR \approx 0.8 \pm 0.2 ~M_\odot~yr}^{-1}$, or ${\rm sSFR \approx 8\times10^{-11} yr^{-1}}$, from the H$\alpha$ line flux. This is consistent, but at the upper end of the uncertainty, with the SFR derived by \texttt{Blast}. 

The third epoch of MeerKAT observations was performed at the UHF-, L-, and S-bands. This allows us to derive a spectral index for the diffuse emission of $\alpha \sim -0.9$, which is consistent with observations of star forming galaxies \citep{2024MNRAS.528.5346A}. Using this spectral index and the relations from \citet{2011ApJ...737...67M}, we find ${\rm SFR} \sim 5~{\rm M}_\odot~{\rm yr}^{-1}$. Given the $\sim$dex scatter in both relations used, and because the radio probes a longer epoch of star formation, the radio and H$\alpha$ derived SFR are broadly consistent. This consistency is further evidence that the diffuse radio emission detected by MeerKAT is consistent with star formation in the past $\sim150$ Myr.

Qualitatively, the combined evidence from \texttt{Blast}, our radio observations, and our H$\alpha$ measurements indicate a consistently high SFR over the past $\sim100$ Myr ($\sim 1 ~{\rm M_\odot~yr}^{-1}$) for the host galaxy of SN\,2025ulz. The SFRs computed here make the SN\,2025ulz host consistent with the larger population of CCSN hosts presented in \citet{qin_linking_2024, nugent_characterizing_2025} but, alone, does not necessarily rule out a BNS origin \citep[\textit{e.g.},][]{belczynski_likelihood_2017}.

\subsection{The Radio Observations Rule Out a SGRB-like On-Axis Jet}\label{sec:radio-discussion}
\begin{figure}
    \centering
    \includegraphics[width=\linewidth]{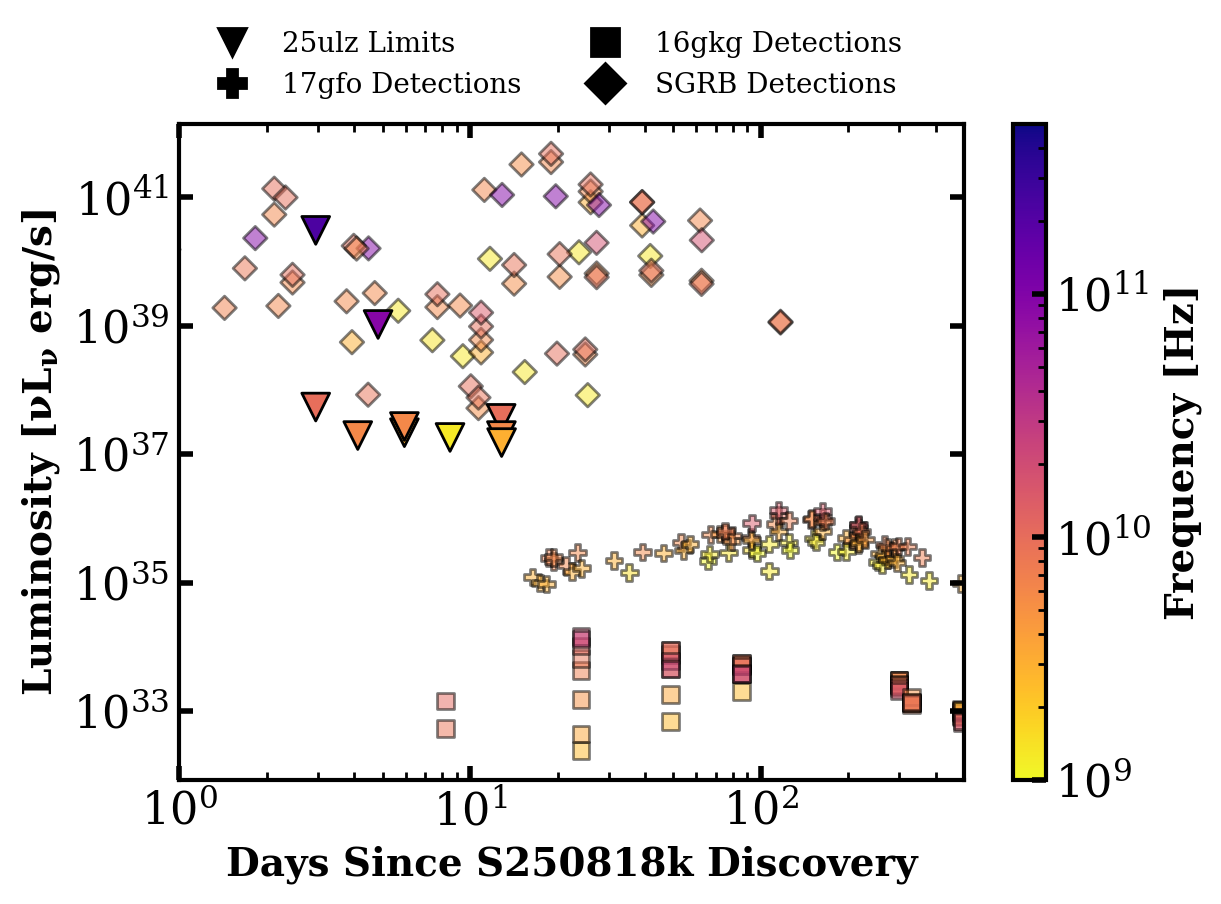}
    \caption{The radio light curves of SN\,2025ulz (downward triangles as limits) as compared to SN\,2016gkg \citep[squares; ][]{a_j_radio_2022}, a sample of short SGRB radio detections \citep[diamonds; ][]{2005Natur.438..988B,2006ApJ...650..261S,2014ApJ...780..118F,2015ApJ...815..102F,2019ApJ...883...48L,2021ApJ...906..127F,2022ApJ...935L..11L,2022ApJ...940...56F,2024ApJ...962....5N,2024ApJ...970..139S,2024Natur.626..737L,2025ApJ...982...42S,2025arXiv250715940D,GCN16815,GCN38189,GCN40966,GCN41038,GCN41046,GCN41419,GCN41060,GCN35097,GCN9958,GCN41455}, and AT\,2017gfo \citep[plus signs; ][]{2017ApJ...848L..21A,2017Sci...358.1579H,2017ApJ...850L..21K,2021ApJ...922..154M,2017Sci...358.1579H,2018ApJ...868L..11M,2018Natur.554..207M,2018Natur.561..355M,2018ApJ...867...57R,2018Natur.554..207M,2018ApJ...868L..11M,2018ApJ...858L..15D,2018ApJ...868L..11M,2018ApJ...868L..11M,2018ApJ...856L..18M,2020MNRAS.494.5110B,2018ApJ...858L..15D,2019Sci...363..968G,2018ApJ...863L..18A,2019MNRAS.489.1919T,2019ApJ...886L..17H}. The color of the point is the representative frequency of the receiver used for the observation. Our radio limits on SN\,2025ulz rule out an SGRB-like on-axis jet but do not rule out SN IIb-like or late-rising AT2017gfo-like radio emission. }
    \label{fig:radio-lc}
\end{figure}
In contrast to the low-resolution ($\sim10''$) MeerKAT observations, our higher-resolution ($\sim 1''$) GMRT, VLA, NOEMA, and SMA observations probe the existence of compact radio emission components. We do not detect radio emission at the transient position with any of these telescopes and we show the light curve of the radio upper limits in \autoref{fig:radio-lc}. For comparison, we show the radio data for AT\,2017gfo, SN\,2016gkg, and a sample of SGRBs from the literature. 

These early-time radio upper limits on SN\,2025ulz rule out an on-axis jet like those seen in SGRBs, which would be expected for a face-on BNS merger \citep[\textit{e.g.},][]{schroeder_long-lived_2025}. However, our limits are still consistent with a BNS merger with an off-axis jet that only brightens in the radio after $\sim 10$ days, like AT\,2017gfo. Our radio limits are also consistent with the less luminous radio emission produced following a CCSN and, in particular, is consistent with the luminosity of the SN\,IIb 2016gkg (see \autoref{fig:radio-lc}).

\section{A Search for Additional Kilonovae Candidates} \label{sec:search}

\begin{figure*}
    \centering
    \includegraphics[width=0.9\linewidth]{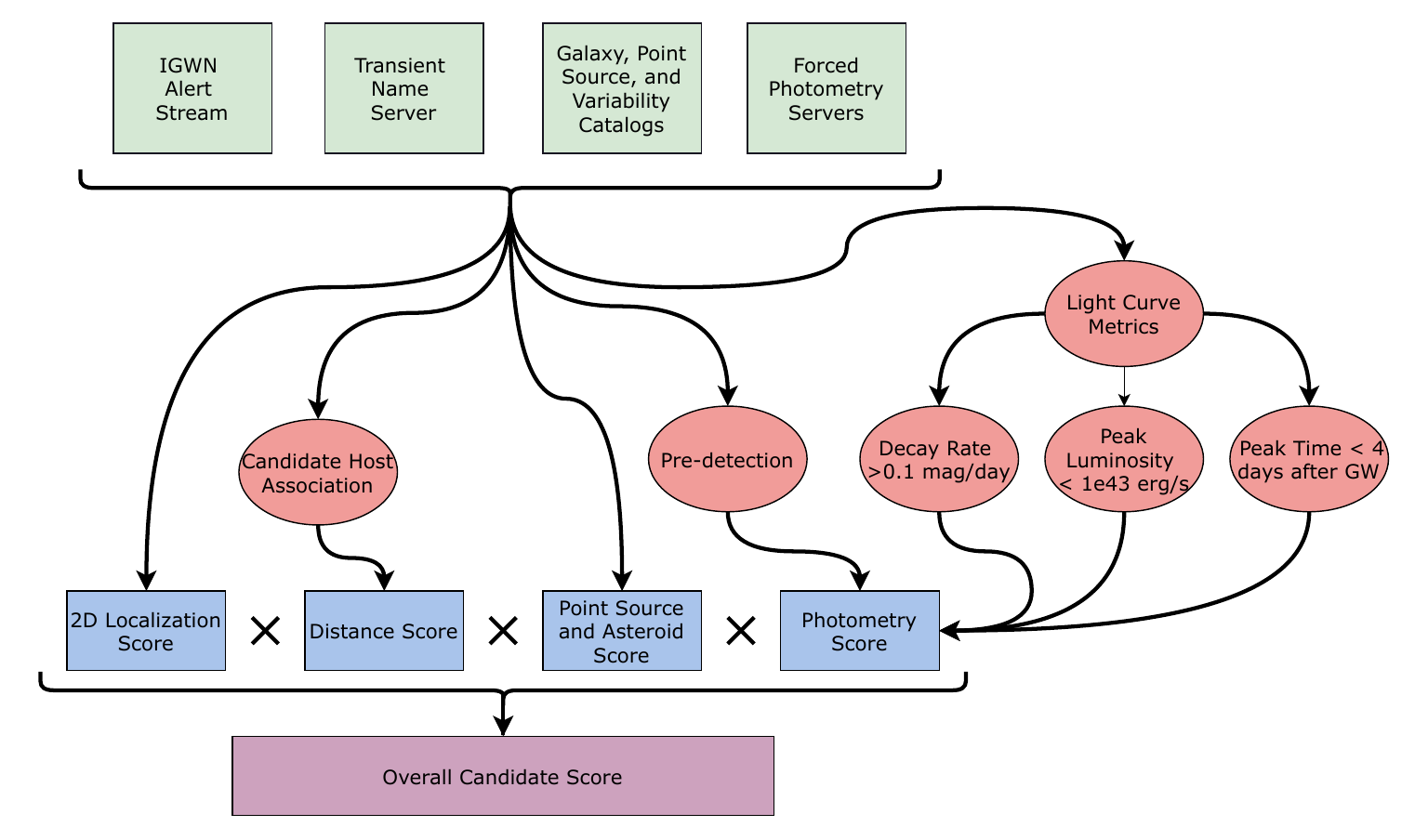}
    \caption{Flow diagram outlining our current candidate scoring algorithm. In general, it takes in IGWN alerts, TNS transients, various static catalogs, and forced photometry and derives multiple ``subscores'' from the data, eventually multiplying them together to obtain a total overall score for a candidate.}
    \label{fig:scoring-diagram}
\end{figure*}

Given the overwhelming photometric and spectroscopic evidence that SN\,2025ulz is a type IIb SNe and not a counterpart to S250818k, we search the TNS for other transients (1) with coordinates inside the 95\% localization region for S250818k from the IGWN data stream and (2) reported within one week of the discovery of S250818k. This search reveals 121 electromagnetic candidate counterparts, one of which is SN\,2025ulz. \added{This suggests that the KN search after S250818k \citep[and likely other GW events, see][]{ackley_localization_2025} would have benefited from follow-up of additional candidates.}

Expanding on the method in \citet{rastinejad_systematic_2022}, we vet these candidates to determine the likelihood of association with S250818k. For each step in their vetting process, we develop a scoring algorithm that quantifies our confidence in the association of a transient with the GW event. Our scoring procedure consists of four components: 
\begin{enumerate}[noitemsep]
    \item Score from the 2D localization (\autoref{sec:2d_score})
    \item Score from the distance localization (\autoref{sec:3d_score})
    \item Score from comparison to known asteroids and point source/variability catalogs (\autoref{sec:mpc_score})
    \item Score from comparing the photometric evolution to models and existing observations (\autoref{sec:phot_score})
\end{enumerate}
Finally, we take the resulting score from each of these components and multiply them to obtain a total score between 0 and 1, such that a higher score generally indicates a more probable candidate. 

A summary of this scoring is shown in \autoref{fig:scoring-diagram} and details are presented in the following subsections. For reference, we validated this scoring algorithm against AT\,2017gfo and found that it has a score of ${\rm S}\approx0.55$, depending on if we use the photometric redshift (from the best matching host, ${\rm S}=0.54$) or spectroscopic redshift (${\rm S}=0.56$). This score of $\sim0.55$ for the known kilonova counterpart to a BNS merger suggests that a candidate need not have a score of $\sim1$ to be worth consideration. 

\subsection{2D Localization Score} \label{sec:2d_score}
The IGWN localization map for GW events provides a probability density in each HEALPix tile describing the likelihood that the GW event originated from that tile. These probability densities are normalized such that integrating over the entire sky provides a localization probability of 1.0. We extract the probability density from the HEALPix tile where each candidate is located and perform a 2D integration over the localization region with a higher probability density \citep{singer_going_2016, singer_supplement_2016}. This cumulative probability is the score that the GW event originated from the candidate coordinates\footnote{Since we use the cumulative probability, the score is less dependent on the NSIDE property of the HEALPix map, as compared to using the probability density.}, and is the 2D localization score ($\rm S_{2D}$).

\subsection{Distance Score} \label{sec:3d_score}
Deriving a distance score requires distances for both the GW signal and the candidate electromagnetic counterpart. The distance to the GW event is given as a function of localization tile, where each tile in the localization map has an associated distance and uncertainty \citep{singer_going_2016, singer_supplement_2016}. We use the distance in the tile corresponding to the candidate coordinates as the distance to the GW event when computing the distance score for this candidate.

Finding the distance to the candidate is more difficult, since the redshift is typically not known. Therefore, we first query galaxy redshift catalogs\footnote{These include DESI (Dark Energy Spectroscopic Instrument) Early Data Release \citep{desi_collaboration_early_2024}, GLADE \citep{Dalya+18}, GWGC (Gravitational Wave Galaxy Catalog) \citep{White+11_GWGC}, Hecate \citep{Kovlakas+21_HECATE}, LS DR10 (Legacy Survey Data Release 10) \citep{Zhou+23_LSDR10}, Pan-STARRS (PS1) Galaxy catalog \citep{Beck+21}, and SDSS DR12 (Sloan Digital Sky Survey Data Release 12) photo-$z$ catalog \citep{Alam+15}.} for all potential host galaxies within $2'$~($\approx150$ kpc at $260$ Mpc) of the candidate. \added{Note that we expect this host galaxy search to be relatively complete at $z\leq0.3$, the distance scale that is relevant for this work, because LS DR10 is $>90\%$ complete out to this redshift \citep{li_photometric_2024}. }For each galaxy match found, we compute the probability of chance coincidence \citep[$P_{\rm cc}$; ][]{bloom_observed_2002}. We select all galaxies with $P_{\rm cc} < 0.1$ as potentially associated with the candidate. If no galaxies have $P_{\rm cc} < 0.1$, we select the galaxy with the minimum $P_{\rm cc}$ as the most likely host.

After finding the most probable host galaxies for the candidate, we extract the redshift and uncertainties from the respective catalog. From this we derive a distance probability distribution for each host galaxy. The distance score ${\rm S_{dist}}$ is then the integrated joint probability of the host distance distribution and the GW distance distribution, derived along the line of sight to the transient (see \autoref{eq:3dscore}). Since we apply this approach to all host galaxies potentially associated with a candidate, and since the photometric redshifts of these galaxies may have unknown systematic uncertainties, we downweight the distance score by selecting the highest score among all potential host galaxies for this candidate. More details on the method used for this calculation are given in \autoref{app:3d}.

\subsection{Minor Planet and Point Source Catalog Comparison} \label{sec:mpc_score}
We perform a cone search on the ASAS-SN (All-Sky Automated Survey for Supernovae) Variable Star catalog X \citep{shappee_man_2014, christy_asas-sn_2023}, Gaia Data Release 3 Variable Star catalog \citep{gaia_collaboration_gaia_2023}, and Pan-STARRS (PS1) Point Source catalog \citep{Beck+21} for any targets within 2 arcseconds of the candidate. If a match is found then it indicates, \textit{e.g.}, a flaring star rather than a KN, and we assign a point source score of ${\rm S_{PS}} = 0$. Otherwise, if no match is found in these catalogs, we assign ${\rm S_{PS}} = 1$.

We also pull the file of asteroid ephemerides from the Minor Planet Center (MPC)\footnote{\url{https://www.minorplanetcenter.net/}}. Using simple 2 body evolution around the Sun we find all asteroids that are approximately within 10 degrees of the candidate location at the time of the first photometric detection. For any asteroid found in this initial search we perform a full $n$-body evolution to find the more precise location at the time of the first photometric detection of the candidate. This is implemented using the {\tt kete} Python package \citep{dahlen_kete_2025}. If a match is found within $25''$~of the candidate, it receives an MPC score of ${\rm S_{MPC}} = 0$, otherwise it receives a score ${\rm S_{MPC}} = 1$. ${\rm S_{MPC}}\times{\rm S_{PS}}$ is the final score for this portion of the analysis.

\subsection{Photometry Scoring} \label{sec:phot_score}
We score candidates with available photometry by comparing basic metrics from their light curve properties to both AT\,2017gfo and KN models (see below). Throughout the photometry scoring we choose a minimum score of 0.1. This, in effect, prevents a single photometric anomaly in the light curve from completely ruling out the candidate. This is a way to ``soften'' the effects of the assumption that all KNe look like the existing models. 

We first check for pre-detections in the ATLAS forced photometry for the $181$ days prior to GW event S250818k. We only consider $5\sigma$ detections with at least 2 other $5\sigma$ detections within the range $\pm 5$ days. If any pre-detections are discovered, we assign a score of ${\rm S_{pre}} = 0.1$, otherwise ${\rm S_{pre}} = 1$.

For candidates with more than one detection after the GW discovery date, we also compare the rise time, decay rate, and peak luminosity to KNe models. To extract these properties from the candidate light curves we first gather all optical photometry in the $g$, $r$, $i$, ATLAS-$c$, or ATLAS-$o$ filters\footnote{Since we are trying to derive the properties of the bulk evolution of the light curve it is safe to assume that these evolve relatively similarly.}. We then fit the light curve with both a power law and a broken power law. 
We only fit a broken power law to the data if more than 5 photometric detections are available; otherwise, the model does not have enough constraining power.

After fitting both models to the photometry, we choose the model with the lowest Akaike Information Criterion (AIC)\footnote{The AIC score takes the model complexity into account for model comparison to help ensure we are not overfitting. We also tried the Bayesian Information Criterion and it did not make a difference.}. From the light curve model we derive a decay rate and a rise time. KN models show that, in all cases, the light curve should decay at a rate faster than 0.1 mag~day$^{-1}$ \citep[\textit{e.g.}, ][]{kasen_kilonova_2015, kasen_origin_2017, kilpatrick_gravity_2021, rastinejad_systematic_2022}. Therefore, for any decay rate $< 0.1$ mag~day$^{-1}$ we assign a decay rate score of ${\rm S_{DR}} = 0.1$, otherwise ${\rm S_{DR}} = 1$. Similarly, no KN models have a rise time greater than 4 days long \citep[\textit{e.g.}, ][]{kasen_kilonova_2015, kasen_origin_2017, kilpatrick_gravity_2021, rastinejad_systematic_2022}. We therefore require the peak of the light curve to occur less than 4 days after the GW event. If the peak occurs after 4 days the candidate receives a score ${\rm S_{RT}} = 0.1$, otherwise ${\rm S_{RT}} = 1$. 

Finally, we also compute the maximum luminosity using the maximum observed flux in the light curve (not the maximum flux of the model) at the distance to the GW event along the line of sight to the candidate. Since the maximum mass of a neutron star sets an upper limit on the maximum outflow mass, we can conservatively assume that a KN should not have a luminosity $\nu L_\nu >10^{43}~{\rm erg~s}^{-1}$ \citep[\textit{e.g.}, ][]{kasen_kilonova_2015, kasen_origin_2017, kilpatrick_gravity_2021, rastinejad_systematic_2022}. Any candidate with $\nu L_\nu >10^{43}~{\rm erg~s}^{-1}$ receives a score ${\rm S_{L}} = 0.1$, otherwise ${\rm S_{L}} = 1$. To obtain a final photometry score, we multiply all of the previously discussed photometry subscores, to obtain ${\rm S_P} = {\rm S_{PD} \times S_{DR} \times S_{RT} \times S_{L}}$. 

\begin{deluxetable*}{lccp{1cm}p{1cm}p{1cm}p{1cm}p{2cm}p{1cm}p{1cm}p{1cm}}
    \tablecaption{Top Candidate Scores and SN\,2025ulz (Full Table in \autoref{app:candidates})\label{tab:candidates-summary}}
    \tablehead{
    \colhead{TNS Name} & \colhead{RA} & \colhead{Dec} & \colhead{Overall Score} & \colhead{2D Score} & \colhead{Point Source Score} & \colhead{Distance Score} & \colhead{Peak Luminosity} & \colhead{Time of Peak\tablenotemark{a}} & \colhead{Decay Rate\tablenotemark{a}} & \colhead{Pre-Detection Score} \\ 
    \colhead{} & \colhead{(deg)} & \colhead{(deg)} & \colhead{} & \colhead{} & \colhead{} & \colhead{} & \colhead{(erg~s$^{-1}$)} & \colhead{(days)} & \colhead{(mag~day$^{-1}$)} & \colhead{}
    }
    \startdata
    AT2025usl & 237.1445 & 32.2938 & 0.73 & 0.85 & 1.00 & 0.86 & $7.42 \times 10^{41}$ & --- & --- & 1 \\
    AT2025uuf & 310.1678 & 64.6115 & 0.51 & 0.59 & 1.00 & 0.87 & $5.11 \times 10^{41}$ & --- & --- & 1 \\
    AT2025uus & 244.2946 & 39.6485 & 0.41 & 0.97 & 1.00 & 0.42 & $9.88 \times 10^{41}$ & --- & --- & 1 \\
    AT2025uua & 261.4404 & 51.9287 & 0.39 & 0.39 & 1.00 & 1.00 & $8.90 \times 10^{41}$ & --- & --- & 1 \\
    AT2025uow & 53.3796 & -30.0963 & 0.38 & 0.45 & 1.00 & 0.87 & $2.38 \times 10^{42}$ & 0.05 & 0.13 & 1 \\
    AT2025usk & 241.2050 & 35.7908 & 0.35 & 0.99 & 1.00 & 0.35 & $1.04 \times 10^{42}$ & --- & --- & 1 \\
    AT2025uvu & 236.7145 & 29.9994 & 0.31 & 0.76 & 1.00 & 0.42 & $5.68 \times 10^{41}$ & --- & --- & 1 \\
    AT2025wfs & 236.5714 & 31.8389 & 0.30 & 0.83 & 1.00 & 0.36 & $3.44 \times 10^{41}$ & --- & --- & 1 \\
    AT2025uxu & 270.6914 & 56.4917 & 0.24 & 0.30 & 1.00 & 0.80 & $4.97 \times 10^{41}$ & 0.16 & 0.25 & 1 \\
    AT2025uut & 283.5697 & 60.0326 & 0.15 & 0.52 & 1.00 & 0.29 & $6.85 \times 10^{41}$ & --- & --- & 1 \\
AT2025utu & 301.9844 & 61.3465 & 0.15 & 0.25 & 1.00 & 0.63 & $6.34 \times 10^{41}$ & 0.37 & 0.33 & 1 \\
AT2025usn & 237.6131 & 30.3337 & 0.12 & 0.64 & 1.00 & 0.20 & $1.04 \times 10^{42}$ & --- & --- & 1 \\
AT2025uuc & 264.9940 & 53.3377 & 0.10 & 0.35 & 1.00 & 0.30 & $8.79 \times 10^{41}$ & --- & --- & 1 \\
AT2025uog & 58.3270 & -33.1844 & 0.09 & 0.52 & 1.00 & 0.18 & $1.60 \times 10^{42}$ & 0.05 & 0.20 & 1 \\
AT2025uul & 237.0139 & 29.7992 & 0.09 & 0.64 & 1.00 & 0.15 & $7.60 \times 10^{41}$ & --- & --- & 1 \\
AT2025uur & 236.7506 & 30.5273 & 0.09 & 0.81 & 1.00 & 0.12 & $6.46 \times 10^{41}$ & --- & --- & 1 \\
AT2025uzu & 244.2891 & 41.7935 & 0.09 & 0.53 & 1.00 & 0.17 & $7.50 \times 10^{41}$ & 0.14 & 0.12 & 1 \\
AT2025unm & 247.0348 & 42.0573 & 0.08 & 0.91 & 1.00 & 0.95 & $4.94 \times 10^{41}$ & 0.12 & 0.02 & 1 \\
    {$\vdots$} & & & & & {$\vdots$}& & & & & {$\vdots$} \\
    SN2025ulz & 237.9758 & 30.9024 & 0.00 & 0.66 & 1.00 & 0.68 & $2.12 \times 10^{42}$ & 26.76 & 3.07 & 1 \\
    \enddata
    \tablenotetext{a}{An empty row indicates that there was only one public photometry point, which is not enough to fit the light curve and compute this value. We therefore do not consider a score from this value when computing the total score.}
\end{deluxetable*}

\subsection{Application to S250818k}
In the case of the BNS GW event S250818k, we find 121 candidates with the scores summarized in \autoref{tab:candidates-summary} (and the full score information is in \autoref{tab:candidates}). The locations of these candidates with respect to the GW localization map is shown in \autoref{fig:skymap}. SN\,2025ulz receives a score of ${\rm S}\approx 0.0011$ as a result of both its distance (${\rm S_{dist}} \approx 0.16$) and photometric evolution (${\rm S_{phot}} \approx 0.01$). We discuss SN\,2025ulz in more detail in \autoref{sec:25ulz-score-time}. 

\begin{figure}
    \centering
    \includegraphics[width=\linewidth]{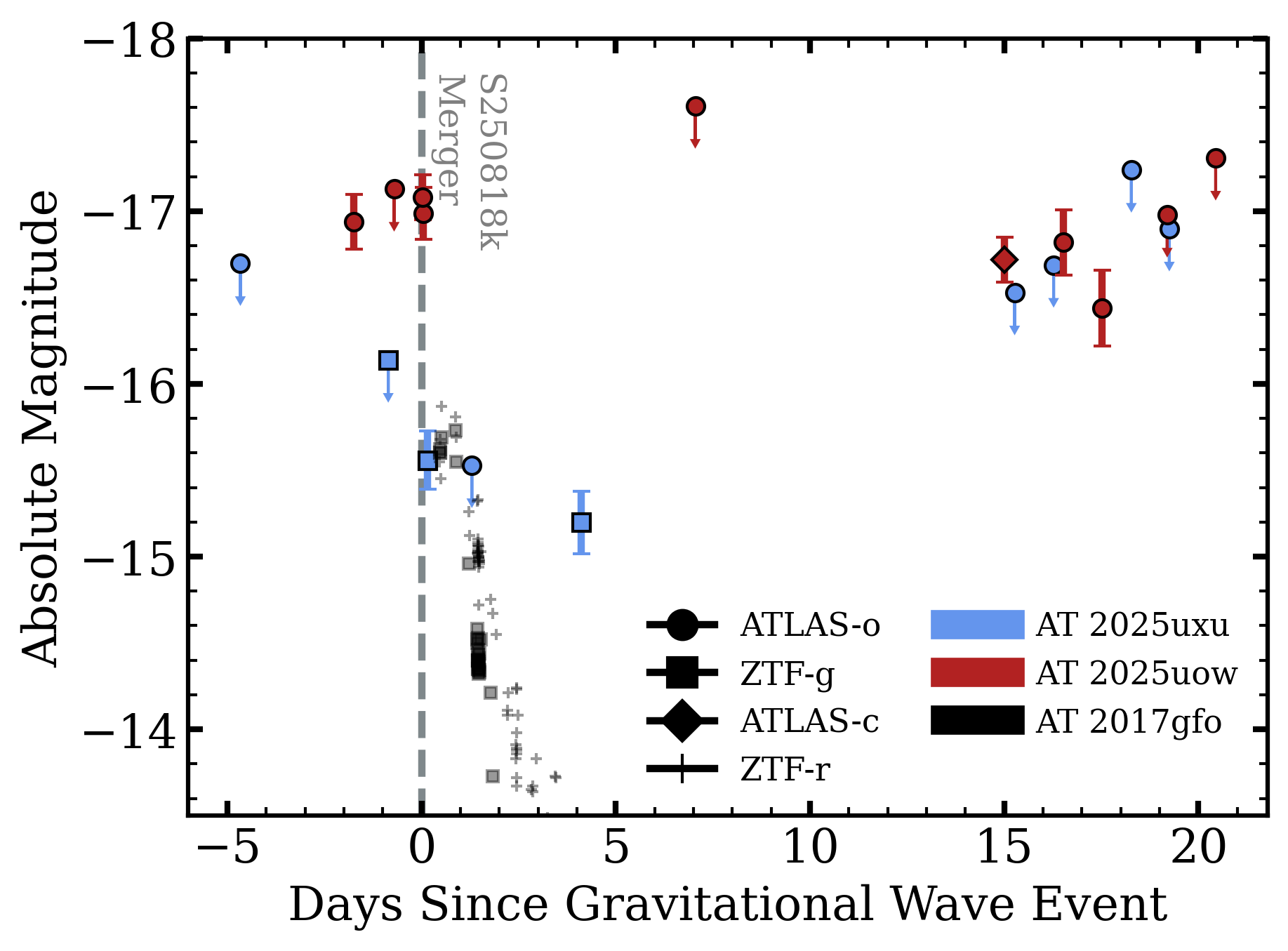}
    \caption{Light curves of the publicly available data from ATLAS and TNS on AT\,2025uxu (blue) and AT\,2025uow (red) as compared to AT\,2017gfo \citep[black;][]{villar_combined_2017}. We scale the observed magnitudes of AT\,2025uxu and AT\,2025uow to the mean distance in the corresponding HEALPix tile in the localization map of S250818k. The merger date associated with S250818k is shown as a grey dashed vertical line. Different markers indicate different telescope filters.}
    \label{fig:cand-lcs}
\end{figure}

We briefly discuss the other high scoring candidates given in \autoref{tab:candidates-summary}. $41/121\approx34\%$ of the candidates have a score $S\geq0.01$ \added{and,} of those 41, only 18 have more than one photometry point. We visually inspect those 18 transients to identify a subset of interest to discuss in more detail. In particular, we ignore candidates with poor host association (\textit{i.e.}, the host used for the distance score is clearly not related to the transient position, \textit{e.g.}, AT\,2025utu), the best matching host galaxy has a distance $>1000$ Mpc (\textit{e.g.}, AT\,2025uog)\footnote{The transients that fall into this category typically lie in a HEALPix tile with an unusually high distance uncertainty and have a photometric distance with an unusually high uncertainty. This combination results in the transient not being ruled out by its distance during the automated scoring.}, and/or those that have multiple photometry points but all in different filters (\textit{e.g.}, AT\,2025vag). We find two promising candidates, \added{AT\,2025uow \citep{2025TNSTR3286....1T} and AT\,2025uxu \citep{2025TNSTR3323....1S}}, but none with enough information to unambiguously prove whether or not they are KNe\added{, suggesting a more comprehensive follow-up strategy of transients found in the localization volume of a GW event would be beneficial for the community}. The light curves of both of these candidates are in \autoref{fig:cand-lcs}.

AT\,2025uow has a total score of 0.38, mostly due to its $S_{\rm 2D} = 0.45$. The light curve reveals a rapid decline at a rate of $0.13$ mag~day$^{-1}$, a red color (based off of only one epoch with an ATLAS c and o filter observation within 1 day of each other), and a peak luminosity $\nu L_\nu \approx 2\times10^{42}$ erg~s$^{-1}$. All three of these photometric properties are consistent with KNe models. However, there is one significant ATLAS predetection at $\delta t \approx -2$ days. Based on this predetection, the light curve appears to {\it peak}, rather than begin to rise, near the discovery of S250818k, making it unlikely that this is a KN associated with a BNS merger.

AT\,2025uxu has a score of 0.24, primarily because of its 2D localization score of 0.3, and only two photometric detections in the ZTF $g$-band following the discovery of S250818k. These two detections fade at a rate of $\sim 0.25$ mag~day$^{-1}$ and have a peak luminosity $\nu L_\nu \approx 5\times10^{41}$ erg~s$^{-1}$, both consistent with KNe models. However, we have no color information and the ATLAS upper limits from before the discovery of S250818k are non-constraining. Therefore, given this lack of information, we hesitate to draw any conclusions for this object.

\citet{gillanders_pan-starrs_2025} finds seven other candidates associated with S250818k, all of which they rule out. Based on our algorithm, six of these candidates, with the exception of AT\,2025uuf, receive a score $S\approx0$ (\autoref{tab:candidates}). AT\,2025uuf has the second highest score (${\rm S} = 0.51$) of all of our candidates. However, only the Pan-STARRS photometry reported to TNS is publicly available, and the ATLAS limits are non-constraining based on the Pan-STARRS detection. Therefore, we \added{defer to the conclusion of \citet{gillanders_pan-starrs_2025} that AT\,2025uuf is inconsistent with a KN}.  

\subsection{Temporal Evolution of the SN\,2025\lowercase{ulz} Score}\label{sec:25ulz-score-time}
\begin{figure*}
    \centering
    \includegraphics[width=\linewidth]{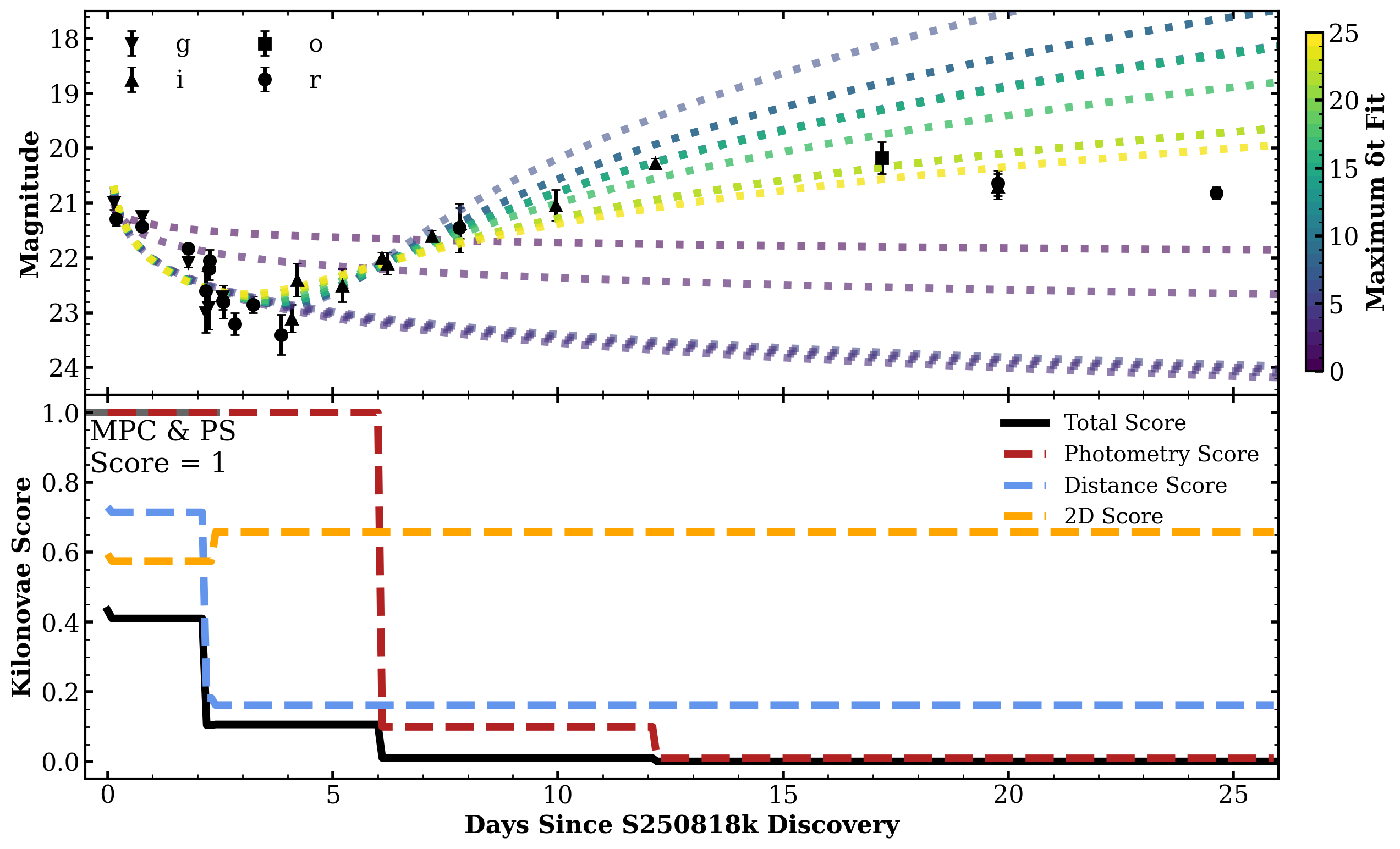}
    \caption{{\it Top: } The best fit light curve after including different ranges of data. The color of the line represents the maximum $\delta t$ days since discovery that was used when selecting data to fit. {\it Bottom:} The change in total score (red), photometry score (blue), distance score (black), and 2D score (orange) over time as we include additional information/measurements of SN\,2025ulz. The distance score drops after a spectroscopic redshift is measured at $\delta t \approx 2.6$ days. After the light curve turns over, the photometry score drops, since this turnover is inconsistent with KN models and the AT\,2017gfo light curve.}
    \label{fig:score-over-time}
\end{figure*}

The public GCNs for the follow-up of SN\,2025ulz provide a unique opportunity to apply this scoring algorithm as a function of time. In this way, we can demonstrate how the score changes as more information is found on the transient. To do this we account for:
\begin{enumerate}
    \item The release of new IGWN alerts with updated skymaps. This affects both the 2D score and distance score.
    \item The measurement of a spectroscopic redshift by \citet{2025GCN.41436....1K} of SN\,2025ulz at $\sim$2.6 days after the discovery of the GW event. This affects the distance score.
    \item The announcements of new photometric measurements of SN\,2025ulz. We do this by fitting the data binned in one day increments. Note that we do not include the Pan-STARRS light curve from \citet{gillanders_pan-starrs_2025} because it was not made public until after the time period considered here.
\end{enumerate}
The minor planet and point source scores are held constant at ${\rm S_{PS} = S_{MPC} = 1}$ since SN\,2025ulz is not found to spatially coincide with any known point sources or asteroids.

In the bottom panel of \autoref{fig:score-over-time} we show the evolution of the 2D, distance, photometry, and total score. In the top panel we show the evolution of the light curve models used for extracting the photometry score. The distance score is strongly affected by the measurement of a spectroscopic redshift because the scoring algorithm switches from using a photometric redshift with large, asymmetric uncertainties to a spectroscopic redshift with small uncertainties, which we set to $\delta z = 10^{-3}$. This, in effect, treats the spectroscopic redshift distance measurement as a delta function, lowering the distance score from ${\rm S_{dist}} = 0.71 \rightarrow 0.16$. 

The photometry score for SN\,2025ulz is initially very high with a score ${\rm S_P} = 1$. This is because the light curve of SN\,2025ulz initially fades quickly and peaks soon after the GW discovery date, as is expected for a KN. However, after the observations on day 6, the score drops to ${\rm S_P} = 0.1$ because, after we include the day 6 photometry, the AIC score for the broken power law drops below the single power law score, indicating a better fit. This indicates that the light curve has begun to rise again and the decay rate score drops to ${\rm S_{DR}} = 0.1$. At day 10 the photometry score drops again because the most recent photometry point is now more luminous than the first observation. This changes the time of peak to 10 days after the GW discovery, outside our conservative range of 0 to 4 days, and also drops the rise time score to ${\rm S_{\rm RT}}=0.1$.

\begin{figure*}
    \centering
    \includegraphics[width=\linewidth]{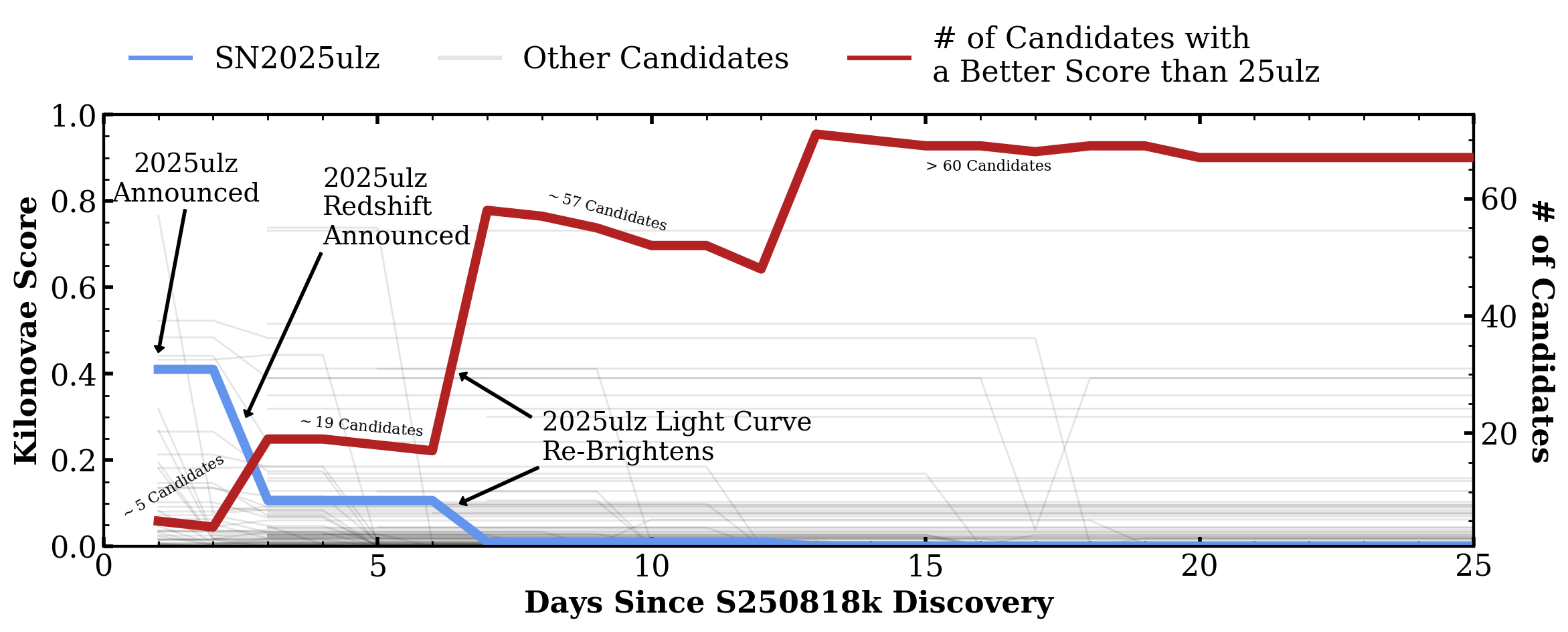}
    \caption{The evolution of the SN\,2025ulz score (blue) as compared to the evolution of the scores of the rest of our candidates (grey). The red line shows the number of candidates with a better score than SN\,2025ulz over time (right y-axis). The labels near the red line are the number of candidates with a score greater than SN\,2025ulz at that time.}
    \label{fig:all-cands-score-over-time}
\end{figure*}

In \autoref{fig:all-cands-score-over-time} we show the evolution of the SN\,2025ulz score compared to the other candidates, and we show the number of candidates with a score greater than SN\,2025ulz as a function of time. At the time SN\,2025ulz was announced, there were 5 transients discovered with a \added{marginally} higher score than SN\,2025ulz, \added{suggesting that additional candidates were also initially consistent with a KN and would have benefited from additional follow-up}. The SN\,2025ulz score drops as more observations are included, and thus more candidates have a higher score than SN\,2025ulz. This increase in the number of candidates with a higher score than SN\,2025ulz over time is likely caused by the lack of follow-up for these candidates, most of which are surely not KNe. With a more comprehensive follow-up strategy of multiple promising candidates the astronomical community could have continuously updated the list of promising KN candidates until each was ruled out or the true counterpart was found. Applying a similar algorithm, in real time, during future KN searches will \added{assist with the community coordination of a comprehensive follow-up strategy}.

\section{Summary \& Conclusions} \label{sec:conclusions}
On 2025-08-20 the LVK collaboration announced the discovery of S250818k, a GW event with a 29\% probability of being a BNS merger and a 71\% probability of being terrestrial. Soon after, the transient SN\,2025ulz was discovered at the 66\% localization contour of S250818k. This prompted follow-up observations across the electromagnetic spectrum to determine if SN\,2025ulz was consistent with KN emission following the potential BNS merger. 

In this paper, we present and analyze observations of  SN\,2025ulz spanning from the ultraviolet to radio wavelengths. We also compare these observations against a novel scoring algorithm to determine the likelihood that a specific transient is associated with a BNS merger GW signal. Based on these observations, we find SN\,2025ulz to be consistent with a SN\,IIb, and inconsistent with both KN models and the KN AT\,2017gfo. A summary of our findings are: 
\begin{itemize}
    \item In the first $\sim 5$ days after the discovery of S250818k, SN\,2025ulz faded rapidly and reddened as it faded in the optical (\autoref{fig:optical-lc}). This is both consistent with the broader population of SN\,IIb and reminiscent of AT\,2017gfo. However, starting $\sim 6$ days after the discovery of S250818k, the SN\,2025ulz optical light curve begins to re-brighten, eventually reaching a peak luminosity at $\gtrsim 25$ days. This optical re-brightening to a more luminous second peak is consistent with an SN\,IIb and inconsistent with both AT\,2017gfo and KN models.
    \item At $\delta t \sim 2.6$ days after discovery, we find that SN\,2025ulz has a combined score  of ${\rm S} \sim 0.45$ (\autoref{fig:score-over-time}). After the measurement of the spectroscopic redshift at $\delta t \approx 2.6$ days this score drops to ${\rm S} \sim 0.16$ because the spectroscopic redshift puts the distance $\sim1.7\sigma$ above the best-fit GW distance along that line of sight. 
    \item By applying our photometry scoring algorithm to SN\,2025ulz over time we find that the early ($\delta t < 5$ days) optical light curve evolution is consistent with KN models. But, similar to our qualitative discussion above, as soon as the optical light curve begins to re-brighten around $\delta t\sim 6$ days, it is no longer consistent with a KN, and the score drops (\autoref{fig:score-over-time}).
    \item Modeling of the initial decline in the SN\,2025ulz light curve shows that it is consistent with the shock cooling expected from a SN\,IIb (\autoref{fig:sw_rsg_bsg}). Additionally, the envelope mass and progenitor radius derived from this model are consistent with the rest of the SN\,IIb population, and most similar to SN\,2016gkg (\autoref{fig:menv_r_phase_space}).
    \item The optical spectrum of SN\,2025ulz (\autoref{fig:binospec}) has a broad P Cygni H$\alpha$ line with a velocity of $\sim 15{,}600$ km~s$^{-1}$. The strength and velocity of this H$\alpha$ line is consistent with both the SN\,IIb SN\,2022hnt and SN\,2016gkg at nearly the same phase. If SN\,2025ulz were a KN we would expect a featureless spectrum, like AT\,2017gfo, and a higher  outflow velocity than measured from the H$\alpha$ line.
    \item We find no evidence for transient radio emission originating from SN\,2025ulz (\autoref{fig:radio-lc}). Our radio limits rule out an on-axis jet, like those seen in SGRBs, as would be expected for a face-on KN. Our radio observations do not rule out an off-axis KN like AT\,2017gfo or an SN\,IIb like SN\,2016gkg.
    \item We derive a SFR from the host H$\alpha$ emission and diffuse radio emission from the MeerKAT observations. Both are consistent with ${\rm SFR} \sim 1~{\rm M_\odot~yr}^{-1}$, indicating that the host galaxy is actively star forming. This finding is consistent with the core collapse SN interpretation of SN\,2025ulz.
\end{itemize}

When this abundance of evidence is considered in its entirety, we find that an SN\,IIb is the most favorable interpretation of SN\,2025ulz. Given this, we also apply our scoring algorithm to the 120 other transients reported to the TNS that are within the 95\% region of the S250818k contour map. By doing this KN search, we find (1) $4-5$ candidates more promising than SN\,2025ulz at $\delta t < 2.6$ days (when the redshift was measured); (2) $19$ candidates more promising than SN\,2025ulz at $2.6 < \delta t \lesssim 6$ days; and (3) $\gtrsim 50$ candidates more promising than SN\,2025ulz at $\delta t \gtrsim 6$ days (\autoref{fig:all-cands-score-over-time}). After further inspection, we find that none of these potential transients appear to be an unambiguous electromagnetic counterpart associated with S250818k --- although some simply do not have enough {\it publicly available} photometry to draw any conclusions. In the future, a coordinated follow-up observation strategy that continues to observe all promising candidates will help shield our KNe searches against impostor SNe\,IIb.

With this in mind, we are building a publicly accessible website for automatically vetting candidate counterparts to poorly localized events (\textit{e.g.}, GWs, neutrinos, poorly localized GRBs, etc.) using the quantitative scoring methodology introduced in this work. This Tool for Rapid Object Vetting and Examination (TROVE) (\url{https://astro-trove.github.io/}), will include both a web application and application programming interface (API) for scoring possible counterparts to poorly localized events. TROVE is under active development and will be made publicly available in the next $\sim$year. In the future, the TROVE candidate vetting will assist with community coordination of follow-up of poorly localized events, working to prevent future contaminating transients from dominating the KN search. 

\begin{acknowledgments}
Time-domain research by the University of Arizona team and D.J.S. is supported by National Science Foundation (NSF) grants 2108032, 2308181, 2407566, and 2432036 and the Heising-Simons Foundation under grant \#2020-1864. N.F. acknowledges support from the National Science Foundation Graduate Research Fellowship Program under Grant No. DGE-2137419.  C.D.K. gratefully acknowledges support from the NSF through AST-2432037, the HST Guest Observer Program through HST-SNAP-17070 and HST-GO-17706, and from JWST Archival Research through JWST-AR-6241 and JWST-AR-5441. KDA and CTC acknowledge support provided by the NSF through award AST-2307668. KDA gratefully acknowledges support from the Alfred P. Sloan Foundation. J.R. was supported by NASA through the NASA Hubble Fellowship grant \#HST-HF2-51587.001-A awarded by the Space Telescope Science Institute, which is operated by the Association of Universities for Research in Astronomy, Inc., for NASA, under contract NAS5-26555.
KAB is supported by an LSST-DA Catalyst Fellowship; this publication was thus made possible through the support of Grant 62192 from the John Templeton Foundation to LSST-DA. The Berger Time-Domain group at Harvard is supported by NSF and NASA, including AST-2206110.

The work of JEA is supported by NOIRLab, which is managed by the Association of Universities for Research in Astronomy (AURA) under a cooperative agreement with the U.S. National Science Foundation. DLC acknowledges support from the Science and Technology Facilities Council (STFC) grant number ST/X001121/1. KJD acknowledges support from NSF Grant PHY-2308986 and the Heising Simons Foundation grant \# 2022-3927. She also respectfully acknowledges that the University of Arizona is home to the O'odham and the Yaqui. She respects and honors the ancestral caretakers of the land, from time immemorial until now, and into the future. This work is supported by the National Science Foundation under Cooperative Agreement PHY-2019786 (The NSF AI Institute for Artificial Intelligence and Fundamental Interactions, http://iaifi.org/). D.H. is supported by NASA grant HST-GO-17770.002-A. R.M. acknowledges support by the National Science Foundation under award No. AST-2224255. VP is in part supported by NASA grant 80NSSC24K0771 and NSF grant PHY-2145421 to the University of Arizona. M.R. acknowledges support from NASA (ATP: 80NSSC24K0932). N.V. acknowledges funding from the National Sciences and Engineering Research Council of Canada (NSERC) Postdoctoral Fellowship 599555.

Some observations reported here were obtained at the MMT Observatory, a joint facility of the University of Arizona and the Smithsonian Institution. We thank the Harvard and Northwestern teams for coordinating a joint MMT/Binospec observation with the University of Arizona.

Some of the data presented herein were obtained at Keck Observatory, which is a private 501(c)3 non-profit organization operated as a scientific partnership among the California Institute of Technology, the University of California, and the National Aeronautics and Space Administration. The Observatory was made possible by the generous financial support of the W. M. Keck Foundation. The authors wish to recognize and acknowledge the very significant cultural role and reverence that the summit of Maunakea has always had within the Native Hawaiian community. We are most fortunate to have the opportunity to conduct observations from this mountain.

This research is based on observations made with the NASA/ESA Hubble Space Telescope obtained from the Space Telescope Science Institute, which is operated by the Association of Universities for Research in Astronomy, Inc., under NASA contract NAS 5–26555. These observations are associated with program(s) GO-17450 and GO-17805 (PI: Troja).

The National Radio Astronomy Observatory (NRAO) is a facility of the National Science Foundation operated under cooperative agreement by Associated Universities, Inc. GMRT observations were obtained for this study. We thank the staff of the GMRT who have made these observations possible. This work is based on observations carried out under project number S25CS with the IRAM NOEMA Interferometer. IRAM is supported by INSU/CNRS (France), MPG (Germany), and IGN (Spain). The Submillimeter Array is a joint project between the Smithsonian Astrophysical Observatory and the Academia Sinica Institute of Astronomy and Astrophysics and is funded by the Smithsonian Institution and the Academia Sinica. We recognize that Maunakea is a culturally important site for the indigenous Hawaiian people; we are privileged to study the cosmos from its summit.

\end{acknowledgments}

\facilities{HST, Keck:I (LRIS), MMT (Binospec), T80, GMRT, MeerKAT, VLA, NOEMA, SMA}

\software{
Astropy \citep{astropy1, astropy2, astropy3},
\texttt{astropy-healpix} \citep{astropy3},
Astroquery \citep{ginsburg_astroquery_2019},
Astro-SCRAPPY \citep{mccully_astropy_2018},
Beautiful Soup \citep{richardson_beautiful_2023},
Binospec IDL Pipeline \citep{Kansky2019},
CASA \citep{mcmullin_casa_2007, casa_team_casa_2022},
\texttt{crispy-bootstrap4} \citep{smith_crispy-bootstrap4_2022},
\texttt{dateutil} \citep{niemeyer_dateutil_2021},
Django \citep{django_software_foundation_django_2023},
\texttt{django-bootstrap4} \citep{verheul_django-bootstrap4_2021},
\texttt{django-crispy-forms} \citep{araujo_django-crispy-forms_2023},
Django Extensions \citep{trier_django_2023},
Django Filter \citep{gibson_django_2021},
\texttt{django-gravatar} \citep{waddington_django-gravatar_2020},
\texttt{django-guardian} \citep{balcerzak_django-guardian_2021},
Django REST Framework \citep{christie_django_2022},
\texttt{django-webpack-loader} \citep{lone_django-webpack-loader_2022},
\texttt{dustmaps} \citet{2018JOSS....3..695M},
\texttt{fastavro} \citep{tebeka_fastavro_2022},
Flask \citep{pallets_projects_flask_2022},
Flask-SQLAlchemy \citep{pallets_projects_flask-sqlalchemy_2021},
\texttt{fundamentals} \citep{young_fundamentals_2023},
\texttt{gracedb-sdk} \citep{singer_gracedb-sdk_2022},
HEALPix Alchemy \citep{singer_healpix_2022},
\texttt{healpy} \citep{zonca_healpy_2019},
Hop Client \citep{godwin_hop_2022},
Hopskotch \citep{scimma_project_hopskotch_2023},
IRAF \citep{Tody1986, Tody1993},
Light Curve Fitting \citep{hosseinzadeh_light_2022},
\texttt{ligo.skymap} \citep{singer_rapid_2016},
\texttt{lmfit} \citep{newville_lmfit_2023},
Apache Kafka \citep{apache_software_foundation_apache_2023},
Matplotlib \citep{hunter_matplotlib:_2007},
MOCPy \citep{baumann_cds-astro_2023},
NumPy \citep{harris_array_2020},
Paramiko \citep{forcier_paramiko_2023},
Photutils \citep{bradley_astropy_2022},
Pillow \citep{murray_python-pillow_2023},
\texttt{plotly.py} \citep{plotly_plotly_2023},
PostgreSQL \citep{postgresql_global_development_group_postgresql_2022},
Psycopg \citep{di_gregorio_psycopg_2023},
Python-Markdown \citep{stienstra_python-markdown_2023},
Q3C \citep{koposov_q3c_2006},
Requests \citep{reitz_requests_2023},
SAGUARO TOM \citep{hosseinzadeh_saguaro_2023},
SASSy Q3C Models \citep{daly_sassy_2023},
SciPy \citep{virtanen_scipy_2020},
\texttt{setuptools} \citep{python_packaging_authority_setuptools_2023},
\texttt{sip\_tpv} \citep{shupe_more_2012},
Source Extractor \citep{bertin_sextractor:_1996,bertin_sextractor_2010},
\texttt{specutils} \citep{earl_astropy_2023},
SQLAlchemy \citep{bayer_sqlalchemy_2012},
SWarp \citep{swarp},
TOM Toolkit \citep{collom_tom_2020,lindstrom_tom_2022},
\texttt{tom-alertstreams} \citep{tom_toolkit_project_tom-alertstreams_2023},
\texttt{tom\_nonlocalizedevents} \citep{tom_toolkit_project_tom_nonlocalizedevents_2023},
\texttt{urllib3} \citep{petrov_urllib3_2023},
\texttt{voevent-parse} \citep{staley_voevent-parse_2014},
Watchdog \citep{mangalapilly_watchdog_2023}
}

\bibliography{main}{}

\begin{thebibliography}{}
\expandafter\ifx\csname natexlab\endcsname\relax\def\natexlab#1{#1}\fi
\providecommand{\url}[1]{\href{#1}{#1}}
\providecommand{\dodoi}[1]{doi:~\href{http://doi.org/#1}{\nolinkurl{#1}}}
\providecommand{\doeprint}[1]{\href{http://ascl.net/#1}{\nolinkurl{http://ascl.net/#1}}}
\providecommand{\doarXiv}[1]{\href{https://arxiv.org/abs/#1}{\nolinkurl{https://arxiv.org/abs/#1}}}

\bibitem[{N. A.~J. {et~al.}(2022)A.~J., Chandra, Krishna, \&
  Anupama}]{a_j_radio_2022}
A.~J., N., Chandra, P., Krishna, A., \& Anupama, G.~C. 2022,
  \bibinfo{title}{Radio {Evolution} of a {Type} {IIb} {Supernova} {SN}
  2016gkg,} The Astrophysical Journal, 934, 186,
  \dodoi{10.3847/1538-4357/ac7c1e}

\bibitem[{K. Ackley(2025)Ackley}]{ackley_localization_2025}
Ackley, K. 2025, From {Localization} to {Discovery}: {Bayesian} {Ranking} of
  {Electromagnetic} {Counterparts} to {Gravitational}-{Wave} {Events}, arXiv,
  \dodoi{10.48550/arXiv.2510.15836}

\bibitem[{K. {Ackley} {et~al.}(2020){Ackley}, {Amati}, {Barbieri}, {Bauer},
  {Benetti}, {Bernardini}, {Bhirombhakdi}, {Botticella}, {Branchesi},
  {Brocato}, {Bruun}, {Bulla}, {Campana}, {Cappellaro}, {Castro-Tirado},
  {Chambers}, {Chaty}, {Chen}, {Ciolfi}, {Coleiro}, {Copperwheat}, {Covino},
  {Cutter}, {D'Ammando}, {D'Avanzo}, {De Cesare}, {D'Elia}, {Della Valle},
  {Denneau}, {De Pasquale}, {Dhillon}, {Dyer}, {Elias-Rosa}, {Evans},
  {Eyles-Ferris}, {Fiore}, {Fraser}, {Fruchter}, {Fynbo}, {Galbany}, {Gall},
  {Galloway}, {Getman}, {Ghirlanda}, {Gillanders}, {Gomboc}, {Gompertz},
  {Gonz{\'a}lez-Fern{\'a}ndez}, {Gonz{\'a}lez-Gait{\'a}n}, {Grado}, {Greco},
  {Gromadzki}, {Groot}, {Guti{\'e}rrez}, {Heikkil{\"a}}, {Heintz}, {Hjorth},
  {Hu}, {Huber}, {Inserra}, {Izzo}, {Japelj}, {Jerkstrand}, {Jin}, {Jonker},
  {Kankare}, {Kann}, {Kennedy}, {Kim}, {Klose}, {Kool}, {Kotak},
  {Kuncarayakti}, {Lamb}, {Leloudas}, {Levan}, {Longo}, {Lowe}, {Lyman},
  {Magnier}, {Maguire}, {Maiorano}, {Mandel}, {Mapelli}, {Mattila}, {McBrien},
  {Melandri}, {Micha{\l}owski}, {Milvang-Jensen}, {Moran}, {Nicastro},
  {Nicholl}, {Nicuesa Guelbenzu}, {Nuttal}, {Oates}, {O'Brien}, {Onori},
  {Palazzi}, {Patricelli}, {Perego}, {Torres}, {Perley}, {Pian}, {Pignata},
  {Piranomonte}, {Poshyachinda}, {Possenti}, {Pumo}, {Quirola-V{\'a}squez},
  {Ragosta}, {Ramsay}, {Rau}, {Rest}, {Reynolds}, {Rosetti}, {Rossi},
  {Rosswog}, {Sabha}, {Sagu{\'e}s Carracedo}, {Salafia}, {Salmon},
  {Salvaterra}, {Savaglio}, {Sbordone}, {Schady}, {Schipani}, {Schultz},
  {Schweyer}, {Smartt}, {Smith}, {Smith}, {Sollerman}, {Srivastav}, {Stanway},
  {Starling}, {Steeghs}, {Stratta}, {Stubbs}, {Tanvir}, {Testa}, {Thrane},
  {Tonry}, {Turatto}, {Ulaczyk}, {van der Horst}, {Vergani}, {Walton},
  {Watson}, {Wiersema}, {Wiik}, {Wyrzykowski}, {Yang}, {Yi}, \&
  {Young}}]{Ackley+20}
{Ackley}, K., {Amati}, L., {Barbieri}, C., {et~al.} 2020,
  \bibinfo{title}{{Observational constraints on the optical and near-infrared
  emission from the neutron star-black hole binary merger candidate
  S190814bv},} \aap, 643, A113, \dodoi{10.1051/0004-6361/202037669}

\bibitem[{T. Ahumada {et~al.}(2024)Ahumada, Anand, Coughlin, Gupta, Kasliwal,
  Karambelkar, Stein, Waratkar, Swain, Jegou~du Laz, Anumarlapudi, Andreoni,
  Bulla, Srinivasaragavan, Toivonen, Wold, Bellm, Cenko, Kaplan, Sollerman,
  Bhalerao, Perley, Salgundi, Suresh, Hinds, Reusch, Necker, Cook, Pletskova,
  Singer, Banerjee, Barna, Copperwheat, Healy, Kiendrebeogo, Kumar, Kumar,
  Pezzella, Sagués-Carracedo, Sravan, Bloom, Chen, Graham, Helou, Laher,
  Mahabal, Purdum, Anupama, Barway, Basu, Raman, \&
  Roychowdhury}]{ahumada_searching_2024}
Ahumada, T., Anand, S., Coughlin, M.~W., {et~al.} 2024,
  \bibinfo{title}{Searching for {Gravitational} {Wave} {Optical} {Counterparts}
  with the {Zwicky} {Transient} {Facility}: {Summary} of {O4a},} Publications
  of the Astronomical Society of the Pacific, 136, 114201,
  \dodoi{10.1088/1538-3873/ad8265}

\bibitem[{T. Ahumada {et~al.}(2025)Ahumada, Anand, Bulla, Gupta, Kasliwal,
  Stein, Karambelkar, Bellm, Jegou~du Laz, Coughlin, Andreoni, Banerjee,
  Bochenek, Hinds, Hu, Palmese, Perley, Pletskova, Salgundi, Singh, Sollerman,
  Swain, Wold, Bhalerao, Cenko, Cook, Copperwheat, Graham, Kaplan, Singer,
  Sravan, Busmann, Gassert, Gruen, Sommer, Zhang, Amsellem, Cabrera, Hall,
  Kunnumkai, O'Connor, Barna, Fontinele~Nunes, Toivonen, Sasli, Masci, Chen,
  Dekany, Purdum, Le-Calloch, Anupama, \& Barway}]{ahumada_ligovirgokagra_2025}
Ahumada, T., Anand, S., Bulla, M., {et~al.} 2025, {LIGO}/{Virgo}/{KAGRA}
  neutron star merger candidate {S250206dm}: {Zwicky} {Transient} {Facility}
  observations, arXiv, \dodoi{10.48550/arXiv.2507.00357}

\bibitem[{S. {Alam} {et~al.}(2015){Alam}, {Albareti}, {Allende Prieto},
  {Anders}, {Anderson}, {Anderton}, {Andrews}, {Armengaud}, {Aubourg},
  {Bailey}, {Basu}, {Bautista}, {Beaton}, {Beers}, {Bender}, {Berlind},
  {Beutler}, {Bhardwaj}, {Bird}, {Bizyaev}, {Blake}, {Blanton}, {Blomqvist},
  {Bochanski}, {Bolton}, {Bovy}, {Shelden Bradley}, {Brandt}, {Brauer},
  {Brinkmann}, {Brown}, {Brownstein}, {Burden}, {Burtin}, {Busca}, {Cai},
  {Capozzi}, {Carnero Rosell}, {Carr}, {Carrera}, {Chambers}, {Chaplin},
  {Chen}, {Chiappini}, {Chojnowski}, {Chuang}, {Clerc}, {Comparat}, {Covey},
  {Croft}, {Cuesta}, {Cunha}, {da Costa}, {Da Rio}, {Davenport}, {Dawson}, {De
  Lee}, {Delubac}, {Deshpande}, {Dhital}, {Dutra-Ferreira}, {Dwelly}, {Ealet},
  {Ebelke}, {Edmondson}, {Eisenstein}, {Ellsworth}, {Elsworth}, {Epstein},
  {Eracleous}, {Escoffier}, {Esposito}, {Evans}, {Fan}, {Fern{\'a}ndez-Alvar},
  {Feuillet}, {Filiz Ak}, {Finley}, {Finoguenov}, {Flaherty}, {Fleming},
  {Font-Ribera}, {Foster}, {Frinchaboy}, {Galbraith-Frew}, {Garc{\'\i}a},
  {Garc{\'\i}a-Hern{\'a}ndez}, {Garc{\'\i}a P{\'e}rez}, {Gaulme}, {Ge},
  {G{\'e}nova-Santos}, {Georgakakis}, {Ghezzi}, {Gillespie}, {Girardi},
  {Goddard}, {Gontcho}, {Gonz{\'a}lez Hern{\'a}ndez}, {Grebel}, {Green},
  {Grieb}, {Grieves}, {Gunn}, {Guo}, {Harding}, {Hasselquist}, {Hawley},
  {Hayden}, {Hearty}, {Hekker}, {Ho}, {Hogg}, {Holley-Bockelmann}, {Holtzman},
  {Honscheid}, {Huber}, {Huehnerhoff}, {Ivans}, {Jiang}, {Johnson},
  {Kinemuchi}, {Kirkby}, {Kitaura}, {Klaene}, {Knapp}, {Kneib}, {Koenig},
  {Lam}, {Lan}, {Lang}, {Laurent}, {Le Goff}, {Leauthaud}, {Lee}, {Lee},
  {Licquia}, {Liu}, {Long}, {L{\'o}pez-Corredoira}, {Lorenzo-Oliveira},
  {Lucatello}, {Lundgren}, {Lupton}, {Mack}, {Mahadevan}, {Maia}, {Majewski},
  {Malanushenko}, {Malanushenko}, {Manchado}, {Manera}, {Mao}, {Maraston},
  {Marchwinski}, {Margala}, {Martell}, {Martig}, {Masters}, {Mathur},
  {McBride}, {McGehee}, {McGreer}, {McMahon}, {M{\'e}nard}, {Menzel},
  {Merloni}, {M{\'e}sz{\'a}ros}, {Miller}, {Miralda-Escud{\'e}}, {Miyatake},
  {Montero-Dorta}, {More}, {Morganson}, {Morice-Atkinson}, {Morrison},
  {Mosser}, {Muna}, {Myers}, {Nandra}, {Newman}, {Neyrinck}, {Nguyen},
  {Nichol}, {Nidever}, {Noterdaeme}, {Nuza}, {O'Connell}, {O'Connell},
  {O'Connell}, {Ogando}, {Olmstead}, {Oravetz}, {Oravetz}, {Osumi}, {Owen},
  {Padgett}, {Padmanabhan}, {Paegert}, {Palanque-Delabrouille}, {Pan},
  {Parejko}, {P{\^a}ris}, {Park}, {Pattarakijwanich}, {Pellejero-Ibanez},
  {Pepper}, {Percival}, {P{\'e}rez-Fournon}, {P{\'e}rez-R{\`a}fols},
  {Petitjean}, {Pieri}, {Pinsonneault}, {Porto de Mello}, {Prada}, {Prakash},
  {Price-Whelan}, {Protopapas}, {Raddick}, {Rahman}, {Reid}, {Rich}, {Rix},
  {Robin}, {Rockosi}, {Rodrigues}, {Rodr{\'\i}guez-Torres}, {Roe}, {Ross},
  {Ross}, {Rossi}, {Ruan}, {Rubi{\~n}o-Mart{\'\i}n}, {Rykoff},
  {Salazar-Albornoz}, {Salvato}, {Samushia}, {S{\'a}nchez}, {Santiago},
  {Sayres}, {Schiavon}, {Schlegel}, {Schmidt}, {Schneider}, {Schultheis},
  {Schwope}, {Sc{\'o}ccola}, {Scott}, {Sellgren}, {Seo}, {Serenelli}, {Shane},
  {Shen}, {Shetrone}, {Shu}, {Silva Aguirre}, {Sivarani}, {Skrutskie},
  {Slosar}, {Smith}, {Sobreira}, {Souto}, {Stassun}, {Steinmetz}, {Stello},
  {Strauss}, {Streblyanska}, {Suzuki}, {Swanson}, {Tan}, {Tayar}, {Terrien},
  {Thakar}, {Thomas}, {Thomas}, {Thompson}, {Tinker}, {Tojeiro}, {Troup},
  {Vargas-Maga{\~n}a}, {Vazquez}, {Verde}, {Viel}, {Vogt}, {Wake}, {Wang},
  {Weaver}, {Weinberg}, {Weiner}, {White}, {Wilson}, {Wisniewski},
  {Wood-Vasey}, {Ye`che}, {York}, {Zakamska}, {Zamora}, {Zasowski}, {Zehavi},
  {Zhao}, {Zheng}, {Zhou}, {Zhou}, {Zou}, \& {Zhu}}]{Alam+15}
{Alam}, S., {Albareti}, F.~D., {Allende Prieto}, C., {et~al.} 2015,
  \bibinfo{title}{{The Eleventh and Twelfth Data Releases of the Sloan Digital
  Sky Survey: Final Data from SDSS-III},} \apjs, 219, 12,
  \dodoi{10.1088/0067-0049/219/1/12}

\bibitem[{K.~D. Alexander {et~al.}(2017)Alexander, Berger, Fong, Williams,
  Guidorzi, Margutti, Metzger, Annis, Blanchard, Brout, Brown, Chen, Chornock,
  Cowperthwaite, Drout, Eftekhari, Frieman, Holz, Nicholl, Rest, Sako,
  Soares-Santos, \& Villar}]{alexander_electromagnetic_2017}
Alexander, K.~D., Berger, E., Fong, W., {et~al.} 2017, \bibinfo{title}{The
  {Electromagnetic} {Counterpart} of the {Binary} {Neutron} {Star} {Merger}
  {LIGO}/{Virgo} {GW170817}. {VI}. {Radio} {Constraints} on a {Relativistic}
  {Jet} and {Predictions} for {Late}-time {Emission} from the {Kilonova}
  {Ejecta},} The Astrophysical Journal Letters, 848, L21,
  \dodoi{10.3847/2041-8213/aa905d}

\bibitem[{K.~D. {Alexander} {et~al.}(2017){Alexander}, {Berger}, {Fong},
  {Williams}, {Guidorzi}, {Margutti}, {Metzger}, {Annis}, {Blanchard}, {Brout},
  {Brown}, {Chen}, {Chornock}, {Cowperthwaite}, {Drout}, {Eftekhari},
  {Frieman}, {Holz}, {Nicholl}, {Rest}, {Sako}, {Soares-Santos}, \&
  {Villar}}]{2017ApJ...848L..21A}
{Alexander}, K.~D., {Berger}, E., {Fong}, W., {et~al.} 2017,
  \bibinfo{title}{{The Electromagnetic Counterpart of the Binary Neutron Star
  Merger LIGO/Virgo GW170817. VI. Radio Constraints on a Relativistic Jet and
  Predictions for Late-time Emission from the Kilonova Ejecta},} \apjl, 848,
  L21, \dodoi{10.3847/2041-8213/aa905d}

\bibitem[{K.~D. {Alexander} {et~al.}(2018){Alexander}, {Margutti}, {Blanchard},
  {Fong}, {Berger}, {Hajela}, {Eftekhari}, {Chornock}, {Cowperthwaite},
  {Giannios}, {Guidorzi}, {Kathirgamaraju}, {MacFadyen}, {Metzger}, {Nicholl},
  {Sironi}, {Villar}, {Williams}, {Xie}, \& {Zrake}}]{2018ApJ...863L..18A}
{Alexander}, K.~D., {Margutti}, R., {Blanchard}, P.~K., {et~al.} 2018,
  \bibinfo{title}{{A Decline in the X-Ray through Radio Emission from GW170817
  Continues to Support an Off-axis Structured Jet},} \apjl, 863, L18,
  \dodoi{10.3847/2041-8213/aad637}

\bibitem[{K.~D. {Alexander} {et~al.}(2021){Alexander}, {Schroeder}, {Paterson},
  {Fong}, {Cowperthwaite}, {Gomez}, {Margalit}, {Margutti}, {Berger},
  {Blanchard}, {Chornock}, {Eftekhari}, {Laskar}, {Metzger}, {Nicholl},
  {Villar}, \& {Williams}}]{Alexander+21}
{Alexander}, K.~D., {Schroeder}, G., {Paterson}, K., {et~al.} 2021,
  \bibinfo{title}{{A Late-time Galaxy-targeted Search for the Radio Counterpart
  of GW190814},} \apj, 923, 66, \dodoi{10.3847/1538-4357/ac281a}

\bibitem[{F. {An} {et~al.}(2024){An}, {Vaccari}, {Best}, {Ocran},
  {Ishwara-Chandra}, {Taylor}, {Leslie}, {R{\"o}ttgering}, {Kondapally},
  {Haskell}, {Collier}, \& {Bonato}}]{2024MNRAS.528.5346A}
{An}, F., {Vaccari}, M., {Best}, P.~N., {et~al.} 2024, \bibinfo{title}{{Radio
  spectral properties of star-forming galaxies between 150 and 5000 MHz in the
  ELAIS-N1 field},} \mnras, 528, 5346, \dodoi{10.1093/mnras/stae364}

\bibitem[{J. {An} {et~al.}(2025{\natexlab{a}}){An}, {Liu}, {Zhu}, {Jiang},
  {He}, {Xu}, {Fu}, \& {Liu}}]{2025GCN.41503....1A}
{An}, J., {Liu}, X., {Zhu}, Z.~P., {et~al.} 2025{\natexlab{a}},
  \bibinfo{title}{{LIGO/Virgo/KAGRA S250818k: JinShan optical observations of
  AT2025ulz},} GRB Coordinates Network, 41503, 1

\bibitem[{J. {An} {et~al.}(2025{\natexlab{b}}){An}, {Malesani}, {Xu}, {Liu},
  {Zhu}, {Schneider}, {De Pasquale}, {D'Elia}, {Thakur}, {Pieterse}, {Levan},
  {Tanvir}, \& {Stargate Collaboration}}]{GCN40966}
{An}, J., {Malesani}, D.~B., {Xu}, D., {et~al.} 2025{\natexlab{b}},
  \bibinfo{title}{{EP250704a / GRB 250704B: VLT/X-shooter spectroscopic
  redshift z = 0.661},} GRB Coordinates Network, 40966, 1

\bibitem[{S. {Anand} {et~al.}(2021){Anand}, {Coughlin}, {Kasliwal}, {Bulla},
  {Ahumada}, {Sagu{\'e}s Carracedo}, {Almualla}, {Andreoni}, {Stein},
  {Foucart}, {Singer}, {Sollerman}, {Bellm}, {Bolin}, {Caballero-Garc{\'\i}a},
  {Castro-Tirado}, {Cenko}, {De}, {Dekany}, {Duev}, {Feeney}, {Fremling},
  {Goldstein}, {Golkhou}, {Graham}, {Guessoum}, {Hankins}, {Hu}, {Kong},
  {Kool}, {Kulkarni}, {Kumar}, {Laher}, {Masci}, {Mr{\'o}z}, {Nissanke},
  {Porter}, {Reusch}, {Riddle}, {Rosnet}, {Rusholme}, {Serabyn},
  {S{\'a}nchez-Ram{\'\i}rez}, {Rigault}, {Shupe}, {Smith}, {Soumagnac},
  {Walters}, \& {Valeev}}]{Anand+21}
{Anand}, S., {Coughlin}, M.~W., {Kasliwal}, M.~M., {et~al.} 2021,
  \bibinfo{title}{{Optical follow-up of the neutron star-black hole mergers
  S200105ae and S200115j},} Nature Astronomy, 5, 46,
  \dodoi{10.1038/s41550-020-1183-3}

\bibitem[{G.~E. {Anderson} {et~al.}(2024){Anderson}, {Belkin}, {Leung}, {van
  der Horst}, {Rhodes}, {Gulati}, {Chastain}, {Gompertz}, \& {ATCA PanRadio GRB
  Collaboration}}]{GCN38189}
{Anderson}, G.~E., {Belkin}, S., {Leung}, J.~K., {et~al.} 2024,
  \bibinfo{title}{{GRB 241105A: ATCA Radio Upper Limits},} GRB Coordinates
  Network, 38189, 1

\bibitem[{I. {Andreoni} {et~al.}(2019){Andreoni}, {Goldstein}, {Anand},
  {Coughlin}, {Singer}, {Ahumada}, {Medford}, {Kool}, {Webb}, {Bulla}, {Bloom},
  {Kasliwal}, {Nugent}, {Bagdasaryan}, {Barnes}, {Cook}, {Cooke}, {Duev},
  {Fremling}, {Gatkine}, {Golkhou}, {Kong}, {Mahabal},
  {Mart{\'\i}nez-Palomera}, {Tao}, \& {Zhang}}]{Andreoni+19}
{Andreoni}, I., {Goldstein}, D.~A., {Anand}, S., {et~al.} 2019,
  \bibinfo{title}{{GROWTH on S190510g: DECam Observation Planning and Follow-up
  of a Distant Binary Neutron Star Merger Candidate},} \apjl, 881, L16,
  \dodoi{10.3847/2041-8213/ab3399}

\bibitem[{I. {Andreoni} {et~al.}(2020){Andreoni}, {Goldstein}, {Kasliwal},
  {Nugent}, {Zhou}, {Newman}, {Bulla}, {Foucart}, {Hotokezaka}, {Nakar},
  {Nissanke}, {Raaijmakers}, {Bloom}, {De}, {Jencson}, {Ward}, {Ahumada},
  {Anand}, {Buckley}, {Caballero-Garc{\'\i}a}, {Castro-Tirado}, {Copperwheat},
  {Coughlin}, {Cenko}, {Gromadzki}, {Hu}, {Karambelkar}, {Perley}, {Sharma},
  {Valeev}, {Cook}, {Fremling}, {Kumar}, {Taggart}, {Bagdasaryan}, {Cooke},
  {Dahiwale}, {Dhawan}, {Dobie}, {Gatkine}, {Golkhou}, {Goobar}, {Chaves},
  {Hankins}, {Kaplan}, {Kong}, {Kool}, {Mohite}, {Sollerman}, {Tzanidakis},
  {Webb}, \& {Zhang}}]{Andreoni+20}
{Andreoni}, I., {Goldstein}, D.~A., {Kasliwal}, M.~M., {et~al.} 2020,
  \bibinfo{title}{{GROWTH on S190814bv: Deep Synoptic Limits on the
  Optical/Near-infrared Counterpart to a Neutron Star─Black Hole Merger},}
  \apj, 890, 131, \dodoi{10.3847/1538-4357/ab6a1b}

\bibitem[{C. {Angulo} {et~al.}(2025){Angulo}, {Watson}, {Dornic}, {Basa},
  {Lee}, {S{\'a}nchez {\'A}lvarez}, {Akl}, {Antier}, {Atteia}, {Becerra},
  {Butler}, {Ducoin}, {Fortin}, {Garc{\'\i}a Garc{\'\i}a}, {Gill}, {Globus},
  {Ocelotl L{\'o}pez}, {L{\'o}pez-C{\'a}mara}, {Magnani}, {Moreno M{\'e}ndez},
  {Pereyra}, {Avo Rakotondrainibe}, {Schneider}, \& {de Ugarte
  Postigo}}]{2025GCN.41518....1A}
{Angulo}, C., {Watson}, A.~M., {Dornic}, D., {et~al.} 2025,
  \bibinfo{title}{{LIGO/Virgo/KAGRA S250818k: COLIBR{\'I} confirmation of
  rebrightening of AT2025ulz},} GRB Coordinates Network, 41518, 1

\bibitem[{S. {Antier} {et~al.}(2025){Antier}, {Pillas}, {Akl}, {Karpov},
  {Coughlin}, {Hello}, {Jacquesson}, \& {ZTF
  Collaboration}}]{2025GCN.41519....1A}
{Antier}, S., {Pillas}, M., {Akl}, D., {et~al.} 2025,
  \bibinfo{title}{{LIGO/Virgo/KAGRA S250818k: CFH/MegaCam follow-up observation
  for AT2025ulz},} GRB Coordinates Network, 41519, 1

\bibitem[{S. {Antier} {et~al.}(2020{\natexlab{a}}){Antier}, {Agayeva},
  {Aivazyan}, {Alishov}, {Arbouch}, {Baransky}, {Barynova}, {Bai}, {Basa},
  {Beradze}, {Bertin}, {Berthier}, {Bla{\v{z}}ek}, {Bo{\"e}r}, {Burkhonov},
  {Burrell}, {Cailleau}, {Chabert}, {Chen}, {Christensen}, {Coleiro},
  {Cordier}, {Corre}, {Coughlin}, {Coward}, {Crisp}, {Delattre}, {Dietrich},
  {Ducoin}, {Duverne}, {Marchal-Duval}, {Gendre}, {Eymar}, {Fock-Hang}, {Han},
  {Hello}, {Howell}, {Inasaridze}, {Ismailov}, {Kann}, {Kapanadze}, {Klotz},
  {Kochiashvili}, {Lachaud}, {Leroy}, {Le Van Su}, {Lin}, {Li}, {Lognone},
  {Marron}, {Mo}, {Moore}, {Natsvlishvili}, {Noysena}, {Perrigault}, {Peyrot},
  {Samadov}, {Sadibekova}, {Simon}, {Stachie}, {Teng}, {Thierry}, {Th{\"o}ne},
  {Tillayev}, {Turpin}, {de Ugarte Postigo}, {Vachier}, {Vardosanidze},
  {Vasylenko}, {Vidadi}, {Wang}, {Wang}, {Wei}, {Yan}, {Zhang}, {Zhang}, \&
  {Zhang}}]{Antier+20a}
{Antier}, S., {Agayeva}, S., {Aivazyan}, V., {et~al.} 2020{\natexlab{a}},
  \bibinfo{title}{{The first six months of the Advanced LIGO's and Advanced
  Virgo's third observing run with GRANDMA},} \mnras, 492, 3904,
  \dodoi{10.1093/mnras/stz3142}

\bibitem[{S. {Antier} {et~al.}(2020{\natexlab{b}}){Antier}, {Agayeva},
  {Almualla}, {Awiphan}, {Baransky}, {Barynova}, {Beradze}, {Bla{\v{z}}ek},
  {Bo{\"e}r}, {Burkhonov}, {Christensen}, {Coleiro}, {Corre}, {Coughlin},
  {Crisp}, {Dietrich}, {Ducoin}, {Duverne}, {Marchal-Duval}, {Gendre},
  {Gokuldass}, {Eggenstein}, {Eymar}, {Hello}, {Howell}, {Ismailov}, {Kann},
  {Karpov}, {Klotz}, {Kochiashvili}, {Lachaud}, {Leroy}, {Lin}, {Li},
  {Ma{\v{s}}ek}, {Mo}, {Menard}, {Morris}, {Noysena}, {Orange}, {Prouza},
  {Rattanamala}, {Sadibekova}, {Saint-Gelais}, {Serrau}, {Simon}, {Stachie},
  {Th{\"o}ne}, {Tillayev}, {Turpin}, {Postigo}, {Vasylenko}, {Vidadi}, {Was},
  {Wang}, {Zhang}, {Zhang}, \& {Zhang}}]{Antier+20b}
{Antier}, S., {Agayeva}, S., {Almualla}, M., {et~al.} 2020{\natexlab{b}},
  \bibinfo{title}{{GRANDMA observations of advanced LIGO's and advanced Virgo's
  third observational campaign},} \mnras, 497, 5518,
  \dodoi{10.1093/mnras/staa1846}

\bibitem[{ {Apache Software Foundation}(2023){Apache Software
  Foundation}}]{apache_software_foundation_apache_2023}
{Apache Software Foundation}. 2023, Apache {{Kafka}} v3.5.1,
  \url{https://kafka.apache.org/}

\bibitem[{M. Araujo(2023)Araujo}]{araujo_django-crispy-forms_2023}
Araujo, M. 2023, django-crispy-forms v2.0,, GitHub
  \url{https://github.com/django-crispy-forms/django-crispy-forms}

\bibitem[{I. {Arcavi} {et~al.}(2011){Arcavi}, {Gal-Yam}, {Yaron}, {Sternberg},
  {Rabinak}, {Waxman}, {Kasliwal}, {Quimby}, {Ofek}, {Horesh}, {Kulkarni},
  {Filippenko}, {Silverman}, {Cenko}, {Li}, {Bloom}, {Sullivan}, {Nugent},
  {Poznanski}, {Gorbikov}, {Fulton}, {Howell}, {Bersier}, {Riou},
  {Lamotte-Bailey}, {Griga}, {Cohen}, {Hachinger}, {Polishook}, {Xu},
  {Ben-Ami}, {Manulis}, {Walker}, {Maguire}, {Pan}, {Matheson}, {Mazzali},
  {Pian}, {Fox}, {Gehrels}, {Law}, {James}, {Marchant}, {Smith}, {Mottram},
  {Barnsley}, {Kandrashoff}, \& {Clubb}}]{Arcavi2011dh}
{Arcavi}, I., {Gal-Yam}, A., {Yaron}, O., {et~al.} 2011, \bibinfo{title}{{SN
  2011dh: Discovery of a Type IIb Supernova from a Compact Progenitor in the
  Nearby Galaxy M51},} \apjl, 742, L18, \dodoi{10.1088/2041-8205/742/2/L18}

\bibitem[{I. Arcavi {et~al.}(2017{\natexlab{a}})Arcavi, Hosseinzadeh, Howell,
  McCully, Poznanski, Kasen, Barnes, Zaltzman, Vasylyev, Maoz, \&
  Valenti}]{arcavi_optical_2017}
Arcavi, I., Hosseinzadeh, G., Howell, D.~A., {et~al.} 2017{\natexlab{a}},
  \bibinfo{title}{Optical emission from a kilonova following a
  gravitational-wave-detected neutron-star merger,} Nature, 551, 64,
  \dodoi{10.1038/nature24291}

\bibitem[{I. Arcavi {et~al.}(2017{\natexlab{b}})Arcavi, Hosseinzadeh, Brown,
  Smartt, Valenti, Tartaglia, Piro, Sanchez, Nicholls, Monard, Howell, McCully,
  Sand, Tonry, Denneau, Stalder, Heinze, Rest, Smith, \&
  Bishop}]{arcavi_constraints_2017}
Arcavi, I., Hosseinzadeh, G., Brown, P.~J., {et~al.} 2017{\natexlab{b}},
  \bibinfo{title}{Constraints on the {Progenitor} of {SN} 2016gkg from {Its}
  {Shock}-cooling {Light} {Curve},} The Astrophysical Journal Letters, 837, L2,
  \dodoi{10.3847/2041-8213/aa5be1}

\bibitem[{I. {Arcavi} {et~al.}(2017){Arcavi}, {Hosseinzadeh}, {Brown},
  {Smartt}, {Valenti}, {Tartaglia}, {Piro}, {Sanchez}, {Nicholls}, {Monard},
  {Howell}, {McCully}, {Sand}, {Tonry}, {Denneau}, {Stalder}, {Heinze}, {Rest},
  {Smith}, \& {Bishop}}]{Arcavi2017}
{Arcavi}, I., {Hosseinzadeh}, G., {Brown}, P.~J., {et~al.} 2017,
  \bibinfo{title}{{Constraints on the Progenitor of SN 2016gkg from Its
  Shock-cooling Light Curve},} \apjl, 837, L2, \dodoi{10.3847/2041-8213/aa5be1}

\bibitem[{P. {Armstrong} {et~al.}(2021){Armstrong}, {Tucker}, {Rest},
  {Ridden-Harper}, {Zenati}, {Piro}, {Hinton}, {Lidman}, {Margheim}, {Narayan},
  {Shaya}, {Garnavich}, {Kasen}, {Villar}, {Zenteno}, {Arcavi}, {Drout},
  {Foley}, {Wheeler}, {Anais}, {Campillay}, {Coulter}, {Dimitriadis}, {Jones},
  {Kilpatrick}, {Mu{\~n}oz-Elgueta}, {Rojas-Bravo}, {Vargas-Gonz{\'a}lez},
  {Bulger}, {Chambers}, {Huber}, {Lowe}, {Magnier}, {Shappee}, {Smartt},
  {Smith}, {Barclay}, {Barentsen}, {Dotson}, {Gully-Santiago}, {Hedges},
  {Howell}, {Cody}, {Auchettl}, {B{\'o}di}, {Bogn{\'a}r}, {Brimacombe},
  {Brown}, {Cseh}, {Galbany}, {Hiramatsu}, {Holoien}, {Howell}, {Jha},
  {K{\"o}nyves-T{\'o}th}, {Kriskovics}, {McCully}, {Milne}, {Mu{\~n}oz}, {Pan},
  {P{\'a}l}, {Sai}, {S{\'a}rneczky}, {Smith}, {S{\'o}dor}, {Szab{\'o}},
  {Szak{\'a}ts}, {Valenti}, {Vink{\'o}}, {Wang}, {Zhang}, \&
  {Zsidi}}]{Armstrong_2017jgh}
{Armstrong}, P., {Tucker}, B.~E., {Rest}, A., {et~al.} 2021,
  \bibinfo{title}{{SN2017jgh: a high-cadence complete shock cooling light curve
  of a SN IIb with the Kepler telescope},} \mnras, 507, 3125,
  \dodoi{10.1093/mnras/stab2138}

\bibitem[{ {Astropy Collaboration} {et~al.}(2013){Astropy Collaboration},
  {Robitaille}, {Tollerud}, {Greenfield}, {Droettboom}, {Bray}, {Aldcroft},
  {Davis}, {Ginsburg}, {Price-Whelan}, {Kerzendorf}, {Conley}, {Crighton},
  {Barbary}, {Muna}, {Ferguson}, {Grollier}, {Parikh}, {Nair}, {Unther},
  {Deil}, {Woillez}, {Conseil}, {Kramer}, {Turner}, {Singer}, {Fox}, {Weaver},
  {Zabalza}, {Edwards}, {Azalee Bostroem}, {Burke}, {Casey}, {Crawford},
  {Dencheva}, {Ely}, {Jenness}, {Labrie}, {Lim}, {Pierfederici}, {Pontzen},
  {Ptak}, {Refsdal}, {Servillat}, \& {Streicher}}]{astropy1}
{Astropy Collaboration}, {Robitaille}, T.~P., {Tollerud}, E.~J., {et~al.} 2013,
  \bibinfo{title}{{Astropy: A community Python package for astronomy},} \aap,
  558, A33, \dodoi{10.1051/0004-6361/201322068}

\bibitem[{ {Astropy Collaboration} {et~al.}(2018){Astropy Collaboration},
  {Price-Whelan}, {Sip{\H{o}}cz}, {G{\"u}nther}, {Lim}, {Crawford}, {Conseil},
  {Shupe}, {Craig}, {Dencheva}, {Ginsburg}, {Vand erPlas}, {Bradley},
  {P{\'e}rez-Su{\'a}rez}, {de Val-Borro}, {Aldcroft}, {Cruz}, {Robitaille},
  {Tollerud}, {Ardelean}, {Babej}, {Bach}, {Bachetti}, {Bakanov}, {Bamford},
  {Barentsen}, {Barmby}, {Baumbach}, {Berry}, {Biscani}, {Boquien}, {Bostroem},
  {Bouma}, {Brammer}, {Bray}, {Breytenbach}, {Buddelmeijer}, {Burke},
  {Calderone}, {Cano Rodr{\'\i}guez}, {Cara}, {Cardoso}, {Cheedella}, {Copin},
  {Corrales}, {Crichton}, {D'Avella}, {Deil}, {Depagne}, {Dietrich}, {Donath},
  {Droettboom}, {Earl}, {Erben}, {Fabbro}, {Ferreira}, {Finethy}, {Fox},
  {Garrison}, {Gibbons}, {Goldstein}, {Gommers}, {Greco}, {Greenfield},
  {Groener}, {Grollier}, {Hagen}, {Hirst}, {Homeier}, {Horton}, {Hosseinzadeh},
  {Hu}, {Hunkeler}, {Ivezi{\'c}}, {Jain}, {Jenness}, {Kanarek}, {Kendrew},
  {Kern}, {Kerzendorf}, {Khvalko}, {King}, {Kirkby}, {Kulkarni}, {Kumar},
  {Lee}, {Lenz}, {Littlefair}, {Ma}, {Macleod}, {Mastropietro}, {McCully},
  {Montagnac}, {Morris}, {Mueller}, {Mumford}, {Muna}, {Murphy}, {Nelson},
  {Nguyen}, {Ninan}, {N{\"o}the}, {Ogaz}, {Oh}, {Parejko}, {Parley}, {Pascual},
  {Patil}, {Patil}, {Plunkett}, {Prochaska}, {Rastogi}, {Reddy Janga},
  {Sabater}, {Sakurikar}, {Seifert}, {Sherbert}, {Sherwood-Taylor}, {Shih},
  {Sick}, {Silbiger}, {Singanamalla}, {Singer}, {Sladen}, {Sooley},
  {Sornarajah}, {Streicher}, {Teuben}, {Thomas}, {Tremblay}, {Turner},
  {Terr{\'o}n}, {van Kerkwijk}, {de la Vega}, {Watkins}, {Weaver}, {Whitmore},
  {Woillez}, {Zabalza}, \& {Astropy Contributors}}]{astropy2}
{Astropy Collaboration}, {Price-Whelan}, A.~M., {Sip{\H{o}}cz}, B.~M., {et~al.}
  2018, \bibinfo{title}{{The Astropy Project: Building an Open-science Project
  and Status of the v2.0 Core Package},} \aj, 156, 123,
  \dodoi{10.3847/1538-3881/aabc4f}

\bibitem[{ {Astropy Collaboration} {et~al.}(2022){Astropy Collaboration},
  {Price-Whelan}, {Lim}, {Earl}, {Starkman}, {Bradley}, {Shupe}, {Patil},
  {Corrales}, {Brasseur}, {N{"o}the}, {Donath}, {Tollerud}, {Morris},
  {Ginsburg}, {Vaher}, {Weaver}, {Tocknell}, {Jamieson}, {van Kerkwijk},
  {Robitaille}, {Merry}, {Bachetti}, {G{"u}nther}, {Aldcroft},
  {Alvarado-Montes}, {Archibald}, {B{'o}di}, {Bapat}, {Barentsen}, {Baz{'a}n},
  {Biswas}, {Boquien}, {Burke}, {Cara}, {Cara}, {Conroy}, {Conseil}, {Craig},
  {Cross}, {Cruz}, {D'Eugenio}, {Dencheva}, {Devillepoix}, {Dietrich},
  {Eigenbrot}, {Erben}, {Ferreira}, {Foreman-Mackey}, {Fox}, {Freij}, {Garg},
  {Geda}, {Glattly}, {Gondhalekar}, {Gordon}, {Grant}, {Greenfield}, {Groener},
  {Guest}, {Gurovich}, {Handberg}, {Hart}, {Hatfield-Dodds}, {Homeier},
  {Hosseinzadeh}, {Jenness}, {Jones}, {Joseph}, {Kalmbach}, {Karamehmetoglu},
  {Ka{l}uszy{'n}ski}, {Kelley}, {Kern}, {Kerzendorf}, {Koch}, {Kulumani},
  {Lee}, {Ly}, {Ma}, {MacBride}, {Maljaars}, {Muna}, {Murphy}, {Norman},
  {O'Steen}, {Oman}, {Pacifici}, {Pascual}, {Pascual-Granado}, {Patil},
  {Perren}, {Pickering}, {Rastogi}, {Roulston}, {Ryan}, {Rykoff}, {Sabater},
  {Sakurikar}, {Salgado}, {Sanghi}, {Saunders}, {Savchenko}, {Schwardt},
  {Seifert-Eckert}, {Shih}, {Jain}, {Shukla}, {Sick}, {Simpson},
  {Singanamalla}, {Singer}, {Singhal}, {Sinha}, {Sip{H{o}}cz}, {Spitler},
  {Stansby}, {Streicher}, {{{S}}umak}, {Swinbank}, {Taranu}, {Tewary},
  {Tremblay}, {Val-Borro}, {Van Kooten}, {Vasovi{'c}}, {Verma}, {de Miranda
  Cardoso}, {Williams}, {Wilson}, {Winkel}, {Wood-Vasey}, {Xue}, {Yoachim},
  {Zhang}, {Zonca}, \& {Astropy Project Contributors}}]{astropy3}
{Astropy Collaboration}, {Price-Whelan}, A.~M., {Lim}, P.~L., {et~al.} 2022,
  \bibinfo{title}{{The Astropy Project: Sustaining and Growing a
  Community-oriented Open-source Project and the Latest Major Release (v5.0) of
  the Core Package},} \apj, 935, 167, \dodoi{10.3847/1538-4357/ac7c74}

\bibitem[{L. Balcerzak(2021)Balcerzak}]{balcerzak_django-guardian_2021}
Balcerzak, L. 2021, django-guardian v2.4.0,, GitHub
  \url{https://github.com/django-guardian/django-guardian}

\bibitem[{S. {Banerjee} {et~al.}(2025){Banerjee}, {Botticella}, {Brennan},
  {Cappellaro}, {Chen}, {D'Avanzo}, {D'Elia}, {de Pasquale}, {Eyles-Ferris},
  {Fraser}, {Gillanders}, {Gompertz}, {Habeeb}, {Izzo}, {Jonker}, {Levan},
  {Bj{\o}rn Malesani}, {Martin-Carrillo}, {Nicholl}, {Oates}, {Piranomonte},
  {Piro}, {Rossi}, {Sharan Salafia}, {Sarin}, {Schulze}, {Singh}, {Smartt},
  {Sneppen}, {Sollerman}, {Steeghs}, {Tanvir}, {Thakur}, \& {Engrave
  Collaboration}}]{2025GCN.41532....1B}
{Banerjee}, S., {Botticella}, M.-T., {Brennan}, S.~J., {et~al.} 2025,
  \bibinfo{title}{{LIGO/Virgo/KAGRA S250818k: ENGRAVE observations of SN
  2025ulz as a type II supernova},} GRB Coordinates Network, 41532, 1

\bibitem[{T. Barna {et~al.}(2025)Barna, Fremling, Ahumada, Andreoni, Banerjee,
  Bloom, Bulla, Chen, Coughlin, Dietrich, Hall, Junell, Rusholme, Sollerman, \&
  Sravan}]{barna_iib_2025}
Barna, T., Fremling, C., Ahumada, T., {et~al.} 2025, \bibinfo{title}{{IIb} or
  not {IIb}: {A} {Catalog} of {ZTF} {Kilonova} {Imposters},} Publications of
  the Astronomical Society of the Pacific, 137, 084105,
  \dodoi{10.1088/1538-3873/adf578}

\bibitem[{J. Barnes \& D. Kasen(2013)Barnes \& Kasen}]{barnes_effect_2013}
Barnes, J., \& Kasen, D. 2013, \bibinfo{title}{Effect of a {High} {Opacity} on
  the {Light} {Curves} of {Radioactively} {Powered} {Transients} from {Compact}
  {Object} {Mergers},} The Astrophysical Journal, 775, 18,
  \dodoi{10.1088/0004-637X/775/1/18}

\bibitem[{M. Baumann {et~al.}(2023)Baumann, Manon, Pineau, Boch, Deil, Cecconi,
  Stewart, \& {odidev}}]{baumann_cds-astro_2023}
Baumann, M., Manon, Pineau, F.-X., {et~al.} 2023, cds-astro/mocpy v0.13.0,,
  Zenodo \dodoi{10.5281/zenodo.8297730}

\bibitem[{M. Bayer(2012)Bayer}]{bayer_sqlalchemy_2012}
Bayer, M. 2012, \bibinfo{title}{{{SQLAlchemy}},} in The architecture of open
  source applications volume {{II}}: {{Structure}}, scale, and a few more
  fearless hacks, ed. A.~Brown \& G.~Wilson ({Mountain View}: {aosabook.org}).
\newblock \url{http://aosabook.org/en/sqlalchemy.html}

\bibitem[{R.~L. {Becerra} {et~al.}(2025{\natexlab{a}}){Becerra}, {Troja}, \&
  {Dichiara}}]{2025GCN.41528....1B}
{Becerra}, R.~L., {Troja}, E., \& {Dichiara}, S. 2025{\natexlab{a}},
  \bibinfo{title}{{LIGO/Virgo/KAGRA S250818k: Swift Observations of AT 2025ulz
  - Second Epoch},} GRB Coordinates Network, 41528, 1

\bibitem[{R.~L. {Becerra} {et~al.}(2025{\natexlab{b}}){Becerra}, {Watson}, \&
  {Troja}}]{2025GCN.41502....1B}
{Becerra}, R.~L., {Watson}, A.~M., \& {Troja}, E. 2025{\natexlab{b}},
  \bibinfo{title}{{LIGO/Virgo/KAGRA S250818k: GTC/OSIRIS Optical Detection of
  AT2025ulz},} GRB Coordinates Network, 41502, 1

\bibitem[{R.~L. {Becerra} {et~al.}(2025{\natexlab{c}}){Becerra}, {Yang},
  {Watson}, {Troja}, \& {Lee}}]{2025GCN.41544....1B}
{Becerra}, R.~L., {Yang}, Y., {Watson}, A.~M., {Troja}, E., \& {Lee}, W.~H.
  2025{\natexlab{c}}, \bibinfo{title}{{LIGO/Virgo/KAGRA S250818k: GTC/OSIRIS
  Confirmation of Rebrightening of AT2025ulz},} GRB Coordinates Network, 41544,
  1

\bibitem[{R.~L. {Becerra} {et~al.}(2021){Becerra}, {Dichiara}, {Watson},
  {Troja}, {Butler}, {Pereyra}, {Moreno M{\'e}ndez}, {De Colle}, {Lee},
  {Kutyrev}, \& {L{\'o}pez}}]{Becerra+21}
{Becerra}, R.~L., {Dichiara}, S., {Watson}, A.~M., {et~al.} 2021,
  \bibinfo{title}{{DDOTI observations of gravitational-wave sources discovered
  in O3},} \mnras, 507, 1401, \dodoi{10.1093/mnras/stab2086}

\bibitem[{R. {Beck} {et~al.}(2021){Beck}, {Szapudi}, {Flewelling}, {Holmberg},
  {Magnier}, \& {Chambers}}]{Beck+21}
{Beck}, R., {Szapudi}, I., {Flewelling}, H., {et~al.} 2021,
  \bibinfo{title}{{PS1-STRM: neural network source classification and
  photometric redshift catalogue for PS1 3{\ensuremath{\pi}} DR1},} \mnras,
  500, 1633, \dodoi{10.1093/mnras/staa2587}

\bibitem[{A. {Becker}(2015){Becker}}]{hotpants}
{Becker}, A. 2015, {HOTPANTS: High Order Transform of PSF ANd Template
  Subtraction},, Astrophysics Source Code Library, record ascl:1504.004

\bibitem[{K. Belczynski {et~al.}(2017)Belczynski, Ryu, Perna, Berti, Tanaka, \&
  Bulik}]{belczynski_likelihood_2017}
Belczynski, K., Ryu, T., Perna, R., {et~al.} 2017, \bibinfo{title}{On the
  likelihood of detecting gravitational waves from {Population} {III} compact
  object binaries,} Monthly Notices of the Royal Astronomical Society, 471,
  4702, \dodoi{10.1093/mnras/stx1759}

\bibitem[{P.~J. Benson {et~al.}(1994)Benson, Herbst, Salzer, Vinton, Hanson,
  Ratcliff, Winkler, Elmegreen, Chromey, Strom, Balonek, \&
  Elmegreen}]{benson_light_1994}
Benson, P.~J., Herbst, W., Salzer, J.~J., {et~al.} 1994, \bibinfo{title}{Light
  {Curves} of {SN} {1993J} {From} {The} {Keck} {Northeast} {Astronomy}
  {Consortium},} The Astronomical Journal, 107, 1453, \dodoi{10.1086/116958}

\bibitem[{E. {Berger} {et~al.}(2005){Berger}, {Price}, {Cenko}, {Gal-Yam},
  {Soderberg}, {Kasliwal}, {Leonard}, {Cameron}, {Frail}, {Kulkarni}, {Murphy},
  {Krzeminski}, {Piran}, {Lee}, {Roth}, {Moon}, {Fox}, {Harrison}, {Persson},
  {Schmidt}, {Penprase}, {Rich}, {Peterson}, \& {Cowie}}]{2005Natur.438..988B}
{Berger}, E., {Price}, P.~A., {Cenko}, S.~B., {et~al.} 2005,
  \bibinfo{title}{{The afterglow and elliptical host galaxy of the short
  {\ensuremath{\gamma}}-ray burst GRB 050724},} \nat, 438, 988,
  \dodoi{10.1038/nature04238}

\bibitem[{M.~C. {Bersten} {et~al.}(2012){Bersten}, {Benvenuto}, {Nomoto},
  {Ergon}, {Folatelli}, {Sollerman}, {Benetti}, {Botticella}, {Fraser},
  {Kotak}, {Maeda}, {Ochner}, \& {Tomasella}}]{B2012}
{Bersten}, M.~C., {Benvenuto}, O.~G., {Nomoto}, K., {et~al.} 2012,
  \bibinfo{title}{{The Type IIb Supernova 2011dh from a Supergiant
  Progenitor},} \apj, 757, 31, \dodoi{10.1088/0004-637X/757/1/31}

\bibitem[{E. {Bertin}(2010){Bertin}}]{swarp}
{Bertin}, E. 2010, {SWarp: Resampling and Co-adding FITS Images Together},,
  Astrophysics Source Code Library, record ascl:1010.068

\bibitem[{E. Bertin \& S. Arnouts(1996)Bertin \&
  Arnouts}]{bertin_sextractor:_1996}
Bertin, E., \& Arnouts, S. 1996, \bibinfo{title}{{{SExtractor}}: {{Software}}
  for source extraction.,} A\&AS, 117, 393, \dodoi{10.1051/aas:1996164}

\bibitem[{E. Bertin \& S. Arnouts(2010)Bertin \&
  Arnouts}]{bertin_sextractor_2010}
Bertin, E., \& Arnouts, S. 2010, {{SExtractor}}: {{Source Extractor}},,
  Astrophysics Source Code Library \doeprint{1010.064}

\bibitem[{D. {Bhakta} {et~al.}(2021){Bhakta}, {Mooley}, {Corsi},
  {Balasubramanian}, {Dobie}, {Frail}, {Hallinan}, {Kaplan}, {Myers}, \&
  {Singer}}]{Bhakta+21}
{Bhakta}, D., {Mooley}, K.~P., {Corsi}, A., {et~al.} 2021, \bibinfo{title}{{The
  JAGWAR Prowls LIGO/Virgo O3 Paper I: Radio Search of a Possible
  Multimessenger Counterpart of the Binary Black Hole Merger Candidate
  S191216ap},} \apj, 911, 77, \dodoi{10.3847/1538-4357/abeaa8}

\bibitem[{A. Bhattacharyya(1943)Bhattacharyya}]{bhattacharyya_1943}
Bhattacharyya, A. 1943, \bibinfo{title}{On Some Sets of Sufficient Conditions
  Leading to the Normal Bivariate Distribution,} Sankhyā: The Indian Journal
  of Statistics (1933-1960), 6, 399.
\newblock \url{http://www.jstor.org/stable/25047806}

\bibitem[{J.~S. Bloom {et~al.}(2002)Bloom, Kulkarni, \&
  Djorgovski}]{bloom_observed_2002}
Bloom, J.~S., Kulkarni, S.~R., \& Djorgovski, S.~G. 2002, \bibinfo{title}{The
  {Observed} {Offset} {Distribution} of {Gamma}-{Ray} {Bursts} from {Their}
  {Host} {Galaxies}: {A} {Robust} {Clue} to the {Nature} of the {Progenitors},}
  The Astronomical Journal, 123, 1111, \dodoi{10.1086/338893}

\bibitem[{L. Bradley {et~al.}(2022)Bradley, Sip{\H o}cz, Robitaille, Tollerud,
  Vin{\'i}cius, Deil, Barbary, Wilson, Busko, Donath, G{\"u}nther, Cara, Lim,
  Me{\ss}linger, Conseil, Bostroem, Droettboom, Bray, Bratholm, Barentsen,
  Craig, Rathi, Pascual, Perren, Georgiev, {de Val-Borro}, Kerzendorf, Bach,
  Quint, \& Souchereau}]{bradley_astropy_2022}
Bradley, L., Sip{\H o}cz, B., Robitaille, T., {et~al.} 2022,
  Astropy/{{Photutils}} v1.5.0,, Zenodo \dodoi{10.5281/zenodo.6825092}

\bibitem[{J.~W. {Broderick} {et~al.}(2020){Broderick}, {Shimwell}, {Gourdji},
  {Rowlinson}, {Nissanke}, {Hotokezaka}, {Jonker}, {Tasse}, {Hardcastle},
  {Oonk}, {Fender}, {Wijers}, {Shulevski}, {Stewart}, {ter Veen}, {Moss}, {van
  der Wiel}, {Nichols}, {Piette}, {Bell}, {Carbone}, {Corbel}, {Eisl{\"o}ffel},
  {Grie{\ss}meier}, {Keane}, {Law}, {Mu{\~n}oz-Darias}, {Pietka}, {Serylak},
  {van der Horst}, {van Leeuwen}, {Wijnands}, {Zarka}, {Anderson}, {Bentum},
  {Blaauw}, {Brouw}, {Br{\"u}ggen}, {Ciardi}, {de Vos}, {Duscha}, {Fallows},
  {Franzen}, {Garrett}, {Gunst}, {Hoeft}, {H{\"o}randel}, {Iacobelli},
  {J{\"u}tte}, {Koopmans}, {Krankowski}, {Maat}, {Mann}, {Mulder}, {Nelles},
  {Paas}, {Pandey-Pommier}, {Pekal}, {Reich}, {R{\"o}ttgering}, {Schwarz},
  {Smirnov}, {Soida}, {Toribio}, {van Haarlem}, {van Weeren}, {Vocks},
  {Wucknitz}, \& {Zucca}}]{2020MNRAS.494.5110B}
{Broderick}, J.~W., {Shimwell}, T.~W., {Gourdji}, K., {et~al.} 2020,
  \bibinfo{title}{{LOFAR 144-MHz follow-up observations of GW170817},} \mnras,
  494, 5110, \dodoi{10.1093/mnras/staa950}

\bibitem[{G. {Bruni} {et~al.}(2025{\natexlab{a}}){Bruni}, {Piro}, {Gianfagna},
  \& {Thakur}}]{2025GCN.41500....1B}
{Bruni}, G., {Piro}, L., {Gianfagna}, G., \& {Thakur}, A.~L.
  2025{\natexlab{a}}, \bibinfo{title}{{LIGO/Virgo/KAGRA S250818k: 3 GHz MeerKAT
  observations of AT2025ulz},} GRB Coordinates Network, 41500, 1

\bibitem[{G. {Bruni} {et~al.}(2025{\natexlab{b}}){Bruni}, {Piro}, {Gianfagna},
  \& {Thakur}}]{2025GCN.41594....1B}
{Bruni}, G., {Piro}, L., {Gianfagna}, G., \& {Thakur}, A.~L.
  2025{\natexlab{b}}, \bibinfo{title}{{LIGO/Virgo/KAGRA S250818k: MeerKAT
  detection of an increase in radio flux from AT2025ulz},} GRB Coordinates
  Network, 41594, 1

\bibitem[{G. {Bruni} {et~al.}(2025{\natexlab{c}}){Bruni}, {Piro}, {Gianfagna},
  \& {Thakur}}]{2025GCN.42032....1B}
{Bruni}, G., {Piro}, L., {Gianfagna}, G., \& {Thakur}, A.~L.
  2025{\natexlab{c}}, \bibinfo{title}{{LIGO/Virgo/KAGRA S250818k: further
  MeerKAT monitoring},} GRB Coordinates Network, 42032, 1

\bibitem[{G. {Bruni} {et~al.}(2025{\natexlab{d}}){Bruni}, {Chandra}, {Christy},
  {Kale}, {Laskar}, {Mohnani}, {Das}, {Resmi}, {Ricci}, \&
  {Troja}}]{2025GCN.41577....1B}
{Bruni}, G., {Chandra}, P., {Christy}, C., S.~C., {et~al.} 2025{\natexlab{d}},
  \bibinfo{title}{{LIGO/Virgo/KAGRA S250818k: uGMRT 1.3 GHz observations of
  AT2025ulz},} GRB Coordinates Network, 41577, 1

\bibitem[{M. {Busmann} {et~al.}(2025){Busmann}, {Hall}, {Gruen}, {O'Connor}, \&
  {Palmese}}]{2025GCN.41421....1B}
{Busmann}, M., {Hall}, X.~J., {Gruen}, D., {O'Connor}, B., \& {Palmese}, A.
  2025, \bibinfo{title}{{LIGO/Virgo/KAGRA S250818k: FTW optical and NIR
  observations of AT 2025ulz},} GRB Coordinates Network, 41421, 1

\bibitem[{ {CASA Team} {et~al.}(2022){CASA Team}, Bean, Bhatnagar, Castro,
  Donovan~Meyer, Emonts, Garcia, Garwood, Golap, Gonzalez~Villalba, Harris,
  Hayashi, Hoskins, Hsieh, Jagannathan, Kawasaki, Keimpema, Kettenis, Lopez,
  Marvil, Masters, McNichols, Mehringer, Miel, Moellenbrock, Montesino,
  Nakazato, Ott, Petry, Pokorny, Raba, Rau, Schiebel, Schweighart, Sekhar,
  Shimada, Small, Steeb, Sugimoto, Suoranta, Tsutsumi, van Bemmel, Verkouter,
  Wells, Xiong, Szomoru, Griffith, Glendenning, \& Kern}]{casa_team_casa_2022}
{CASA Team}, Bean, B., Bhatnagar, S., {et~al.} 2022, \bibinfo{title}{{CASA},
  the {Common} {Astronomy} {Software} {Applications} for {Radio} {Astronomy},}
  Publications of the Astronomical Society of the Pacific, 134, 114501,
  \dodoi{10.1088/1538-3873/ac9642}

\bibitem[{K.~C. {Chambers} {et~al.}(2016){Chambers}, {Magnier}, {Metcalfe},
  {Flewelling}, {Huber}, {Waters}, {Denneau}, {Draper}, {Farrow}, {Finkbeiner},
  {Holmberg}, {Koppenhoefer}, {Price}, {Rest}, {Saglia}, {Schlafly}, {Smartt},
  {Sweeney}, {Wainscoat}, {Burgett}, {Chastel}, {Grav}, {Heasley}, {Hodapp},
  {Jedicke}, {Kaiser}, {Kudritzki}, {Luppino}, {Lupton}, {Monet}, {Morgan},
  {Onaka}, {Shiao}, {Stubbs}, {Tonry}, {White}, {Ba{\~n}ados}, {Bell},
  {Bender}, {Bernard}, {Boegner}, {Boffi}, {Botticella}, {Calamida},
  {Casertano}, {Chen}, {Chen}, {Cole}, {Deacon}, {Frenk}, {Fitzsimmons},
  {Gezari}, {Gibbs}, {Goessl}, {Goggia}, {Gourgue}, {Goldman}, {Grant},
  {Grebel}, {Hambly}, {Hasinger}, {Heavens}, {Heckman}, {Henderson}, {Henning},
  {Holman}, {Hopp}, {Ip}, {Isani}, {Jackson}, {Keyes}, {Koekemoer}, {Kotak},
  {Le}, {Liska}, {Long}, {Lucey}, {Liu}, {Martin}, {Masci}, {McLean}, {Mindel},
  {Misra}, {Morganson}, {Murphy}, {Obaika}, {Narayan}, {Nieto-Santisteban},
  {Norberg}, {Peacock}, {Pier}, {Postman}, {Primak}, {Rae}, {Rai}, {Riess},
  {Riffeser}, {Rix}, {R{\"o}ser}, {Russel}, {Rutz}, {Schilbach}, {Schultz},
  {Scolnic}, {Strolger}, {Szalay}, {Seitz}, {Small}, {Smith}, {Soderblom},
  {Taylor}, {Thomson}, {Taylor}, {Thakar}, {Thiel}, {Thilker}, {Unger},
  {Urata}, {Valenti}, {Wagner}, {Walder}, {Walter}, {Watters}, {Werner},
  {Wood-Vasey}, \& {Wyse}}]{PanSTARRS}
{Chambers}, K.~C., {Magnier}, E.~A., {Metcalfe}, N., {et~al.} 2016,
  \bibinfo{title}{{The Pan-STARRS1 Surveys},} arXiv e-prints, arXiv:1612.05560,
  \dodoi{10.48550/arXiv.1612.05560}

\bibitem[{K.~C. {Chambers} {et~al.}(2025){Chambers}, {Boer}, {Fairlamb},
  {Huber}, {Lin}, {Lowe}, {Magnier}, {Minguez}, {Paek}, {Schultz}, {Smith},
  {Wainscoat}, {Smartt}, {Smith}, {Srivastav}, {Young}, {Fulton}, {Nicholl},
  {Moore}, {Weston}, \& {Angus}}]{2025TNSTR3300....1C}
{Chambers}, K.~C., {Boer}, T.~D., {Fairlamb}, J., {et~al.} 2025,
  \bibinfo{title}{{Pan-STARRS Transient Discovery Report for 2025-08-21},}
  Transient Name Server Discovery Report, 2025-3300, 1

\bibitem[{S.-W. {Chang} {et~al.}(2021){Chang}, {Onken}, {Wolf}, {Luvaul},
  {M{\"o}ller}, {Scalzo}, {Schmidt}, {Scott}, {Sura}, \& {Yuan}}]{Chang+21}
{Chang}, S.-W., {Onken}, C.~A., {Wolf}, C., {et~al.} 2021,
  \bibinfo{title}{{SkyMapper optical follow-up of gravitational wave triggers:
  Alert science data pipeline and LIGO/Virgo O3 run},} \pasa, 38, e024,
  \dodoi{10.1017/pasa.2021.17}

\bibitem[{S.~S. Chaudhary {et~al.}(2024)Chaudhary, Toivonen, Waratkar, Mo,
  Chatterjee, Antier, Brockill, Coughlin, Essick, Ghosh, Morisaki, Baral,
  Baylor, Adhikari, Brady, Davies, Canton, Cavaglià, Creighton, Choudhary,
  Chu, Clearwater, Davis, Dent, Drago, Ewing, Godwin, Guo, Hanna, Huxford,
  Harry, Katsavounidis, Kovalam, Li, Magee, Marx, Meacher, Messick,
  Morice-Atkinson, Pace, Pietri, Piotrzkowski, Roy, Sachdev, Singer, Singh,
  Szczepanczyk, Tang, Trevor, Tsukada, Villa-Ortega, Wen, \&
  Wysocki}]{chaudhary_low-latency_2024}
Chaudhary, S.~S., Toivonen, A., Waratkar, G., {et~al.} 2024,
  \bibinfo{title}{Low-latency gravitational wave alert products and their
  performance at the time of the fourth {LIGO}-{Virgo}-{KAGRA} observing run,}
  Proceedings of the National Academy of Sciences, 121, e2316474121,
  \dodoi{10.1073/pnas.2316474121}

\bibitem[{Y.-X. Chen \& B.~D. Metzger(2025)Chen \&
  Metzger}]{chen_gravitational_2025}
Chen, Y.-X., \& Metzger, B.~D. 2025, \bibinfo{title}{Gravitational
  {Instability} and {Fragmentation} in {Collapsar} {Disks} {Supports} the
  {Formation} of {Subsolar} {Neutron} {Stars},} The Astrophysical Journal, 991,
  L22, \dodoi{10.3847/2041-8213/ae045d}

\bibitem[{R.~A. Chevalier \& A.~M. Soderberg(2010)Chevalier \&
  Soderberg}]{chevalier_type_2010}
Chevalier, R.~A., \& Soderberg, A.~M. 2010, \bibinfo{title}{Type {IIb}
  {Supernovae} with {Compact} and {Extended} {Progenitors},} The Astrophysical
  Journal, 711, L40, \dodoi{10.1088/2041-8205/711/1/L40}

\bibitem[{R.~A. {Chevalier} \& A.~M. {Soderberg}(2010){Chevalier} \&
  {Soderberg}}]{CS2010}
{Chevalier}, R.~A., \& {Soderberg}, A.~M. 2010, \bibinfo{title}{{Type IIb
  Supernovae with Compact and Extended Progenitors},} \apjl, 711, L40,
  \dodoi{10.1088/2041-8205/711/1/L40}

\bibitem[{ {CHIME/FRB Collaboration} {et~al.}(2025){CHIME/FRB Collaboration},
  {Abbott}, {Amouyal}, {Andersen}, {Andrew}, {Bandura}, {Bhardwaj}, {Bhopi},
  {Bhusare}, {Brar}, {Cai}, {Cassanelli}, {Chatterjee}, {Cliche}, {Cook},
  {Curtin}, {Davies-Velie}, {Dobbs}, {Dong}, {Dong}, {Eadie}, {Eftekhari},
  {Fong}, {Fonseca}, {Gaensler}, {Gusinskaia}, {Hessels}, {Hewitt}, {Huang},
  {Jain}, {Joseph}, {Kahinga}, {Kaspi}, {Khan}, {Kharel}, {Lanman}, {L'Argent},
  {Lazda}, {Leung}, {Main}, {Mas-Ribas}, {Masui}, {McGregor}, {McKinven},
  {Mena-Parra}, {Michilli}, {Mulyk}, {Ng}, {Nimmo}, {Pandhi}, {Patil},
  {Pearlman}, {Pen}, {Pleunis}, {Prochaska}, {Rafiei-Ravandi}, {Ransom},
  {Sachdeva}, {Sammons}, {Sand}, {Scholz}, {Shah}, {Shin}, {Siegel}, {Simha},
  {Smith}, {Stairs}, {Stenning}, {Wang}, {Boles}, {Cognard}, {Dijkema},
  {Filippenko}, {Gawro{\'n}ski}, {Herrmann}, {Kilpatrick}, {Kirsten}, {Knabel},
  {Ould-Boukattine}, {Paugnat}, {Puchalska}, {Sheu}, {Suresh}, {Tohuvavohu},
  {Treu}, \& {Zheng}}]{CHIME_FRB_2025}
{CHIME/FRB Collaboration}, {Abbott}, T.~C., {Amouyal}, D., {et~al.} 2025,
  \bibinfo{title}{{FRB 20250316A: A Brilliant and Nearby One-off Fast Radio
  Burst Localized to 13 pc Precision},} \apjl, 989, L48,
  \dodoi{10.3847/2041-8213/adf62f}

\bibitem[{T. Christie(2022)Christie}]{christie_django_2022}
Christie, T. 2022, Django {{REST Framework}} v3.14.0,, GitHub
  \url{https://github.com/encode/django-rest-framework}

\bibitem[{C.~T. Christy {et~al.}(2023)Christy, Jayasinghe, Stanek, Kochanek,
  Thompson, Shappee, Holoien, Prieto, Dong, \& Giles}]{christy_asas-sn_2023}
Christy, C.~T., Jayasinghe, T., Stanek, K.~Z., {et~al.} 2023,
  \bibinfo{title}{The {ASAS}-{SN} catalogue of variable stars {X}: discovery of
  116 000 new variable stars using {G}-band photometry,} Monthly Notices of the
  Royal Astronomical Society, 519, 5271, \dodoi{10.1093/mnras/stac3801}

\bibitem[{J.~C. {Clemens} {et~al.}(2004){Clemens}, {Crain}, \&
  {Anderson}}]{Clemens04}
{Clemens}, J.~C., {Crain}, J.~A., \& {Anderson}, R. 2004, \bibinfo{title}{{The
  Goodman spectrograph},} in Society of Photo-Optical Instrumentation Engineers
  (SPIE) Conference Series, Vol. 5492, Ground-based Instrumentation for
  Astronomy, ed. A.~F.~M. {Moorwood} \& M.~{Iye}, 331--340,
  \dodoi{10.1117/12.550069}

\bibitem[{D. Collom {et~al.}(2020)Collom, Lindstrom, Riba, Street, McCully, \&
  Bowman}]{collom_tom_2020}
Collom, D., Lindstrom, L., Riba, A., {et~al.} 2020, The {{TOM Toolkit}}
  v2.0.0,, Zenodo \dodoi{10.5281/zenodo.4437764}

\bibitem[{M.~W. Coughlin {et~al.}(2018)Coughlin {et~al.}}]{Coughlin:2018lta}
Coughlin, M.~W., {et~al.} 2018, \bibinfo{title}{{Optimizing searches for
  electromagnetic counterparts of gravitational wave triggers},} Mon. Not. Roy.
  Astron. Soc., 478, 692, \dodoi{10.1093/mnras/sty1066}

\bibitem[{M.~W. {Coughlin} {et~al.}(2019){Coughlin}, {Ahumada}, {Anand}, {De},
  {Hankins}, {Kasliwal}, {Singer}, {Bellm}, {Andreoni}, {Cenko}, {Cooke},
  {Copperwheat}, {Dugas}, {Jencson}, {Perley}, {Yu}, {Bhalerao}, {Kumar},
  {Bloom}, {Anupama}, {Ashley}, {Bagdasaryan}, {Biswas}, {Buckley}, {Burdge},
  {Cook}, {Cromer}, {Cunningham}, {D'A{\`\i}}, {Dekany}, {Delacroix},
  {Dichiara}, {Duev}, {Dutta}, {Feeney}, {Frederick}, {Gatkine}, {Ghosh},
  {Goldstein}, {Golkhou}, {Goobar}, {Graham}, {Hanayama}, {Horiuchi}, {Hung},
  {Jha}, {Kong}, {Giomi}, {Kaplan}, {Karambelkar}, {Kowalski}, {Kulkarni},
  {Kupfer}, {Masci}, {Mazzali}, {Moore}, {Mogotsi}, {Neill}, {Ngeow},
  {Mart{\'\i}nez-Palomera}, {La Parola}, {Pavana}, {Ofek}, {Patil}, {Riddle},
  {Rigault}, {Rusholme}, {Serabyn}, {Shupe}, {Sharma}, {Singh}, {Sollerman},
  {Soon}, {Staats}, {Taggart}, {Tan}, {Travouillon}, {Troja}, {Waratkar}, \&
  {Yatsu}}]{Coughlin+19}
{Coughlin}, M.~W., {Ahumada}, T., {Anand}, S., {et~al.} 2019,
  \bibinfo{title}{{GROWTH on S190425z: Searching Thousands of Square Degrees to
  Identify an Optical or Infrared Counterpart to a Binary Neutron Star Merger
  with the Zwicky Transient Facility and Palomar Gattini-IR},} \apjl, 885, L19,
  \dodoi{10.3847/2041-8213/ab4ad8}

\bibitem[{M.~W. Coughlin {et~al.}(2019)Coughlin {et~al.}}]{Coughlin:2019qkn}
Coughlin, M.~W., {et~al.} 2019, \bibinfo{title}{{Optimizing multitelescope
  observations of gravitational-wave counterparts},} Mon. Not. Roy. Astron.
  Soc., 489, 5775, \dodoi{10.1093/mnras/stz2485}

\bibitem[{D.~A. Coulter {et~al.}(2017)Coulter, Foley, Kilpatrick, Drout, Piro,
  Shappee, Siebert, Simon, Ulloa, Kasen, Madore, Murguia-Berthier, Pan,
  Prochaska, Ramirez-Ruiz, Rest, \& Rojas-Bravo}]{coulter_swope_2017}
Coulter, D.~A., Foley, R.~J., Kilpatrick, C.~D., {et~al.} 2017,
  \bibinfo{title}{Swope {Supernova} {Survey} 2017a ({SSS17a}), the optical
  counterpart to a gravitational wave source,} Science, 358, 1556,
  \dodoi{10.1126/science.aap9811}

\bibitem[{P.~S. Cowperthwaite {et~al.}(2017)Cowperthwaite, Berger, Villar,
  Metzger, Nicholl, Chornock, Blanchard, Fong, Margutti, Soares-Santos,
  Alexander, Allam, Annis, Brout, Brown, Butler, Chen, Diehl, Doctor, Drout,
  Eftekhari, Farr, Finley, Foley, Frieman, Fryer, García-Bellido, Gill,
  Guillochon, Herner, Holz, Kasen, Kessler, Marriner, Matheson, Neilsen,
  Quataert, Palmese, Rest, Sako, Scolnic, Smith, Tucker, Williams, Balbinot,
  Carlin, Cook, Durret, Li, Lopes, Lourenço, Marshall, Medina, Muir, Muñoz,
  Sauseda, Schlegel, Secco, Vivas, Wester, Zenteno, Zhang, Abbott, Banerji,
  Bechtol, Benoit-Lévy, Bertin, Buckley-Geer, Burke, Capozzi, Carnero~Rosell,
  Carrasco~Kind, Castander, Crocce, Cunha, D’Andrea, Costa, Davis, DePoy,
  Desai, Dietrich, Drlica-Wagner, Eifler, Evrard, Fernandez, Flaugher, Fosalba,
  Gaztanaga, Gerdes, Giannantonio, Goldstein, Gruen, Gruendl, Gutierrez,
  Honscheid, Jain, James, Jeltema, Johnson, Johnson, Kent, Krause, Kron, Kuehn,
  Nuropatkin, Lahav, Lima, Lin, Maia, March, Martini, McMahon, Menanteau,
  Miller, Miquel, Mohr, Neilsen, Nichol, Ogando, Plazas, Roe, Romer, Roodman,
  Rykoff, Sanchez, Scarpine, Schindler, Schubnell, Sevilla-Noarbe, Smith,
  Smith, Sobreira, Suchyta, Swanson, Tarle, Thomas, Thomas, Troxel, Vikram,
  Walker, Wechsler, Weller, Yanny, \&
  Zuntz}]{cowperthwaite_electromagnetic_2017}
Cowperthwaite, P.~S., Berger, E., Villar, V.~A., {et~al.} 2017,
  \bibinfo{title}{The {Electromagnetic} {Counterpart} of the {Binary} {Neutron}
  {Star} {Merger} {LIGO}/{Virgo} {GW170817}. {II}. {UV}, {Optical}, and
  {Near}-infrared {Light} {Curves} and {Comparison} to {Kilonova} {Models},}
  The Astrophysical Journal Letters, 848, L17, \dodoi{10.3847/2041-8213/aa8fc7}

\bibitem[{D. Dahlen {et~al.}(2025)Dahlen, Kwon, Masiero, Spahr, \&
  Mainzer}]{dahlen_kete_2025}
Dahlen, D., Kwon, Y.~G., Masiero, J.~R., Spahr, T., \& Mainzer, A.~K. 2025,
  Kete: {Predicting} {Known} {Minor} {Bodies} in {Images}, arXiv,
  \dodoi{10.48550/arXiv.2509.04666}

\bibitem[{P.~N. Daly {et~al.}(2023)Daly, Bostroem, \&
  Hosseinzadeh}]{daly_sassy_2023}
Daly, P.~N., Bostroem, K.~A., \& Hosseinzadeh, G. 2023, {{SASSy Q3C Models}}
  v1.4.0,, Zenodo \dodoi{10.5281/zenodo.8436176}

\bibitem[{G. {D{\'a}lya} {et~al.}(2018){D{\'a}lya}, {Galg{\'o}czi}, {Dobos},
  {Frei}, {Heng}, {Macas}, {Messenger}, {Raffai}, \& {de Souza}}]{Dalya+18}
{D{\'a}lya}, G., {Galg{\'o}czi}, G., {Dobos}, L., {et~al.} 2018,
  \bibinfo{title}{{GLADE: A galaxy catalogue for multimessenger searches in the
  advanced gravitational-wave detector era},} \mnras, 479, 2374,
  \dodoi{10.1093/mnras/sty1703}

\bibitem[{P. Darc {et~al.}(2025)Darc, Bom, Kilpatrick, Santos, Fraga,
  Rodríguez-Ramírez, Coulter, Mendes~de Oliveira, Kanaan, Ribeiro, Schoenell,
  \& Lacerda}]{darc_long-term_2025}
Darc, P., Bom, C., Kilpatrick, C., {et~al.} 2025, \bibinfo{title}{Long-term
  optical follow up of {S231206cc}: {Multimodel} constraints on {BBH} merger
  emission in {AGN} disks,} Physical Review D, 112, 063019,
  \dodoi{10.1103/6rg8-2xxz}

\bibitem[{K.~K. Das {et~al.}(2023)Das, Kasliwal, Fremling, Yang, Schulze,
  Sollerman, Sit, De, Tzanidakis, Perley, Anand, Andreoni, Barbarino, Brudge,
  Drake, Gal-Yam, Laher, Karambelkar, Kulkarni, Masci, Medford, Polin, Reedy,
  Riddle, Sharma, Smith, Yan, Yang, \& Yao}]{das_probing_2023}
Das, K.~K., Kasliwal, M.~M., Fremling, C., {et~al.} 2023,
  \bibinfo{title}{Probing the {Low}-mass {End} of {Core}-collapse {Supernovae}
  {Using} a {Sample} of {Strongly}-stripped {Calcium}-rich {Type} {IIb}
  {Supernovae} from the {Zwicky} {Transient} {Facility},} The Astrophysical
  Journal, 959, 12, \dodoi{10.3847/1538-4357/acfeeb}

\bibitem[{K.~K. {Das} {et~al.}(2023){Das}, {Kasliwal}, {Fremling}, {Yang},
  {Schulze}, {Sollerman}, {Sit}, {De}, {Tzanidakis}, {Perley}, {Anand},
  {Andreoni}, {Barbarino}, {Brudge}, {Drake}, {Gal-Yam}, {Laher},
  {Karambelkar}, {Kulkarni}, {Masci}, {Medford}, {Polin}, {Reedy}, {Riddle},
  {Sharma}, {Smith}, {Yan}, {Yang}, \& {Yao}}]{Das2023}
{Das}, K.~K., {Kasliwal}, M.~M., {Fremling}, C., {et~al.} 2023,
  \bibinfo{title}{{Probing the Low-mass End of Core-collapse Supernovae Using a
  Sample of Strongly-stripped Calcium-rich Type IIb Supernovae from the Zwicky
  Transient Facility},} \apj, 959, 12, \dodoi{10.3847/1538-4357/acfeeb}

\bibitem[{T. {de Jaeger} {et~al.}(2022){de Jaeger}, {Shappee}, {Kochanek},
  {Stanek}, {Beacom}, {Holoien}, {Thompson}, {Franckowiak}, \&
  {Holmbo}}]{deJaeger+22}
{de Jaeger}, T., {Shappee}, B.~J., {Kochanek}, C.~S., {et~al.} 2022,
  \bibinfo{title}{{ASAS-SN search for optical counterparts of
  gravitational-wave events from the third observing run of Advanced
  LIGO/Virgo},} \mnras, 509, 3427, \dodoi{10.1093/mnras/stab3141}

\bibitem[{S. {de Wet} {et~al.}(2021){de Wet}, {Groot}, {Bloemen}, {Le Poole},
  {Klein-Wolt}, {K{\"o}rding}, {McBride}, {Paterson}, {Pieterse}, {Vreeswijk},
  \& {Woudt}}]{de_Wet+21}
{de Wet}, S., {Groot}, P.~J., {Bloemen}, S., {et~al.} 2021,
  \bibinfo{title}{{GW190814 follow-up with the optical telescope MeerLICHT},}
  \aap, 649, A72, \dodoi{10.1051/0004-6361/202040231}

\bibitem[{ {DESI Collaboration} {et~al.}(2024){DESI Collaboration}, Adame,
  Aguilar, Ahlen, Alam, Aldering, Alexander, Alfarsy, Allende~Prieto, Alvarez,
  Alves, Anand, Andrade-Oliveira, Armengaud, Asorey, Avila, Aviles, Bailey,
  Balaguera-Antolínez, Ballester, Baltay, Bault, Bautista, Behera, Beltran,
  BenZvi, Beraldo~e Silva, Bermejo-Climent, Berti, Besuner, Beutler, Bianchi,
  Blake, Blum, Bolton, Brieden, Brodzeller, Brooks, Brown, Buckley-Geer,
  Burtin, Cabayol-Garcia, Cai, Canning, Cardiel-Sas, Carnero~Rosell, Castander,
  Cervantes-Cota, Chabanier, Chaussidon, Chaves-Montero, Chen, Chen, Chuang,
  Claybaugh, Cole, Cooper, Cuceu, Davis, Dawson, de~Belsunce, de~la Cruz, de~la
  Macorra, Della~Costa, de~Mattia, Demina, Demirbozan, DeRose, Dey, Dey,
  Dhungana, Ding, Ding, Doel, Doshi, Douglass, Edge, Eftekharzadeh, Eisenstein,
  Elliott, Ereza, Escoffier, Fagrelius, Fan, Fanning, Fawcett, Ferraro,
  Flaugher, Font-Ribera, Forero-Romero, Forero-Sánchez, Frenk, Gänsicke,
  García, García-Bellido, Garcia-Quintero, Garrison, Gil-Marín, Golden-Marx,
  Gontcho A~Gontcho, Gonzalez-Morales, Gonzalez-Perez, Gordon, Graur, Green,
  Gruen, Guy, Hadzhiyska, Hahn, Han, Hanif, Herrera-Alcantar, Honscheid, Hou,
  Howlett, Huterer, Iršič, Ishak, Jacques, Jana, Jiang, Jimenez, Jing,
  Joudaki, Joyce, Jullo, Juneau, Karaçaylı, Karim, Kehoe, Kent, Khederlarian,
  Kim, Kirkby, Kisner, Kitaura, Kizhuprakkat, Kneib, Koposov, Kovács, Kremin,
  Krolewski, L'Huillier, Lahav, Lambert, Lamman, Lan, Landriau, Lang, Lange,
  Lasker, Leauthaud, Le~Guillou, Levi, Li, Linder, Lyons, Magneville, Manera,
  Manser, Margala, Martini, McDonald, Medina, Medina-Varela, Meisner,
  Mena-Fernández, Meneses-Rizo, Mezcua, Miquel, Montero-Camacho, Moon, Moore,
  Moustakas, Mueller, Mundet, Muñoz-Gutiérrez, Myers, Nadathur, Napolitano,
  Neveux, Newman, Nie, Nikutta, Niz, Norberg, Noriega, Paillas,
  Palanque-Delabrouille, Palmese, Pan, Parkinson, Penmetsa, Percival,
  Pérez-Fernández, Pérez-Ràfols, Pieri, Poppett, Porredon, \&
  Pothier}]{desi_collaboration_early_2024}
{DESI Collaboration}, Adame, A.~G., Aguilar, J., {et~al.} 2024,
  \bibinfo{title}{The {Early} {Data} {Release} of the {Dark} {Energy}
  {Spectroscopic} {Instrument},} The Astronomical Journal, 168, 58,
  \dodoi{10.3847/1538-3881/ad3217}

\bibitem[{F. Di~Gregorio \& D. Varrazzo(2023)Di~Gregorio \&
  Varrazzo}]{di_gregorio_psycopg_2023}
Di~Gregorio, F., \& Varrazzo, D. 2023, Psycopg v2.9.6,, GitHub
  \url{https://github.com/psycopg/psycopg2}

\bibitem[{S. {Dichiara} {et~al.}(2021){Dichiara}, {Becerra}, {Chase}, {Troja},
  {Lee}, {Watson}, {Butler}, {O'Connor}, {Pereyra}, {L{\'o}pez}, {Lien},
  {Gottlieb}, \& {Kutyrev}}]{Dichiara+21}
{Dichiara}, S., {Becerra}, R.~L., {Chase}, E.~A., {et~al.} 2021,
  \bibinfo{title}{{Constraints on the Electromagnetic Counterpart of the
  Neutron-star-Black-hole Merger GW200115},} \apjl, 923, L32,
  \dodoi{10.3847/2041-8213/ac4259}

\bibitem[{ {Dimple} {et~al.}(2025){Dimple}, {Gompertz}, {Levan}, {Malesani},
  {Laskar}, {Bala}, {Chrimes}, {Heintz}, {Izzo}, {Lamb}, {O'Neill}, {Palmerio},
  {Saccardi}, {Anderson}, {De Barra}, {Huang}, {Kumar}, {Li}, {McBreen},
  {Mukherjee}, {Oates}, {Pathak}, {Qiu}, {Roberts}, {Sonawane}, {Veres},
  {Ackley}, {Han}, {Julakanti}, {Wang}, {D'Avanzo}, {Martin-Carrillo},
  {Ravasio}, {Rossi}, {Tanvir}, {Anderson}, {Arabsalmani}, {Belkin}, {Breton},
  {Brivio}, {Burns}, {Casares}, {Campana}, {Chastain}, {D'Elia}, {Dhillon},
  {Dyer}, {Fynbo}, {Galloway}, {Gulati}, {Godson}, {Goodwin}, {Gromadzki},
  {Hartmann}, {Jakobsson}, {Killestein}, {Kotak}, {Leung}, {Lyman}, {Melandri},
  {Mattila}, {McGee}, {Morley}, {Mukherjee}, {Muller-Bravo}, {Noysena},
  {Nuttall}, {O'Brien}, {De Pasquale}, {Pignata}, {Pollacco}, {Pugliese},
  {Ramsay}, {Sahu}, {Salvaterra}, {Schady}, {Schneider}, {Steeghs}, {Starling},
  {Tsalapatas}, {Ulaczyk}, {van der Horst}, {Wang}, {Wiersema}, {Worssam},
  {Wortley}, {Xiong}, \& {Zafar}}]{2025arXiv250715940D}
{Dimple}, {Gompertz}, B.~P., {Levan}, A.~J., {et~al.} 2025,
  \bibinfo{title}{{GRB 241105A: A test case for GRB classification and rapid
  r-process nucleosynthesis channels},} arXiv e-prints, arXiv:2507.15940,
  \dodoi{10.48550/arXiv.2507.15940}

\bibitem[{ {Django Software Foundation}(2023){Django Software
  Foundation}}]{django_software_foundation_django_2023}
{Django Software Foundation}. 2023, Django v4.2,, GitHub
  \url{https://github.com/django/django/}

\bibitem[{D. {Dobie} {et~al.}(2018){Dobie}, {Kaplan}, {Murphy}, {Lenc},
  {Mooley}, {Lynch}, {Corsi}, {Frail}, {Kasliwal}, \&
  {Hallinan}}]{2018ApJ...858L..15D}
{Dobie}, D., {Kaplan}, D.~L., {Murphy}, T., {et~al.} 2018, \bibinfo{title}{{A
  Turnover in the Radio Light Curve of GW170817},} \apjl, 858, L15,
  \dodoi{10.3847/2041-8213/aac105}

\bibitem[{D. {Dobie} {et~al.}(2019){Dobie}, {Stewart}, {Murphy}, {Lenc},
  {Wang}, {Kaplan}, {Andreoni}, {Banfield}, {Brown}, {Corsi}, {De},
  {Goldstein}, {Hallinan}, {Hotan}, {Hotokezaka}, {Jaodand}, {Karambelkar},
  {Kasliwal}, {McConnell}, {Mooley}, {Moss}, {Newman}, {Perley}, {Prakash},
  {Pritchard}, {Sadler}, {Sharma}, {Ward}, {Whiting}, \& {Zhou}}]{Dobie+19}
{Dobie}, D., {Stewart}, A., {Murphy}, T., {et~al.} 2019, \bibinfo{title}{{An
  ASKAP Search for a Radio Counterpart to the First High-significance Neutron
  Star-Black Hole Merger LIGO/Virgo S190814bv},} \apjl, 887, L13,
  \dodoi{10.3847/2041-8213/ab59db}

\bibitem[{D. {Dobie} {et~al.}(2021){Dobie}, {Stewart}, {Hotokezaka}, {Murphy},
  {Kaplan}, {Buckley}, {Cooke}, {Ho}, {Lenc}, {Leung}, {Gromadzki}, {O'Brien},
  {Pintaldi}, {Pritchard}, {Wang}, \& {Wang}}]{Dobie+21}
{Dobie}, D., {Stewart}, A., {Hotokezaka}, K., {et~al.} 2021, \bibinfo{title}{{A
  comprehensive search for the radio counterpart of GW190814 with the
  Australian Square Kilometre Array Pathfinder},} \mnras,
  \dodoi{10.1093/mnras/stab3628}

\bibitem[{A. {Dolphin}(2016){Dolphin}}]{dolphot}
{Dolphin}, A. 2016, {DOLPHOT: Stellar photometry},, Astrophysics Source Code
  Library, record ascl:1608.013 \doeprint{1608.013}

\bibitem[{A. Domínguez {et~al.}(2013)Domínguez, Siana, Henry, Scarlata,
  Bedregal, Malkan, Atek, Ross, Colbert, Teplitz, Rafelski, McCarthy, Bunker,
  Hathi, Dressler, Martin, \& Masters}]{dominguez_dust_2013}
Domínguez, A., Siana, B., Henry, A.~L., {et~al.} 2013, \bibinfo{title}{{DUST}
  {EXTINCTION} {FROM} {BALMER} {DECREMENTS} {OF} {STAR}-{FORMING} {GALAXIES}
  {AT} 0.75 ⩽ \textit{z} ⩽ 1.5 {WITH} \textit{{HUBBLE} {SPACE} {TELESCOPE}}
  /{WIDE}-{FIELD}-{CAMERA} 3 {SPECTROSCOPY} {FROM} {THE} {WFC3} {INFRARED}
  {SPECTROSCOPIC} {PARALLEL} {SURVEY},} The Astrophysical Journal, 763, 145,
  \dodoi{10.1088/0004-637X/763/2/145}

\bibitem[{Y. {Dong} {et~al.}(2024){Dong}, {Eftekhari}, {Fong}, {Bhandari},
  {Berger}, {Ould-Boukattine}, {Hessels}, {Sridhar}, {Reines}, {Margalit},
  {Darling}, {Gordon}, {Greene}, {Kilpatrick}, {Marcote}, {Metzger}, {Nimmo},
  {Nugent}, {Paragi}, \& {Williams}}]{Dong24}
{Dong}, Y., {Eftekhari}, T., {Fong}, W., {et~al.} 2024, \bibinfo{title}{{A
  Radio Study of Persistent Radio Sources in Nearby Dwarf Galaxies:
  Implications for Fast Radio Bursts},} \apj, 973, 133,
  \dodoi{10.3847/1538-4357/ad6568}

\bibitem[{M.~R. Drout {et~al.}(2017)Drout, Piro, Shappee, Kilpatrick, Simon,
  Contreras, Coulter, Foley, Siebert, Morrell, Boutsia, Di~Mille, Holoien,
  Kasen, Kollmeier, Madore, Monson, Murguia-Berthier, Pan, Prochaska,
  Ramirez-Ruiz, Rest, Adams, Alatalo, Bañados, Baughman, Beers, Bernstein,
  Bitsakis, Campillay, Hansen, Higgs, Ji, Maravelias, Marshall, Bidin, Prieto,
  Rasmussen, Rojas-Bravo, Strom, Ulloa, Vargas-González, Wan, \&
  Whitten}]{drout_light_2017}
Drout, M.~R., Piro, A.~L., Shappee, B.~J., {et~al.} 2017, \bibinfo{title}{Light
  curves of the neutron star merger {GW170817}/{SSS17a}: {Implications} for
  r-process nucleosynthesis,} Science, 358, 1570,
  \dodoi{10.1126/science.aaq0049}

\bibitem[{N. Earl {et~al.}(2023)Earl, Tollerud, O'Steen, {brechmos},
  Kerzendorf, Busko, {shaileshahuja}, D'Avella, Robitaille, Ginsburg, Lim,
  Homeier, Sip{\H o}cz, Averbukh, Tocknell, Cherinka, Ogaz, Geda, Davies,
  Conroy, G{\"u}nther, Barbary, Foster, Droettboom, Nguyen, Bray, Casey,
  Teuben, Crawford, \& Ferguson}]{earl_astropy_2023}
Earl, N., Tollerud, E., O'Steen, R., {et~al.} 2023, astropy/specutils v1.11.0,,
  Zenodo \dodoi{10.5281/zenodo.8049033}

\bibitem[{W.~E. East {et~al.}(2015)East, Paschalidis, \&
  Pretorius}]{east_eccentric_2015}
East, W.~E., Paschalidis, V., \& Pretorius, F. 2015, \bibinfo{title}{Eccentric
  mergers of black holes with spinning neutron stars,} The Astrophysical
  Journal, 807, L3, \dodoi{10.1088/2041-8205/807/1/L3}

\bibitem[{A. Elmhamdi {et~al.}(2006)Elmhamdi, Danziger, Branch, Leibundgut,
  Baron, \& Kirshner}]{elmhamdi_hydrogen_2006}
Elmhamdi, A., Danziger, I.~J., Branch, D., {et~al.} 2006,
  \bibinfo{title}{Hydrogen and helium traces in type {Ib}-c supernovae,}
  Astronomy and Astrophysics, 450, 305, \dodoi{10.1051/0004-6361:20054366}

\bibitem[{D. {Fabricant} {et~al.}(2019){Fabricant}, {Fata}, {Epps}, {Gauron},
  {Mueller}, {Zajac}, {Amato}, {Barberis}, {Bergner}, {Brennan}, {Brown},
  {Chilingarian}, {Geary}, {Kradinov}, {McLeod}, {Smith}, \&
  {Woods}}]{Fabricant2019}
{Fabricant}, D., {Fata}, R., {Epps}, H., {et~al.} 2019,
  \bibinfo{title}{{Binospec: A Wide-field Imaging Spectrograph for the MMT},}
  \pasp, 131, 075004, \dodoi{10.1088/1538-3873/ab1d78}

\bibitem[{J.~R. Farah {et~al.}(2025)Farah, Howell, Hiramatsu, McCully, Andrews,
  Newsome, Padilla~Gonzalez, Pellegrino, Berger, Blanchard, Gomez, Kumar,
  Bostroem, Ni, Gagliano, \& Ravi}]{farah_when_2025}
Farah, J.~R., Howell, D.~A., Hiramatsu, D., {et~al.} 2025, When {IIb} {Ceases}
  {To} {Be}: {Bridging} the {Gap} {Between} {IIb} and {Short}-plateau
  {Supernovae}, arXiv, \dodoi{10.48550/arXiv.2509.12470}

\bibitem[{J.~R. {Farah} {et~al.}(2025){Farah}, {Howell}, {Terreran}, {Irani},
  {Morag}, {Pellegrino}, {McCully}, {Newsome}, {Padilla Gonzalez}, {Bostroem},
  \& {Hosseinzadeh}}]{F2025_2022hnt}
{Farah}, J.~R., {Howell}, D.~A., {Terreran}, G., {et~al.} 2025,
  \bibinfo{title}{{Shock-cooling Constraints via Early-time Observations of the
  Type IIb SN 2022hnt},} arXiv e-prints, arXiv:2501.17221,
  \dodoi{10.48550/arXiv.2501.17221}

\bibitem[{H.~A. {Flewelling} {et~al.}(2020){Flewelling}, {Magnier}, {Chambers},
  {Heasley}, {Holmberg}, {Huber}, {Sweeney}, {Waters}, {Calamida}, {Casertano},
  {Chen}, {Farrow}, {Hasinger}, {Henderson}, {Long}, {Metcalfe}, {Narayan},
  {Nieto-Santisteban}, {Norberg}, {Rest}, {Saglia}, {Szalay}, {Thakar},
  {Tonry}, {Valenti}, {Werner}, {White}, {Denneau}, {Draper}, {Hodapp},
  {Jedicke}, {Kaiser}, {Kudritzki}, {Price}, {Wainscoat}, {Chastel}, {McLean},
  {Postman}, \& {Shiao}}]{Flewelling20}
{Flewelling}, H.~A., {Magnier}, E.~A., {Chambers}, K.~C., {et~al.} 2020,
  \bibinfo{title}{{The Pan-STARRS1 Database and Data Products},} \apjs, 251, 7,
  \dodoi{10.3847/1538-4365/abb82d}

\bibitem[{W. {Fong} {et~al.}(2015){Fong}, {Berger}, {Margutti}, \&
  {Zauderer}}]{2015ApJ...815..102F}
{Fong}, W., {Berger}, E., {Margutti}, R., \& {Zauderer}, B.~A. 2015,
  \bibinfo{title}{{A Decade of Short-duration Gamma-Ray Burst Broadband
  Afterglows: Energetics, Circumburst Densities, and Jet Opening Angles},}
  \apj, 815, 102, \dodoi{10.1088/0004-637X/815/2/102}

\bibitem[{W. {Fong} {et~al.}(2025){Fong}, {Gordon}, {Levan}, {Tanvir}, {Dong},
  {Suresh}, \& {Liu}}]{GCN41419}
{Fong}, W., {Gordon}, A.~C., {Levan}, A.~J., {et~al.} 2025,
  \bibinfo{title}{{GRB 250818B: Keck redshift of the optical afterglow},} GRB
  Coordinates Network, 41419, 1

\bibitem[{W. {Fong} {et~al.}(2014){Fong}, {Berger}, {Metzger}, {Margutti},
  {Chornock}, {Migliori}, {Foley}, {Zauderer}, {Lunnan}, {Laskar}, {Desch},
  {Meech}, {Sonnett}, {Dickey}, {Hedlund}, \& {Harding}}]{2014ApJ...780..118F}
{Fong}, W., {Berger}, E., {Metzger}, B.~D., {et~al.} 2014,
  \bibinfo{title}{{Short GRB 130603B: Discovery of a Jet Break in the Optical
  and Radio Afterglows, and a Mysterious Late-time X-Ray Excess},} \apj, 780,
  118, \dodoi{10.1088/0004-637X/780/2/118}

\bibitem[{W. Fong {et~al.}(2017)Fong, Berger, Blanchard, Margutti,
  Cowperthwaite, Chornock, Alexander, Metzger, Villar, Nicholl, Eftekhari,
  Williams, Annis, Brout, Brown, Chen, Doctor, Diehl, Holz, Rest, Sako, \&
  Soares-Santos}]{fong_electromagnetic_2017}
Fong, W., Berger, E., Blanchard, P.~K., {et~al.} 2017, \bibinfo{title}{The
  {Electromagnetic} {Counterpart} of the {Binary} {Neutron} {Star} {Merger}
  {LIGO}/{Virgo} {GW170817}. {VIII}. {A} {Comparison} to {Cosmological}
  {Short}-duration {Gamma}-{Ray} {Bursts},} The Astrophysical Journal Letters,
  848, L23, \dodoi{10.3847/2041-8213/aa9018}

\bibitem[{W. {Fong} {et~al.}(2021){Fong}, {Laskar}, {Rastinejad}, {Escorial},
  {Schroeder}, {Barnes}, {Kilpatrick}, {Paterson}, {Berger}, {Metzger}, {Dong},
  {Nugent}, {Strausbaugh}, {Blanchard}, {Goyal}, {Cucchiara}, {Terreran},
  {Alexander}, {Eftekhari}, {Fryer}, {Margalit}, {Margutti}, \&
  {Nicholl}}]{2021ApJ...906..127F}
{Fong}, W., {Laskar}, T., {Rastinejad}, J., {et~al.} 2021, \bibinfo{title}{{The
  Broadband Counterpart of the Short GRB 200522A at z = 0.5536: A Luminous
  Kilonova or a Collimated Outflow with a Reverse Shock?},} \apj, 906, 127,
  \dodoi{10.3847/1538-4357/abc74a}

\bibitem[{W.-f. {Fong} {et~al.}(2022){Fong}, {Nugent}, {Dong}, {Berger},
  {Paterson}, {Chornock}, {Levan}, {Blanchard}, {Alexander}, {Andrews}, {Cobb},
  {Cucchiara}, {Fox}, {Fryer}, {Gordon}, {Kilpatrick}, {Lunnan}, {Margutti},
  {Miller}, {Milne}, {Nicholl}, {Perley}, {Rastinejad}, {Escorial},
  {Schroeder}, {Smith}, {Tanvir}, \& {Terreran}}]{2022ApJ...940...56F}
{Fong}, W.-f., {Nugent}, A.~E., {Dong}, Y., {et~al.} 2022,
  \bibinfo{title}{{Short GRB Host Galaxies. I. Photometric and Spectroscopic
  Catalogs, Host Associations, and Galactocentric Offsets},} \apj, 940, 56,
  \dodoi{10.3847/1538-4357/ac91d0}

\bibitem[{J. Forcier(2023)Forcier}]{forcier_paramiko_2023}
Forcier, J. 2023, Paramiko 3.1.0,, GitHub
  \url{https://github.com/paramiko/paramiko}

\bibitem[{N. Franz {et~al.}(2025)Franz, Alexander, Gomez, Christy, Laskar, van
  Velzen, Earl, Gezari, Karmen, Margutti, Pearson, Villar, \&
  Zabludoff}]{franz_open_2025}
Franz, N., Alexander, K.~D., Gomez, S., {et~al.} 2025, The {Open}
  {mulTiwavelength} {Transient} {Event} {Repository} ({OTTER}):
  {Infrastructure} {Release} and {Tidal} {Disruption} {Event} {Catalog}, arXiv,
  \dodoi{10.48550/arXiv.2509.05405}

\bibitem[{J. {Freeburn} {et~al.}(2025){Freeburn}, {O'Connor}, {Hall},
  {Busmann}, {Andreoni}, {Palmese}, {Gruen}, {Hu}, {Cabrera}, {Kunnumkai}, \&
  {Amsellem}}]{2025GCN.41507....1F}
{Freeburn}, J., {O'Connor}, B., {Hall}, X.~J., {et~al.} 2025,
  \bibinfo{title}{{LIGO/Virgo/KAGRA S250818k: Rebrightening detected with
  Gemini/GMOS},} GRB Coordinates Network, 41507, 1

\bibitem[{W.~L. Freedman {et~al.}(2020)Freedman, Madore, Hoyt, Jang, Beaton,
  Lee, Monson, Neeley, \& Rich}]{freedman_calibration_2020}
Freedman, W.~L., Madore, B.~F., Hoyt, T., {et~al.} 2020,
  \bibinfo{title}{Calibration of the {Tip} of the {Red} {Giant} {Branch},} The
  Astrophysical Journal, 891, 57, \dodoi{10.3847/1538-4357/ab7339}

\bibitem[{D. Frostig {et~al.}(2025)Frostig, Karambelkar, Stein, Lourie,
  Kasliwal, Simcoe, Bulla, Ahumada, Mo, Purdum, Juneau, Malonis, \&
  Fűrész}]{frostig_winter_2025}
Frostig, D., Karambelkar, V.~R., Stein, R.~D., {et~al.} 2025,
  \bibinfo{title}{{WINTER} on {S250206dm}: {A} {Near}-infrared {Search} for an
  {Electromagnetic} {Counterpart} to a {Gravitational}-wave {Event},}
  Publications of the Astronomical Society of the Pacific, 137, 074203,
  \dodoi{10.1088/1538-3873/ade478}

\bibitem[{ {Gaia Collaboration} {et~al.}(2023{\natexlab{a}}){Gaia
  Collaboration}, {Vallenari}, {Brown}, {Prusti}, {de Bruijne}, {Arenou},
  {Babusiaux}, {Biermann}, {Creevey}, {Ducourant}, {Evans}, {Eyer}, {Guerra},
  {Hutton}, {Jordi}, {Klioner}, {Lammers}, {Lindegren}, {Luri}, {Mignard},
  {Panem}, {Pourbaix}, {Randich}, {Sartoretti}, {Soubiran}, {Tanga}, {Walton},
  {Bailer-Jones}, {Bastian}, {Drimmel}, {Jansen}, {Katz}, {Lattanzi}, {van
  Leeuwen}, {Bakker}, {Cacciari}, {Casta{\~n}eda}, {De Angeli}, {Fabricius},
  {Fouesneau}, {Fr{\'e}mat}, {Galluccio}, {Guerrier}, {Heiter}, {Masana},
  {Messineo}, {Mowlavi}, {Nicolas}, {Nienartowicz}, {Pailler}, {Panuzzo},
  {Riclet}, {Roux}, {Seabroke}, {Sordo}, {Th{\'e}venin}, {Gracia-Abril},
  {Portell}, {Teyssier}, {Altmann}, {Andrae}, {Audard}, {Bellas-Velidis},
  {Benson}, {Berthier}, {Blomme}, {Burgess}, {Busonero}, {Busso},
  {C{\'a}novas}, {Carry}, {Cellino}, {Cheek}, {Clementini}, {Damerdji},
  {Davidson}, {de Teodoro}, {Nu{\~n}ez Campos}, {Delchambre}, {Dell'Oro},
  {Esquej}, {Fern{\'a}ndez-Hern{\'a}ndez}, {Fraile}, {Garabato},
  {Garc{\'\i}a-Lario}, {Gosset}, {Haigron}, {Halbwachs}, {Hambly}, {Harrison},
  {Hern{\'a}ndez}, {Hestroffer}, {Hodgkin}, {Holl}, {Jan{\ss}en}, {Jevardat de
  Fombelle}, {Jordan}, {Krone-Martins}, {Lanzafame}, {L{\"o}ffler}, {Marchal},
  {Marrese}, {Moitinho}, {Muinonen}, {Osborne}, {Pancino}, {Pauwels},
  {Recio-Blanco}, {Reyl{\'e}}, {Riello}, {Rimoldini}, {Roegiers}, {Rybizki},
  {Sarro}, {Siopis}, {Smith}, {Sozzetti}, {Utrilla}, {van Leeuwen}, {Abbas},
  {{\'A}brah{\'a}m}, {Abreu Aramburu}, {Aerts}, {Aguado}, {Ajaj},
  {Aldea-Montero}, {Altavilla}, {{\'A}lvarez}, {Alves}, {Anders}, {Anderson},
  {Anglada Varela}, {Antoja}, {Baines}, {Baker}, {Balaguer-N{\'u}{\~n}ez},
  {Balbinot}, {Balog}, {Barache}, {Barbato}, {Barros}, {Barstow},
  {Bartolom{\'e}}, {Bassilana}, {Bauchet}, {Becciani}, {Bellazzini},
  {Berihuete}, {Bernet}, {Bertone}, {Bianchi}, {Binnenfeld}, {Blanco-Cuaresma},
  {Blazere}, {Boch}, {Bombrun}, {Bossini}, {Bouquillon}, {Bragaglia},
  {Bramante}, {Breedt}, {Bressan}, {Brouillet}, {Brugaletta}, {Bucciarelli},
  {Burlacu}, {Butkevich}, {Buzzi}, {Caffau}, {Cancelliere}, {Cantat-Gaudin},
  {Carballo}, {Carlucci}, {Carnerero}, {Carrasco}, {Casamiquela}, {Castellani},
  {Castro-Ginard}, {Chaoul}, {Charlot}, {Chemin}, {Chiaramida}, {Chiavassa},
  {Chornay}, {Comoretto}, {Contursi}, {Cooper}, {Cornez}, {Cowell}, {Crifo},
  {Cropper}, {Crosta}, {Crowley}, {Dafonte}, {Dapergolas}, {David}, {David},
  {de Laverny}, {De Luise}, \& {De March}}]{gaiadr3}
{Gaia Collaboration}, {Vallenari}, A., {Brown}, A.~G.~A., {et~al.}
  2023{\natexlab{a}}, \bibinfo{title}{{Gaia Data Release 3. Summary of the
  content and survey properties},} \aap, 674, A1,
  \dodoi{10.1051/0004-6361/202243940}

\bibitem[{ {Gaia Collaboration} {et~al.}(2023{\natexlab{b}}){Gaia
  Collaboration}, Vallenari, Brown, Prusti, de~Bruijne, Arenou, Babusiaux,
  Biermann, Creevey, Ducourant, Evans, Eyer, Guerra, Hutton, Jordi, Klioner,
  Lammers, Lindegren, Luri, Mignard, Panem, Pourbaix, Randich, Sartoretti,
  Soubiran, Tanga, Walton, Bailer-Jones, Bastian, Drimmel, Jansen, Katz,
  Lattanzi, van Leeuwen, Bakker, Cacciari, Castañeda, De~Angeli, Fabricius,
  Fouesneau, Frémat, Galluccio, Guerrier, Heiter, Masana, Messineo, Mowlavi,
  Nicolas, Nienartowicz, Pailler, Panuzzo, Riclet, Roux, Seabroke, Sordo,
  Thévenin, Gracia-Abril, Portell, Teyssier, Altmann, Andrae, Audard,
  Bellas-Velidis, Benson, Berthier, Blomme, Burgess, Busonero, Busso, Cánovas,
  Carry, Cellino, Cheek, Clementini, Damerdji, Davidson, de~Teodoro,
  Nuñez~Campos, Delchambre, Dell'Oro, Esquej, Fernández-Hernández, Fraile,
  Garabato, García-Lario, Gosset, Haigron, Halbwachs, Hambly, Harrison,
  Hernández, Hestroffer, Hodgkin, Holl, Janßen, Jevardat~de Fombelle, Jordan,
  Krone-Martins, Lanzafame, Löffler, Marchal, Marrese, Moitinho, Muinonen,
  Osborne, Pancino, Pauwels, Recio-Blanco, Reylé, Riello, Rimoldini, Roegiers,
  Rybizki, Sarro, Siopis, Smith, Sozzetti, Utrilla, van Leeuwen, Abbas,
  Ábrahám, Abreu~Aramburu, Aerts, Aguado, Ajaj, Aldea-Montero, Altavilla,
  Álvarez, Alves, Anders, Anderson, Anglada~Varela, Antoja, Baines, Baker,
  Balaguer-Núñez, Balbinot, Balog, Barache, Barbato, Barros, Barstow,
  Bartolomé, Bassilana, Bauchet, Becciani, Bellazzini, Berihuete, Bernet,
  Bertone, Bianchi, Binnenfeld, Blanco-Cuaresma, Blazere, Boch, Bombrun,
  Bossini, Bouquillon, Bragaglia, Bramante, Breedt, Bressan, Brouillet,
  Brugaletta, Bucciarelli, Burlacu, Butkevich, Buzzi, Caffau, Cancelliere,
  Cantat-Gaudin, Carballo, Carlucci, Carnerero, Carrasco, Casamiquela,
  Castellani, Castro-Ginard, Chaoul, Charlot, Chemin, Chiaramida, Chiavassa,
  Chornay, Comoretto, Contursi, Cooper, Cornez, Cowell, Crifo, Cropper, Crosta,
  Crowley, Dafonte, Dapergolas, David, David, de~Laverny, De~Luise, \&
  De~March}]{gaia_collaboration_gaia_2023}
{Gaia Collaboration}, Vallenari, A., Brown, A. G.~A., {et~al.}
  2023{\natexlab{b}}, \bibinfo{title}{Gaia {Data} {Release} 3. {Summary} of the
  content and survey properties,} Astronomy and Astrophysics, 674, A1,
  \dodoi{10.1051/0004-6361/202243940}

\bibitem[{A. {Garcia} {et~al.}(2020){Garcia}, {Morgan}, {Herner}, {Palmese},
  {Soares-Santos}, {Annis}, {Brout}, {Vivas}, {Drlica-Wagner}, {Santana-Silva},
  {Tucker}, {Allam}, {Wiesner}, {Garc{\'\i}a-Bellido}, {Gill}, {Sako},
  {Kessler}, {Davis}, {Scolnic}, {Casares}, {Chen}, {Conselice}, {Cooke},
  {Doctor}, {Foley}, {Horvath}, {Howell}, {Kilpatrick}, {Lidman}, {Olivares
  E.}, {Paz-Chinch{\'o}n}, {Pineda-G.}, {Quirola-V{\'a}squez}, {Rest},
  {Sherman}, {Abbott}, {Aguena}, {Avila}, {Bertin}, {Bhargava}, {Brooks},
  {Burke}, {Carnero Rosell}, {Carrasco Kind}, {Carretero}, {Costanzi}, {da
  Costa}, {Desai}, {Diehl}, {Dietrich}, {Doel}, {Everett}, {Flaugher},
  {Fosalba}, {Friedel}, {Frieman}, {Gaztanaga}, {Gerdes}, {Gruen}, {Gruendl},
  {Gschwend}, {Gutierrez}, {Hinton}, {Hollowood}, {Honscheid}, {James},
  {Kuehn}, {Kuropatkin}, {Lahav}, {Lima}, {Maia}, {March}, {Marshall},
  {Menanteau}, {Miquel}, {Ogando}, {Plazas}, {Romer}, {Roodman}, {Sanchez},
  {Scarpine}, {Schubnell}, {Serrano}, {Sevilla-Noarbe}, {Smith}, {Suchyta},
  {Swanson}, {Tarle}, {Thomas}, {Varga}, {Walker}, {Weller}, \& {DES
  Collaboration}}]{Garcia+20}
{Garcia}, A., {Morgan}, R., {Herner}, K., {et~al.} 2020, \bibinfo{title}{{A
  DESGW Search for the Electromagnetic Counterpart to the LIGO/Virgo
  Gravitational-wave Binary Neutron Star Merger Candidate S190510g},} \apj,
  903, 75, \dodoi{10.3847/1538-4357/abb823}

\bibitem[{G. {Ghirlanda} {et~al.}(2019){Ghirlanda}, {Salafia}, {Paragi},
  {Giroletti}, {Yang}, {Marcote}, {Blanchard}, {Agudo}, {An}, {Bernardini},
  {Beswick}, {Branchesi}, {Campana}, {Casadio}, {Chassande-Mottin}, {Colpi},
  {Covino}, {D'Avanzo}, {D'Elia}, {Frey}, {Gawronski}, {Ghisellini}, {Gurvits},
  {Jonker}, {van Langevelde}, {Melandri}, {Moldon}, {Nava}, {Perego},
  {Perez-Torres}, {Reynolds}, {Salvaterra}, {Tagliaferri}, {Venturi},
  {Vergani}, \& {Zhang}}]{2019Sci...363..968G}
{Ghirlanda}, G., {Salafia}, O.~S., {Paragi}, Z., {et~al.} 2019,
  \bibinfo{title}{{Compact radio emission indicates a structured jet was
  produced by a binary neutron star merger},} Science, 363, 968,
  \dodoi{10.1126/science.aau8815}

\bibitem[{C. Gibson(2021)Gibson}]{gibson_django_2021}
Gibson, C. 2021, Django {{Filter}} v21.1,, GitHub
  \url{https://github.com/carltongibson/django-filter}

\bibitem[{J.~H. Gillanders {et~al.}(2025)Gillanders, Huber, Nicholl, Smartt,
  Smith, Chambers, Young, Tweddle, Srivastav, Fulton, Stoppa, Paek, Aamer,
  Alarcon, Andersson, Aryan, Auchettl, Chen, Boer, Kong, Licandro, Lowe,
  Magill, Magnier, Minguez, Moore, Pignata, Rest, Serra-Ricart, Shappee, Smith,
  Tucker, \& Wainscoat}]{gillanders_pan-starrs_2025}
Gillanders, J.~H., Huber, M.~E., Nicholl, M., {et~al.} 2025, Pan-{STARRS}
  follow-up of the gravitational-wave event {S250818k} and the lightcurve of
  {SN} 2025ulz, arXiv, \dodoi{10.48550/arXiv.2510.01142}

\bibitem[{J.~H. {Gillanders} {et~al.}(2025{\natexlab{a}}){Gillanders}, {Huber},
  {Chambers}, {Smartt}, {Smith}, {Srivastav}, {Stoppa}, {Stevance}, {Tweddle},
  {Nicholl}, {Young}, {Aamer}, {Angus}, {Fulton}, {Magill}, {McCollum},
  {Moore}, {Sim}, {Weston}, {Sheng}, {Chen}, {Shingles}, {Ramsden}, {Schultz},
  {de Boer}, {Fairlamb}, {Lin}, {Lowe}, {Magnier}, {Minguez}, {Paek}, {Smith},
  {Wainscoat}, {Rest}, \& {Stubbs}}]{2025GCN.41540....1G}
{Gillanders}, J.~H., {Huber}, M.~E., {Chambers}, K.~C., {et~al.}
  2025{\natexlab{a}}, \bibinfo{title}{{LIGO/Virgo/KAGRA S250818k: Pan-STARRS
  imaging confirms re-brightening of SN2025ulz},} GRB Coordinates Network,
  41540, 1

\bibitem[{J.~H. {Gillanders} {et~al.}(2025{\natexlab{b}}){Gillanders}, {Huber},
  {Chambers}, {Smartt}, {Smith}, {Srivastav}, {Stoppa}, {Stevance}, {Tweddle},
  {Nicholl}, {Young}, {Aamer}, {Angus}, {Fulton}, {Magill}, {McCollum},
  {Moore}, {Sim}, {Weston}, {Sheng}, {Chen}, {Shingles}, {Ramsden}, {Schultz},
  {de Boer}, {Fairlamb}, {Lin}, {Lowe}, {Magnier}, {Minguez}, {Paek}, {Smith},
  {Wainscoat}, {Rest}, \& {Stubbs}}]{2025GCN.41454....1G}
{Gillanders}, J.~H., {Huber}, M.~E., {Chambers}, K.~C., {et~al.}
  2025{\natexlab{b}}, \bibinfo{title}{{LIGO/Virgo/KAGRA S250818k: Pan-STARRS
  grizy-band imaging and photometry of AT2025ulz},} GRB Coordinates Network,
  41454, 1

\bibitem[{A. Ginsburg {et~al.}(2019)Ginsburg, Sip{\H o}cz, Brasseur,
  Cowperthwaite, Craig, Deil, Guillochon, Guzman, Liedtke, Lian~Lim, Lockhart,
  Mommert, Morris, Norman, Parikh, Persson, Robitaille, Segovia, Singer,
  Tollerud, {de Val-Borro}, Valtchanov, Woillez, {Astroquery Collaboration}, \&
  {a subset of astropy Collaboration}}]{ginsburg_astroquery_2019}
Ginsburg, A., Sip{\H o}cz, B.~M., Brasseur, C.~E., {et~al.} 2019,
  \bibinfo{title}{astroquery: {{An Astronomical Web-querying Package}} in
  {{Python}},} AJ, 157, 98, \dodoi{10.3847/1538-3881/aafc33}

\bibitem[{P. Godwin(2022)Godwin}]{godwin_hop_2022}
Godwin, P. 2022, Hop {{Client}} v0.8.0,, GitHub
  \url{https://github.com/scimma/hop-client}

\bibitem[{A. Goldstein {et~al.}(2017)Goldstein, Veres, Burns, Briggs, Hamburg,
  Kocevski, Wilson-Hodge, Preece, Poolakkil, Roberts, Hui, Connaughton,
  Racusin, Kienlin, Canton, Christensen, Littenberg, Siellez, Blackburn,
  Broida, Bissaldi, Cleveland, Gibby, Giles, Kippen, McBreen, McEnery, Meegan,
  Paciesas, \& Stanbro}]{goldstein_ordinary_2017}
Goldstein, A., Veres, P., Burns, E., {et~al.} 2017, \bibinfo{title}{An
  {Ordinary} {Short} {Gamma}-{Ray} {Burst} with {Extraordinary} {Implications}:
  {Fermi}-{GBM} {Detection} of {GRB} {170817A},} The Astrophysical Journal
  Letters, 848, L14, \dodoi{10.3847/2041-8213/aa8f41}

\bibitem[{D.~A. {Goldstein} {et~al.}(2019){Goldstein}, {Andreoni}, {Nugent},
  {Kasliwal}, {Coughlin}, {Anand}, {Bloom}, {Mart{\'\i}nez-Palomera}, {Zhang},
  {Ahumada}, {Bagdasaryan}, {Cooke}, {De}, {Duev}, {Fremling}, {Gatkine},
  {Graham}, {Ofek}, {Singer}, \& {Yan}}]{Goldstein+19}
{Goldstein}, D.~A., {Andreoni}, I., {Nugent}, P.~E., {et~al.} 2019,
  \bibinfo{title}{{GROWTH on S190426c: Real-time Search for a Counterpart to
  the Probable Neutron Star-Black Hole Merger using an Automated Difference
  Imaging Pipeline for DECam},} \apjl, 881, L7,
  \dodoi{10.3847/2041-8213/ab3046}

\bibitem[{S. {Gomez} {et~al.}(2019){Gomez}, {Hosseinzadeh}, {Cowperthwaite},
  {Villar}, {Berger}, {Gardner}, {Alexander}, {Blanchard}, {Chornock}, {Drout},
  {Eftekhari}, {Fong}, {Gill}, {Margutti}, {Nicholl}, {Paterson}, \&
  {Williams}}]{Gomez+19}
{Gomez}, S., {Hosseinzadeh}, G., {Cowperthwaite}, P.~S., {et~al.} 2019,
  \bibinfo{title}{{A Galaxy-targeted Search for the Optical Counterpart of the
  Candidate NS-BH Merger S190814bv with Magellan},} \apjl, 884, L55,
  \dodoi{10.3847/2041-8213/ab4ad5}

\bibitem[{B.~P. {Gompertz} {et~al.}(2020){Gompertz}, {Levan}, \&
  {Tanvir}}]{Gompertz+20}
{Gompertz}, B.~P., {Levan}, A.~J., \& {Tanvir}, N.~R. 2020, \bibinfo{title}{{A
  Search for Neutron Star-Black Hole Binary Mergers in the Short Gamma-Ray
  Burst Population},} \apj, 895, 58, \dodoi{10.3847/1538-4357/ab8d24}

\bibitem[{G. {Green}(2018){Green}}]{2018JOSS....3..695M}
{Green}, G. 2018, \bibinfo{title}{{dustmaps: A Python interface for maps of
  interstellar dust},} The Journal of Open Source Software, 3, 695,
  \dodoi{10.21105/joss.00695}

\bibitem[{J. Guillochon {et~al.}(2018)Guillochon, Nicholl, Villar, Mockler,
  Narayan, Mandel, Berger, \& Williams}]{guillochon_mosfit_2018}
Guillochon, J., Nicholl, M., Villar, V.~A., {et~al.} 2018,
  \bibinfo{title}{{MOSFiT}: {Modular} {Open} {Source} {Fitter} for
  {Transients},} The Astrophysical Journal Supplement Series, 236, 6,
  \dodoi{10.3847/1538-4365/aab761}

\bibitem[{A. {Hajela} {et~al.}(2019){Hajela}, {Margutti}, {Alexander},
  {Kathirgamaraju}, {Baldeschi}, {Guidorzi}, {Giannios}, {Fong}, {Wu},
  {MacFadyen}, {Paggi}, {Berger}, {Blanchard}, {Chornock}, {Coppejans},
  {Cowperthwaite}, {Eftekhari}, {Gomez}, {Hosseinzadeh}, {Laskar}, {Metzger},
  {Nicholl}, {Paterson}, {Radice}, {Sironi}, {Terreran}, {Villar}, {Williams},
  {Xie}, \& {Zrake}}]{2019ApJ...886L..17H}
{Hajela}, A., {Margutti}, R., {Alexander}, K.~D., {et~al.} 2019,
  \bibinfo{title}{{Two Years of Nonthermal Emission from the Binary Neutron
  Star Merger GW170817: Rapid Fading of the Jet Afterglow and First Constraints
  on the Kilonova Fastest Ejecta},} \apjl, 886, L17,
  \dodoi{10.3847/2041-8213/ab5226}

\bibitem[{X.~J. {Hall} {et~al.}(2025){Hall}, {Busmann}, {Gruen}, {O'Connor}, \&
  {Palmese}}]{2025GCN.41433....1H}
{Hall}, X.~J., {Busmann}, M., {Gruen}, D., {O'Connor}, B., \& {Palmese}, A.
  2025, \bibinfo{title}{{LIGO/Virgo/KAGRA S250818k: FTW Fast reddening of AT
  2025ulz},} GRB Coordinates Network, 41433, 1

\bibitem[{G. {Hallinan} {et~al.}(2017){Hallinan}, {Corsi}, {Mooley},
  {Hotokezaka}, {Nakar}, {Kasliwal}, {Kaplan}, {Frail}, {Myers}, {Murphy},
  {De}, {Dobie}, {Allison}, {Bannister}, {Bhalerao}, {Chandra}, {Clarke},
  {Giacintucci}, {Ho}, {Horesh}, {Kassim}, {Kulkarni}, {Lenc}, {Lockman},
  {Lynch}, {Nichols}, {Nissanke}, {Palliyaguru}, {Peters}, {Piran}, {Rana},
  {Sadler}, \& {Singer}}]{2017Sci...358.1579H}
{Hallinan}, G., {Corsi}, A., {Mooley}, K.~P., {et~al.} 2017, \bibinfo{title}{{A
  radio counterpart to a neutron star merger},} Science, 358, 1579,
  \dodoi{10.1126/science.aap9855}

\bibitem[{C.~R. Harris {et~al.}(2020)Harris, Millman, {van der Walt}, Gommers,
  Virtanen, Cournapeau, Wieser, Taylor, Berg, Smith, Kern, Picus, Hoyer, {van
  Kerkwijk}, Brett, Haldane, {del R{\'i}o}, Wiebe, Peterson,
  {G{\'e}rard-Marchant}, Sheppard, Reddy, Weckesser, Abbasi, Gohlke, \&
  Oliphant}]{harris_array_2020}
Harris, C.~R., Millman, K.~J., {van der Walt}, S.~J., {et~al.} 2020,
  \bibinfo{title}{Array programming with {{NumPy}},} Natur, 585, 357,
  \dodoi{10.1038/s41586-020-2649-2}

\bibitem[{G. Hosseinzadeh {et~al.}(2024)Hosseinzadeh, Bostroem, Ben-Ami, \&
  Gomez}]{hosseinzadeh_2024_11405219}
Hosseinzadeh, G., Bostroem, K.~A., Ben-Ami, T., \& Gomez, S. 2024, Light Curve
  Fitting v0.10.0, v0.10.0 Zenodo, \dodoi{10.5281/zenodo.11405219}

\bibitem[{G. Hosseinzadeh \& S. Gomez(2022)Hosseinzadeh \&
  Gomez}]{hosseinzadeh_light_2022}
Hosseinzadeh, G., \& Gomez, S. 2022, Light {{Curve Fitting}} v0.7.0,, Zenodo
  \dodoi{10.5281/zenodo.4312178}

\bibitem[{G. Hosseinzadeh {et~al.}(2023)Hosseinzadeh, Rastinejad, \&
  Shrestha}]{hosseinzadeh_saguaro_2023}
Hosseinzadeh, G., Rastinejad, J., \& Shrestha, M. 2023, {{SAGUARO Target}} and
  {{Observation Manager}} v1.0.0,, Zenodo \dodoi{10.5281/zenodo.8436090}

\bibitem[{G. {Hosseinzadeh} {et~al.}(2019){Hosseinzadeh}, {Cowperthwaite},
  {Gomez}, {Villar}, {Nicholl}, {Margutti}, {Berger}, {Chornock}, {Paterson},
  {Fong}, {Savchenko}, {Short}, {Alexander}, {Blanchard}, {Braga}, {Calkins},
  {Cartier}, {Coppejans}, {Eftekhari}, {Laskar}, {Ly}, {Patton}, {Pelisoli},
  {Reichart}, {Terreran}, \& {Williams}}]{Hosseinzadeh+19}
{Hosseinzadeh}, G., {Cowperthwaite}, P.~S., {Gomez}, S., {et~al.} 2019,
  \bibinfo{title}{{Follow-up of the Neutron Star Bearing Gravitational-wave
  Candidate Events S190425z and S190426c with MMT and SOAR},} \apjl, 880, L4,
  \dodoi{10.3847/2041-8213/ab271c}

\bibitem[{G. {Hosseinzadeh} {et~al.}(2024){Hosseinzadeh}, {Paterson},
  {Rastinejad}, {Shrestha}, {Daly}, {Lundquist}, {Sand}, {Fong}, {Bostroem},
  {Hall}, {Wyatt}, {Gibbs}, {Christensen}, {Lindstrom}, {Nation}, {Chatelain},
  \& {McCully}}]{2024ApJ...964...35H}
{Hosseinzadeh}, G., {Paterson}, K., {Rastinejad}, J.~C., {et~al.} 2024,
  \bibinfo{title}{{SAGUARO: Time-domain Infrastructure for the Fourth
  Gravitational-wave Observing Run and Beyond},} \apj, 964, 35,
  \dodoi{10.3847/1538-4357/ad2170}

\bibitem[{L. Hu {et~al.}(2025)Hu, Cabrera, Palmese, Freeburn, Bulla, Andreoni,
  Hall, O'Connor, Amsellem, Bom, Busmann, Fabà, Gassert, Kalabalik, Kunnumkai,
  Gruen, Santana-Silva, Santos, Ahumada, Carney, Coughlin, Chen, Ford, Holz,
  Kasliwal, Magaña~Hernandez, Mihalenko, Perna, Riffeser, Ries, Schnappinger,
  Schmidt, Sommer, Teague, Vega, Volchansky, Wang, \& Zhang}]{hu_kilonova_2025}
Hu, L., Cabrera, T., Palmese, A., {et~al.} 2025, \bibinfo{title}{Kilonova
  {Constraints} for the {LIGO}/{Virgo}/{KAGRA} {Neutron} {Star} {Merger}
  {Candidate} {S250206dm}: {GW}-{MMADS} {Observations},} The Astrophysical
  Journal, 990, L46, \dodoi{10.3847/2041-8213/adfd49}

\bibitem[{J.~D. Hunter(2007)Hunter}]{hunter_matplotlib:_2007}
Hunter, J.~D. 2007, \bibinfo{title}{Matplotlib: {{A 2D}} graphics environment,}
  CSE, 9, 90, \dodoi{10.1109/MCSE.2007.55}

\bibitem[{D.~O. Jones {et~al.}(2024)Jones, McGill, Manning, Gagliano, Wang,
  Coulter, Foley, Narayan, Villar, Braff, Engel, Farias, Lai, Loertscher,
  Kutcka, Thorp, \& Vazquez}]{jones_blast_2024}
Jones, D.~O., McGill, P., Manning, T.~A., {et~al.} 2024, Blast: a {Web}
  {Application} for {Characterizing} the {Host} {Galaxies} of {Astrophysical}
  {Transients}, arXiv, \dodoi{10.48550/arXiv.2410.17322}

\bibitem[{J. {Kansky} {et~al.}(2019){Kansky}, {Chilingarian}, {Fabricant},
  {Matthews}, {Moran}, {Paegert}, {Duane Gibson}, {Porter}, \&
  {Roll}}]{Kansky2019}
{Kansky}, J., {Chilingarian}, I., {Fabricant}, D., {et~al.} 2019,
  \bibinfo{title}{{Binospec Software System},} \pasp, 131, 075005,
  \dodoi{10.1088/1538-3873/ab1ceb}

\bibitem[{V. {Karambelkar} {et~al.}(2025){Karambelkar}, {Kasliwal}, {Hall},
  {Ztf Collaboration}, \& {Growth Collaboration}}]{2025GCN.41436....1K}
{Karambelkar}, V., {Kasliwal}, M.~M., {Hall}, X.~J., {Ztf Collaboration}, \&
  {Growth Collaboration}. 2025, \bibinfo{title}{{LIGO/Virgo/KAGRA S250818k:
  Keck I LRIS spectroscopy of ZTF25abjmnps (AT2025ulz)},} GRB Coordinates
  Network, 41436, 1

\bibitem[{D. Kasen {et~al.}(2015)Kasen, Fernández, \&
  Metzger}]{kasen_kilonova_2015}
Kasen, D., Fernández, R., \& Metzger, B.~D. 2015, \bibinfo{title}{Kilonova
  light curves from the disc wind outflows of compact object mergers,} Monthly
  Notices of the Royal Astronomical Society, 450, 1777,
  \dodoi{10.1093/mnras/stv721}

\bibitem[{D. Kasen {et~al.}(2017)Kasen, Metzger, Barnes, Quataert, \&
  Ramirez-Ruiz}]{kasen_origin_2017}
Kasen, D., Metzger, B., Barnes, J., Quataert, E., \& Ramirez-Ruiz, E. 2017,
  \bibinfo{title}{Origin of the heavy elements in binary neutron-star mergers
  from a gravitational-wave event,} Nature, 551, 80,
  \dodoi{10.1038/nature24453}

\bibitem[{M.~M. Kasliwal {et~al.}(2017)Kasliwal, Nakar, Singer, Kaplan, Cook,
  Van~Sistine, Lau, Fremling, Gottlieb, Jencson, Adams, Feindt, Hotokezaka,
  Ghosh, Perley, Yu, Piran, Allison, Anupama, Balasubramanian, Bannister,
  Bally, Barnes, Barway, Bellm, Bhalerao, Bhattacharya, Blagorodnova, Bloom,
  Brady, Cannella, Chatterjee, Cenko, Cobb, Copperwheat, Corsi, De, Dobie,
  Emery, Evans, Fox, Frail, Frohmaier, Goobar, Hallinan, Harrison, Helou,
  Hinderer, Ho, Horesh, Ip, Itoh, Kasen, Kim, Kuin, Kupfer, Lynch, Madsen,
  Mazzali, Miller, Mooley, Murphy, Ngeow, Nichols, Nissanke, Nugent, Ofek, Qi,
  Quimby, Rosswog, Rusu, Sadler, Schmidt, Sollerman, Steele, Williamson, Xu,
  Yan, Yatsu, Zhang, \& Zhao}]{kasliwal_illuminating_2017}
Kasliwal, M.~M., Nakar, E., Singer, L.~P., {et~al.} 2017,
  \bibinfo{title}{Illuminating gravitational waves: {A} concordant picture of
  photons from a neutron star merger,} Science, 358, 1559,
  \dodoi{10.1126/science.aap9455}

\bibitem[{M.~M. {Kasliwal} {et~al.}(2020){Kasliwal}, {Anand}, {Ahumada},
  {Stein}, {Carracedo}, {Andreoni}, {Coughlin}, {Singer}, {Kool}, {De},
  {Kumar}, {AlMualla}, {Yao}, {Bulla}, {Dobie}, {Reusch}, {Perley}, {Cenko},
  {Bhalerao}, {Kaplan}, {Sollerman}, {Goobar}, {Copperwheat}, {Bellm},
  {Anupama}, {Corsi}, {Nissanke}, {Agudo}, {Bagdasaryan}, {Barway}, {Belicki},
  {Bloom}, {Bolin}, {Buckley}, {Burdge}, {Burruss}, {Caballero-Garc{\'\i}a},
  {Cannella}, {Castro-Tirado}, {Cook}, {Cooke}, {Cunningham}, {Dahiwale},
  {Deshmukh}, {Dichiara}, {Duev}, {Dutta}, {Feeney}, {Franckowiak},
  {Frederick}, {Fremling}, {Gal-Yam}, {Gatkine}, {Ghosh}, {Goldstein},
  {Golkhou}, {Graham}, {Graham}, {Hankins}, {Helou}, {Hu}, {Ip}, {Jaodand},
  {Karambelkar}, {Kong}, {Kowalski}, {Khandagale}, {Kulkarni}, {Kumar},
  {Laher}, {Li}, {Mahabal}, {Masci}, {Miller}, {Mogotsi}, {Mohite}, {Mooley},
  {Mroz}, {Newman}, {Ngeow}, {Oates}, {Patil}, {Pandey}, {Pavana}, {Pian},
  {Riddle}, {S{\'a}nchez-Ram{\'\i}rez}, {Sharma}, {Singh}, {Smith},
  {Soumagnac}, {Taggart}, {Tan}, {Tzanidakis}, {Troja}, {Valeev}, {Walters},
  {Waratkar}, {Webb}, {Yu}, {Zhang}, {Zhou}, \& {Zolkower}}]{Kasliwal+20}
{Kasliwal}, M.~M., {Anand}, S., {Ahumada}, T., {et~al.} 2020,
  \bibinfo{title}{{Kilonova Luminosity Function Constraints Based on Zwicky
  Transient Facility Searches for 13 Neutron Star Merger Triggers during O3},}
  \apj, 905, 145, \dodoi{10.3847/1538-4357/abc335}

\bibitem[{M.~M. {Kasliwal} {et~al.}(2025){Kasliwal}, {Karambelkar}, {Fremling},
  {Ahumada}, {Hall}, {Perley}, {Anand}, {Liu}, {Das}, {Bhalerao}, {Swain},
  {Saikia}, {Ztf Collaboration}, \& {Growth
  Collaboration}}]{2025GCN.41538....1K}
{Kasliwal}, M.~M., {Karambelkar}, V., {Fremling}, C., {et~al.} 2025,
  \bibinfo{title}{{LIGO/Virgo/KAGRA S250818k: Continued Keck I LRIS
  spectroscopy of ZTF25abjmnps (AT2025ulz)},} GRB Coordinates Network, 41538, 1

\bibitem[{R.~C. Kennicutt \& N.~J. Evans(2012)Kennicutt \&
  Evans}]{kennicutt_star_2012}
Kennicutt, R.~C., \& Evans, N.~J. 2012, \bibinfo{title}{Star {Formation} in the
  {Milky} {Way} and {Nearby} {Galaxies},} Annual Review of Astronomy and
  Astrophysics, 50, 531, \dodoi{10.1146/annurev-astro-081811-125610}

\bibitem[{R.~C. Kennicutt(1998)Kennicutt}]{kennicutt_star_1998}
Kennicutt, Jr., R.~C. 1998, \bibinfo{title}{{STAR} {FORMATION} {IN} {GALAXIES}
  {ALONG} {THE} {HUBBLE} {SEQUENCE},} Annual Review of Astronomy and
  Astrophysics, 36, 189, \dodoi{10.1146/annurev.astro.36.1.189}

\bibitem[{C.~D. {Kilpatrick}(2021){Kilpatrick}}]{hst123}
{Kilpatrick}, C.~D. 2021, {charliekilpatrick/hst123: hst123}, v1.0.0 Zenodo,
  \dodoi{10.5281/zenodo.5573941}

\bibitem[{C.~D. Kilpatrick {et~al.}(2017)Kilpatrick, Foley, Kasen,
  Murguia-Berthier, Ramirez-Ruiz, Coulter, Drout, Piro, Shappee, Boutsia,
  Contreras, Di~Mille, Madore, Morrell, Pan, Prochaska, Rest, Rojas-Bravo,
  Siebert, Simon, \& Ulloa}]{kilpatrick_electromagnetic_2017}
Kilpatrick, C.~D., Foley, R.~J., Kasen, D., {et~al.} 2017,
  \bibinfo{title}{Electromagnetic evidence that {SSS17a} is the result of a
  binary neutron star merger,} Science, 358, 1583,
  \dodoi{10.1126/science.aaq0073}

\bibitem[{C.~D. Kilpatrick {et~al.}(2021)Kilpatrick, Coulter, Arcavi, Brink,
  Dimitriadis, Filippenko, Foley, Howell, Jones, Kasen, Makler, Piro,
  Rojas-Bravo, Sand, Swift, Tucker, Zheng, Allam, Annis, Antilen, Bachmann,
  Bloom, Bom, Bostroem, Brout, Burke, Butler, Butner, Campillay, Clever,
  Conselice, Cooke, Dage, de~Carvalho, de~Jaeger, Desai, Garcia,
  Garcia-Bellido, Gill, Girish, Hallakoun, Herner, Hiramatsu, Holz, Huber,
  Kawash, McCully, Medallon, Metzger, Modak, Morgan, Muñoz, Muñoz-Elgueta,
  Murakami, Felipe~Olivares, Palmese, Patra, Pereira, Pessi, Pineda-Garcia,
  Quirola-Vásquez, Ramirez-Ruiz, Rembold, Rest, Rodríguez, Santana-Silva,
  Sherman, Siebert, Smith, Smith, Soares-Santos, Stacey, Stahl, Strader,
  Strasburger, Sunseri, Tinyanont, Tucker, Ulloa, Valenti, Vasylyev, Wiesner,
  \& Zhang}]{kilpatrick_gravity_2021}
Kilpatrick, C.~D., Coulter, D.~A., Arcavi, I., {et~al.} 2021,
  \bibinfo{title}{The {Gravity} {Collective}: {A} {Search} for the
  {Electromagnetic} {Counterpart} to the {Neutron} {Star}-{Black} {Hole}
  {Merger} {GW190814},} The Astrophysical Journal, 923, 258,
  \dodoi{10.3847/1538-4357/ac23c6}

\bibitem[{C.~D. {Kilpatrick} {et~al.}(2021){Kilpatrick}, {Drout}, {Auchettl},
  {Dimitriadis}, {Foley}, {Jones}, {DeMarchi}, {French}, {Gall}, {Hjorth},
  {Jacobson-Gal{\'a}n}, {Margutti}, {Piro}, {Ramirez-Ruiz}, {Rest}, \&
  {Rojas-Bravo}}]{Kilpatrick_2019yvr}
{Kilpatrick}, C.~D., {Drout}, M.~R., {Auchettl}, K., {et~al.} 2021,
  \bibinfo{title}{{A cool and inflated progenitor candidate for the Type Ib
  supernova 2019yvr at 2.6 yr before explosion},} \mnras, 504, 2073,
  \dodoi{10.1093/mnras/stab838}

\bibitem[{S. {Kim} {et~al.}(2017){Kim}, {Schulze}, {Resmi},
  {Gonz{\'a}lez-L{\'o}pez}, {Higgins}, {Ishwara-Chandra}, {Bauer}, {de
  Gregorio-Monsalvo}, {De Pasquale}, {de Ugarte Postigo}, {Kann},
  {Mart{\'\i}n}, {Oates}, {Starling}, {Tanvir}, {Buchner}, {Campana}, {Cano},
  {Covino}, {Fruchter}, {Fynbo}, {Hartmann}, {Hjorth}, {Jakobsson}, {Levan},
  {Malesani}, {Micha{\l}owski}, {Milvang-Jensen}, {Misra}, {O'Brien},
  {S{\'a}nchez-Ram{\'\i}rez}, {Th{\"o}ne}, {Watson}, \&
  {Wiersema}}]{2017ApJ...850L..21K}
{Kim}, S., {Schulze}, S., {Resmi}, L., {et~al.} 2017, \bibinfo{title}{{ALMA and
  GMRT Constraints on the Off-axis Gamma-Ray Burst 170817A from the Binary
  Neutron Star Merger GW170817},} \apjl, 850, L21,
  \dodoi{10.3847/2041-8213/aa970b}

\bibitem[{S. Koposov \& O. Bartunov(2006)Koposov \&
  Bartunov}]{koposov_q3c_2006}
Koposov, S., \& Bartunov, O. 2006, \bibinfo{title}{{{Q3C}}, {{Quad Tree Cube}}
  -- {{The}} new {{Sky-indexing Concept}} for {{Huge Astronomical Catalogues}}
  and its {{Realization}} for {{Main Astronomical Queries}} ({{Cone Search}}
  and {{Xmatch}}) in {{Open Source Database PostgreSQL}},} ASPC, 351, 735.
\newblock \url{https://ui.adsabs.harvard.edu/abs/2006ASPC..351..735K}

\bibitem[{K. {Kovlakas} {et~al.}(2021){Kovlakas}, {Zezas}, {Andrews},
  {Basu-Zych}, {Fragos}, {Hornschemeier}, {Kouroumpatzakis}, {Lehmer}, \&
  {Ptak}}]{Kovlakas+21_HECATE}
{Kovlakas}, K., {Zezas}, A., {Andrews}, J.~J., {et~al.} 2021,
  \bibinfo{title}{{The Heraklion Extragalactic Catalogue (HECATE): a
  value-added galaxy catalogue for multimessenger astrophysics},} \mnras, 506,
  1896, \dodoi{10.1093/mnras/stab1799}

\bibitem[{G.~P. {Lamb} {et~al.}(2019){Lamb}, {Tanvir}, {Levan}, {de Ugarte
  Postigo}, {Kawaguchi}, {Corsi}, {Evans}, {Gompertz}, {Malesani}, {Page},
  {Wiersema}, {Rosswog}, {Shibata}, {Tanaka}, {van der Horst}, {Cano}, {Fynbo},
  {Fruchter}, {Greiner}, {Heintz}, {Higgins}, {Hjorth}, {Izzo}, {Jakobsson},
  {Kann}, {O'Brien}, {Perley}, {Pian}, {Pugliese}, {Starling}, {Th{\"o}ne},
  {Watson}, {Wijers}, \& {Xu}}]{2019ApJ...883...48L}
{Lamb}, G.~P., {Tanvir}, N.~R., {Levan}, A.~J., {et~al.} 2019,
  \bibinfo{title}{{Short GRB 160821B: A Reverse Shock, a Refreshed Shock, and a
  Well-sampled Kilonova},} \apj, 883, 48, \dodoi{10.3847/1538-4357/ab38bb}

\bibitem[{T. {Laskar} {et~al.}(2022){Laskar}, {Escorial}, {Schroeder}, {Fong},
  {Berger}, {Veres}, {Bhandari}, {Rastinejad}, {Kilpatrick}, {Tohuvavohu},
  {Margutti}, {Alexander}, {DeLaunay}, {Kennea}, {Nugent}, {Paterson}, \&
  {Williams}}]{2022ApJ...935L..11L}
{Laskar}, T., {Escorial}, A.~R., {Schroeder}, G., {et~al.} 2022,
  \bibinfo{title}{{The First Short GRB Millimeter Afterglow: The Wide-angled
  Jet of the Extremely Energetic SGRB 211106A},} \apjl, 935, L11,
  \dodoi{10.3847/2041-8213/ac8421}

\bibitem[{A.~J. {Levan} {et~al.}(2009){Levan}, {Fynbo}, {Hjorth}, {Malesani},
  {D'Avanzo}, \& {D'Elia}}]{GCN9958}
{Levan}, A.~J., {Fynbo}, J.~P.~U., {Hjorth}, J., {et~al.} 2009,
  \bibinfo{title}{{GRB 090927: VLT redshift.},} GRB Coordinates Network, 9958,
  1

\bibitem[{A.~J. {Levan} {et~al.}(2024){Levan}, {Gompertz}, {Salafia}, {Bulla},
  {Burns}, {Hotokezaka}, {Izzo}, {Lamb}, {Malesani}, {Oates}, {Ravasio}, {Rouco
  Escorial}, {Schneider}, {Sarin}, {Schulze}, {Tanvir}, {Ackley}, {Anderson},
  {Brammer}, {Christensen}, {Dhillon}, {Evans}, {Fausnaugh}, {Fong},
  {Fruchter}, {Fryer}, {Fynbo}, {Gaspari}, {Heintz}, {Hjorth}, {Kennea},
  {Kennedy}, {Laskar}, {Leloudas}, {Mandel}, {Martin-Carrillo}, {Metzger},
  {Nicholl}, {Nugent}, {Palmerio}, {Pugliese}, {Rastinejad}, {Rhodes}, {Rossi},
  {Saccardi}, {Smartt}, {Stevance}, {Tohuvavohu}, {van der Horst}, {Vergani},
  {Watson}, {Barclay}, {Bhirombhakdi}, {Breedt}, {Breeveld}, {Brown},
  {Campana}, {Chrimes}, {D'Avanzo}, {D'Elia}, {De Pasquale}, {Dyer},
  {Galloway}, {Garbutt}, {Green}, {Hartmann}, {Jakobsson}, {Kerry},
  {Kouveliotou}, {Langeroodi}, {Le Floc'h}, {Leung}, {Littlefair}, {Munday},
  {O'Brien}, {Parsons}, {Pelisoli}, {Sahman}, {Salvaterra}, {Sbarufatti},
  {Steeghs}, {Tagliaferri}, {Th{\"o}ne}, {de Ugarte Postigo}, \&
  {Kann}}]{2024Natur.626..737L}
{Levan}, A.~J., {Gompertz}, B.~P., {Salafia}, O.~S., {et~al.} 2024,
  \bibinfo{title}{{Heavy-element production in a compact object merger observed
  by JWST},} \nat, 626, 737, \dodoi{10.1038/s41586-023-06759-1}

\bibitem[{C. Li {et~al.}(2024)Li, Zhang, Cui, Wei, Zhang, Zhao, Wu, Tao, Li,
  Wang, \& Kang}]{li_photometric_2024}
Li, C., Zhang, Y., Cui, C., {et~al.} 2024, \bibinfo{title}{A {Photometric}
  {Redshift} {Catalogue} of {Galaxies} from the {DESI} {Legacy} {Imaging}
  {Surveys} {DR10},} The Astronomical Journal, 168, 233,
  \dodoi{10.3847/1538-3881/ad7c52}

\bibitem[{L.-X. Li \& B. Paczyński(1998)Li \& Paczyński}]{li_transient_1998}
Li, L.-X., \& Paczyński, B. 1998, \bibinfo{title}{Transient {Events} from
  {Neutron} {Star} {Mergers},} The Astrophysical Journal, 507, L59,
  \dodoi{10.1086/311680}

\bibitem[{R.~Z. {Li} {et~al.}(2025){Li}, {Xu}, {Sun}, {Li}, {Liu}, {Yuan},
  {Zhang}, \& {Einstein Probe Team}}]{2025GCN.41460....1L}
{Li}, R.~Z., {Xu}, X.~P., {Sun}, H., {et~al.} 2025,
  \bibinfo{title}{{LIGO/Virgo/KAGRA S250818k: EP/FXT observation of AT
  2025ulz},} GRB Coordinates Network, 41460, 1

\bibitem[{ {LIGO Scientific Collaboration} \&  {Virgo Collaboration}(2017){LIGO
  Scientific Collaboration} \& {Virgo
  Collaboration}}]{ligo_scientific_collaboration_and_virgo_collaboration_gw170817_2017}
{LIGO Scientific Collaboration}, \& {Virgo Collaboration}. 2017,
  \bibinfo{title}{{GW170817: Observation of Gravitational Waves from a Binary
  Neutron Star Inspiral},} \prl, 119, 161101,
  \dodoi{10.1103/PhysRevLett.119.161101}

\bibitem[{ {LIGO Scientific Collaboration} \&  {Virgo Collaboration}(2018){LIGO
  Scientific Collaboration} \& {Virgo
  Collaboration}}]{the_ligo_scientific_collaboration_and_the_virgo_collaboration_gw170817_2018}
{LIGO Scientific Collaboration}, \& {Virgo Collaboration}. 2018,
  \bibinfo{title}{{GW170817}: {Measurements} of {Neutron} {Star} {Radii} and
  {Equation} of {State},} Physical Review Letters, 121, 161101,
  \dodoi{10.1103/PhysRevLett.121.161101}

\bibitem[{ {LIGO Scientific Collaboration} {et~al.}(2025{\natexlab{a}}){LIGO
  Scientific Collaboration}, {Virgo Collaboration}, \& {KAGRA
  Collaboration}}]{gcn41437}
{LIGO Scientific Collaboration}, {Virgo Collaboration}, \& {KAGRA
  Collaboration}. 2025{\natexlab{a}}, \bibinfo{title}{{LIGO/Virgo/KAGRA
  S250818k: Properties of the low-significance GW compact binary merger
  candidate potentially associated with AT 2025ulz},} GRB Coordinates Network,
  41437, 1

\bibitem[{ {LIGO Scientific Collaboration} {et~al.}(2025{\natexlab{b}}){LIGO
  Scientific Collaboration}, {Virgo Collaboration}, \& {KAGRA
  Collaboration}}]{gcn41440}
{LIGO Scientific Collaboration}, {Virgo Collaboration}, \& {KAGRA
  Collaboration}. 2025{\natexlab{b}}, \bibinfo{title}{{LIGO/Virgo/KAGRA
  S250818k: Updated Sky localization and EM Bright Classification},} GRB
  Coordinates Network, 41440, 1

\bibitem[{ {LIGO Scientific Collaboration} {et~al.}(2017{\natexlab{a}}){LIGO
  Scientific Collaboration}, {Virgo Collaboration},
  {et~al.}}]{abbott_multi-messenger_2017}
{LIGO Scientific Collaboration}, {Virgo Collaboration}, {et~al.}
  2017{\natexlab{a}}, \bibinfo{title}{{Multi-messenger Observations of a Binary
  Neutron Star Merger},} \apjl, 848, L12, \dodoi{10.3847/2041-8213/aa91c9}

\bibitem[{ {LIGO Scientific Collaboration} {et~al.}(2017{\natexlab{b}}){LIGO
  Scientific Collaboration}, {Virgo Collaboration},
  {et~al.}}]{abbott_gravitational_2017}
{LIGO Scientific Collaboration}, {Virgo Collaboration}, {et~al.}
  2017{\natexlab{b}}, \bibinfo{title}{{Gravitational Waves and Gamma-Rays from
  a Binary Neutron Star Merger: GW170817 and GRB 170817A},} \apjl, 848, L13,
  \dodoi{10.3847/2041-8213/aa920c}

\bibitem[{W. Lindstrom {et~al.}(2022)Lindstrom, Chatelain, Collom, Riba,
  Street, McCully, \& Bowman}]{lindstrom_tom_2022}
Lindstrom, W., Chatelain, J., Collom, D., {et~al.} 2022, {{TOM Toolkit}}:
  {{Target}} and {{Observation Manager Toolkit}},, Astrophysics Source Code
  Library \doeprint{2208.004}

\bibitem[{V.~M. Lipunov {et~al.}(2017)Lipunov, Gorbovskoy, Kornilov, Tyurina,
  Balanutsa, Kuznetsov, Vlasenko, Kuvshinov, Gorbunov, Buckley, Krylov,
  Podesta, Lopez, Podesta, Levato, Saffe, Mallamachi, Potter, Budnev, Gress,
  Ishmuhametova, Vladimirov, Zimnukhov, Yurkov, Sergienko, Gabovich, Rebolo,
  Serra-Ricart, Israelyan, Chazov, Wang, Tlatov, \&
  Panchenko}]{lipunov_master_2017}
Lipunov, V.~M., Gorbovskoy, E., Kornilov, V.~G., {et~al.} 2017,
  \bibinfo{title}{{MASTER} {Optical} {Detection} of the {First} {LIGO}/{Virgo}
  {Neutron} {Star} {Binary} {Merger} {GW170817},} The Astrophysical Journal
  Letters, 850, L1, \dodoi{10.3847/2041-8213/aa92c0}

\bibitem[{Z.~Y. {Liu} {et~al.}(2025){Liu}, {Xu}, {Meng}, {Jiang}, {Zhao},
  {Cai}, {Dai}, {Fan}, {Jiang}, {Kong}, {Wang}, {Fan}, {Geng}, {Jin}, {Wu}, \&
  {WFST Collaboration}}]{2025GCN.41461....1L}
{Liu}, Z.~Y., {Xu}, Z.~L., {Meng}, D.~Z., {et~al.} 2025,
  \bibinfo{title}{{LIGO/Virgo/KAGRA S250818k: WFST pre-discovery limits and
  follow-up observations of AT 2025ulz},} GRB Coordinates Network, 41461, 1

\bibitem[{O. Lone(2022)Lone}]{lone_django-webpack-loader_2022}
Lone, O. 2022, django-webpack-loader v1.6.0,, GitHub
  \url{https://github.com/django-webpack/django-webpack-loader}

\bibitem[{M.~J. {Lundquist} {et~al.}(2019{\natexlab{a}}){Lundquist},
  {Paterson}, {Fong}, {Sand}, {Andrews}, {Shivaei}, {Daly}, {Valenti}, {Yang},
  {Christensen}, {Gibbs}, {Shelly}, {Wyatt}, {Eskandari}, {Kuhn}, {Amaro},
  {Arcavi}, {Behroozi}, {Butler}, {Chomiuk}, {Corsi}, {Drout}, {Egami}, {Fan},
  {Foley}, {Frye}, {Gabor}, {Green}, {Grier}, {Guzman}, {Hamden}, {Howell},
  {Jannuzi}, {Kelly}, {Milne}, {Moe}, {Nugent}, {Olszewski}, {Palazzi},
  {Paschalidis}, {Psaltis}, {Reichart}, {Rest}, {Rossi}, {Schroeder}, {Smith},
  {Smith}, {Spekkens}, {Strader}, {Stark}, {Trilling}, {Veillet}, {Wagner},
  {Weiner}, {Wheeler}, {Williams}, \& {Zabludoff}}]{Lundquist+19}
{Lundquist}, M.~J., {Paterson}, K., {Fong}, W., {et~al.} 2019{\natexlab{a}},
  \bibinfo{title}{{Searches after Gravitational Waves Using ARizona
  Observatories (SAGUARO): System Overview and First Results from Advanced
  LIGO/Virgo{\textquoteright}s Third Observing Run},} \apjl, 881, L26,
  \dodoi{10.3847/2041-8213/ab32f2}

\bibitem[{M.~J. {Lundquist} {et~al.}(2019{\natexlab{b}}){Lundquist},
  {Paterson}, {Fong}, {Sand}, {Andrews}, {Shivaei}, {Daly}, {Valenti}, {Yang},
  {Christensen}, {Gibbs}, {Shelly}, {Wyatt}, {Eskandari}, {Kuhn}, {Amaro},
  {Arcavi}, {Behroozi}, {Butler}, {Chomiuk}, {Corsi}, {Drout}, {Egami}, {Fan},
  {Foley}, {Frye}, {Gabor}, {Green}, {Grier}, {Guzman}, {Hamden}, {Howell},
  {Jannuzi}, {Kelly}, {Milne}, {Moe}, {Nugent}, {Olszewski}, {Palazzi},
  {Paschalidis}, {Psaltis}, {Reichart}, {Rest}, {Rossi}, {Schroeder}, {Smith},
  {Smith}, {Spekkens}, {Strader}, {Stark}, {Trilling}, {Veillet}, {Wagner},
  {Weiner}, {Wheeler}, {Williams}, \& {Zabludoff}}]{2019ApJ...881L..26L}
{Lundquist}, M.~J., {Paterson}, K., {Fong}, W., {et~al.} 2019{\natexlab{b}},
  \bibinfo{title}{{Searches after Gravitational Waves Using ARizona
  Observatories (SAGUARO): System Overview and First Results from Advanced
  LIGO/Virgo{\textquoteright}s Third Observing Run},} \apjl, 881, L26,
  \dodoi{10.3847/2041-8213/ab32f2}

\bibitem[{S. {Makhathini} {et~al.}(2021){Makhathini}, {Mooley}, {Brightman},
  {Hotokezaka}, {Nayana}, {Intema}, {Dobie}, {Lenc}, {Perley}, {Fremling},
  {Mold{\`o}n}, {Lazzati}, {Kaplan}, {Balasubramanian}, {Brown}, {Carbone},
  {Chandra}, {Corsi}, {Camilo}, {Deller}, {Frail}, {Murphy}, {Murphy}, {Nakar},
  {Smirnov}, {Beswick}, {Fender}, {Hallinan}, {Heywood}, {Kasliwal}, {Lee},
  {Lu}, {Rana}, {Perkins}, {White}, {J{\'o}zsa}, {Hugo}, \&
  {Kamphuis}}]{2021ApJ...922..154M}
{Makhathini}, S., {Mooley}, K.~P., {Brightman}, M., {et~al.} 2021,
  \bibinfo{title}{{The Panchromatic Afterglow of GW170817: The Full Uniform
  Data Set, Modeling, Comparison with Previous Results, and Implications},}
  \apj, 922, 154, \dodoi{10.3847/1538-4357/ac1ffc}

\bibitem[{D.~B. {Malesani} {et~al.}(2025){Malesani}, {Boye}, {Izzo},
  {Leloudas}, {An}, {Liu}, {Xu}, {Fraser}, {Brennan}, {Broe Bendsten}, {Koch},
  {Wagner}, {Magaard Knudsen}, {Hein Pedersen}, {Fynbo}, {Holmberg Rasmussen},
  {Valeckas}, \& {De Pasquale}}]{2025GCN.41492....1M}
{Malesani}, D.~B., {Boye}, A., {Izzo}, L., {et~al.} 2025,
  \bibinfo{title}{{LIGO/Virgo/KAGRA S250818k: NOT optical observations of
  AT2025ulz},} GRB Coordinates Network, 41492, 1

\bibitem[{Y. Mangalapilly(2023)Mangalapilly}]{mangalapilly_watchdog_2023}
Mangalapilly, Y. 2023, Watchdog v3.0.0,, GitHub
  \url{https://github.com/gorakhargosh/watchdog}

\bibitem[{R. Margutti {et~al.}(2017)Margutti, Berger, Fong, Guidorzi,
  Alexander, Metzger, Blanchard, Cowperthwaite, Chornock, Eftekhari, Nicholl,
  Villar, Williams, Annis, Brown, Chen, Doctor, Frieman, Holz, Sako, \&
  Soares-Santos}]{margutti_electromagnetic_2017}
Margutti, R., Berger, E., Fong, W., {et~al.} 2017, \bibinfo{title}{The
  {Electromagnetic} {Counterpart} of the {Binary} {Neutron} {Star} {Merger}
  {LIGO}/{Virgo} {GW170817}. {V}. {Rising} {X}-{Ray} {Emission} from an
  {Off}-axis {Jet},} The Astrophysical Journal Letters, 848, L20,
  \dodoi{10.3847/2041-8213/aa9057}

\bibitem[{R. {Margutti} {et~al.}(2018){Margutti}, {Alexander}, {Xie}, {Sironi},
  {Metzger}, {Kathirgamaraju}, {Fong}, {Blanchard}, {Berger}, {MacFadyen},
  {Giannios}, {Guidorzi}, {Hajela}, {Chornock}, {Cowperthwaite}, {Eftekhari},
  {Nicholl}, {Villar}, {Williams}, \& {Zrake}}]{2018ApJ...856L..18M}
{Margutti}, R., {Alexander}, K.~D., {Xie}, X., {et~al.} 2018,
  \bibinfo{title}{{The Binary Neutron Star Event LIGO/Virgo GW170817 160 Days
  after Merger: Synchrotron Emission across the Electromagnetic Spectrum},}
  \apjl, 856, L18, \dodoi{10.3847/2041-8213/aab2ad}

\bibitem[{T. Matheson {et~al.}(2000)Matheson, Filippenko, Ho, Barth, \&
  Leonard}]{matheson_detailed_2000}
Matheson, T., Filippenko, A.~V., Ho, L.~C., Barth, A.~J., \& Leonard, D.~C.
  2000, \bibinfo{title}{Detailed {Analysis} of {Early} to {Late}-{Time}
  {Spectra} of {Supernova1993J},} The Astronomical Journal, 120, 1499,
  \dodoi{10.1086/301519}

\bibitem[{K. {Matthews} {et~al.}(2002){Matthews}, {Neugebauer}, {Armus}, \&
  {Soifer}}]{2002AJ....123..753M}
{Matthews}, K., {Neugebauer}, G., {Armus}, L., \& {Soifer}, B.~T. 2002,
  \bibinfo{title}{{Early Near-Infrared Observations of SN 1993J},} \aj, 123,
  753, \dodoi{10.1086/338646}

\bibitem[{O.~R. McBrien {et~al.}(2021)McBrien, Smartt, Huber, Rest, Chambers,
  Barbieri, Bulla, Jha, Gromadzki, Srivastav, Smith, Young, McLaughlin,
  Inserra, Nicholl, Fraser, Maguire, Chen, Wevers, Anderson, Müller-Bravo,
  Olivares E., Kankare, Gal-Yam, \& Waters}]{mcbrien_ps15cey_2021}
McBrien, O.~R., Smartt, S.~J., Huber, M.~E., {et~al.} 2021,
  \bibinfo{title}{{PS15cey} and {PS17cke}: prospective candidates from the
  {Pan}-{STARRS} {Search} for kilonovae,} Monthly Notices of the Royal
  Astronomical Society, 500, 4213, \dodoi{10.1093/mnras/staa3361}

\bibitem[{C. McCully {et~al.}(2018)McCully, Crawford, Kovacs, Tollerud, Betts,
  Bradley, Craig, Turner, Streicher, Sipocz, Robitaille, \&
  Deil}]{mccully_astropy_2018}
McCully, C., Crawford, S., Kovacs, G., {et~al.} 2018, astropy/astroscrappy
  v1.0.5,, Zenodo \dodoi{10.5281/zenodo.1482019}

\bibitem[{J.~P. McMullin {et~al.}(2007)McMullin, Waters, Schiebel, Young, \&
  Golap}]{mcmullin_casa_2007}
McMullin, J.~P., Waters, B., Schiebel, D., Young, W., \& Golap, K. 2007,
  \bibinfo{title}{{CASA} {Architecture} and {Applications},} 376, 127.
\newblock \url{https://ui.adsabs.harvard.edu/abs/2007ASPC..376..127M}

\bibitem[{B.~D. Metzger(2017)Metzger}]{metzger_kilonovae_2017}
Metzger, B.~D. 2017, \bibinfo{title}{Kilonovae,} Living Reviews in Relativity,
  20, 3, \dodoi{10.1007/s41114-017-0006-z}

\bibitem[{B.~D. Metzger {et~al.}(2024)Metzger, Hui, \&
  Cantiello}]{metzger_fragmentation_2024}
Metzger, B.~D., Hui, L., \& Cantiello, M. 2024, \bibinfo{title}{Fragmentation
  in {Gravitationally} {Unstable} {Collapsar} {Disks} and {Subsolar} {Neutron}
  {Star} {Mergers},} The Astrophysical Journal, 971, L34,
  \dodoi{10.3847/2041-8213/ad6990}

\bibitem[{B.~D. {Metzger} {et~al.}(2010){Metzger}, {Mart{\'\i}nez-Pinedo},
  {Darbha}, {Quataert}, {Arcones}, {Kasen}, {Thomas}, {Nugent}, {Panov}, \&
  {Zinner}}]{Metzger10}
{Metzger}, B.~D., {Mart{\'\i}nez-Pinedo}, G., {Darbha}, S., {et~al.} 2010,
  \bibinfo{title}{{Electromagnetic counterparts of compact object mergers
  powered by the radioactive decay of r-process nuclei},} \mnras, 406, 2650,
  \dodoi{10.1111/j.1365-2966.2010.16864.x}

\bibitem[{D. {Milisavljevic} {et~al.}(2013){Milisavljevic}, {Margutti},
  {Soderberg}, {Pignata}, {Chomiuk}, {Fesen}, {Bufano}, {Sanders}, {Parrent},
  {Parker}, {Mazzali}, {Pian}, {Pickering}, {Buckley}, {Crawford}, {Gulbis},
  {Hettlage}, {Hooper}, {Nordsieck}, {O'Donoghue}, {Husser}, {Potter},
  {Kniazev}, {Kotze}, {Romero-Colmenero}, {Vaisanen}, {Wolf}, {Bietenholz},
  {Bartel}, {Fransson}, {Walker}, {Brunthaler}, {Chakraborti}, {Levesque},
  {MacFadyen}, {Drescher}, {Bock}, {Marples}, {Anderson}, {Benetti},
  {Reichart}, \& {Ivarsen}}]{Dan2013}
{Milisavljevic}, D., {Margutti}, R., {Soderberg}, A.~M., {et~al.} 2013,
  \bibinfo{title}{{Multi-wavelength Observations of Supernova 2011ei:
  Time-dependent Classification of Type IIb and Ib Supernovae and Implications
  for Their Progenitors},} \apj, 767, 71, \dodoi{10.1088/0004-637X/767/1/71}

\bibitem[{G. {Mo} {et~al.}(2025){Mo}, {Stein}, {Kasliwal}, {Karambelkar},
  {Frostig}, {Lourie}, \& {Simcoe}}]{2025GCN.41456....1M}
{Mo}, G., {Stein}, R., {Kasliwal}, M., {et~al.} 2025,
  \bibinfo{title}{{LIGO/Virgo/KAGRA S250818k: WINTER J-band observations of AT
  2025ulz},} GRB Coordinates Network, 41456, 1

\bibitem[{K.~P. {Mooley} {et~al.}(2018{\natexlab{a}}){Mooley}, {Frail},
  {Dobie}, {Lenc}, {Corsi}, {De}, {Nayana}, {Makhathini}, {Heywood}, {Murphy},
  {Kaplan}, {Chandra}, {Smirnov}, {Nakar}, {Hallinan}, {Camilo}, {Fender},
  {Goedhart}, {Groot}, {Kasliwal}, {Kulkarni}, \&
  {Woudt}}]{2018ApJ...868L..11M}
{Mooley}, K.~P., {Frail}, D.~A., {Dobie}, D., {et~al.} 2018{\natexlab{a}},
  \bibinfo{title}{{A Strong Jet Signature in the Late-time Light Curve of
  GW170817},} \apjl, 868, L11, \dodoi{10.3847/2041-8213/aaeda7}

\bibitem[{K.~P. {Mooley} {et~al.}(2018{\natexlab{b}}){Mooley}, {Nakar},
  {Hotokezaka}, {Hallinan}, {Corsi}, {Frail}, {Horesh}, {Murphy}, {Lenc},
  {Kaplan}, {de}, {Dobie}, {Chandra}, {Deller}, {Gottlieb}, {Kasliwal},
  {Kulkarni}, {Myers}, {Nissanke}, {Piran}, {Lynch}, {Bhalerao}, {Bourke},
  {Bannister}, \& {Singer}}]{2018Natur.554..207M}
{Mooley}, K.~P., {Nakar}, E., {Hotokezaka}, K., {et~al.} 2018{\natexlab{b}},
  \bibinfo{title}{{A mildly relativistic wide-angle outflow in the neutron-star
  merger event GW170817},} \nat, 554, 207, \dodoi{10.1038/nature25452}

\bibitem[{K.~P. {Mooley} {et~al.}(2018{\natexlab{c}}){Mooley}, {Deller},
  {Gottlieb}, {Nakar}, {Hallinan}, {Bourke}, {Frail}, {Horesh}, {Corsi}, \&
  {Hotokezaka}}]{2018Natur.561..355M}
{Mooley}, K.~P., {Deller}, A.~T., {Gottlieb}, O., {et~al.} 2018{\natexlab{c}},
  \bibinfo{title}{{Superluminal motion of a relativistic jet in the
  neutron-star merger GW170817},} \nat, 561, 355,
  \dodoi{10.1038/s41586-018-0486-3}

\bibitem[{A. {Morales-Garoffolo} {et~al.}(2014){Morales-Garoffolo},
  {Elias-Rosa}, {Benetti}, {Taubenberger}, {Cappellaro}, {Pastorello},
  {Klauser}, {Valenti}, {Howerton}, {Ochner}, {Schramm}, {Siviero},
  {Tartaglia}, \& {Tomasella}}]{morales2014_2013df}
{Morales-Garoffolo}, A., {Elias-Rosa}, N., {Benetti}, S., {et~al.} 2014,
  \bibinfo{title}{{SN 2013df, a double-peaked IIb supernova from a compact
  progenitor and an extended H envelope},} \mnras, 445, 1647,
  \dodoi{10.1093/mnras/stu1837}

\bibitem[{A. {Morales-Garoffolo} {et~al.}(2015){Morales-Garoffolo},
  {Elias-Rosa}, {Bersten}, {Jerkstrand}, {Taubenberger}, {Benetti},
  {Cappellaro}, {Kotak}, {Pastorello}, {Bufano}, {Dom{\'\i}nguez}, {Ergon},
  {Fraser}, {Gao}, {Garc{\'\i}a}, {Howell}, {Isern}, {Smartt}, {Tomasella}, \&
  {Valenti}}]{Morales2015_2011fu}
{Morales-Garoffolo}, A., {Elias-Rosa}, N., {Bersten}, M., {et~al.} 2015,
  \bibinfo{title}{{SN 2011fu: a type IIb supernova with a luminous
  double-peaked light curve},} \mnras, 454, 95, \dodoi{10.1093/mnras/stv1972}

\bibitem[{R. {Morgan} {et~al.}(2020){Morgan}, {Soares-Santos}, {Annis},
  {Herner}, {Garcia}, {Palmese}, {Drlica-Wagner}, {Kessler},
  {Garc{\'\i}a-Bellido}, {Bachmann}, {Sherman}, {Allam}, {Bechtol}, {Bom},
  {Brout}, {Butler}, {Butner}, {Cartier}, {Chen}, {Conselice}, {Cook}, {Davis},
  {Doctor}, {Farr}, {Figueiredo}, {Finley}, {Foley}, {Galarza}, {Gill},
  {Gruendl}, {Holz}, {Kuropatkin}, {Lidman}, {Lin}, {Malik}, {Mann},
  {Marriner}, {Marshall}, {Mart{\'\i}nez-V{\'a}zquez}, {Meza}, {Neilsen},
  {Nicolaou}, {Olivares E.}, {Paz-Chinch{\'o}n}, {Points},
  {Quirola-V{\'a}squez}, {Rodriguez}, {Sako}, {Scolnic}, {Smith}, {Sobreira},
  {Tucker}, {Vivas}, {Wiesner}, {Wood}, {Yanny}, {Zenteno}, {Abbott}, {Aguena},
  {Avila}, {Bertin}, {Bhargava}, {Brooks}, {Burke}, {Carnero Rosell}, {Carrasco
  Kind}, {Carretero}, {da Costa}, {Costanzi}, {De Vicente}, {Desai}, {Diehl},
  {Doel}, {Eifler}, {Everett}, {Flaugher}, {Frieman}, {Gaztanaga}, {Gerdes},
  {Gruen}, {Gschwend}, {Gutierrez}, {Hartley}, {Hinton}, {Hollowood},
  {Honscheid}, {James}, {Kuehn}, {Lahav}, {Lima}, {Maia}, {March}, {Miquel},
  {Ogando}, {Plazas}, {Roodman}, {Sanchez}, {Scarpine}, {Schubnell}, {Serrano},
  {Sevilla-Noarbe}, {Suchyta}, \& {Tarle}}]{Morgan+20}
{Morgan}, R., {Soares-Santos}, M., {Annis}, J., {et~al.} 2020,
  \bibinfo{title}{{Constraints on the Physical Properties of GW190814 through
  Simulations Based on DECam Follow-up Observations by the Dark Energy
  Survey},} \apj, 901, 83, \dodoi{10.3847/1538-4357/abafaa}

\bibitem[{A. Murguia-Berthier {et~al.}(2021)Murguia-Berthier, Ramirez-Ruiz,
  Colle, Janiuk, Rosswog, \& Lee}]{murguia-berthier_fate_2021}
Murguia-Berthier, A., Ramirez-Ruiz, E., Colle, F.~D., {et~al.} 2021,
  \bibinfo{title}{The {Fate} of the {Merger} {Remnant} in {GW170817} and {Its}
  {Imprint} on the {Jet} {Structure},} The Astrophysical Journal, 908, 152,
  \dodoi{10.3847/1538-4357/abd08e}

\bibitem[{E.~J. {Murphy} {et~al.}(2011){Murphy}, {Condon}, {Schinnerer},
  {Kennicutt}, {Calzetti}, {Armus}, {Helou}, {Turner}, {Aniano}, {Beir{\~a}o},
  {Bolatto}, {Brandl}, {Croxall}, {Dale}, {Donovan Meyer}, {Draine},
  {Engelbracht}, {Hunt}, {Hao}, {Koda}, {Roussel}, {Skibba}, \&
  {Smith}}]{2011ApJ...737...67M}
{Murphy}, E.~J., {Condon}, J.~J., {Schinnerer}, E., {et~al.} 2011,
  \bibinfo{title}{{Calibrating Extinction-free Star Formation Rate Diagnostics
  with 33 GHz Free-free Emission in NGC 6946},} \apj, 737, 67,
  \dodoi{10.1088/0004-637X/737/2/67}

\bibitem[{A. Murray {et~al.}(2023)Murray, van Kemenade, {wiredfool},
  Clark~(Alex), Karpinsky, Baranovi{\v c}, Gohlke, Dufresne, Yay295, DWesl,
  Schmidt, Kopachev, Houghton, Mani, Landey, Ware, {vashek}, Piolie, Douglas,
  T, Caro, Martinez, Kossouho, Lahd, Lee, Brown, Tonnhofer, \&
  Bonfill}]{murray_python-pillow_2023}
Murray, A., van Kemenade, H., {wiredfool}, {et~al.} 2023,
  python-pillow/{{Pillow}} v10.0.1,, Zenodo \dodoi{10.5281/zenodo.8349181}

\bibitem[{M. {Nakajima} {et~al.}(2025){Nakajima}, {Negoro}, {Takagi}, {Kawai},
  {Mihara}, {Sugita}, {Serino}, {Kawakubo}, {Hiramatsu}, {Kondo}, \& {MAXI
  Team}}]{2025GCN.41410....1N}
{Nakajima}, M., {Negoro}, H., {Takagi}, K., {et~al.} 2025,
  \bibinfo{title}{{LIGO/Virgo/KAGRA S250818k: Coverage and upper limits from
  MAXI/GSC observations},} GRB Coordinates Network, 41410, 1

\bibitem[{A.~J. {Nayana} \& P. {Chandra}(2014){Nayana} \& {Chandra}}]{GCN16815}
{Nayana}, A.~J., \& {Chandra}, P. 2014, \bibinfo{title}{{Possible radio
  detection of GRB 140903A with the GMRT.},} GRB Coordinates Network, 16815, 1

\bibitem[{M. Newville {et~al.}(2023)Newville, Otten, Nelson, Stensitzki,
  Ingargiola, Allan, Fox, Carter, Micha{\l}, Osborn, Pustakhod, {lneuhaus},
  Weigand, Aristov, Glenn, Deil, {mgunyho}, Mark, Hansen, Pasquevich, Foks,
  Zobrist, Frost, Stuermer, {azelcer}, Polloreno, Persaud, Nielsen, Pompili, \&
  Eendebak}]{newville_lmfit_2023}
Newville, M., Otten, R., Nelson, A., {et~al.} 2023, lmfit/lmfit-py v1.2.2,,
  Zenodo \dodoi{10.5281/zenodo.8145703}

\bibitem[{M. {Nicholl} {et~al.}(2025){Nicholl}, {Young}, {Aamer}, {Angus},
  {Fulton}, {Magill}, {McCollum}, {Moore}, {Sim}, {Weston}, {Sheng}, {Smartt},
  {Smith}, {Gillanders}, {Srivastav}, {Stevance}, {Stoppa}, {Tweddle},
  {Shingles}, {Ramsden}, {Chambers}, {Huber}, {Schultz}, {de Boer}, {Fairlamb},
  {Lin}, {Lowe}, {Magnier}, {Minguez}, {Paek}, {Smith}, {Wainscoat}, {Chen},
  {Rest}, \& {Stubbs}}]{2025GCN.41439....1N}
{Nicholl}, M., {Young}, D.~R., {Aamer}, A., {et~al.} 2025,
  \bibinfo{title}{{LIGO/Virgo/KAGRA S250818k: Pan-STARRS pre-detection limits
  for AT2025ulz},} GRB Coordinates Network, 41439, 1

\bibitem[{G. Niemeyer {et~al.}(2021)Niemeyer, Pievil{\"a}inen, {de Leeuw}, \&
  Ganssle}]{niemeyer_dateutil_2021}
Niemeyer, G., Pievil{\"a}inen, T., {de Leeuw}, Y., \& Ganssle, P. 2021,
  dateutil/dateutil v2.8.2,, GitHub \url{https://github.com/dateutil/dateutil}

\bibitem[{A.~E. {Nugent} {et~al.}(2024){Nugent}, {Fong}, {Castrejon}, {Leja},
  {Zevin}, \& {Ji}}]{2024ApJ...962....5N}
{Nugent}, A.~E., {Fong}, W.-f., {Castrejon}, C., {et~al.} 2024,
  \bibinfo{title}{{A Population of Short-duration Gamma-Ray Bursts with Dwarf
  Host Galaxies},} \apj, 962, 5, \dodoi{10.3847/1538-4357/ad17c0}

\bibitem[{A.~E. Nugent {et~al.}(2025)Nugent, Villar, Gagliano, Jones, Horowicz,
  Soto, Wang, \& Margalit}]{nugent_characterizing_2025}
Nugent, A.~E., Villar, V.~A., Gagliano, A., {et~al.} 2025, Characterizing
  {Supernovae} {Host} {Galaxies} with {FrankenBlast}: {A} {Scalable} {Tool} for
  {Transient} {Host} {Galaxy} {Association}, {Photometry}, and {Stellar}
  {Population} {Modeling}, arXiv, \dodoi{10.48550/arXiv.2509.08874}

\bibitem[{S.~R. {Oates} {et~al.}(2021){Oates}, {Marshall}, {Breeveld}, {Kuin},
  {Brown}, {De Pasquale}, {Evans}, {Fenney}, {Gronwall}, {Kennea}, {Klingler},
  {Page}, {Siegel}, {Tohuvavohu}, {Ambrosi}, {Barthelmy}, {Beardmore},
  {Bernardini}, {Campana}, {Caputo}, {Cenko}, {Cusumano}, {D'A{\`\i}},
  {D'Avanzo}, {D'Elia}, {Giommi}, {Hartmann}, {Krimm}, {Laha}, {Malesani},
  {Melandri}, {Nousek}, {O'Brien}, {Osborne}, {Pagani}, {Page}, {Palmer},
  {Perri}, {Racusin}, {Sakamoto}, {Sbarufatti}, {Schlieder}, {Tagliaferri}, \&
  {Troja}}]{Oates+21}
{Oates}, S.~R., {Marshall}, F.~E., {Breeveld}, A.~A., {et~al.} 2021,
  \bibinfo{title}{{Swift/UVOT follow-up of gravitational wave alerts in the O3
  era},} \mnras, 507, 1296, \dodoi{10.1093/mnras/stab2189}

\bibitem[{B. {O'Connor} {et~al.}(2025){O'Connor}, {Freeburn}, {Hall},
  {Busmann}, {Andreoni}, {Palmese}, {Gruen}, {Hu}, {Cabrera}, {Kunnumkai}, \&
  {Amsellem}}]{2025GCN.41452....1O}
{O'Connor}, B., {Freeburn}, J., {Hall}, X.~J., {et~al.} 2025,
  \bibinfo{title}{{LIGO/Virgo/KAGRA S250818k: Multi-band Gemini GMOS
  Detections},} GRB Coordinates Network, 41452, 1

\bibitem[{T. {Ohgami} {et~al.}(2021){Ohgami}, {Tominaga}, {Utsumi}, {Niino},
  {Tanaka}, {Banerjee}, {Hamasaki}, {Yoshida}, {Terai}, {Takagi}, {Morokuma},
  {Sasada}, {Akitaya}, {Yasuda}, {Yanagisawa}, \& {Ohsawa}}]{Ohgami+21}
{Ohgami}, T., {Tominaga}, N., {Utsumi}, Y., {et~al.} 2021,
  \bibinfo{title}{{Optical follow-up observation for GW event S190510g using
  Subaru/Hyper Suprime-Cam},} \pasj, 73, 350, \dodoi{10.1093/pasj/psab002}

\bibitem[{J.~B. {Oke} {et~al.}(1995){Oke}, {Cohen}, {Carr}, {Cromer},
  {Dingizian}, {Harris}, {Labrecque}, {Lucinio}, {Schaal}, {Epps}, \&
  {Miller}}]{lris}
{Oke}, J.~B., {Cohen}, J.~G., {Carr}, M., {et~al.} 1995, \bibinfo{title}{{The
  Keck Low-Resolution Imaging Spectrometer},} \pasp, 107, 375,
  \dodoi{10.1086/133562}

\bibitem[{G.~S.~H. Paek {et~al.}(2025)Paek, Im, Jeong, Chang, Hur, Hong, Kim,
  Lee, Lee, Lee, Jung, Kim, Lee, Lee, \& Kim}]{paek_gecko_2025}
Paek, G. S.~H., Im, M., Jeong, M., {et~al.} 2025, \bibinfo{title}{{GECKO}
  {Follow}-up {Observations} of the {Binary} {Neutron} {Star}–{Black} {Hole}
  {Merger} {Candidate} {S230518h},} The Astrophysical Journal, 981, 38,
  \dodoi{10.3847/1538-4357/adaf99}

\bibitem[{ {Pallets Projects}(2021){Pallets
  Projects}}]{pallets_projects_flask-sqlalchemy_2021}
{Pallets Projects}. 2021, Flask-{{SQLAlchemy}} v2.5.1,, GitHub
  \url{https://github.com/pallets-eco/flask-sqlalchemy}

\bibitem[{ {Pallets Projects}(2022){Pallets
  Projects}}]{pallets_projects_flask_2022}
{Pallets Projects}. 2022, Flask v2.2.2,, GitHub
  \url{https://github.com/pallets/flask}

\bibitem[{P.~T.~H. Pang {et~al.}(2023)Pang, Dietrich, Coughlin, Bulla, Tews,
  Almualla, Barna, Kiendrebeogo, Kunert, Mansingh, Reed, Sravan, Toivonen,
  Antier, VandenBerg, Heinzel, Nedora, Salehi, Sharma, Somasundaram, \&
  Broeck}]{Pang:2022rzc}
Pang, P. T.~H., Dietrich, T., Coughlin, M.~W., {et~al.} 2023,
  \bibinfo{title}{An updated nuclear-physics and multi-messenger astrophysics
  framework for binary neutron star mergers,} Nature Communications, 14, 8352,
  \dodoi{10.1038/s41467-023-43932-6}

\bibitem[{A. {Pastorello} {et~al.}(2008){Pastorello}, {Kasliwal}, {Crockett},
  {Valenti}, {Arbour}, {Itagaki}, {Kaspi}, {Gal-Yam}, {Smartt}, {Griffith},
  {Maguire}, {Ofek}, {Seymour}, {Stern}, \& {Wiethoff}}]{Pastorello2008}
{Pastorello}, A., {Kasliwal}, M.~M., {Crockett}, R.~M., {et~al.} 2008,
  \bibinfo{title}{{The Type IIb SN 2008ax: spectral and light curve
  evolution},} \mnras, 389, 955, \dodoi{10.1111/j.1365-2966.2008.13618.x}

\bibitem[{K. {Paterson} {et~al.}(2021{\natexlab{a}}){Paterson}, {Lundquist},
  {Rastinejad}, {Fong}, {Sand}, {Andrews}, {Amaro}, {Eskandari}, {Wyatt},
  {Daly}, {Bradley}, {Zhou-Wright}, {Valenti}, {Yang}, {Christensen}, {Gibbs},
  {Shelly}, {Bilinski}, {Chomiuk}, {Corsi}, {Drout}, {Foley}, {Gabor},
  {Garnavich}, {Grier}, {Hamden}, {Krantz}, {Olszewski}, {Paschalidis},
  {Reichart}, {Rest}, {Smith}, {Strader}, {Trilling}, {Veillet}, {Wagner},
  {Weiner}, \& {Zabludoff}}]{Paterson+21}
{Paterson}, K., {Lundquist}, M.~J., {Rastinejad}, J.~C., {et~al.}
  2021{\natexlab{a}}, \bibinfo{title}{{Searches after Gravitational Waves Using
  ARizona Observatories (SAGUARO): Observations and Analysis from Advanced
  LIGO/Virgo's Third Observing Run},} \apj, 912, 128,
  \dodoi{10.3847/1538-4357/abeb71}

\bibitem[{K. {Paterson} {et~al.}(2021{\natexlab{b}}){Paterson}, {Lundquist},
  {Rastinejad}, {Fong}, {Sand}, {Andrews}, {Amaro}, {Eskandari}, {Wyatt},
  {Daly}, {Bradley}, {Zhou-Wright}, {Valenti}, {Yang}, {Christensen}, {Gibbs},
  {Shelly}, {Bilinski}, {Chomiuk}, {Corsi}, {Drout}, {Foley}, {Gabor},
  {Garnavich}, {Grier}, {Hamden}, {Krantz}, {Olszewski}, {Paschalidis},
  {Reichart}, {Rest}, {Smith}, {Strader}, {Trilling}, {Veillet}, {Wagner},
  {Weiner}, \& {Zabludoff}}]{2021ApJ...912..128P}
{Paterson}, K., {Lundquist}, M.~J., {Rastinejad}, J.~C., {et~al.}
  2021{\natexlab{b}}, \bibinfo{title}{{Searches after Gravitational Waves Using
  ARizona Observatories (SAGUARO): Observations and Analysis from Advanced
  LIGO/Virgo's Third Observing Run},} \apj, 912, 128,
  \dodoi{10.3847/1538-4357/abeb71}

\bibitem[{J. Pearson {et~al.}(2023)Pearson, Hosseinzadeh, Sand, Andrews,
  Jencson, Dong, Bostroem, Valenti, Janzen, Retamal, Lundquist, Wyatt, Amaro,
  Burke, Howell, McCully, Hiramatsu, Jha, Smith, Haislip, Kouprianov, Reichart,
  Yang, \& Rho}]{pearson_circumstellar_2023}
Pearson, J., Hosseinzadeh, G., Sand, D.~J., {et~al.} 2023,
  \bibinfo{title}{Circumstellar {Medium} {Interaction} in {SN} 2018lab, {A}
  {Low}-luminosity {Type} {IIP} {Supernova} {Observed} with {TESS},} The
  Astrophysical Journal, 945, 107, \dodoi{10.3847/1538-4357/acb8a9}

\bibitem[{C. {Pellegrino} {et~al.}(2023){Pellegrino}, {Hiramatsu}, {Arcavi},
  {Howell}, {Bostroem}, {Brown}, {Burke}, {Elias-Rosa}, {Itagaki}, {Kaneda},
  {McCully}, {Modjaz}, {Padilla Gonzalez}, {Pritchard}, \&
  {Yesmin}}]{Pellegrino_2023_SN2020bio}
{Pellegrino}, C., {Hiramatsu}, D., {Arcavi}, I., {et~al.} 2023,
  \bibinfo{title}{{SN 2020bio: A Double-peaked, H-poor Type IIb Supernova with
  Evidence of Circumstellar Interaction},} \apj, 954, 35,
  \dodoi{10.3847/1538-4357/ace595}

\bibitem[{A. Petrov {et~al.}(2023)Petrov, Larson, \&
  Pradet}]{petrov_urllib3_2023}
Petrov, A., Larson, S.~M., \& Pradet, Q. 2023, urllib3/urllib3 v1.26.15,,
  GitHub \url{https://github.com/urllib3/urllib3}

\bibitem[{E. Pian {et~al.}(2017)Pian, D'Avanzo, Benetti, Branchesi, Brocato,
  Campana, Cappellaro, Covino, D'Elia, Fynbo, Getman, Ghirlanda, Ghisellini,
  Grado, Greco, Hjorth, Kouveliotou, Levan, Limatola, Malesani, Mazzali,
  Melandri, Møller, Nicastro, Palazzi, Piranomonte, Rossi, Salafia, Selsing,
  Stratta, Tanaka, Tanvir, Tomasella, Watson, Yang, Amati, Antonelli, Ascenzi,
  Bernardini, Boër, Bufano, Bulgarelli, Capaccioli, Casella, Castro-Tirado,
  Chassande-Mottin, Ciolfi, Copperwheat, Dadina, De~Cesare, di~Paola, Fan,
  Gendre, Giuffrida, Giunta, Hunt, Israel, Jin, Kasliwal, Klose, Lisi, Longo,
  Maiorano, Mapelli, Masetti, Nava, Patricelli, Perley, Pescalli, Piran,
  Possenti, Pulone, Razzano, Salvaterra, Schipani, Spera, Stamerra, Stella,
  Tagliaferri, Testa, Troja, Turatto, Vergani, \&
  Vergani}]{pian_spectroscopic_2017}
Pian, E., D'Avanzo, P., Benetti, S., {et~al.} 2017,
  \bibinfo{title}{Spectroscopic identification of r-process nucleosynthesis in
  a double neutron-star merger,} Nature, 551, 67, \dodoi{10.1038/nature24298}

\bibitem[{M. {Pillas} {et~al.}(2025){Pillas}, {Antier}, {Ackley}, {Ahumada},
  {Akl}, {de Almeida}, {Anand}, {Andrade}, {Andreoni}, {Bostroem}, {Bulla},
  {Burns}, {Cabrera}, {Chang}, {Choi}, {O'Connor}, {Coughlin}, {Corradi},
  {Gibbs}, {Dietrich}, {Dornic}, {Ducoin}, {Duverne}, {Eggenstein}, {Freeberg},
  {Dyer}, {Fausnaugh}, {Fong}, {Foucart}, {Frostig}, {Guessoum}, {Gupta},
  {Hello}, {Hosseinzadeh}, {Hu}, {Hussenot-Desenonges}, {Im}, {Jayaraman},
  {Jeong}, {Karambelkar}, {Kasliwal}, {Kim}, {Kilpatrick}, {Kochiashvili},
  {Karpov}, {Kunnumkai}, {Lamoureux}, {Lee}, {Lourie}, {Lyman}, {Ma{\v{s}}ek},
  {Magnani}, {Mo}, {Molham}, {Nitz}, {Nicholl}, {Navarete}, {Noysena},
  {O'Neill}, {Paek}, {Palmese}, {Poggiani}, {Pradier}, {Pyshna}, {Rajabov},
  {Rastinejad}, {Sand}, {Shawhan}, {Shrestha}, {Simcoe}, {Smartt}, {Steeghs},
  {Stein}, {Stevance}, {Takey}, {Sun}, {Toivonen}, {Turpin}, {Ulaczyk}, {Wold},
  \& {Wouters}}]{Pillas25}
{Pillas}, M., {Antier}, S., {Ackley}, K., {et~al.} 2025,
  \bibinfo{title}{{Limits on the ejecta mass during the search for kilonovae
  associated with neutron star-black hole mergers: A case study of S230518h,
  GW230529, S230627c and the low-significance candidate S240422ed},} \prd, 112,
  083002, \dodoi{10.1103/6ld6-95xh}

\bibitem[{A.~L. Piro {et~al.}(2017)Piro, Muhleisen, Arcavi, Sand, Tartaglia, \&
  Valenti}]{piro_numerically_2017}
Piro, A.~L., Muhleisen, M., Arcavi, I., {et~al.} 2017,
  \bibinfo{title}{Numerically {Modeling} the {First} {Peak} of the {Type} {IIb}
  {SN} 2016gkg,} The Astrophysical Journal, 846, 94,
  \dodoi{10.3847/1538-4357/aa8595}

\bibitem[{ {Plotly}(2023){Plotly}}]{plotly_plotly_2023}
{Plotly}. 2023, plotly.py v5.14.1,, GitHub
  \url{https://github.com/plotly/plotly.py}

\bibitem[{P. Podsiadlowski {et~al.}(1993)Podsiadlowski, Hsu, Joss, \&
  Ross}]{podsiadlowski_progenitor_1993}
Podsiadlowski, P., Hsu, J. J.~L., Joss, P.~C., \& Ross, R.~R. 1993,
  \bibinfo{title}{The progenitor of supernova {1993J}: a stripped supergiant in
  a binary system?} Nature, 364, 509, \dodoi{10.1038/364509a0}

\bibitem[{ {PostgreSQL Global Development Group}(2022){PostgreSQL Global
  Development Group}}]{postgresql_global_development_group_postgresql_2022}
{PostgreSQL Global Development Group}. 2022, {{PostgreSQL}} v14.6,
  \url{https://www.postgresql.org/}

\bibitem[{A.~S. {Pozanenko} {et~al.}(2020){Pozanenko}, {Minaev}, {Grebenev}, \&
  {Chelovekov}}]{Pozanenko+20}
{Pozanenko}, A.~S., {Minaev}, P.~Y., {Grebenev}, S.~A., \& {Chelovekov}, I.~V.
  2020, \bibinfo{title}{{Observation of the Second LIGO/Virgo Event Connected
  with a Binary Neutron Star Merger S190425z in the Gamma-Ray Range},}
  Astronomy Letters, 45, 710, \dodoi{10.1134/S1063773719110057}

\bibitem[{J.~X. {Prochaska} {et~al.}(2020{\natexlab{a}}){Prochaska}, {Hennawi},
  {Westfall}, {Cooke}, {Wang}, {Hsyu}, {Davies}, {Farina}, \&
  Pelliccia}]{pypeit:joss_pub}
{Prochaska}, J.~X., {Hennawi}, J.~F., {Westfall}, K.~B., {et~al.}
  2020{\natexlab{a}}, \bibinfo{title}{PypeIt: The Python Spectroscopic Data
  Reduction Pipeline,} Journal of Open Source Software, 5, 2308,
  \dodoi{10.21105/joss.02308}

\bibitem[{J.~X. {Prochaska} {et~al.}(2020{\natexlab{b}}){Prochaska}, {Hennawi},
  {Cooke}, {Westfall}, {Wang}, {EmAstro}, {Tiffanyhsyu}, {Wasserman},
  {Villaume}, {Marijana777}, {Schindler}, {Young}, {Simha}, {Wilde}, {Tejos},
  {Isbell}, {Fl{\"o}rs}, {Sandford}, {Vasovi{\'c}}, {Betts}, \&
  {Holden}}]{pypeit:zenodo}
{Prochaska}, J.~X., {Hennawi}, J., {Cooke}, R., {et~al.} 2020{\natexlab{b}},
  {pypeit/PypeIt: Release 1.0.0}, v1.0.0 Zenodo, \dodoi{10.5281/zenodo.3743493}

\bibitem[{ {Python Packaging Authority}(2023){Python Packaging
  Authority}}]{python_packaging_authority_setuptools_2023}
{Python Packaging Authority}. 2023, setuptools v67.6.1,, GitHub
  \url{https://github.com/pypa/setuptools}

\bibitem[{Y.-J. Qin \& A. Zabludoff(2024)Qin \& Zabludoff}]{qin_linking_2024}
Qin, Y.-J., \& Zabludoff, A. 2024, \bibinfo{title}{Linking transients to their
  host galaxies – {II}. {A} comparison of host galaxy properties and rate
  dependencies across supernova types,} Monthly Notices of the Royal
  Astronomical Society, 533, 3517, \dodoi{10.1093/mnras/stae1921}

\bibitem[{D. Radice {et~al.}(2018)Radice, Perego, Zappa, \&
  Bernuzzi}]{radice_gw170817_2018}
Radice, D., Perego, A., Zappa, F., \& Bernuzzi, S. 2018,
  \bibinfo{title}{{GW170817}: {Joint} {Constraint} on the {Neutron} {Star}
  {Equation} of {State} from {Multimessenger} {Observations},} The
  Astrophysical Journal Letters, 852, L29, \dodoi{10.3847/2041-8213/aaa402}

\bibitem[{J.~C. Rastinejad {et~al.}(2022)Rastinejad, Paterson, Fong, Sand,
  Lundquist, Hosseinzadeh, Christensen, Daly, Gibbs, Hall, Shelly, \&
  Yang}]{rastinejad_systematic_2022}
Rastinejad, J.~C., Paterson, K., Fong, W., {et~al.} 2022, \bibinfo{title}{A
  {Systematic} {Exploration} of {Kilonova} {Candidates} from {Neutron} {Star}
  {Mergers} during the {Third} {Gravitational}-wave {Observing} {Run},} The
  Astrophysical Journal, 927, 50, \dodoi{10.3847/1538-4357/ac4d34}

\bibitem[{J.~C. {Rastinejad} {et~al.}(2022){Rastinejad}, {Paterson}, {Fong},
  {Sand}, {Lundquist}, {Hosseinzadeh}, {Christensen}, {Daly}, {Gibbs}, {Hall},
  {Shelly}, \& {Yang}}]{2022ApJ...927...50R}
{Rastinejad}, J.~C., {Paterson}, K., {Fong}, W., {et~al.} 2022,
  \bibinfo{title}{{A Systematic Exploration of Kilonova Candidates from Neutron
  Star Mergers during the Third Gravitational-wave Observing Run},} \apj, 927,
  50, \dodoi{10.3847/1538-4357/ac4d34}

\bibitem[{J.~C. {Rastinejad} {et~al.}(2025){Rastinejad}, {Levan}, {Jonker},
  {Kilpatrick}, {Fryer}, {Sarin}, {Gompertz}, {Liu}, {Eyles-Ferris}, {Fong},
  {Burns}, {Gillanders}, {Mandel}, {Malesani}, {O'Brien}, {Tanvir}, {Ackley},
  {Aryan}, {Bauer}, {Bloemen}, {de Boer}, {Bom}, {Chac{\'o}n}, {Chambers},
  {Chen}, {Chrimes}, {van Dalen}, {D'Elia}, {De Pasquale}, {Fulton}, {Groot},
  {Gupta}, {Hartmann}, {van Hoof}, {Huber}, {Izzo}, {Jacobson-Galan},
  {Jakobsson}, {Kong}, {Laskar}, {Lowe}, {Magnier}, {Maiorano},
  {Martin-Carrillo}, {Mas-Ribas}, {Mata S{\'a}nchez}, {Nicholl}, {Nixon},
  {Oates}, {Paek}, {Palmerio}, {Paris}, {Pieterse}, {Pugliese}, {Quirola
  Vasquez}, {van Roestel}, {Rossi}, {Rouco Escorial}, {Salvaterra},
  {Schneider}, {Smartt}, {Smith}, {Smith}, {Srivastav}, {Torres}, {Ventura},
  {Vreeswijk}, {Wainscoat}, {Yang}, \& {Yang}}]{Rastinejad25}
{Rastinejad}, J.~C., {Levan}, A.~J., {Jonker}, P.~G., {et~al.} 2025,
  \bibinfo{title}{{EP 250108a/SN 2025kg: Observations of the Most Nearby
  Broad-line Type Ic Supernova Following an Einstein Probe Fast X-Ray
  Transient},} \apjl, 988, L13, \dodoi{10.3847/2041-8213/ade7f9}

\bibitem[{A. {Reguitti} {et~al.}(2025){Reguitti}, {Pastorello}, {Smartt},
  {Valerin}, {Pignata}, {Campana}, {Chen}, {Sankar. K.}, {Moran}, {Mazzali},
  {Duarte}, {Salmaso}, {Anderson}, {Ashall}, {Benetti}, {Gromadzki},
  {Gutierrez}, {Humina}, {Inserra}, {Kankare}, {Kravtsov}, {Muller-Bravo},
  {Pessi}, {Young}, {Chambers}, {de Boer}, {Gao}, {Huber}, {Lin}, {Lowe},
  {Magnier}, {Minguez}, {Smith}, {Smith}, {Srivastav}, {Wainscoat}, \&
  {Benedet}}]{Regutti2025}
{Reguitti}, A., {Pastorello}, A., {Smartt}, S.~J., {et~al.} 2025,
  \bibinfo{title}{{SN 2024abfo: a partially stripped SN II from a white
  supergiant},} arXiv e-prints, arXiv:2503.03851,
  \dodoi{10.48550/arXiv.2503.03851}

\bibitem[{K. Reitz(2023)Reitz}]{reitz_requests_2023}
Reitz, K. 2023, Requests v2.28.2,, GitHub \url{https://github.com/psf/requests}

\bibitem[{L. {Resmi} {et~al.}(2018){Resmi}, {Schulze}, {Ishwara-Chandra},
  {Misra}, {Buchner}, {De Pasquale}, {S{\'a}nchez-Ram{\'\i}rez}, {Klose},
  {Kim}, {Tanvir}, \& {O'Brien}}]{2018ApJ...867...57R}
{Resmi}, L., {Schulze}, S., {Ishwara-Chandra}, C.~H., {et~al.} 2018,
  \bibinfo{title}{{Low-frequency View of GW170817/GRB 170817A with the Giant
  Metrewave Radio Telescope},} \apj, 867, 57, \dodoi{10.3847/1538-4357/aae1a6}

\bibitem[{A. {Rest} {et~al.}(2005){Rest}, {Stubbs}, {Becker}, {Miknaitis},
  {Miceli}, {Covarrubias}, {Hawley}, {Smith}, {Suntzeff}, {Olsen}, {Prieto},
  {Hiriart}, {Welch}, {Cook}, {Nikolaev}, {Huber}, {Prochtor}, {Clocchiatti},
  {Minniti}, {Garg}, {Challis}, {Keller}, \& {Schmidt}}]{Rest05}
{Rest}, A., {Stubbs}, C., {Becker}, A.~C., {et~al.} 2005,
  \bibinfo{title}{{Testing LMC Microlensing Scenarios: The Discrimination Power
  of the SuperMACHO Microlensing Survey},} \apj, 634, 1103,
  \dodoi{10.1086/497060}

\bibitem[{L. {Rhodes} {et~al.}(2023){Rhodes}, {Anderson}, {van der Horst},
  {Leung}, {Ryder}, {Gulati}, {Chastain}, \& {PanRadio GRB
  Collaboration}}]{GCN35097}
{Rhodes}, L., {Anderson}, G.~E., {van der Horst}, A.~J., {et~al.} 2023,
  \bibinfo{title}{{GRB 231117A: ATCA detection of a radio counterpart},} GRB
  Coordinates Network, 35097, 1

\bibitem[{L. {Rhodes} {et~al.}(2025){Rhodes}, {Smirnov}, {Mooley}, \&
  {Woudt}}]{2025GCN.41666....1R}
{Rhodes}, L., {Smirnov}, O., {Mooley}, K., \& {Woudt}, P. 2025,
  \bibinfo{title}{{LIGO/VIRGO/KAGRA S250818k: No evidence for radio variability
  of AT2025ulz in MeerKAT 3 GHz data},} GRB Coordinates Network, 41666, 1

\bibitem[{R. {Ricci} {et~al.}(2025{\natexlab{a}}){Ricci}, {Becerra}, {Troja},
  \& {ERC BHianca Team}}]{GCN41046}
{Ricci}, R., {Becerra}, R.~L., {Troja}, E., \& {ERC BHianca Team}.
  2025{\natexlab{a}}, \bibinfo{title}{{EP250704a/GRB 250704B: 10 GHz VLA
  detection},} GRB Coordinates Network, 41046, 1

\bibitem[{R. {Ricci} {et~al.}(2025{\natexlab{b}}){Ricci}, {Yadav}, \&
  {Troja}}]{2025GCN.41464....1R}
{Ricci}, R., {Yadav}, M., \& {Troja}, E. 2025{\natexlab{b}},
  \bibinfo{title}{{LIGO/Virgo/KAGRA S250818k: 10 GHz VLA observations of
  AT2025ulz},} GRB Coordinates Network, 41464, 1

\bibitem[{R. {Ricci} {et~al.}(2025{\natexlab{c}}){Ricci}, {Yadav}, \&
  {Troja}}]{GCN41455}
{Ricci}, R., {Yadav}, M., \& {Troja}, E. 2025{\natexlab{c}},
  \bibinfo{title}{{GRB 250818B: 10 GHz VLA detection},} GRB Coordinates
  Network, 41455, 1

\bibitem[{R. {Ricci} {et~al.}(2025{\natexlab{d}}){Ricci}, {Yadav}, {Troja}, \&
  {ERC BHianca Team}}]{2025GCN.41542....1R}
{Ricci}, R., {Yadav}, M., {Troja}, E., \& {ERC BHianca Team}.
  2025{\natexlab{d}}, \bibinfo{title}{{LIGO/Virgo/KAGRA S250818k: VLA upper
  limits on AT2025ulz},} GRB Coordinates Network, 41542, 1

\bibitem[{L. Richardson(2023)Richardson}]{richardson_beautiful_2023}
Richardson, L. 2023, Beautiful {{Soup}} v4.12.2,, Launchpad
  \url{https://launchpad.net/beautifulsoup/}

\bibitem[{M.~W. Richmond {et~al.}(1994)Richmond, Treffers, Filippenko, Paik,
  Leibundgut, Schulman, \& Cox}]{richmond_ubvri_1994}
Richmond, M.~W., Treffers, R.~R., Filippenko, A.~V., {et~al.} 1994,
  \bibinfo{title}{{UBVRI} {Photometry} of {SN} {1993J} in {M81}: {The} {First}
  120 {Days},} The Astronomical Journal, 107, 1022, \dodoi{10.1086/116915}

\bibitem[{M.~W. {Richmond} {et~al.}(1994){Richmond}, {Treffers}, {Filippenko},
  {Paik}, {Leibundgut}, {Schulman}, \& {Cox}}]{R1994}
{Richmond}, M.~W., {Treffers}, R.~R., {Filippenko}, A.~V., {et~al.} 1994,
  \bibinfo{title}{{UBVRI Photometry of SN 1993J in M81: The First 120 Days},}
  \aj, 107, 1022, \dodoi{10.1086/116915}

\bibitem[{N. {Rudolf} {et~al.}(2016){Rudolf}, {G{\"u}nther}, {Schneider}, \&
  {Schmitt}}]{telluric_1}
{Rudolf}, N., {G{\"u}nther}, H.~M., {Schneider}, P.~C., \& {Schmitt},
  J.~H.~M.~M. 2016, \bibinfo{title}{{Modelling telluric line spectra in the
  optical and infrared with an application to VLT/X-Shooter spectra},} \aap,
  585, A113, \dodoi{10.1051/0004-6361/201322749}

\bibitem[{A. {Salgundi}(2025){Salgundi}}]{2025TNSTR3323....1S}
{Salgundi}, A. 2025, \bibinfo{title}{{ZTF Transient Discovery Report for
  2025-08-22},} Transient Name Server Discovery Report, 2025-3323, 1

\bibitem[{A. {Santos} {et~al.}(2025){Santos}, {Bom}, {Kilpatrick},
  {Santana-Silva}, {Darc}, {Teixeira}, {Mendes de Oliveira}, \& {STEP
  Collaboration}}]{2025GCN.41501....1S}
{Santos}, A., {Bom}, C.~R., {Kilpatrick}, C.~D., {et~al.} 2025,
  \bibinfo{title}{{LIGO/Virgo/Kagra S250818k: STEP/T80N upper limits on
  2025ulz},} GRB Coordinates Network, 41501, 1

\bibitem[{A. {Santos} {et~al.}(2024){Santos}, {Kilpatrick}, {Bom}, {Darc},
  {Herpich}, {Lacerda}, {Sartori}, {Alvarez-Candal}, {Mendes de Oliveira},
  {Kanaan}, {Ribeiro}, \& {Schoenell}}]{Santos24}
{Santos}, A., {Kilpatrick}, C.~D., {Bom}, C.~R., {et~al.} 2024,
  \bibinfo{title}{{The S-PLUS Transient Extension Program: imaging pipeline,
  transient identification, and survey optimization for multimessenger
  astronomy},} \mnras, 529, 59, \dodoi{10.1093/mnras/stae466}

\bibitem[{N. {Sapir} \& E. {Waxman}(2017){Sapir} \& {Waxman}}]{SW2017}
{Sapir}, N., \& {Waxman}, E. 2017, \bibinfo{title}{{UV/Optical Emission from
  the Expanding Envelopes of Type II Supernovae},} \apj, 838, 130,
  \dodoi{10.3847/1538-4357/aa64df}

\bibitem[{N. Sarin {et~al.}(2024)Sarin, Hübner, Omand, Setzer, Schulze,
  Adhikari, Sagués-Carracedo, Galaudage, Wallace, Lamb, \&
  Lin}]{sarin_redback_2024}
Sarin, N., Hübner, M., Omand, C. M.~B., {et~al.} 2024,
  \bibinfo{title}{redback: a {Bayesian} inference software package for
  electromagnetic transients,} Monthly Notices of the Royal Astronomical
  Society, 531, 1203, \dodoi{10.1093/mnras/stae1238}

\bibitem[{V. Savchenko {et~al.}(2017)Savchenko, Ferrigno, Kuulkers, Bazzano,
  Bozzo, Brandt, Chenevez, Courvoisier, Diehl, Domingo, Hanlon, Jourdain, von
  Kienlin, Laurent, Lebrun, Lutovinov, Martin-Carrillo, Mereghetti, Natalucci,
  Rodi, Roques, Sunyaev, \& Ubertini}]{savchenko_integral_2017}
Savchenko, V., Ferrigno, C., Kuulkers, E., {et~al.} 2017,
  \bibinfo{title}{{INTEGRAL} {Detection} of the {First} {Prompt} {Gamma}-{Ray}
  {Signal} {Coincident} with the {Gravitational}-wave {Event} {GW170817},} The
  Astrophysical Journal Letters, 848, L15, \dodoi{10.3847/2041-8213/aa8f94}

\bibitem[{P.~L. {Schechter} {et~al.}(1993){Schechter}, {Mateo}, \&
  {Saha}}]{dophot}
{Schechter}, P.~L., {Mateo}, M., \& {Saha}, A. 1993, \bibinfo{title}{{DoPHOT, A
  CCD Photometry Program: Description and Tests},} \pasp, 105, 1342,
  \dodoi{10.1086/133316}

\bibitem[{E.~F. Schlafly \& D.~P. Finkbeiner(2011)Schlafly \&
  Finkbeiner}]{schlafly_measuring_2011}
Schlafly, E.~F., \& Finkbeiner, D.~P. 2011, \bibinfo{title}{{MEASURING}
  {REDDENING} {WITH} {SLOAN} {DIGITAL} {SKY} {SURVEY} {S}℡{LAR} {SPECTRA}
  {AND} {RECALIBRATING} {SFD},} The Astrophysical Journal, 737, 103,
  \dodoi{10.1088/0004-637X/737/2/103}

\bibitem[{D.~J. Schlegel {et~al.}(1998)Schlegel, Finkbeiner, \&
  Davis}]{schlegel_maps_1998}
Schlegel, D.~J., Finkbeiner, D.~P., \& Davis, M. 1998, \bibinfo{title}{Maps of
  {Dust} {Infrared} {Emission} for {Use} in {Estimation} of {Reddening} and
  {Cosmic} {Microwave} {Background} {Radiation} {Foregrounds},} The
  Astrophysical Journal, 500, 525, \dodoi{10.1086/305772}

\bibitem[{G. {Schroeder} {et~al.}(2025{\natexlab{a}}){Schroeder}, {Rastinejad},
  {Fong}, \& {Laskar}}]{GCN41038}
{Schroeder}, G., {Rastinejad}, J., {Fong}, W., \& {Laskar}, T.
  2025{\natexlab{a}}, \bibinfo{title}{{GRB 250704B / EP250704a: VLA radio
  detection},} GRB Coordinates Network, 41038, 1

\bibitem[{G. {Schroeder} {et~al.}(2025{\natexlab{b}}){Schroeder}, {Rhodes},
  {Fong}, {Laskar}, \& {Berger}}]{GCN41060}
{Schroeder}, G., {Rhodes}, L., {Fong}, W., {Laskar}, T., \& {Berger}, E.
  2025{\natexlab{b}}, \bibinfo{title}{{GRB 250704B / EP250704a: 1.3 GHz MeerKAT
  Detection},} GRB Coordinates Network, 41060, 1

\bibitem[{G. {Schroeder} {et~al.}(2024){Schroeder}, {Rhodes}, {Laskar},
  {Nugent}, {Rouco Escorial}, {Rastinejad}, {Fong}, {van der Horst}, {Veres},
  {Alexander}, {Andersson}, {Berger}, {Blanchard}, {Chastain}, {Christensen},
  {Fender}, {Green}, {Groot}, {Heywood}, {Horesh}, {Izzo}, {Kilpatrick},
  {K{\"o}rding}, {Lien}, {Malesani}, {McBride}, {Mooley}, {Rowlinson}, {Sears},
  {Stappers}, {Tanvir}, {Vergani}, {Wijers}, {Williams-Baldwin}, \&
  {Woudt}}]{2024ApJ...970..139S}
{Schroeder}, G., {Rhodes}, L., {Laskar}, T., {et~al.} 2024, \bibinfo{title}{{A
  Radio Flare in the Long-lived Afterglow of the Distant Short GRB 210726A:
  Energy Injection or a Reverse Shock from Shell Collisions?},} \apj, 970, 139,
  \dodoi{10.3847/1538-4357/ad49ab}

\bibitem[{G. {Schroeder} {et~al.}(2025{\natexlab{c}}){Schroeder}, {Fong},
  {Kilpatrick}, {Rouco Escorial}, {Laskar}, {Nugent}, {Rastinejad},
  {Alexander}, {Berger}, {Brink}, {Chornock}, {de Bom}, {Dong}, {Eftekhari},
  {Filippenko}, {Fuentes-Carvajal}, {Jacobson-Gal{\'a}n}, {Malkan}, {Margutti},
  {Pearson}, {Rhodes}, {Salinas}, {Sand}, {Santana-Silva}, {Santos}, {Sears},
  {Shrestha}, {Smith}, {Webb}, {de Wet}, \& {Yang}}]{2025ApJ...982...42S}
{Schroeder}, G., {Fong}, W.-f., {Kilpatrick}, C.~D., {et~al.}
  2025{\natexlab{c}}, \bibinfo{title}{{The Long-lived Broadband Afterglow of
  Short Gamma-Ray Burst 231117A and the Growing Radio-detected Short Gamma-Ray
  Burst Population},} \apj, 982, 42, \dodoi{10.3847/1538-4357/ada9e5}

\bibitem[{G. Schroeder {et~al.}(2025)Schroeder, Fong, Kilpatrick,
  Rouco~Escorial, Laskar, Nugent, Rastinejad, Alexander, Berger, Brink,
  Chornock, de~Bom, Dong, Eftekhari, Filippenko, Fuentes-Carvajal,
  Jacobson-Galán, Malkan, Margutti, Pearson, Rhodes, Salinas, Sand,
  Santana-Silva, Santos, Sears, Shrestha, Smith, Webb, de~Wet, \&
  Yang}]{schroeder_long-lived_2025}
Schroeder, G., Fong, W.-f., Kilpatrick, C.~D., {et~al.} 2025,
  \bibinfo{title}{The {Long}-lived {Broadband} {Afterglow} of {Short}
  {Gamma}-{Ray} {Burst} {231117A} and the {Growing} {Radio}-detected {Short}
  {Gamma}-{Ray} {Burst} {Population},} The Astrophysical Journal, 982, 42,
  \dodoi{10.3847/1538-4357/ada9e5}

\bibitem[{ {SCiMMA Project}(2023){SCiMMA
  Project}}]{scimma_project_hopskotch_2023}
{SCiMMA Project}. 2023, Hopskotch, \url{https://scimma.org/hopskotch.html}

\bibitem[{B.~J. Shappee {et~al.}(2014)Shappee, Prieto, Grupe, Kochanek, Stanek,
  De~Rosa, Mathur, Zu, Peterson, Pogge, Komossa, Im, Jencson, Holoien, Basu,
  Beacom, Szczygieł, Brimacombe, Adams, Campillay, Choi, Contreras, Dietrich,
  Dubberley, Elphick, Foale, Giustini, Gonzalez, Hawkins, Howell, Hsiao, Koss,
  Leighly, Morrell, Mudd, Mullins, Nugent, Parrent, Phillips, Pojmanski,
  Rosing, Ross, Sand, Terndrup, Valenti, Walker, \& Yoon}]{shappee_man_2014}
Shappee, B.~J., Prieto, J.~L., Grupe, D., {et~al.} 2014, \bibinfo{title}{The
  {Man} behind the {Curtain}: {X}-{Rays} {Drive} the {UV} through {NIR}
  {Variability} in the 2013 {Active} {Galactic} {Nucleus} {Outburst} in {NGC}
  2617,} The Astrophysical Journal, 788, 48, \dodoi{10.1088/0004-637X/788/1/48}

\bibitem[{L. {Shingles} {et~al.}(2021){Shingles}, {Smith}, {Young}, {Smartt},
  {Tonry}, {Denneau}, {Heinze}, {Weiland}, {Flewelling}, {Stalder},
  {Clocchiatti}, {F{\"o}rster}, {Pignata}, {Rest}, {Anderson}, {Stubbs}, \&
  {Erasmus}}]{2021TNSAN...7....1S}
{Shingles}, L., {Smith}, K.~W., {Young}, D.~R., {et~al.} 2021,
  \bibinfo{title}{{Release of the ATLAS Forced Photometry server for public
  use},} Transient Name Server AstroNote, 7, 1

\bibitem[{M. {Shrestha} {et~al.}(2024){Shrestha}, {Pearson}, {Wyatt}, {Sand},
  {Hosseinzadeh}, {Bostroem}, {Andrews}, {Dong}, {Hoang}, {Janzen}, {Jencson},
  {Lundquist}, {Mehta}, {Retamal}, {Valenti}, {Rastinejad}, {Daly}, {Porter},
  {Hinz}, {Self}, {Weiner}, {Williams}, {Hiramatsu}, {Howell}, {McCully},
  {Gonzalez}, {Pellegrino}, {Terreran}, {Newsome}, {Farah}, {Itagaki}, {Jha},
  {Kwok}, {Smith}, {Schwab}, {Rho}, \& {Yang}}]{Shrestha2024}
{Shrestha}, M., {Pearson}, J., {Wyatt}, S., {et~al.} 2024,
  \bibinfo{title}{{Evidence of Weak Circumstellar Medium Interaction in the
  Type II SN 2023axu},} \apj, 961, 247, \dodoi{10.3847/1538-4357/ad11e1}

\bibitem[{D.~L. Shupe {et~al.}(2012)Shupe, Laher, {Storrie-Lombardi}, Surace,
  Grillmair, Levitan, \& Sesar}]{shupe_more_2012}
Shupe, D.~L., Laher, R.~R., {Storrie-Lombardi}, L., {et~al.} 2012,
  \bibinfo{title}{More flexibility in representing geometric distortion in
  astronomical images,} Proc.\ SPIE, 8451, E1M, \dodoi{10.1117/12.925460}

\bibitem[{L. Singer(2022)Singer}]{singer_gracedb-sdk_2022}
Singer, L. 2022, gracedb-sdk v0.1.7,
  \url{https://git.ligo.org/emfollow/gracedb-sdk}

\bibitem[{L.~P. Singer {et~al.}(2022)Singer, Parazin, Coughlin, Bloom,
  {Crellin-Quick}, Goldstein, \& van~der Walt}]{singer_healpix_2022}
Singer, L.~P., Parazin, B., Coughlin, M.~W., {et~al.} 2022,
  \bibinfo{title}{{{HEALPix Alchemy}}: {{Fast All-Sky Geometry}} and {{Image
  Arithmetic}} in a {{Relational Database}} for {{Multimessenger Astronomy
  Brokers}},} AJ, 163, 209, \dodoi{10.3847/1538-3881/ac5ab8}

\bibitem[{L.~P. Singer \& L.~R. Price(2016)Singer \& Price}]{singer_rapid_2016}
Singer, L.~P., \& Price, L.~R. 2016, \bibinfo{title}{Rapid {{Bayesian}}
  position reconstruction for gravitational-wave transients,} PhRvD, 93,
  024013, \dodoi{10.1103/PhysRevD.93.024013}

\bibitem[{L.~P. Singer {et~al.}(2016{\natexlab{a}})Singer, Chen, Holz, Farr,
  Price, Raymond, Cenko, Gehrels, Cannizzo, Kasliwal, Nissanke, Coughlin, Farr,
  Urban, Vitale, Veitch, Graff, Berry, Mohapatra, \&
  Mandel}]{singer_going_2016}
Singer, L.~P., Chen, H.-Y., Holz, D.~E., {et~al.} 2016{\natexlab{a}},
  \bibinfo{title}{{GOING} {THE} {DISTANCE}: {MAPPING} {HOST} {GALAXIES} {OF}
  {LIGO} {AND} {VIRGO} {SOURCES} {IN} {THREE} {DIMENSIONS} {USING} {LOCAL}
  {COSMOGRAPHY} {AND} {TARGETED} {FOLLOW}-{UP},} The Astrophysical Journal
  Letters, 829, L15, \dodoi{10.3847/2041-8205/829/1/L15}

\bibitem[{L.~P. Singer {et~al.}(2016{\natexlab{b}})Singer, Chen, Holz, Farr,
  Price, Raymond, Cenko, Gehrels, Cannizzo, Kasliwal, Nissanke, Coughlin, Farr,
  Urban, Vitale, Veitch, Graff, Berry, Mohapatra, \&
  Mandel}]{singer_supplement_2016}
Singer, L.~P., Chen, H.-Y., Holz, D.~E., {et~al.} 2016{\natexlab{b}},
  \bibinfo{title}{{SUPPLEMENT}: “{GOING} {THE} {DISTANCE}: {MAPPING} {HOST}
  {GALAXIES} {OF} {LIGO} {AND} {VIRGO} {SOURCES} {IN} {THREE} {DIMENSIONS}
  {USING} {LOCAL} {COSMOGRAPHY} {AND} {TARGETED} {FOLLOW}-{UP}” (2016,
  {ApJL}, 829, {L15}),} The Astrophysical Journal Supplement Series, 226, 10,
  \dodoi{10.3847/0067-0049/226/1/10}

\bibitem[{S.~J. Smartt {et~al.}(2017)Smartt, Chen, Jerkstrand, Coughlin,
  Kankare, Sim, Fraser, Inserra, Maguire, Chambers, Huber, Krühler, Leloudas,
  Magee, Shingles, Smith, Young, Tonry, Kotak, Gal-Yam, Lyman, Homan, Agliozzo,
  Anderson, Angus, Ashall, Barbarino, Bauer, Berton, Botticella, Bulla, Bulger,
  Cannizzaro, Cano, Cartier, Cikota, Clark, De~Cia, Della~Valle, Denneau,
  Dennefeld, Dessart, Dimitriadis, Elias-Rosa, Firth, Flewelling, Flörs,
  Franckowiak, Frohmaier, Galbany, González-Gaitán, Greiner, Gromadzki,
  Guelbenzu, Gutiérrez, Hamanowicz, Hanlon, Harmanen, Heintz, Heinze,
  Hernandez, Hodgkin, Hook, Izzo, James, Jonker, Kerzendorf, Klose,
  Kostrzewa-Rutkowska, Kowalski, Kromer, Kuncarayakti, Lawrence, Lowe, Magnier,
  Manulis, Martin-Carrillo, Mattila, McBrien, Müller, Nordin, O'Neill, Onori,
  Palmerio, Pastorello, Patat, Pignata, Podsiadlowski, Pumo, Prentice, Rau,
  Razza, Rest, Reynolds, Roy, Ruiter, Rybicki, Salmon, Schady, Schultz,
  Schweyer, Seitenzahl, Smith, Sollerman, Stalder, Stubbs, Sullivan, Szegedi,
  Taddia, Taubenberger, Terreran, van Soelen, Vos, Wainscoat, Walton, Waters,
  Weiland, Willman, Wiseman, Wright, Wyrzykowski, \&
  Yaron}]{smartt_kilonova_2017}
Smartt, S.~J., Chen, T.~W., Jerkstrand, A., {et~al.} 2017, \bibinfo{title}{A
  kilonova as the electromagnetic counterpart to a gravitational-wave source,}
  Nature, 551, 75, \dodoi{10.1038/nature24303}

\bibitem[{D. Smith(2022)Smith}]{smith_crispy-bootstrap4_2022}
Smith, D. 2022, crispy-bootstrap4 v2022.1,, GitHub
  \url{https://github.com/django-crispy-forms/crispy-bootstrap4}

\bibitem[{K.~W. {Smith} {et~al.}(2020){Smith}, {Smartt}, {Young}, {Tonry},
  {Denneau}, {Flewelling}, {Heinze}, {Weiland}, {Stalder}, {Rest}, {Stubbs},
  {Anderson}, {Chen}, {Clark}, {Do}, {F{\"o}rster}, {Fulton}, {Gillanders},
  {McBrien}, {O'Neill}, {Srivastav}, \& {Wright}}]{2020PASP..132h5002S}
{Smith}, K.~W., {Smartt}, S.~J., {Young}, D.~R., {et~al.} 2020,
  \bibinfo{title}{{Design and Operation of the ATLAS Transient Science
  Server},} \pasp, 132, 085002, \dodoi{10.1088/1538-3873/ab936e}

\bibitem[{M. Soares-Santos {et~al.}(2017)Soares-Santos, Holz, Annis, Chornock,
  Herner, Berger, Brout, Chen, Kessler, Sako, Allam, Tucker, Butler, Palmese,
  Doctor, Diehl, Frieman, Yanny, Lin, Scolnic, Cowperthwaite, Neilsen,
  Marriner, Kuropatkin, Hartley, Paz-Chinchón, Alexander, Balbinot, Blanchard,
  Brown, Carlin, Conselice, Cook, Drlica-Wagner, Drout, Durret, Eftekhari,
  Farr, Finley, Foley, Fong, Fryer, García-Bellido, Gill, Gruendl, Hanna,
  Kasen, Li, Lopes, Lourenço, Margutti, Marshall, Matheson, Medina, Metzger,
  Muñoz, Muir, Nicholl, Quataert, Rest, Sauseda, Schlegel, Secco, Sobreira,
  Stebbins, Villar, Vivas, Walker, Wester, Williams, Zenteno, Zhang, Abbott,
  Abdalla, Banerji, Bechtol, Benoit-Lévy, Bertin, Brooks, Buckley-Geer, Burke,
  Rosell, Kind, Carretero, Castander, Crocce, Cunha, D’Andrea, Costa, Davis,
  Desai, Dietrich, Doel, Eifler, Fernandez, Flaugher, Fosalba, Gaztanaga,
  Gerdes, Giannantonio, Goldstein, Gruen, Gschwend, Gutierrez, Honscheid, Jain,
  James, Jeltema, Johnson, Johnson, Kent, Krause, Kron, Kuehn, Kuhlmann, Lahav,
  Lima, Maia, March, McMahon, Menanteau, Miquel, Mohr, Nichol, Nord, Ogando,
  Petravick, Plazas, Romer, Roodman, Rykoff, Sanchez, Scarpine, Schubnell,
  Sevilla-Noarbe, Smith, Smith, Suchyta, Swanson, Tarle, Thomas, Thomas,
  Troxel, Vikram, Wechsler, Weller, Survey, \&
  Collaboration)}]{soares-santos_electromagnetic_2017}
Soares-Santos, M., Holz, D.~E., Annis, J., {et~al.} 2017, \bibinfo{title}{The
  {Electromagnetic} {Counterpart} of the {Binary} {Neutron} {Star} {Merger}
  {LIGO}/{Virgo} {GW170817}. {I}. {Discovery} of the {Optical} {Counterpart}
  {Using} the {Dark} {Energy} {Camera},} The Astrophysical Journal Letters,
  848, L16, \dodoi{10.3847/2041-8213/aa9059}

\bibitem[{A.~M. {Soderberg} {et~al.}(2006){Soderberg}, {Berger}, {Kasliwal},
  {Frail}, {Price}, {Schmidt}, {Kulkarni}, {Fox}, {Cenko}, {Gal-Yam}, {Nakar},
  \& {Roth}}]{2006ApJ...650..261S}
{Soderberg}, A.~M., {Berger}, E., {Kasliwal}, M., {et~al.} 2006,
  \bibinfo{title}{{The Afterglow, Energetics, and Host Galaxy of the Short-Hard
  Gamma-Ray Burst 051221a},} \apj, 650, 261, \dodoi{10.1086/506429}

\bibitem[{N. {Sravan} {et~al.}(2020){Sravan}, {Marchant}, {Kalogera},
  {Milisavljevic}, \& {Margutti}}]{S2020}
{Sravan}, N., {Marchant}, P., {Kalogera}, V., {Milisavljevic}, D., \&
  {Margutti}, R. 2020, \bibinfo{title}{{Progenitors of Type IIb Supernovae. II.
  Observable Properties},} \apj, 903, 70, \dodoi{10.3847/1538-4357/abb8d5}

\bibitem[{T.~D. Staley(2014)Staley}]{staley_voevent-parse_2014}
Staley, T.~D. 2014, voevent-parse: {{Parse}}, manipulate, and generate
  {{VOEvent XML}} packets,, Astrophysics Source Code Library
  \doeprint{1411.003}

\bibitem[{R. {Stein}(2025){Stein}}]{2025TNSTR3264....1S}
{Stein}, R. 2025, \bibinfo{title}{{ZTF Transient Discovery Report for
  2025-08-18},} Transient Name Server Discovery Report, 2025-3264, 1

\bibitem[{R. {Stein} {et~al.}(2025){Stein}, {Ahumada}, {Kasliwal}, {Du Laz},
  {Pathak}, {Swain}, {Salgundi}, {Bhalerao}, {Hall}, {Ztf Collaboration}, \&
  {Growth Collaboration}}]{2025GCN.41414....1S}
{Stein}, R., {Ahumada}, T., {Kasliwal}, M., {et~al.} 2025,
  \bibinfo{title}{{LIGO/Virgo/KAGRA S250818k: Candidates from the Zwicky
  Transient Facility},} GRB Coordinates Network, 41414, 1

\bibitem[{M. Stienstra {et~al.}(2023)Stienstra, Takhteyev, \&
  Limberg}]{stienstra_python-markdown_2023}
Stienstra, M., Takhteyev, Y., \& Limberg, W. 2023, Python-{{Markdown}} v3.4.3,,
  GitHub \url{https://github.com/Python-Markdown/markdown}

\bibitem[{ {STSCI Development Team}(2012){STSCI Development Team}}]{drizzlepac}
{STSCI Development Team}. 2012, {DrizzlePac: HST image software},, Astrophysics
  Source Code Library, record ascl:1212.011 \doeprint{1212.011}

\bibitem[{B.~M. {Subrayan} {et~al.}(2025){Subrayan}, {Sand}, {Bostroem}, {Jha},
  {Ravi}, {Schwab}, {Andrews}, {Hosseinzadeh}, {Valenti}, {Dong}, {Pearson},
  {Shrestha}, {Kwok}, {Hoang}, {Rho}, {Park}, {Yoon}, {Geballe}, {Haislip},
  {Janzen}, {Kouprianov}, {Mehta}, {Meza Retamal}, {Reichart}, {Andrews},
  {Farah}, {Newsome}, {Howell}, \& {McCully}}]{Subrayan_2024uwq}
{Subrayan}, B.~M., {Sand}, D.~J., {Bostroem}, K.~A., {et~al.} 2025,
  \bibinfo{title}{{Early Shock Cooling Observations and Progenitor Constraints
  of Type IIb Supernova SN 2024uwq},} \apjl, 990, L68,
  \dodoi{10.3847/2041-8213/adfe52}

\bibitem[{V. {Swain} {et~al.}(2025){Swain}, {Anupama}, {Sahu}, {Barway}, {Das},
  {Basu}, {Chamoli}, {Singh}, {Tiwari}, {Mohan}, {Saikia}, \&
  {Bhalerao}}]{2025GCN.41837....1S}
{Swain}, V., {Anupama}, G.~C., {Sahu}, D.~K., {et~al.} 2025,
  \bibinfo{title}{{LIGO/Virgo/KAGRA S250818k: HCT optical follow-up},} GRB
  Coordinates Network, 41837, 1

\bibitem[{M. Tanaka \& K. Hotokezaka(2013)Tanaka \&
  Hotokezaka}]{tanaka_radiative_2013}
Tanaka, M., \& Hotokezaka, K. 2013, \bibinfo{title}{Radiative {Transfer}
  {Simulations} of {Neutron} {Star} {Merger} {Ejecta},} The Astrophysical
  Journal, 775, 113, \dodoi{10.1088/0004-637X/775/2/113}

\bibitem[{N.~R. Tanvir {et~al.}(2017)Tanvir, Levan, González-Fernández,
  Korobkin, Mandel, Rosswog, Hjorth, D’Avanzo, Fruchter, Fryer, Kangas,
  Milvang-Jensen, Rosetti, Steeghs, Wollaeger, Cano, Copperwheat, Covino,
  D’Elia, de~Ugarte~Postigo, Evans, Even, Fairhurst, Jaimes, Fontes, Fujii,
  Fynbo, Gompertz, Greiner, Hodosan, Irwin, Jakobsson, Jørgensen, Kann, Lyman,
  Malesani, McMahon, Melandri, O’Brien, Osborne, Palazzi, Perley, Pian,
  Piranomonte, Rabus, Rol, Rowlinson, Schulze, Sutton, Thöne, Ulaczyk, Watson,
  Wiersema, \& Wijers}]{tanvir_emergence_2017}
Tanvir, N.~R., Levan, A.~J., González-Fernández, C., {et~al.} 2017,
  \bibinfo{title}{The {Emergence} of a {Lanthanide}-rich {Kilonova} {Following}
  the {Merger} of {Two} {Neutron} {Stars},} The Astrophysical Journal Letters,
  848, L27, \dodoi{10.3847/2041-8213/aa90b6}

\bibitem[{L. Tartaglia {et~al.}(2017)Tartaglia, Fraser, Sand, Valenti, Smartt,
  McCully, Anderson, Arcavi, Elias-Rosa, Galbany, Gal-Yam, Haislip,
  Hosseinzadeh, Howell, Inserra, Jha, Kankare, Lundqvist, Maguire, Mattila,
  Reichart, Smith, Smith, Stritzinger, Sullivan, Taddia, \&
  Tomasella}]{tartaglia_progenitor_2017}
Tartaglia, L., Fraser, M., Sand, D.~J., {et~al.} 2017, \bibinfo{title}{The
  {Progenitor} and {Early} {Evolution} of the {Type} {IIb} {SN} 2016gkg,} The
  Astrophysical Journal, 836, L12, \dodoi{10.3847/2041-8213/aa5c7f}

\bibitem[{M. Tebeka(2022)Tebeka}]{tebeka_fastavro_2022}
Tebeka, M. 2022, fastavro v1.6.1,, GitHub
  \url{https://github.com/fastavro/fastavro}

\bibitem[{A.~L. {Thakur} {et~al.}(2020){Thakur}, {Dichiara}, {Troja}, {Chase},
  {S{\'a}nchez-Ram{\'\i}rez}, {Piro}, {Fryer}, {Butler}, {Watson}, {Wollaeger},
  {Ambrosi}, {Becerra Gonz{\'a}lez}, {Becerra}, {Bruni}, {Cenko}, {Cusumano},
  {D'A{\`\i}}, {Durbak}, {Fontes}, {Gatkine}, {Hungerford}, {Korobkin},
  {Kutyrev}, {Lee}, {Lotti}, {Minervini}, {Novara}, {La Parola}, {Pereyra},
  {Ricci}, {Tiengo}, \& {Veilleux}}]{Thakur+20}
{Thakur}, A.~L., {Dichiara}, S., {Troja}, E., {et~al.} 2020, \bibinfo{title}{{A
  search for optical and near-infrared counterparts of the compact binary
  merger GW190814},} \mnras, 499, 3868, \dodoi{10.1093/mnras/staa2798}

\bibitem[{D. {Tody}(1986){Tody}}]{Tody1986}
{Tody}, D. 1986, \bibinfo{title}{{The IRAF Data Reduction and Analysis
  System},} in Society of Photo-Optical Instrumentation Engineers (SPIE)
  Conference Series, Vol. 627, Instrumentation in astronomy VI, ed. D.~L.
  {Crawford}, 733, \dodoi{10.1117/12.968154}

\bibitem[{D. {Tody}(1993){Tody}}]{Tody1993}
{Tody}, D. 1993, \bibinfo{title}{{IRAF in the Nineties},} in Astronomical
  Society of the Pacific Conference Series, Vol.~52, Astronomical Data Analysis
  Software and Systems II, ed. R.~J. {Hanisch}, R.~J.~V. {Brissenden}, \&
  J.~{Barnes}, 173

\bibitem[{ {TOM Toolkit Project} {et~al.}(2023){TOM Toolkit Project}, Collom,
  Lindstrom, \& Nation}]{tom_toolkit_project_tom_nonlocalizedevents_2023}
{TOM Toolkit Project}, Collom, D., Lindstrom, L., \& Nation, J. 2023,
  tom\_nonlocalizedevents v0.7.7,, GitHub
  \url{https://github.com/TOMToolkit/tom_nonlocalizedevents}

\bibitem[{ {TOM Toolkit Project} \& W. Lindstrom(2023){TOM Toolkit Project} \&
  Lindstrom}]{tom_toolkit_project_tom-alertstreams_2023}
{TOM Toolkit Project}, \& Lindstrom, W. 2023, tom-alertstreams v0.6.2,, GitHub
  \url{https://github.com/TOMToolkit/tom-alertstreams}

\bibitem[{J. {Tonry} {et~al.}(2025){Tonry}, {Denneau}, {Weiland}, {Siverd},
  {Erasmus}, {Koorts}, {Jordan}, {Suc}, {Alarc{\'o}n}, {Licandro}, {Nichita},
  {Smartt}, {Smith}, {Young}, {Nicholl}, {Fulton}, {McCollum}, {Moore},
  {Weston}, {Sheng}, {Angus}, {Wilson}, {Aamer}, {Magill}, {Ramsden},
  {Shingles}, {Srivastav}, {Gillanders}, {Stevance}, {Cooper}, {Stoppa},
  {Tweddle}, {Rhodes}, {Rest}, {Chen}, {Stubbs}, {Sommer}, \&
  {Schmidt}}]{2025TNSTR3286....1T}
{Tonry}, J., {Denneau}, L., {Weiland}, H., {et~al.} 2025,
  \bibinfo{title}{{ATLAS Transient Discovery Report for 2025-08-20},} Transient
  Name Server Discovery Report, 2025-3286, 1

\bibitem[{J.~L. {Tonry} {et~al.}(2018){Tonry}, {Denneau}, {Heinze}, {Stalder},
  {Smith}, {Smartt}, {Stubbs}, {Weiland}, \& {Rest}}]{2018PASP..130f4505T}
{Tonry}, J.~L., {Denneau}, L., {Heinze}, A.~N., {et~al.} 2018,
  \bibinfo{title}{{ATLAS: A High-cadence All-sky Survey System},} \pasp, 130,
  064505, \dodoi{10.1088/1538-3873/aabadf}

\bibitem[{M. Trier \& B. {van Oostveen}(2023)Trier \& {van
  Oostveen}}]{trier_django_2023}
Trier, M., \& {van Oostveen}, B. 2023, Django {{Extensions}} v3.2.3,, GitHub
  \url{https://github.com/django-extensions/django-extensions}

\bibitem[{E. {Troja} {et~al.}(2025){Troja}, {O'Connor}, \&
  {Becerra}}]{2025GCN.41506....1T}
{Troja}, E., {O'Connor}, B., \& {Becerra}, R.~L. 2025,
  \bibinfo{title}{{LIGO/Virgo/KAGRA S250818k: HST nIR detection of AT
  2025ulz},} GRB Coordinates Network, 41506, 1

\bibitem[{E. {Troja} {et~al.}(2019){Troja}, {van Eerten}, {Ryan}, {Ricci},
  {Burgess}, {Wieringa}, {Piro}, {Cenko}, \& {Sakamoto}}]{2019MNRAS.489.1919T}
{Troja}, E., {van Eerten}, H., {Ryan}, G., {et~al.} 2019, \bibinfo{title}{{A
  year in the life of GW 170817: the rise and fall of a structured jet from a
  binary neutron star merger},} \mnras, 489, 1919,
  \dodoi{10.1093/mnras/stz2248}

\bibitem[{D.~L. Tucker {et~al.}(2022)Tucker, Wiesner, Allam, Soares-Santos,
  Bom, Butner, Garcia, Morgan, Olivares~E., Palmese, Santana-Silva,
  Shrivastava, Annis, García-Bellido, Gill, Herner, Kilpatrick, Makler,
  Sherman, Amara, Lin, Smith, Swann, Arcavi, Bachmann, Bechtol, Berlfein,
  Briceño, Brout, Butler, Cartier, Casares, Chen, Conselice, Contreras, Cook,
  Cooke, Dage, D'Andrea, Davis, de~Carvalho, Diehl, Dietrich, Doctor,
  Drlica-Wagner, Drout, Farr, Finley, Fishbach, Foley, Förster-Burón,
  Fosalba, Friedel, Frieman, Frohmaier, Gruendl, Hartley, Hiramatsu, Holz,
  Howell, Kawash, Kessler, Kuropatkin, Lahav, Lundgren, Lundquist, Malik, Mann,
  Marriner, Marshall, Martínez-Vázquez, McCully, Menanteau, Meza, Narayan,
  Neilsen, Nicolaou, Nichol, Paz-Chinchón, Pereira, Pineda, Points,
  Quirola-Vásquez, Rembold, Rest, Rodriguez, Romer, Sako, Salim, Scolnic,
  Smith, Strader, Sullivan, Swanson, Thomas, Valenti, Varga, Walker, Weller,
  Wood, Yanny, Zenteno, Aguena, Andrade-Oliveira, Bertin, Brooks, Burke,
  Carnero~Rosell, Carrasco~Kind, Carretero, Costanzi, da~Costa, De~Vicente,
  Desai, Everett, Ferrero, Flaugher, Gaztanaga, Gerdes, Gruen, Gschwend,
  Gutierrez, Hinton, Hollowood, Honscheid, James, Kuehn, Lima, Maia, Miquel,
  Ogando, Pieres, Plazas~Malagón, Rodriguez-Monroy, Sanchez, Scarpine,
  Schubnell, Serrano, Sevilla-Noarbe, Suchyta, Tarle, To, \&
  Zhang}]{tucker_soargoodman_2022}
Tucker, D.~L., Wiesner, M.~P., Allam, S.~S., {et~al.} 2022,
  \bibinfo{title}{{SOAR}/{Goodman} {Spectroscopic} {Assessment} of {Candidate}
  {Counterparts} of the {LIGO}/{Virgo} {Event} {GW190814},} The Astrophysical
  Journal, 929, 115, \dodoi{10.3847/1538-4357/ac5b60}

\bibitem[{S.~A. {Usman} {et~al.}(2016){Usman}, {Nitz}, {Harry}, {Biwer},
  {Brown}, {Cabero}, {Capano}, {Dal Canton}, {Dent}, {Fairhurst}, {Kehl},
  {Keppel}, {Krishnan}, {Lenon}, {Lundgren}, {Nielsen}, {Pekowsky}, {Pfeiffer},
  {Saulson}, {West}, \& {Willis}}]{pycbc}
{Usman}, S.~A., {Nitz}, A.~H., {Harry}, I.~W., {et~al.} 2016,
  \bibinfo{title}{{The PyCBC search for gravitational waves from compact binary
  coalescence},} Classical and Quantum Gravity, 33, 215004,
  \dodoi{10.1088/0264-9381/33/21/215004}

\bibitem[{S. Valenti {et~al.}(2017)Valenti, Sand, Yang, Cappellaro, Tartaglia,
  Corsi, Jha, Reichart, Haislip, \& Kouprianov}]{valenti_discovery_2017}
Valenti, S., Sand, D.~J., Yang, S., {et~al.} 2017, \bibinfo{title}{The
  {Discovery} of the {Electromagnetic} {Counterpart} of {GW170817}: {Kilonova}
  {AT} 2017gfo/{DLT17ck},} The Astrophysical Journal, 848, L24,
  \dodoi{10.3847/2041-8213/aa8edf}

\bibitem[{D. Verheul(2021)Verheul}]{verheul_django-bootstrap4_2021}
Verheul, D. 2021, django-bootstrap4 v3.0.1,, GitHub
  \url{https://github.com/zostera/django-bootstrap4}

\bibitem[{N. {Vieira} {et~al.}(2020){Vieira}, {Ruan}, {Haggard}, {Drout},
  {Nynka}, {Boyce}, {Spekkens}, {Safi-Harb}, {Carlberg}, {Fern{\'a}ndez},
  {Piro}, {Afsariardchi}, \& {Moon}}]{Vieira+20}
{Vieira}, N., {Ruan}, J.~J., {Haggard}, D., {et~al.} 2020, \bibinfo{title}{{A
  Deep CFHT Optical Search for a Counterpart to the Possible Neutron Star-Black
  Hole Merger GW190814},} \apj, 895, 96, \dodoi{10.3847/1538-4357/ab917d}

\bibitem[{V.~A. Villar {et~al.}(2017)Villar, Guillochon, Berger, Metzger,
  Cowperthwaite, Nicholl, Alexander, Blanchard, Chornock, Eftekhari, Fong,
  Margutti, \& Williams}]{villar_combined_2017}
Villar, V.~A., Guillochon, J., Berger, E., {et~al.} 2017, \bibinfo{title}{The
  {Combined} {Ultraviolet}, {Optical}, and {Near}-infrared {Light} {Curves} of
  the {Kilonova} {Associated} with the {Binary} {Neutron} {Star} {Merger}
  {GW170817}: {Unified} {Data} {Set}, {Analytic} {Models}, and {Physical}
  {Implications},} The Astrophysical Journal, 851, L21,
  \dodoi{10.3847/2041-8213/aa9c84}

\bibitem[{P. Virtanen {et~al.}(2020)Virtanen, Gommers, Oliphant, Haberland,
  Reddy, Cournapeau, Burovski, Peterson, Weckesser, Bright, {van der Walt},
  Brett, Wilson, Jarrod~Millman, Mayorov, Nelson, Jones, Kern, Larson, Carey,
  Polat, Feng, Moore, {Vand erPlas}, Laxalde, Perktold, Cimrman, Henriksen,
  Quintero, Harris, Archibald, Ribeiro, Pedregosa, {van Mulbregt}, \&
  Contributors}]{virtanen_scipy_2020}
Virtanen, P., Gommers, R., Oliphant, T.~E., {et~al.} 2020,
  \bibinfo{title}{{{SciPy}} 1.0: {{Fundamental Algorithms}} for {{Scientific
  Computing}} in {{Python}},} NatMe, 17, 261, \dodoi{10.1038/s41592-019-0686-2}

\bibitem[{T. Waddington(2020)Waddington}]{waddington_django-gravatar_2020}
Waddington, T. 2020, django-gravatar v1.4.4,, GitHub
  \url{https://github.com/twaddington/django-gravatar}

\bibitem[{A.~M. {Watson} {et~al.}(2020){Watson}, {Butler}, {Lee}, {Becerra},
  {Pereyra}, {Angeles}, {Farah}, {Figueroa}, {G{\'o}nzalez-Buitrago},
  {Quir{\'o}s}, {Ru{\'\i}z-D{\'\i}az-Soto}, {Tejada de Vargas}, {Tinoco}, \&
  {Wolfram}}]{Watson+20}
{Watson}, A.~M., {Butler}, N.~R., {Lee}, W.~H., {et~al.} 2020,
  \bibinfo{title}{{Limits on the electromagnetic counterpart to S190814bv},}
  \mnras, 492, 5916, \dodoi{10.1093/mnras/staa161}

\bibitem[{D.~J. {White} {et~al.}(2011){White}, {Daw}, \&
  {Dhillon}}]{White+11_GWGC}
{White}, D.~J., {Daw}, E.~J., \& {Dhillon}, V.~S. 2011, \bibinfo{title}{{A list
  of galaxies for gravitational wave searches},} Classical and Quantum Gravity,
  28, 085016, \dodoi{10.1088/0264-9381/28/8/085016}

\bibitem[{S.~E. Woosley {et~al.}(1994)Woosley, Eastman, Weaver, \&
  Pinto}]{woosley_sn_1994}
Woosley, S.~E., Eastman, R.~G., Weaver, T.~A., \& Pinto, P.~A. 1994,
  \bibinfo{title}{{SN} {1993J}: {A} {Type} {IIb} {Supernova},} The
  Astrophysical Journal, 429, 300, \dodoi{10.1086/174319}

\bibitem[{S.~E. {Woosley} {et~al.}(1994){Woosley}, {Eastman}, {Weaver}, \&
  {Pinto}}]{Woosley1994a}
{Woosley}, S.~E., {Eastman}, R.~G., {Weaver}, T.~A., \& {Pinto}, P.~A. 1994,
  \bibinfo{title}{{SN 1993J: A Type IIb Supernova},} \apj, 429, 300,
  \dodoi{10.1086/174319}

\bibitem[{O. {Yaron} \& A. {Gal-Yam}(2012){Yaron} \&
  {Gal-Yam}}]{2012PASP..124..668Y}
{Yaron}, O., \& {Gal-Yam}, A. 2012, \bibinfo{title}{{WISeREP{\textemdash}An
  Interactive Supernova Data Repository},} \pasp, 124, 668,
  \dodoi{10.1086/666656}

\bibitem[{S.-C. {Yoon} {et~al.}(2017){Yoon}, {Dessart}, \&
  {Clocchiatti}}]{Yoon2017}
{Yoon}, S.-C., {Dessart}, L., \& {Clocchiatti}, A. 2017, \bibinfo{title}{{Type
  Ib and IIb Supernova Progenitors in Interacting Binary Systems},} \apj, 840,
  10, \dodoi{10.3847/1538-4357/aa6afe}

\bibitem[{D. Young(2023)Young}]{young_fundamentals_2023}
Young, D. 2023, fundamentals v2.4.1,, Zenodo \dodoi{10.5281/zenodo.8037510}

\bibitem[{R. {Zhou} {et~al.}(2023){Zhou}, {Ferraro}, {White}, {DeRose},
  {Sailer}, {Aguilar}, {Ahlen}, {Bailey}, {Brooks}, {Claybaugh}, {Dawson}, {de
  la Macorra}, {Dey}, {Doel}, {Font-Ribera}, {Forero-Romero}, {Gontcho A
  Gontcho}, {Guy}, {Kremin}, {Lambert}, {Le Guillou}, {Levi}, {Magneville},
  {Manera}, {Meisner}, {Miquel}, {Moustakas}, {Myers}, {Newman}, {Nie},
  {Percival}, {Rezaie}, {Rossi}, {Sanchez}, {Schlegel}, {Schubnell}, {Seo},
  {Tarl{\'e}}, \& {Zhou}}]{Zhou+23_LSDR10}
{Zhou}, R., {Ferraro}, S., {White}, M., {et~al.} 2023, \bibinfo{title}{{DESI
  luminous red galaxy samples for cross-correlations},} \jcap, 2023, 097,
  \dodoi{10.1088/1475-7516/2023/11/097}

\bibitem[{A. Zonca {et~al.}(2019)Zonca, Singer, Lenz, Reinecke, Rosset, Hivon,
  \& Gorski}]{zonca_healpy_2019}
Zonca, A., Singer, L., Lenz, D., {et~al.} 2019, \bibinfo{title}{healpy: equal
  area pixelization and spherical harmonics transforms for data on the sphere
  in {{Python}},} JOSS, 4, 1298, \dodoi{10.21105/joss.01298}

\end{thebibliography}
\bibliographystyle{aasjournalv7}

\onecolumngrid 
\appendix

\section{Detailed Method for Deriving the Distance Score}\label{app:3d}
We begin with the following parameters:

\newcommand{\gwdist}{\overline{D_{\rm GW}}}
\newcommand{\gwsig}{\sigma_{\rm GW}}
\newcommand{\canddist}{\overline{D_{\rm C}}}
\newcommand{\candsig}{\sigma_{\rm C}}
\newcommand{\candsigpos}{\sigma_{\rm C}^{+}}
\newcommand{\candsigneg}{\sigma_{\rm C}^{-}}

\begin{itemize}
    \item $ \gwdist\rightarrow$ The mean distance to the gravitational wave event at the position of the candidate as given by the IGWN alerts.
    \item $\gwsig\rightarrow$ The symmetric uncertainty on $\gwdist$ reported in the IGWN alert.
    \item $\canddist \rightarrow$ The mean derived luminosity distance to the host galaxy with the lowest $P_{\rm cc}$ score.
    \item $\candsigpos$ and $\candsigneg \rightarrow$ The positive and negative uncertainties on $\canddist$, respectively. In the case of a $\canddist$ derived from a spectroscopic redshift or a photometric redshift with symmetric uncertainties $\candsig = \candsigpos = \candsigneg$.
\end{itemize}

For the GW event the uncertainties are symmetric so, assuming the uncertainties are Gaussian, the probability distribution of the distance to the GW event is given by equation \autoref{eq:gw-pdf}.
\begin{align}
    {\rm P_{\rm GW}(D)} = \frac{1}{\gwsig \sqrt{2\pi}} \exp\left[-\frac{1}{2}\left(\frac{D - \gwdist}{\gwsig}\right)^2\right] \label{eq:gw-pdf}
\end{align}
Since the uncertainty on the distance to the candidate event is potentially asymmetric, we use a split normal distribution (\autoref{eq:cand-pdf}) to describe the probability distribution.
\begin{align}
    {\rm P_{\rm C}(D)} = \sqrt{\frac{2}{\pi(\candsigpos + \candsigneg)^2}} \times
    \begin{cases}
        \exp\left[-\frac{1}{2}\left(\frac{D - \canddist}{\candsigpos}\right)^2\right], & D > \canddist \\ \\
        \exp\left[-\frac{1}{2}\left(\frac{D - \canddist}{\candsigneg}\right)^2\right], & D \leq \canddist
    \end{cases} \label{eq:cand-pdf}
\end{align}
The normalization factor $2 / [\sqrt{\pi}(\candsigpos + \candsigneg)]$ is derived by integrating the piecewise function from $-\infty \rightarrow\infty$. From these two probability distributions we compute the Bhattacharyya coefficient \citep{bhattacharyya_1943} --- a non-parametric, normalized measure of the distances between the distributions such that 0 indicates no similarity and 1 indicates identical distributions. 
\begin{align}
    {\rm S_{\rm dist}} &= \int_{0}^{\infty} \sqrt{{\rm P_{\rm GW}(D) \times P_{\rm C}(D)}} ~ dD \\
    &= \left[ \pi \gwsig (\candsigpos + \candsigneg)  \right]^{-1/2} \bigint_{0}^{\infty} \left[ \exp\left[-\frac{1}{2}\left(\frac{D - \gwdist}{\gwsig}\right)^2\right] \times     \begin{cases}
        \exp\left[-\frac{1}{2}\left(\frac{D - \canddist}{\candsigpos}\right)^2\right], & D > \canddist \\ \\
        \exp\left[-\frac{1}{2}\left(\frac{D - \canddist}{\candsigneg}\right)^2\right], & D \leq \canddist
    \end{cases} ~~~~~~ \right]^{1/2} {\rm dD} \label{eq:3dscore}
\end{align}
where $S_{\rm dist}$ is the score we assign for the distance association step in our vetting algorithm. In practice, to derive a score for the distance association we integrate equation \autoref{eq:3dscore} numerically using \texttt{numpy.trapezoid} over a range $0 \rightarrow 10,000$ Mpc.

\section{Candidates Table}\label{app:candidates}
All of the candidates are ranked by their score in \autoref{tab:candidates}.
\begin{longtable}{lccp{1cm}p{1cm}p{1cm}p{1cm}p{2cm}p{1cm}p{1cm}p{1cm}}
\caption{Candidate Scores} \\
\toprule
TNS Name & RA & Dec & Overall Score & 2D Score & Point Source Score & 3D Score & Peak Luminosity (erg/s) & Time of Peak (days) & Decay Rate (mag /day) & Pre-Detection Score \\
\midrule
\endfirsthead
\caption[]{Candidate Scores} \\
\toprule
TNS Name & RA & Dec & Overall Score & 2D Score & Point Source Score & 3D Score & Peak Luminosity (erg/s) & Time of Peak (days) & Decay Rate (mag /day) & Pre-Detection Score \\
\midrule
\endhead
\midrule
\multicolumn{11}{r}{Continued on next page} \\
\midrule
\endfoot
\bottomrule
\endlastfoot
AT2025usl & 237.1445 & 32.2938 & 0.73 & 0.85 & 1.00 & 0.86 & $7.42 \times 10^{41}$ & --- & --- & 1 \\
AT2025uuf & 310.1678 & 64.6115 & 0.51 & 0.59 & 1.00 & 0.87 & $5.11 \times 10^{41}$ & --- & --- & 1 \\
AT2025uus & 244.2946 & 39.6485 & 0.41 & 0.97 & 1.00 & 0.42 & $9.88 \times 10^{41}$ & --- & --- & 1 \\
AT2025uua & 261.4404 & 51.9287 & 0.39 & 0.39 & 1.00 & 1.00 & $8.90 \times 10^{41}$ & --- & --- & 1 \\
AT2025uow & 53.3796 & -30.0963 & 0.38 & 0.45 & 1.00 & 0.87 & $2.38 \times 10^{42}$ & 0.05 & 0.13 & 1 \\
AT2025usk & 241.2050 & 35.7908 & 0.35 & 0.99 & 1.00 & 0.35 & $1.04 \times 10^{42}$ & --- & --- & 1 \\
AT2025uvu & 236.7145 & 29.9994 & 0.31 & 0.76 & 1.00 & 0.42 & $5.68 \times 10^{41}$ & --- & --- & 1 \\
AT2025wfs & 236.5714 & 31.8389 & 0.30 & 0.83 & 1.00 & 0.36 & $3.44 \times 10^{41}$ & --- & --- & 1 \\
AT2025uxu & 270.6914 & 56.4917 & 0.24 & 0.30 & 1.00 & 0.80 & $4.97 \times 10^{41}$ & 0.16 & 0.25 & 1 \\
AT2025uut & 283.5697 & 60.0326 & 0.15 & 0.52 & 1.00 & 0.29 & $6.85 \times 10^{41}$ & --- & --- & 1 \\
AT2025utu & 301.9844 & 61.3465 & 0.15 & 0.25 & 1.00 & 0.63 & $6.34 \times 10^{41}$ & 0.37 & 0.33 & 1 \\
AT2025usn & 237.6131 & 30.3337 & 0.12 & 0.64 & 1.00 & 0.20 & $1.04 \times 10^{42}$ & --- & --- & 1 \\
AT2025uuc & 264.9940 & 53.3377 & 0.10 & 0.35 & 1.00 & 0.30 & $8.79 \times 10^{41}$ & --- & --- & 1 \\
AT2025uog & 58.3270 & -33.1844 & 0.09 & 0.52 & 1.00 & 0.18 & $1.60 \times 10^{42}$ & 0.05 & 0.20 & 1 \\
AT2025uul & 237.0139 & 29.7992 & 0.09 & 0.64 & 1.00 & 0.15 & $7.60 \times 10^{41}$ & --- & --- & 1 \\
AT2025uur & 236.7506 & 30.5273 & 0.09 & 0.81 & 1.00 & 0.12 & $6.46 \times 10^{41}$ & --- & --- & 1 \\
AT2025uzu & 244.2891 & 41.7935 & 0.09 & 0.53 & 1.00 & 0.17 & $7.50 \times 10^{41}$ & 0.14 & 0.12 & 1 \\
AT2025unm & 247.0348 & 42.0573 & 0.08 & 0.91 & 1.00 & 0.95 & $4.94 \times 10^{41}$ & 0.12 & 0.02 & 1 \\
AT2025uuk & 262.3959 & 54.6045 & 0.08 & 0.30 & 1.00 & 0.28 & $1.30 \times 10^{42}$ & --- & --- & 1 \\
AT2025war & 298.4335 & 60.7029 & 0.07 & 0.19 & 1.00 & 0.41 & $5.77 \times 10^{41}$ & --- & --- & 1 \\
AT2025uvt & 238.8852 & 31.0098 & 0.07 & 0.59 & 1.00 & 0.12 & $3.08 \times 10^{41}$ & --- & --- & 1 \\
AT2025wek & 237.0823 & 32.7603 & 0.07 & 0.83 & 1.00 & 0.09 & $8.83 \times 10^{41}$ & --- & --- & 1 \\
AT2025uug & 285.5158 & 59.3271 & 0.07 & 0.44 & 1.00 & 0.17 & $6.74 \times 10^{41}$ & --- & --- & 1 \\
AT2025uub & 300.2631 & 61.9816 & 0.06 & 0.49 & 1.00 & 0.13 & $1.80 \times 10^{42}$ & 0.37 & 0.13 & 1 \\
AT2025uya & 248.6400 & 43.9504 & 0.04 & 0.86 & 1.00 & 0.05 & $8.30 \times 10^{41}$ & --- & --- & 1 \\
AT2025uuu & 267.1103 & 56.0134 & 0.04 & 0.28 & 1.00 & 0.16 & $4.60 \times 10^{41}$ & --- & --- & 1 \\
AT2025vag & 232.3371 & 24.4240 & 0.03 & 0.57 & 1.00 & 0.07 & $3.89 \times 10^{41}$ & 0.18 & 0.13 & 1 \\
AT2025vaj & 245.1994 & 42.0982 & 0.03 & 0.81 & 1.00 & 0.04 & $1.06 \times 10^{42}$ & 0.12 & 0.12 & 1 \\
AT2025unj & 242.1796 & 40.0271 & 0.03 & 0.49 & 1.00 & 0.71 & $5.36 \times 10^{41}$ & 0.18 & -1.48 & 1 \\
AT2025urh & 259.7336 & 50.9121 & 0.02 & 0.42 & 1.00 & 0.05 & $4.13 \times 10^{41}$ & 0.20 & 0.20 & 1 \\
AT2025uvs & 269.2303 & 55.3516 & 0.02 & 0.33 & 1.00 & 0.07 & $9.93 \times 10^{41}$ & 1.30 & 2.24 & 1 \\
AT2025utq & 262.9452 & 54.7965 & 0.02 & 0.29 & 1.00 & 0.94 & $1.74 \times 10^{42}$ & 2.30 & 0.44 & 0.10 \\
AT2025uuq & 244.6472 & 37.9425 & 0.02 & 0.71 & 1.00 & 0.04 & $1.75 \times 10^{42}$ & --- & --- & 1 \\
AT2025usm & 236.9292 & 30.5482 & 0.02 & 0.80 & 1.00 & 0.03 & $4.90 \times 10^{41}$ & --- & --- & 1 \\
AT2025unp & 241.0241 & 35.6278 & 0.02 & 0.99 & 1.00 & 0.30 & $9.40 \times 10^{41}$ & 1.12 & -0.01 & 1 \\
AT2025uzf & 255.1251 & 49.6713 & 0.01 & 0.41 & 1.00 & 0.45 & $4.84 \times 10^{41}$ & 0.20 & 0.02 & 1 \\
AT2025vtz & 215.8520 & -1.2628 & 0.01 & 0.05 & 1.00 & 0.36 & $2.84 \times 10^{41}$ & --- & --- & 1 \\
AT2025van & 247.9201 & 40.5134 & 0.01 & 0.43 & 1.00 & 0.03 & $6.46 \times 10^{41}$ & --- & --- & 1 \\
AT2025uph & 297.0605 & 64.6750 & 0.01 & 0.48 & 1.00 & 0.40 & $1.81 \times 10^{42}$ & 0.37 & 0.06 & 1 \\
AT2025uzl & 241.6009 & 37.7818 & 0.01 & 0.95 & 1.00 & 0.18 & $8.20 \times 10^{41}$ & 0.12 & 0.08 & 1 \\
AT2025uzq & 232.9914 & 25.2775 & 0.01 & 0.58 & 1.00 & 0.20 & $7.41 \times 10^{41}$ & 0.14 & 0.00 & 1 \\
AT2025vau & 234.3892 & 29.6586 & 0.00 & 0.50 & 0.00 & --- & --- & --- & --- & 1 \\
AT2025utv & 258.7515 & 51.3671 & 0.00 & 0.50 & 1.00 & 0.34 & $2.23 \times 10^{42}$ & 19.22 & -0.03 & 1 \\
AT2025vam & 235.0854 & 30.5155 & 0.00 & 0.68 & 1.00 & 0.27 & $2.05 \times 10^{42}$ & 11.23 & -0.86 & 1 \\
AT2025uzx & 242.0438 & 38.3653 & 0.00 & 0.92 & 0.00 & --- & --- & --- & --- & 1 \\
AT2025unn & 239.1654 & 33.2560 & 0.00 & 0.92 & 1.00 & 0.09 & $7.43 \times 10^{41}$ & 0.12 & 0.09 & 1 \\
AT2025vaf & 244.7487 & 38.1098 & 0.00 & 0.70 & 1.00 & 0.03 & $4.40 \times 10^{41}$ & 1.11 & -0.09 & 1 \\
AT2025uzk & 239.9571 & 33.5515 & 0.00 & 0.82 & 1.00 & 0.23 & $1.80 \times 10^{42}$ & 11.22 & -0.54 & 1 \\
SN2025uuw & 249.2317 & 44.3800 & 0.00 & 0.84 & 1.00 & 0.87 & $9.53 \times 10^{42}$ & 20.21 & -2.66 & 1 \\
AT2025uzr & 241.4425 & 37.0061 & 0.00 & 0.99 & 1.00 & 0.49 & $1.95 \times 10^{42}$ & 17.22 & -0.32 & 1 \\
AT2025uzh & 239.1768 & 34.1537 & 0.00 & 0.97 & 0.00 & --- & --- & --- & --- & 1 \\
AT2025val & 245.4769 & 40.0921 & 0.00 & 0.96 & 1.00 & 0.00 & $7.44 \times 10^{41}$ & 4.13 & -0.36 & 1 \\
AT2025vab & 241.8167 & 36.9921 & 0.00 & 1.00 & 1.00 & 0.00 & $8.11 \times 10^{41}$ & --- & --- & 1 \\
AT2025vas & 242.6047 & 36.9211 & 0.00 & 0.95 & 1.00 & 0.00 & $4.43 \times 10^{41}$ & 0.17 & -1.51 & 1 \\
AT2025unk & 246.3895 & 40.7304 & 0.00 & 0.89 & 1.00 & 0.00 & $8.41 \times 10^{41}$ & 0.14 & 0.30 & 1 \\
AT2025uzw & 241.2052 & 35.2971 & 0.00 & 0.88 & 1.00 & 0.00 & $5.17 \times 10^{41}$ & 1.12 & -0.17 & 1 \\
AT2025uzy & 231.0306 & 23.8806 & 0.00 & 0.55 & 0.00 & --- & --- & --- & --- & 1 \\
AT2025vat & 233.0136 & 27.5937 & 0.00 & 0.42 & 1.00 & 0.41 & $6.80 \times 10^{41}$ & 4.14 & -0.01 & 1 \\
AT2025vad & 237.2357 & 28.6869 & 0.00 & 0.43 & 1.00 & 0.01 & $5.86 \times 10^{41}$ & 1.13 & -0.13 & 1 \\
AT2025uup & 297.7439 & 61.5513 & 0.00 & 0.44 & 1.00 & 0.06 & $4.63 \times 10^{41}$ & 9.33 & -0.01 & 1 \\
AT2025uxs & 236.0598 & 27.4783 & 0.00 & 0.44 & 1.00 & 0.96 & $8.88 \times 10^{42}$ & 20.19 & -2.67 & 1 \\
AT2025unh & 260.4936 & 52.5422 & 0.00 & 0.44 & 1.00 & 0.19 & $2.41 \times 10^{42}$ & 19.23 & -0.56 & 0.10 \\
AT2025uuv & 311.7227 & 65.1269 & 0.00 & 0.46 & 1.00 & 0.12 & $6.56 \times 10^{41}$ & 2.38 & -0.13 & 1 \\
AT2025uzd & 251.2641 & 44.2983 & 0.00 & 0.46 & 1.00 & 0.96 & $4.14 \times 10^{42}$ & 19.21 & -0.90 & 1 \\
AT2025ung & 251.2002 & 44.2290 & 0.00 & 0.46 & 0.00 & --- & --- & --- & --- & 1 \\
AT2025uuh & 283.6991 & 60.2078 & 0.00 & 0.49 & 0.00 & --- & --- & --- & --- & 1 \\
AT2025utw & 296.9660 & 64.5091 & 0.00 & 0.48 & 1.00 & 0.63 & $7.50 \times 10^{41}$ & 9.33 & -0.37 & 1 \\
AT2025uzv & 232.9374 & 25.0276 & 0.00 & 0.54 & 1.00 & 0.16 & $4.92 \times 10^{41}$ & 4.14 & -0.34 & 1 \\
AT2025unl & 250.1465 & 46.7436 & 0.00 & 0.52 & 1.00 & 0.19 & $5.63 \times 10^{41}$ & 0.19 & -0.27 & 1 \\
AT2025utz & 247.7828 & 40.7123 & 0.00 & 0.51 & 1.00 & 0.00 & $1.30 \times 10^{42}$ & --- & --- & 1 \\
AT2025vaa & 254.1146 & 46.8251 & 0.00 & 0.50 & 1.00 & 0.00 & $3.61 \times 10^{41}$ & 0.20 & 0.03 & 1 \\
AT2025uzz & 238.4062 & 36.1216 & 0.00 & 0.39 & 0.00 & --- & --- & --- & --- & 1 \\
AT2025uzo & 234.0232 & 25.5990 & 0.00 & 0.34 & 1.00 & 0.10 & $1.07 \times 10^{42}$ & 4.14 & -0.34 & 1 \\
AT2025utr & 271.9448 & 56.8702 & 0.00 & 0.32 & 1.00 & 0.83 & $2.44 \times 10^{42}$ & 19.25 & -0.27 & 1 \\
AT2025uzg & 233.7461 & 24.8628 & 0.00 & 0.32 & 1.00 & 0.21 & $5.61 \times 10^{41}$ & 4.14 & -0.40 & 1 \\
AT2025uqe & 257.9832 & 49.1794 & 0.00 & 0.33 & 1.00 & 0.83 & $1.83 \times 10^{42}$ & 15.26 & -0.87 & 1 \\
AT2025uxo & 248.8884 & 41.0161 & 0.00 & 0.33 & 1.00 & 0.33 & $4.10 \times 10^{42}$ & 16.24 & -1.76 & 1 \\
AT2025usy & 268.8021 & 56.0339 & 0.00 & 0.29 & 1.00 & 0.75 & $1.53 \times 10^{42}$ & 19.24 & -1.04 & 1 \\
AT2025utt & 267.5892 & 54.9656 & 0.00 & 0.28 & 1.00 & 0.05 & $1.75 \times 10^{42}$ & 19.24 & -0.26 & 1 \\
AT2025uzi & 244.8820 & 36.8049 & 0.00 & 0.28 & 1.00 & 0.26 & $5.88 \times 10^{41}$ & 4.10 & -0.04 & 1 \\
AT2025vae & 234.2500 & 24.4686 & 0.00 & 0.29 & 1.00 & 0.15 & $7.48 \times 10^{41}$ & 1.13 & -0.39 & 1 \\
AT2025vao & 257.7600 & 48.7028 & 0.00 & 0.27 & 1.00 & 0.02 & $5.70 \times 10^{41}$ & 1.23 & -0.26 & 1 \\
AT2025uno & 231.2551 & 26.2796 & 0.00 & 0.27 & 1.00 & 0.00 & $5.86 \times 10^{41}$ & 0.14 & 0.14 & 1 \\
AT2025vfa & 246.2516 & 45.0537 & 0.00 & 0.26 & 1.00 & 0.11 & $9.01 \times 10^{42}$ & 20.21 & -3.12 & 1 \\
AT2025uui & 239.7301 & 30.9933 & 0.00 & 0.26 & 1.00 & 0.17 & $1.69 \times 10^{42}$ & 11.23 & -0.87 & 1 \\
AT2025uuo & 257.8897 & 51.5603 & 0.00 & 0.35 & 1.00 & 0.02 & $1.63 \times 10^{42}$ & --- & --- & 1 \\
AT2025uud & 313.4742 & 65.1826 & 0.00 & 0.39 & 0.00 & --- & --- & --- & --- & 1 \\
AT2025uzn & 235.2292 & 31.3958 & 0.00 & 0.41 & 0.00 & --- & --- & --- & --- & 1 \\
SN2025vnt & 237.5921 & 29.0246 & 0.00 & 0.42 & 1.00 & 0.98 & $6.51 \times 10^{42}$ & 20.19 & -3.44 & 1 \\
AT2025vie & 273.2270 & 58.9476 & 0.00 & 0.24 & 0.00 & --- & --- & --- & --- & 1 \\
AT2025vay & 230.7166 & 24.8829 & 0.00 & 0.25 & 1.00 & 0.07 & $9.08 \times 10^{41}$ & 4.09 & -0.13 & 1 \\
AT2025vav & 235.3140 & 32.6537 & 0.00 & 0.25 & 1.00 & 0.01 & $4.18 \times 10^{41}$ & 4.14 & -0.08 & 1 \\
AT2025var & 232.6559 & 28.8539 & 0.00 & 0.25 & 1.00 & 0.19 & $6.01 \times 10^{41}$ & 0.14 & 0.06 & 1 \\
AT2025uzs & 240.9930 & 32.4010 & 0.00 & 0.20 & 1.00 & 0.06 & $4.81 \times 10^{41}$ & 0.18 & -4.74 & 1 \\
AT2025vpk & 43.2178 & -11.1210 & 0.00 & 0.24 & 1.00 & 0.52 & $3.82 \times 10^{42}$ & 20.12 & -0.71 & 1 \\
AT2025uzj & 233.5968 & 23.1899 & 0.00 & 0.21 & 1.00 & 0.02 & $7.71 \times 10^{41}$ & 4.09 & -0.13 & 1 \\
AT2025uql & 267.2360 & 57.8759 & 0.00 & 0.22 & 1.00 & 0.28 & $2.43 \times 10^{42}$ & 19.25 & -0.02 & 1 \\
AT2025uzm & 234.2632 & 31.8666 & 0.00 & 0.19 & 1.00 & 0.21 & $7.87 \times 10^{41}$ & 4.14 & -0.13 & 1 \\
AT2025uty & 265.4641 & 52.4306 & 0.00 & 0.18 & 1.00 & 0.00 & $8.40 \times 10^{41}$ & --- & --- & 1 \\
AT2025utx & 251.0454 & 42.4653 & 0.00 & 0.16 & 0.00 & --- & --- & --- & --- & 1 \\
AT2025uor & 58.3948 & -31.3176 & 0.00 & 0.15 & 0.00 & --- & --- & --- & --- & 1 \\
AT2025vac & 239.7555 & 40.2066 & 0.00 & 0.14 & 1.00 & 0.02 & $4.14 \times 10^{41}$ & 4.13 & -0.08 & 1 \\
AT2025vax & 241.0788 & 32.3593 & 0.00 & 0.20 & 1.00 & 0.23 & $9.97 \times 10^{41}$ & 0.23 & -0.78 & 1 \\
AT2025whj & 258.5650 & 53.6545 & 0.00 & 0.20 & 1.00 & 0.49 & $1.69 \times 10^{42}$ & 11.17 & -1.93 & 1 \\
AT2025vqh & 270.4048 & 59.0380 & 0.00 & 0.20 & 1.00 & 0.18 & $2.30 \times 10^{42}$ & 19.25 & -0.95 & 1 \\
AT2025usb & 53.2648 & -31.3207 & 0.00 & 0.13 & 0.00 & --- & --- & --- & --- & 1 \\
AT2025uzp & 229.7613 & 25.6649 & 0.00 & 0.13 & 0.00 & --- & --- & --- & --- & 1 \\
AT2025vap & 265.4473 & 52.2285 & 0.00 & 0.14 & 1.00 & 0.09 & $1.39 \times 10^{42}$ & 7.30 & -0.34 & 1 \\
AT2025upw & 222.0841 & 9.1170 & 0.00 & 0.12 & 1.00 & 0.91 & $5.99 \times 10^{42}$ & 15.95 & -0.46 & 1 \\
AT2025uzt & 236.2462 & 35.7672 & 0.00 & 0.09 & 1.00 & 0.03 & $4.93 \times 10^{41}$ & 0.19 & -1.53 & 1 \\
AT2025wju & 42.7855 & -14.6134 & 0.00 & 0.12 & 1.00 & 0.91 & $2.96 \times 10^{42}$ & 19.34 & -2.12 & 1 \\
AT2025usa & 286.0330 & 63.8107 & 0.00 & 0.12 & 1.00 & 0.07 & $1.02 \times 10^{42}$ & 1.30 & 0.28 & 1 \\
AT2025uzc & 235.9834 & 25.1267 & 0.00 & 0.10 & 0.00 & --- & --- & --- & --- & 1 \\
AT2025vah & 239.4820 & 29.2197 & 0.00 & 0.08 & 1.00 & 0.06 & $7.62 \times 10^{41}$ & 1.13 & -0.13 & 1 \\
AT2025vak & 237.6184 & 38.5752 & 0.00 & 0.08 & 0.00 & --- & --- & --- & --- & 1 \\
AT2025uze & 306.3231 & 67.3743 & 0.00 & 0.09 & 1.00 & 0.16 & $1.54 \times 10^{42}$ & 13.32 & -0.43 & 1 \\
AT2025uxm & 239.2477 & 29.4235 & 0.00 & 0.08 & 0.00 & --- & --- & --- & --- & 1 \\
AT2025uso & 244.2393 & 35.0558 & 0.00 & 0.05 & 1.00 & 0.88 & $9.78 \times 10^{42}$ & 20.21 & -2.55 & 1 \\
AT2025unr & 220.8751 & 13.1996 & 0.00 & 0.06 & 1.00 & 0.62 & $1.75 \times 10^{42}$ & 8.22 & -0.90 & 1 \\
AT2025vai & 242.0910 & 32.7409 & 0.00 & 0.07 & 0.00 & --- & --- & --- & --- & 1 \\
SN2025ulz & 237.9758 & 30.9024 & 0.00 & 0.66 & 1.00 & 0.68 & $2.12 \times 10^{42}$ & 26.76 & -3.07 & 1 \\
\label{tab:candidates}
\end{longtable}

\onecolumngrid
\section{Photometry Tables}\label{app:data}
The optical and ultraviolet photometry is given in \autoref{tab:phot} and the radio photometry is given in \autoref{tab:radio}.

\begin{longtable}{lcrrr} \label{tab:phot} \\
\caption{SN\,2025ulz Ultraviolet, Optical, and Infrared Photometry} \\
\toprule
MJD & Filter & Magnitude & Magnitude Error & Telescope (Source) \\
\midrule
\endfirsthead
\caption[]{SN 2025ulz Photometry} \\
\toprule
MJD & Filter & Magnitude & Magnitude Error & Telescope (Source) \\
\midrule
\endhead
\midrule
\multicolumn{5}{r}{Continued on next page} \\
\midrule
\endfoot
\bottomrule
\endlastfoot
60901.64 & u & $>$22.10 & --- & WFST \citep{2025GCN.41461....1L} \\
60901.67 & g & $>$22.40 & --- & WFST \citep{2025GCN.41461....1L} \\
60903.32 & o & $>$21.48 & --- & ATLAS (This work) \\
60903.32 & o & $>$21.23 & --- & ATLAS (This work) \\
60903.32 & o & 21.61 & 0.35 & ATLAS (This work) \\
60904.22 & r & $>$21.01 & --- & P48 (TNS) \\
60904.23 & y & $>$19.70 & --- & PS \citep{2025GCN.41439....1N} \\
60905.19 & g & 20.99 & 0.13 & P48 (TNS) \\
60905.24 & r & 21.29 & 0.13 & P48 (TNS) \\
60905.82 & r & 21.43 & 0.06 & Wendelstein \citep{2025GCN.41421....1B} \\
60905.82 & g & 21.25 & 0.03 & Wendelstein \citep{2025GCN.41421....1B} \\
60906.13 & J & $>$19.30 & --- & WINTER \citep{2025GCN.41456....1M} \\
60906.84 & r & 21.83 & 0.06 & Wendelstein \citep{2025GCN.41433....1H}\\
60906.84 & g & 22.08 & 0.09 & Wendelstein \citep{2025GCN.41433....1H}\\
60907.23 & r & 22.60 & 0.00 & Gemini \citep{2025GCN.41452....1O} \\
60907.23 & g & 23.00 & 0.00 & Gemini \citep{2025GCN.41452....1O} \\
60907.30 & g & 22.90 & 0.40 & PS \citep{2025GCN.41454....1G} \\
60907.31 & r & 22.20 & 0.20 & PS \citep{2025GCN.41454....1G} \\
60907.31 & r & 22.05 & 0.20 & PS1 (TNS) \\
60907.33 & I & 21.80 & 0.20 & PS \citep{2025GCN.41454....1G} \\
60907.34 & z & 21.40 & 0.20 & PS \citep{2025GCN.41454....1G} \\
60907.60 & g & 22.71 & 0.10 & WFST \citep{2025GCN.41461....1L} \\
60907.61 & r & 22.78 & 0.16 & WFST \citep{2025GCN.41461....1L} \\
60907.63 & r & 22.80 & 0.30 & JinShan \citep{2025GCN.41503....1A} \\
60907.65 & u & 23.24 & 0.24 & WFST \citep{2025GCN.41461....1L} \\
60907.88 & r & 23.20 & 0.20 & NOT \citep{2025GCN.41492....1M} \\
60908.28 & r & 22.85 & 0.15 & CFH \citep{2025GCN.41519....1A} \\
60908.90 & r & 23.40 & --- & GTC \citep{2025GCN.41502....1B} \\
60908.91 & r & $>$21.72 & --- & STEP/T80N \citep{2025GCN.41501....1S} \\
60908.92 & i & $>$21.62 & --- & STEP/T80N \citep{2025GCN.41501....1S} \\
60909.14 & i & 23.10 & 0.25 & COLIBRI \citep{2025GCN.41518....1A} \\
60909.20 & i & $>$22.96 & --- & MMT (This work) \\
60909.21 & r & $>$22.90 & --- & MMT (This work) \\
60909.26 & i & 22.40 & 0.30 & PS \citep{2025GCN.41540....1G} \\
60909.84 & F336W & 23.85 & 0.30 & Hubble WFC3/UVIS \citep[][re-reduced in this work]{2025GCN.41506....1T} \\
60909.94 & F110W & 22.87 & 0.23 & Hubble WFC3/IR \citep[][re-reduced in this work]{2025GCN.41506....1T} \\
60910.17 & F160W & 22.15 & 0.30 & Hubble WFC3/IR \citep[][re-reduced in this work]{2025GCN.41506....1T} \\
60910.26 & i & 22.50 & 0.30 & PS \citep{2025GCN.41540....1G} \\
60911.15 & i & 22.00 & 0.10 & COLIBRI \citep{2025GCN.41518....1A} \\
60911.26 & i & 22.10 & 0.20 & PS \citep{2025GCN.41540....1G} \\
60912.26 & i & 21.60 & 0.10 & PS \citep{2025GCN.41540....1G} \\
60912.86 & r & 21.45 & 0.44 & STEP (This work) \\
60912.88 & i & 21.37 & 0.28 & STEP (This work) \\
60913.90 & F606W & 22.03 & 0.03 & Hubble WFC3/UVIS (PI: Troja; This work) \\
60914.82 & F110W & 21.81 & 0.02 & Hubble WFC3/IR (PI: Troja; This work) \\
60914.99 & r & $>$21.04 & --- & SOAR (This work) \\
60915.00 & i & 21.04 & 0.28 & SOAR (This work) \\
60915.01 & z & $>$20.76 & --- & SOAR (This work) \\
60915.47 & F160W & 21.79 & 0.05 & Hubble WFC3/IR (PI: Troja; This work) \\
60916.28 & o & $>$20.43 & --- & ATLAS (This work) \\
60917.19 & z & 20.40 & 0.08 & MMT (This work) \\
60917.22 & i & 20.28 & 0.09 & MMT (This work) \\
60919.25 & o & $>$17.85 & --- & ATLAS (This work) \\
60922.24 & o & 20.18 & 0.29 & ATLAS (This work) \\
60923.26 & o & $>$20.12 & --- & ATLAS (This work) \\
60924.26 & o & $>$20.15 & --- & ATLAS (This work) \\
60924.82 & r & 20.64 & 0.23 & STEP (This work) \\
60924.83 & i & 20.70 & 0.23 & STEP (This work) \\
60925.26 & o & $>$19.46 & --- & ATLAS (This work) \\
60929.67 & r & 20.82 & 0.11 & HCT \citep{2025GCN.41837....1S} \\
60931.81 & i & 20.41 & 0.18 & T80 (This work) \\
60931.82 & r & 20.77 & 0.21 & T80 (This work) 
\end{longtable}

\begin{table*}
    \centering
        \caption{Summary of radio observations for SN 2025ulz.  The reported uncertainties are 1$\sigma$ statistical errors. Non-detections are reported as
3$\sigma$ upper limits.}
        \begin{tabular}{lcccccp{4cm}}
        \hline\hline
        Telescope & MJD & $\nu_{\rm eff}$ (GHz) & $\theta_{\rm B, major} \times \theta_{\rm B, minor}$  & $\theta_{\rm PA}$ & Flux Density ($\mu$Jy) & Source \\[5pt]
        \hline
        GMRT       & 60913.6  &    1.26  &   $2.5'' \times   2.0'' $&     -5$^\circ$ & $<80$ & \citet{2025GCN.41577....1B}, This work \\[5pt]

        MeerKAT    & 60932  &    0.82  &  $18.3'' \times  14.7'' $&    167$^\circ$ & $185\pm44$ & \citet{2025GCN.42032....1B}, re-reduced in this work \\
        MeerKAT    & 60932  &    1.28  &  $ 8.5'' \times   7.5'' $&     27$^\circ$ & $186\pm18$ & \citet{2025GCN.42032....1B}, re-reduced in this work \\
        MeerKAT    & 60908  &    3.06  &  $ 8.0'' \times   3.7'' $&    163$^\circ$ & $59\pm5$ & \citet{2025GCN.41500....1B}, re-reduced in this work \\
        MeerKAT    & 60915  &    3.06  &  $ 4.5'' \times   3.8'' $&      5$^\circ$ & $83\pm6$ & \citet{2025GCN.41594....1B}, \citet{2025GCN.41666....1R}, re-reduced in this work \\
        MeerKAT    & 60931  &    3.06  &  $ 4.3'' \times   3.9'' $&      7$^\circ$ & $76\pm9$ & \citet{2025GCN.42032....1B}, re-reduced in this work \\[5pt]

        NOEMA      & 60909.9  &   92  &  $ 3.8'' \times   2.7'' $&     54$^\circ$ & $<63$ & This work \\[5pt]

        SMA      & 60908.3  &   225.5  &  $ 3.2'' \times  3.0'' $&     89.1$^\circ$ & $<750$ & This work \\[5pt]

        VLA        & 60917.9  &    3  &  $ 1.7'' \times   1.5'' $&    -81$^\circ$ & $<28$ & This work \\
        VLA        & 60917.9  &    6  &  $ 1.0'' \times   0.8'' $&    -85$^\circ$ & $<20$ & This work \\
        VLA        & 60909.2  &    6  &  $ 1.0'' \times   0.9'' $&     90$^\circ$ & $<18$ & This work \\
        VLA        & 60917.9  &   10  &  $ 0.7'' \times   0.5'' $&    -80$^\circ$ & $<20$ & This work \\[5pt]

        \hline

        VLA & 60911 & 3 & --- & --- & $<40$ & \citet{2025GCN.41542....1R} \\ 
        VLA & 60911 & 6 & --- & --- & $<25$ & \citet{2025GCN.41542....1R} \\ 
        VLA & 60908 & 10 & --- & --- & $<30$ & \citet{2025GCN.41464....1R} \\[5pt]

        \hline\hline
        \end{tabular}
    \label{tab:radio}
\end{table*} 

\end{document}